# Investigating the Nature and Structure of Broad Line Region in Active Galactic Nuclei

A thesis submitted for the degree of
**Doctor of Philoshophy**
in
**Physics**

by

# Vivek Kumar Jha

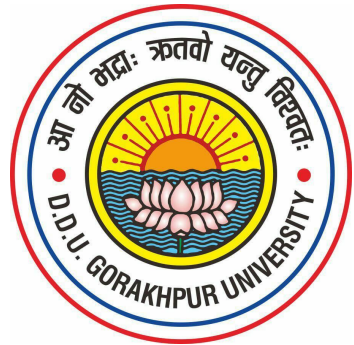

**Deen Dayal Upadhyaya Gorakhpur University**
Gorakhpur, Uttar Pradesh, India; 273009

Research centre:
**Aryabhatta Research Institute of observational sciencES (ARIES)**
Nainital, Uttarakhand, India; 263001

**September, 2023**

*Who really knows? Who will here proclaim it?*
*Whence was it produced? Whence is this creation?*
*Gods came afterwards, with the creation of this universe.*
*Who then knows whence it has arisen?*
*Whether God's will created it, or whether He was mute;*
*Perhaps it formed itself, or perhaps it did not;*
*Only He who is its overseer in highest heaven knows,*
*Only He knows, or perhaps He does not know.*

**- Verse 6-7, Nasadiya sukta: Hymn of creation, Rigveda (10.129)**

# Acknowledgements

This Ph.D. thesis signifies a journey to the world of academic research that I started as a graduate student, out of the University and looking to make sense of the research environment around me. My supervisor, Prof. Hum Chand, has played the leading role in this journey, being a friend, a mentor, and sometimes a life coach. Throughout the duration of my thesis, he has been a constant companion, looking at the minute details of my research work and steering me onto the correct path whenever I felt wayward. His enthusiasm for computer codes, for learning new things, and the capability to manoeuvre multiple jobs without a worry on his face is something I would love to emulate in my life too. The journey had its speed breakers, but it became a smooth ride having a supervisor like him.

Besides my supervisor, I would also like to acknowledge the guidance provided by Dr. Ravi Joshi, who has provided constant support, discussed various research ideas, and ensured I was approaching them correctly. His patience and persistence have been essential in successfully executing the research projects. I appreciate the help provided by Dr. Vineet Ojha for being a helping hand and always playing the role of a very approachable senior cum mentor, whom I could approach with any question. I also acknowledge all other collaborators who have played a significant role in shaping my research career so far.

When I started as a Ph.D. student in early 2018, I met a few individuals in ARIES who taught me not only the fundamentals of research but also valuable life lessons, the most important of them handling the tough times. Right from the beginning, I felt blessed to have the company of senior researchers: Anjasha Gangopadhyay, Mridweeka Singh, Mukesh Kumar Vyas, and Arti Joshi, who played the double role of friends and mentors, even going to the extent of being family members at times. My friends Dimple, Ankur, Amar, and Rahul were not mere friends but became family for the duration I stayed in ARIES. How can we forget the dinners at various restaurants during the chilly winters in Nainital, the movies we watched together, or the weekend treks! I acknowledge Lakshitha for patiently helping me through the tough times I may have faced and the boring discussions I occasionally engaged her in. As time went by, excellent junior fellows joined in this journey. Thanks to Arpit, Tushar, Rahul, Shubham, Gurpreet and all the research fellows for the happy memories. The tea after dinner, walking in the campus after lunch, and the road trip to Munsyari with you people will remain etched in my heart. I am also grateful to have met Pankaj Kushwaha during my stay. Be it a discussion about politics, sports, and our shared area of research- Active galaxies- he always amazes me with his stimulating ideas and immense knowledge about things around us. We went for weekend treks together, and discussing various things during those treks helped me see things from a different perspective.

I want to thank the academic committee of ARIES for reviewing my work annually and providing their valuable feedback on my research progress. I am indebted to the authorities at Deen Dayal Upadhyaya Gorakhpur University (DDUGU) for providing me the opportunity to register for a Ph.D. degree. I am grateful to my co-supervisor, Prof. Shantanu Rastogi, for providing all the help and support and always being available for any queries. Without his constant help and support, the thesis would not have been possible. I acknowledge the warm environment provided by Atul, Vishnu, Prayag, Vaibhav and all other research fellows at DDUGU whenever we visited the department.

I thank DST - SERB for funding a significant part of my Ph.D. under grant number EMR/2016/001723. Without their financial support, the research conducted during this thesis would not have been possible. I acknowledge the help regarding telescope observations provided by the staff at the Devasthal observatory. I acknowledge IUCAA for providing access to their High-Performance Computing (HPC) facility. This thesis used data from multiple public surveys. I want to acknowledge the Zwicky Transient Facility (ZTF) and the Sloan Digital Sky Survey ( SDSS) for providing the extremely valuable datasets publicly. I hope astronomical surveys in the future emulate the generous standards set by these surveys. I also acknowledge the ASTROPY project, the JUPYTER project, and various other PYTHON developers who have devoted a significant amount of their time developing packages available to the public for free. Without the availability of these packages, research would have been much tricky indeed.

Lastly, the COVID-19 pandemic, which raged worldwide for the past 3 years, affected almost every human being on this planet. Through the lockdowns, our lives were disrupted. Moreover, being infected with COVID-19 and admitted to the hospital was a scary experience, which I faced, unfortunately. I acknowledge the help provided by Mohit Joshi and the Director of ARIES, Prof. Dipankar Banerjee, for constantly monitoring my health and arranging all the required medical and quarantine facilities. Fortunately, I recovered with minor complications only, and due to their support, I never felt away from home.

**- Vivek Kumar Jha**

# Abstract

Active galactic nuclei (AGN) are the innermost regions of the active galaxies, which outshine their hosts. Since these objects are located at far distances, even with large telescope apertures and improvements in detector technology, resolving the inner regions of AGN remains a formidable challenge. Nevertheless, understanding the innermost regions of these galaxies is critical in understanding the galaxy evolution and the dynamics of matter in the vicinity of a Supermassive Black Hole (SMBH). In the absence of direct methods to resolve the inner regions of AGN, indirect methods are used. In this thesis, I explored a few important indirect methods to understand the innermost (sub-parsec) regions.

Firstly, we used the reverberation mapping technique to estimate the accretion disk sizes for a sample of AGN with previously known SMBH mass estimates to look for a correlation between the estimated disk size, known SMBH mass, and luminosities. The results show that the accretion disk sizes computed using this method are, on average, 3.9 times larger than the predictions of the Shakura Sunyev (SS) standard disk model. Additionally, we find a weak correlation between the obtained accretion disk sizes and the SMBH mass. Next, we present initial results from a new accretion disk monitoring program titled: 'Investigating the central parsec regions around supermassive black holes'(INTERVAL) to probe the accretion disk structure of Super Eddington Accreting AGN. We report the results for IRAS 04416+121 and find the disk sizes to be about 4 times larger than the SS disk model. In the next step, we calibrated the narrow-band photometric reverberation mapping (PRM) technique to develop tools for a large systematic narrow-band PRM project which can enable the measurement of SMBH mass and BLR sizes for a large number of quasars at a fraction of the telescope's time. To develop an optimum strategy for executing a successful PRM project, we used simulations to test the effect of cadence, variability of the light curves, and the length of light curves in recovering the reverberation lags.

Secondly, we used the method of microvariability observed in the accretion disk continuum to study the dichotomy between AGNs with and without detected jets.

Through a monitoring program spanning 53 sessions with a minimum duration of 3 hours, we demonstrated that the jets can induce significant microvariability in the optical continuum in these AGNs, and the AGNs with confirmed jets are about 3 times more variable on short time scales than the AGNs without a confirmed jet. We also find evidence that Narrow Line Seyfert 1 (NLSy1) galaxies with $\gamma-$ray detection have blazar-like behavior in terms of microvariability.

Finally, we performed statistical analysis on a large sample of low luminosity AGNs using optical spectra to investigate the dynamics of matter in BLR and infer the properties of the NLSy1 galaxies. Based on a sample of 144 NLSy1 and 117 BLSy1 galaxies, we find that the NLSy1 galaxies are more likely to have outflow signatures than their broad-line counterparts hinting toward the disk wind origin of the material in BLR. Further, through a principal component analysis of this data, we find out the principal components for NLSy1 galaxies differ from the BLSy1 galaxies, which points towards the fact that the NLSy1 galaxies are not just a subclass of BLSy1 galaxies but could be occupying their own parameter space.

In summary, we have measured accretion disk sizes for a handful of AGN, which adds to the growing number of AGN with accretion disk size measurement. We have calibrated the PRM technique and developed simulations to execute a successful narrow-band PRM campaign to get the BLR sizes and SMBH mass for a large sample of AGN. We have found evidence that relativistic jets induce microvariability in the AGN optical continuum from the accretion disk. Finally, we have also found that the emission line region in the NLSy1 galaxies is likelier to show outflows through the H$\beta$ emission line as compared to the BLSy1 galaxies.

# Table of contents

# List of figures

# List of tables

# Chapter 1

# Introduction

A fraction of the observed population of galaxies is known as Active Galaxies or, more popularly– Quasars. These objects are different because the central Super Massive Black Hole (SMBH) is accreting mass, and a tremendous amount of energy is released in this process. Thus, the Active Galactic Nuclei (AGN) usually outshine their host galaxies. The presence of quasars is ubiquitous across the night sky. Since these objects are observed at redshifts as high as 7, their study is essential for understanding the evolution of the Universe itself. However, the inner regions of AGN are challenging to resolve with current technology due to the smaller angular projection of these objects. Indirect methods have been widely used to infer the characteristics of matter in the vicinity of the central SMBH. Based on the results obtained through these methods, we know that the SMBH is surrounded by the matter, which forms an accretion disk, a line emitting region known as the broad line region (BLR) is present, further enveloped by a dusty torus, and through the technique of reverberation mapping, we can estimate the mass of the SMBH residing at the centre. Nevertheless, the extent of the accretion disk, the dynamics of matter in the BLR, and the interplay of the disk with the other components remain unknown. Through this thesis, I have used these indirect methods to extract information about the inner region of these objects. This chapter aims to describe the phenomenon of AGN and its components and the physical processes happening in their innermost regions; and then move on to some of the current challenges involved in understanding these regions, which are very difficult to resolve physically. We explore the available techniques to probe these inner regions, which will be used in the following chapters to present new results obtained during this work.

## 1.1 Historical Background

AGN refers to the phase of galaxy evolution during which the emission from the central regions outshines the whole galaxy. The active phase is signified by enhanced accretion by the central SMBH, which in a few cases reaches super Eddington rates (Takeo et al., 2019) (see Figure 1.1 for examples of a few AGN). The AGN phase is characterized by rapid flux variability, emission in all wavelength ranges, and the presence of jets as well as massive outflows (Ulrich et al., 1997, Peterson, 2001, Feruglio et al., 2010). Observed luminosity of AGNs ranges from $10^{40}-10^{47}$ erg/sec or higher (Fabian, 1999). Such luminosities arise from the central part of these galaxies meaning the central regions are compact. AGNs have been detected in all wavelengths and are quite common across the night sky (Kellermann, 2014).

A few nebular objects have been known since the 19th century (see Curtis, 1917, Slipher, 1917). However, it was not certain what these nebulous objects could be. In 1943, Carl Seyfert detected abnormally broad lines in the spectra of 6 such objects (Seyfert, 1943), but the extragalactic origin of such objects remained a puzzle. These objects were confirmed to be extragalactic only when Martin Schmidt (Schmidt, 1963) pointed to the radio observation of 3C 273 and interpreted it as coming from an object where emission lines are redshifted to 0.158. Since then, the knowledge of AGN has grown by leaps and bounds. Sulentic et al. (2012) provide detailed information about the discoveries in the past 50 years or so since AGN were first discovered. These objects were initially known as quasi-stellar sources, sources with a star-like appearance but non-stellar features. The short form of the quasi-stellar object (QSO) − *quasar* caught up, which persists even today. The AGN nomenclature can be confusing, and for no funny reasons, it is known as the AGN zoo (see Table 1 in Padovani et al., 2017). Now, the general convention is to use the term AGN for the entire population, while the term *quasar*/ QSO is typically used for AGN with luminosities over $10^{44}$ ergs/sec (Sazonov & Revnivtsev, 2004). In this thesis, we shall follow the same convention.

It is now believed that most galaxies undergo the AGN phase, which may last up to a few million years (Schmidt et al., 2017). With the availability of all-sky surveys such as the SDSS, it has become possible to catalogue hundreds of thousands of AGNs in the literature, providing us with unique data sets to work on and explore various cosmological models. The total number of confirmed AGN is more than 700,000 in the latest SDSS data release (Lyke et al., 2020, also see Figure 1.2) and the all-sky surveys provide multi-wavelength data for hundreds of thousands of AGN in the databases (see La Mura et al., 2017, for further information). AGN are mainly detected in all-sky surveys based on their unique features; the big blue bump in the SED and the bluer when brighter trend seen in their observations. If the spectrum is available, the AGN can be classified from other galaxies based on their emission line ratios through the Baldwin, Phillips & Terlevich (BPT) diagram (Baldwin et al., 1981, see Figure 1.3). Similarly, multi-wavelength diagnostics differentiate AGN from other stellar and extragalactic objects (e.g., see Padovani et al., 2017).

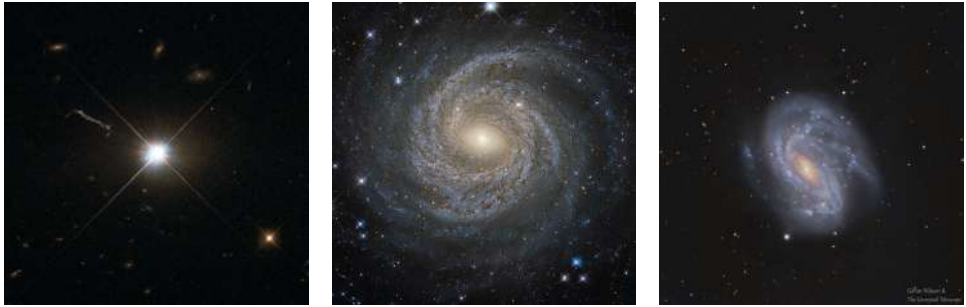

Fig. 1.1 Optical images for 3 AGNs located at various redshifts: On the left, 3C273 is located at a redshift of 0.15, in the middle, NGC 6814 at a redshift of 0.005, and on the right, NGC 4051 is located at a redshift of 0.002. Images obtained from Wikimedia Commons.

With the onset of all-sky optical and IR surveys, many of the quasars at $z \geq 5$ have been known and catalogued (Wang et al., 2017b, Pipien et al., 2018, Ross & Cross, 2020). Recently, the quasars have been discovered up to redshifts as high as $z \geq 7$ (Mortlock et al., 2011, Momjian et al., 2014, Bañados et al., 2021), making them an essential source of information about the early Universe in the era of cosmic re-ionization and offer clues to the black hole growth up to several billion $M\odot$ in the early phase of the universe (Fan et al., 2006, 2019). The discovery of quasars at such early epochs suggests that the growth of SMBH could have started earlier in the Universe with massive seeds. AGN being located further may provide a suitable anchor for matter dynamics in the early universe (Valiante et al., 2017). Further, the quasars at high redshift can be excellent candidates to test cosmological models as the current supernova-based inferences are limited to $z \sim 1.5$ (Risaliti & Lusso, 2015).

In the next section, we briefly describe the various components that constitute an AGN.

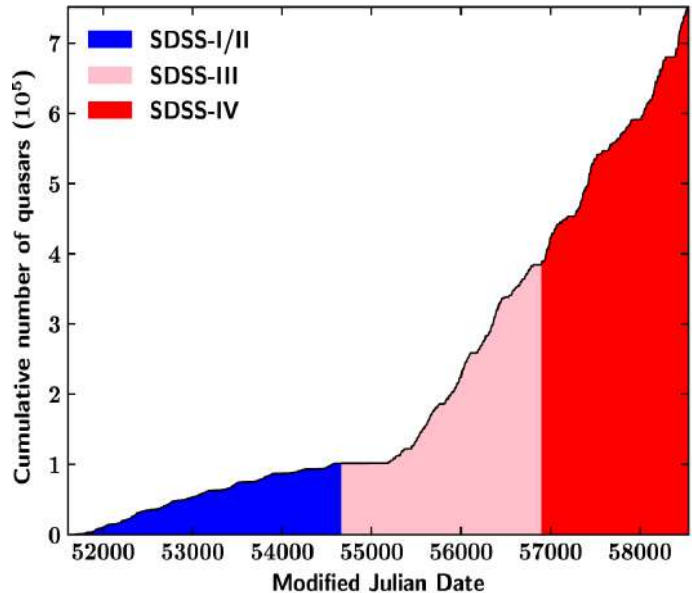

Fig. 1.2 The figure shows the growth in the population of AGN with the ongoing SDSS surveys. The total number of AGNs in the SDSS database has increased to 750,414 in the latest data release- (DR16). Image obtained from: Lyke et al. (2020).

## 1.2 Components of AGN

The fundamental components of AGN are at a scale of subparsec to kiloparsec from the central region (Netzer & Peterson, 1997, Netzer, 2008, 2015). In the centre of an AGN, an accreting SMBH with a mass ranging from $10^6$–$10^9$ M$\odot$ is believed to exist, where M$\odot$ is the mass of the Sun. The black hole is fed by the matter, forming an accretion disk around it which forms a flow. This disk is responsible for the continuum emission in the AGN. Above the accretion disk, a corona-like structure is believed to be present near the black hole and emits in X-ray wavelengths (Haardt & Maraschi, 1991). A relativistic jet may be present, which may induce variability in the continuum and is often the reason behind radio lobe emissions. The Broad Line Region (BLR) exists beyond the accretion disk from where all the broad (of the order of a few 1000 km/s) emission lines observed in the AGN spectra are emitted. Outside the BLR is the dusty torus, which emits infrared wavelengths and can envelop the innermost regions and provides shielding to the accretion disk and the BLR in the equatorial plane. Next to the torus lies the Narrow line Region (NLR). The NLR emits the narrow lines (of the order of a few 100 km/s) seen in the AGN spectrum, and this region is essential to study the AGN host galaxy connection. More information about the innermost regions is detailed in these reviews (Davidson

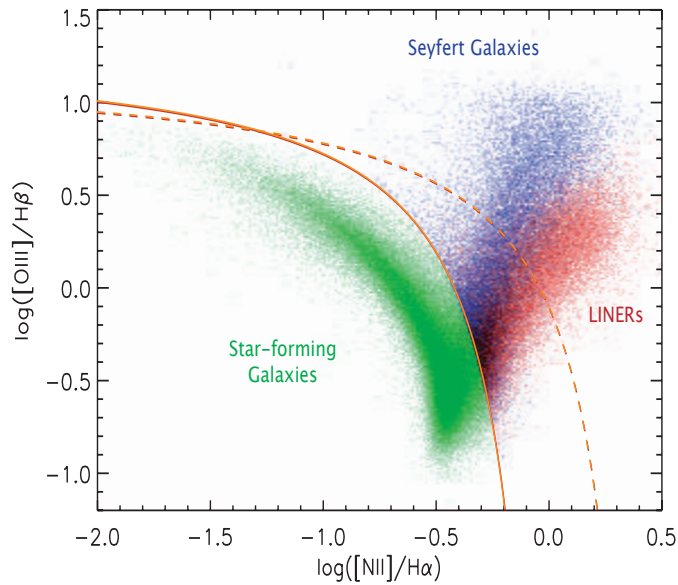

Fig. 1.3 Figure shows the BPT diagram to differentiate the AGN population from the other types of galaxies. The different types of AGN can be identified from the star-forming galaxies based on the emission line ratios [OIII]/H$\beta$ and the [NII]/H$\alpha$. Image obtained from Fosbury et al. (2007).

& Netzer, 1979, Sulentic et al., 2000a, Véron-Cetty & Véron, 2000, Risaliti & Elvis, 2004, Netzer, 2008, Tadhunter, 2008, Padovani et al., 2017). We shall briefly look at the various components and the physical processes in these regions and the role these regions play in generating the AGN Spectral Energy Distribution (SED).

### 1.2.1 The Supermassive Black Hole

The idea that an SMBH might exist in the centre of AGN was based on the observation that the gravitationally bound regions in these objects must be in a very compact region with mass $\sim 10^8$ M$\odot$ or more. This compactness can only be found around a black hole, so a central SMBH was proposed. A black hole is a region of space-time where the gravitational field dominates to an extent where even light cannot escape. The evidence for the presence of a central SMBH most convincingly comes from the shape of Fe k$\alpha$ line, which shows the general relativistic effect attributed to a compact object, most likely an SMBH (Tanaka et al., 1995). The SMBH is known to drive the power in the AGN. The formalism of SMBH was essential for the knowledge of AGN, as calculations yielded that the power needed to generate such luminosities could not be generated by fusion only. Hoyle & Fowler (1963) proposed that gravitational collapsing objects could generate such luminosities. This black hole drives the power of the AGN. The accretion rate can determine the accretion of the matter onto the black hole, which might estimate the age of the AGN in simplest terms. The gravitational radius ($r_g$) for an object with mass M is given as:

$$r_g = \frac{GM}{c^2} \tag{1.1}$$

where G is the Gravitational constant and c is the speed of light. The force due to the radiation pressure of the matter around the SMBH is given as:

$$F_{rad} = \frac{N_e \sigma_t}{4\pi r^2 c} \int_0^\infty L_\nu d\nu \tag{1.2}$$

where $\sigma_t$ is the Thomson scattering cross section, $N_e$ is the number of electrons, $L_\nu$ is the luminosity at a particular frequency $\nu$. In spherical accretion, at a critical luminosity, the radiation force balances the gravitational force. ($F_{grav} = F_{rad}$):

$$\frac{GM\mu m_p}{r^2} = \frac{\sigma_t L_E}{4\pi r^2 c} \tag{1.3}$$

This luminosity is known as the Eddington luminosity, and it is given as:

$$L_E = \frac{4\pi c G M \mu m_p}{\sigma_t} \tag{1.4}$$

Assuming an accretion efficiency of $\eta$, the luminosity can be obtained using the mass in terms of mass accretion rate:

$$L = \dot{M}\eta c^2 \tag{1.5}$$

and

$$L_E = \dot{M}_{edd} c^2 \tag{1.6}$$

gives the Eddington mass accretion rate for such a black hole. The observed luminosities in AGNs require the SMBH masses of $\sim 10^8$ M$\odot$ or higher, even below the Eddington accretion rates.

### 1.2.2 The Accretion Disk

The accreting matter onto the supermassive black hole is expected to form an accretion disk (Salpeter, 1964, Lynden-Bell, 1969). The viscosity of the infalling matter is supposed to drive the disk accretion. The simplest model for the accretion disk is proposed by Shakura & Sunyaev (1973) (SS disk henceforth). According to the standard SS disk model, the effective temperature $T$ at a particular location in the accretion disk $R_{disk}$ is related to the disk size and the SMBH mass as follows:

$$T(R_{disk}) = \left(\frac{3GM_{BH}\dot{M}}{8\pi\sigma(R_{disk})^3}\right)^{1/4} \qquad (1.7)$$

where $\sigma$ denotes the Stefan-Boltzmann constant. As is evident from this equation, the accretion disk size (R) scales inversely with the temperature (T) by a power law of index 4/3.

The optical depth of such a disk is relatively high, and the spectrum of the disk in a particular region is assumed to be a black body where free free and free bound processed and absorption of heavy elements provide the emitted spectrum. In the intermediate regions of the disk, the Thomson scattering dominates, while in the inner regions, the comptonization dominates. The resultant spectrum from the black hole is a black body spectrum with peaks at multiple temperatures. AGN accretion disks can take many forms and may be slim, thin, or thick based on the accretion rates. The emission from the accretion disk peaks in the UV, and there is significant emission in the optical bands. The big blue bump observed in the SED of AGN is an observational feature arising due to the emission peaking in the UV wavelengths (Shang et al., 2005). In the inner regions of the accretion disk, X-ray observations point towards a non-disk element, as the disk cannot go beyond a certain temperature threshold. This emission is believed to originate from a corona, which illuminates the accretion disk, and the emission is reprocessed through UV and optical wavelengths (Sun et al., 2020). This disk reprocessing can be observed and is assumed to drive the variations in the accretion disk and the nearby emission-line regions.

Despite several efforts, the structure of the AGN accretion disk remains poorly understood. The simple SS disk model does not predict the X-ray emission in the AGN spectra. The current understanding has been that the corona up scatters the photons from the accretion disk, and then the reprocessed emission is seen in the UV and optical wavelengths. Recent observations have yielded that the size of the accretion disk is a few times larger than the SS disk predictions (Starkey et al., 2016a, Fausnaugh et al., 2016, Edelson et al., 2015). Moreover, based on the X-ray emission, it has been observed that hard X-rays sometimes lead or lag the soft X-ray emission, which is challenging to disk reprocessing (McHardy et al., 2014). The corona's structure needs a continuous flow of matter from the accretion disk unless it cools down, which is not incorporated in the accretion disk models. Also, the current accretion disk models do not consider the magnetic field, which may affect the dynamics of the accretion disk.

### 1.2.3 The Broad-line Region

The emission lines seen in the optical spectra of AGN are generated in the Broad Line Region (BLR). This region lies in the range of $\sim$ 0.1-1pc from the central SMBH. The temperature of the BLR lines is of the order of $10^3$-$10^4$ K, which is just below the dust sublimation radius (Vestergaard & Peterson, 2006). The boundary of the BLR is assumed to be the dust sublimation radius that separates it from the dusty torus. The nature and structure of this region is a crucial aspect that remains unexplored and unknown at the moment. The emission lines are formed by photoionization of the continuum, which can be observed by correlated continuum and emission line variability. This concept forms the basis for Reverberation mapping. We shall revisit the BLR when we discuss the concept of Reverberation mapping in the subsequent sections. In the BLR, emission lines form due to the recombination and the resonance lines excited by collision. The BLR itself may have a composite structure where the High Ionization Lines (HIL) and the Low Ionization Lines (LIL) arise from distinct regions.

The BLR remains poorly understood. We find that the origin of material in BLR is highly debated, and not a single model has been able to explain the various features seen in the spectra. At the same time, models of the BLR range from clumps of matter disrupted tidally (Wang et al., 2017a), the failed radiation-driven wind (Czerny et al., 2017) to the accretion disk winds (Kollatschny & Zetzl, 2013), self gravitation disks (Wang et al., 2012). Further, what is the structure of the BLR, where is it located, and what can determine its location? Dust sublimation radius can form the outer boundary of the BLR due to the temperature-based arguments. There is no evidence why there can be clouds or clumpy gas in the BLR. Dynamical modelling of the reverberation mapping data for the BLR has yielded steady flow and the possibility of inflow and outflow (Pancoast et al., 2012).

### 1.2.4 The Dusty Torus

The dusty torus envelopes the AGN and mostly re-emits the emission from the accretion disk in the infrared wavelengths. The torus is small, clumpy, well-structured, and a dynamic changing quantity (Mason, 2015). The torus spans up to a few parsecs in size, as has been observed through recent interferometric observations (see Nikutta et al., 2021). Dust particles such as graphite cannot survive in regions with strong ultraviolet (UV) radiation; they evaporate if located in a place where the equilibrium black body temperature is above approximately 1500 K. The radius at which graphite can exist is called the dust sublimation radius, and it is considered to be the starting point of the so-called obscuring/dusty torus (see Almeida et al., 2017). In the unified AGN model, the dusty torus provides the opacity needed to hide the central parts of the AGN if observed at the right angle, leading us to observe seemingly different active galaxies. Obscuration by the dusty torus results in the Type-1 and Type-2 classification of the AGN, with the presence of with and without the BLR emission lines, respectively.

### 1.2.5 The Narrow Line Region

The emission line widths of a few hundred km/s are detected in the AGN spectra. These emission lines are generated in the Narrow Line Region (NLR). As the name suggests, the emission lines from this area are more narrow (of the order of a few 100 km/s) than those from the BLR (of the order of a few 1000 km/s), which is a consequence of the larger distance from the SMBH. The narrow line region (NLR) is made up of interstellar gas, which is less dense than that of the BLR (see Netzer, 2008). The region is also larger and extends much further from the SMBH. The interesting thing about the NLR is that both the permitted and forbidden emission lines can be seen in the NLR spectra. Permitted emission lines occur readily on Earth and other high-density environments. Forbidden emission lines correspond to low probability transitions from meta-stable energy levels in atoms. The low probability leads to a long lifetime for the transitions, and in sufficiently high-density regions, the energy is carried away by atom collisions before the low probability transition may occur. With a more considerable distance to the SMBH, broadening the emission lines is not equally vital in the narrow line region, resulting in both the permitted and forbidden narrow emission lines. A study of the NLR region is crucial in understanding the AGN and host galaxy feedback. Since the NLR is located beyond the tours, the obscuration due to the dust is minimal in this region. Instead, this region has significant outflows related to the AGN and host galaxy feedback. In the [OIII] emission line, blue outliers are present, tracing the outflow of these gases (Berton et al., 2016, Gaur et al., 2019).

### 1.2.6 The Jet

Relativistic jets are launched from the central regions of Active Galactic Nuclei. These jets are highly collimated with minimal opening angles. The motion of the material in these jets is mostly relativistic. The power of the jets lies in $10^{43} - 10^{48}$ erg/sec range (Ghisellini et al., 2014). The emission near the launching point of the jet can be electromagnetic, while on the larger length scales, as it becomes relativistic, inverse comptonization of the photons takes place (Boccardi et al., 2017). The energy gets dissipated through radio emission, giving rise to diffuse radio lobes. The jets are associated with the radio loudness of the AGNs and are usually found in only radio-loud AGNs. The jets can reach up to a length of a few Megaparsecs (Blandford et al., 2019).

The emission in the jets varies at very short time scales, implying that the jet emission region is very small. However, resolving such regions is still challenging due to small angular resolutions. How the matter can eject in the form of a jet and what can be the launching point of the jets in these AGNs is not known clearly. Moreover, the connection of jets with the accretion disk remains an unsolved problem. In the absence of this information, the condition responsible for the launch of relativistic jets remains unknown.

All these components play a unique role in generating the composite AGN Spectral Energy Distribution (SED), which is seen to be contributed by a wide variety of processes (see Figure 1.4). The SED has a unique shape with two humps, one near the optical/UV wavelengths and another near the high energy X-ray to $\gamma-$ray wavelengths. This provides the so-called *double-humped* structure of the AGN SED and is a characteristic of such objects.

## 1.3  AGN classification and the Unification model

AGNs display a wide variety of features in different wavelengths; thus, classifying these objects is no simple task. According to the properties of their optical spectra, AGNs are classified into Types 1 and 2, with Type 1 showing both broad (of the order of a few 1000 km/s) and narrow emission lines (of the order of a few 100 km/s) and the Type 2 showing an absence of broad lines and only narrow emission lines are present. This happens due to the obscuration by the dusty torus of the BLR and the accretion disk in these AGNs (for comprehensive reviews see Tadhunter, 2008, Netzer, 2015, Merloni, 2016, Padovani et al., 2017, and references therein.). Quasars and Seyfert galaxies are classified based on their luminosity, with the quasars having higher luminosity (traditionally, this limit is taken as $10^{44}$ ergs/sec). Blazars are those AGN characterized by rapid optical variability, high optical polarization, large X-ray and $\gamma$-ray luminosity, and a variable, flat-spectrum, superluminal radio core (Véron-Cetty & Véron, 2000). BL Lac objects are the blazars characterized by a strong continuum and very weak or absent emission lines. Optically Violent Variables (OVVs) are another class of Blazars similar to BL Lac objects, with the difference that their spectra show emission lines. LINERs ( Low-Ionization Nuclear Emission-line Regions) are the least luminous and the most common AGNs. Approximately one-third of all nearby galaxies may be classified as LINER galaxies. However, it remains uncertain whether all LINERS are indeed AGN or rather a subset of them represent nuclear star-forming regions. In terms of spectra, the narrow lines arising out of LINERs are clearly distinguished from that of the Seyfert galaxies, showing a lower degree of ionization. In terms of their detection in radio wavelengths, AGNs can be classified as radio-loud and radio-quiet AGNs. However, only about 10 % of the AGN fraction is radio loud. Recently, another distinction- jetted or non-jetted AGN has been proposed by Padovani (2017) instead of the radio loud radio-quiet dichotomy.

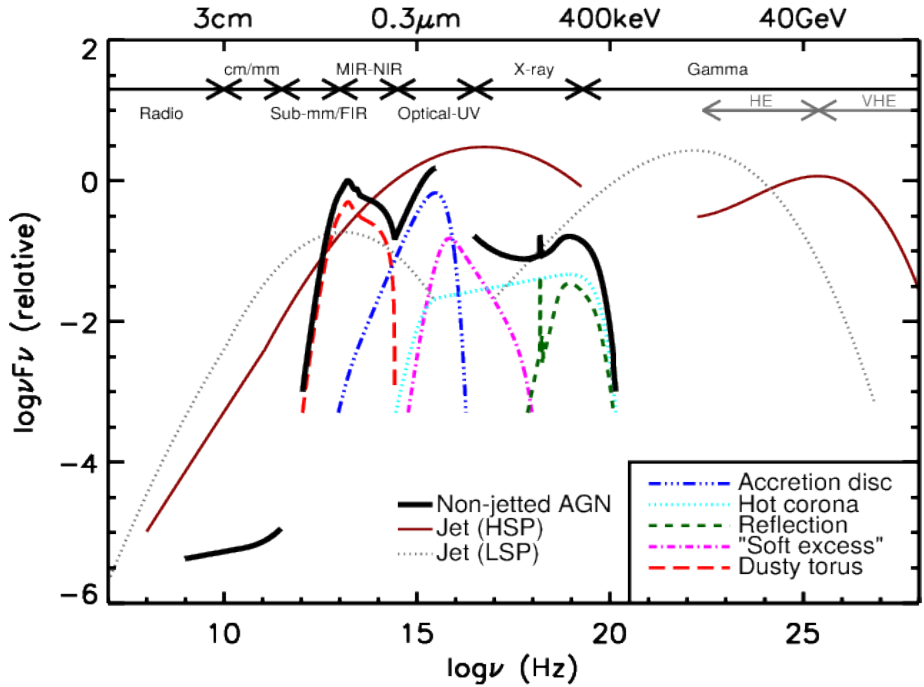

Fig. 1.4 The Spectral Energy Density (SED) of AGN shows the contribution of different components to the different wavelengths. Image adapted from Harrison (2014).

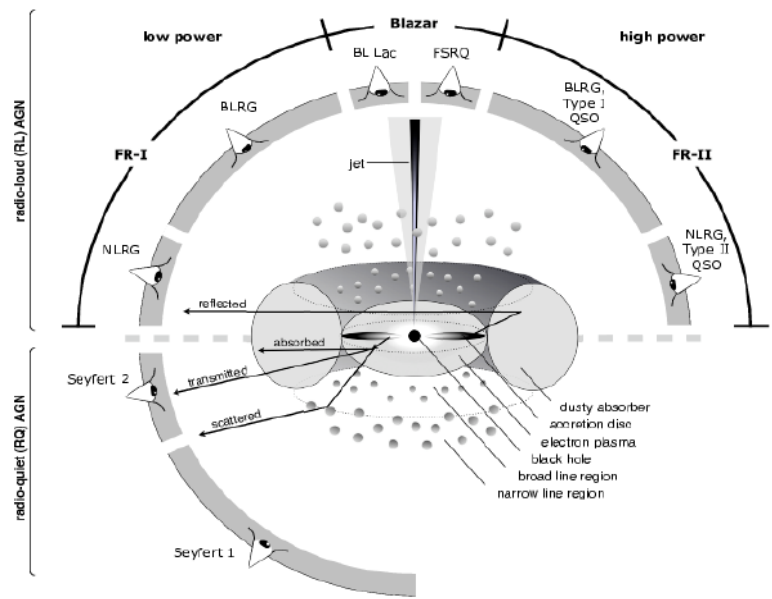

Fig. 1.5 The Unification model proposed by (Urry & Padovani, 1995) aims to unify the different AGN properties based on their orientation with respect to the observer. Image obtained from Beckmann & Shrader (2012).

The AGN population ranges in hundreds of thousands, thanks to the recent advances in telescope technology. The terminology in AGN is confusing, and many methods have been used to unify the AGN population into one. Since these objects are located far away, not every region is observable for all the objects. The classification of AGN into various sub-types is influenced by their orientation with respect to the observer. Different spectral signatures are seen depending on the orientation at which this structure is viewed. In particular, the broad emission line region is only visible to observers with a relatively face-on view. If viewed edge-on, the dusty torus blocks the same, and only narrow emission lines are seen in the AGN spectrum. Multiple components may give rise to multiple observational phenomena; hence the AGN taxonomy consists of a variety of objects. A real breakthrough was the unification model (Urry & Padovani, 1995, Antonucci, 1993), which unified all the AGN into one family and simplified our understanding of the AGN phenomenon by classifying them based on their orientation as shown in Figure 1.5. The Unification model has been successful in arranging the AGN based on their orientation with respect to the observer.

Among the various classes of AGN based on the unification model, the work in this thesis focuses on the study of the Type-1 AGN population only. The Type 1 AGN are the only ones with little or no obscuration of the BLR and the accretion disk; thus, these AGN allow us to study these regions through their observations with the ground-based telescopes as outlined in the subsequent sections.

## 1.4 Tools to probe the Innermost Regions

### 1.4.1 Variability

As mentioned above, the sub-pc regions involve the innermost regions, the accretion disk, and the BLR. Being too distant, even in the closest of AGNs, these regions are too compact to resolve spatially. However, the fact that most of the AGNs are variable at all time scales can be exploited to understand and temporally resolve the innermost regions of these objects. Variability on shorter time scales has smaller amplitudes. Sometimes quasi-periodic variability is also observed in AGN (Gierliński et al., 2008). There is a more rapid variation in the X-ray than in the other bands, possibly due to the relativistic motion of the particles in proximity to the SMBH. The causes of the variability in AGNs have been proposed to be the accretion disk hotspots, magnetic re-connection, accretion disk corona interplay, and the disk jet interconnection ( see Ulrich et al., 1997). The AGN variability has been recently modelled using stochastic processes. The Damped Random Walk (DRW) model proposed by Kelly et al. (2009) has been successful for a large number of AGN (MacLeod et al., 2010, Zu et al., 2016); however, the limitations of the DRW model have also been pointed out in a few recent works (e.g., see Kasliwal et al., 2015, Kozłowski, 2017).

**Reverberation mapping**

Since the innermost regions of the AGN cannot be resolved by any current or near-future observational facilities, time resolution substitutes spatial resolution. Reverberation Mapping (RM) (Bahcall et al., 1972, Blandford & McKee, 1982) is based on the premise that the line emitting regions are driven by the intrinsic variation from the accretion disk itself, and thus a correlated variability can be observed, which results in a lag. The innermost regions of AGN, although challenging to resolve spatially, are resolved temporally using the technique of RM. RM has been applied to get the BLR sizes (Peterson & Horne, 2004), the torus sizes (see Koshida, 2015, Mandal et al., 2018, Minezaki et al., 2019, etc.) and more recently, it has also been applied to get the accretion disk sizes (Starkey et al., 2016a, Cackett et al., 2020) (see Figure 1.6). The earlier RM campaigns centred on getting the BLR sizes using the H$\beta$ emission line RM, and the emission reverberation mapping has been very helpful in resolving the BLR and estimating the size and structure of the region. The most important result from the RM experiments is that the AGN's SMBH masses can be estimated based on scaling relations. The scaling for SMBH remains the only method to estimate the SMBH masses even at high redshifts assuming all the AGNs produce similar spectra.

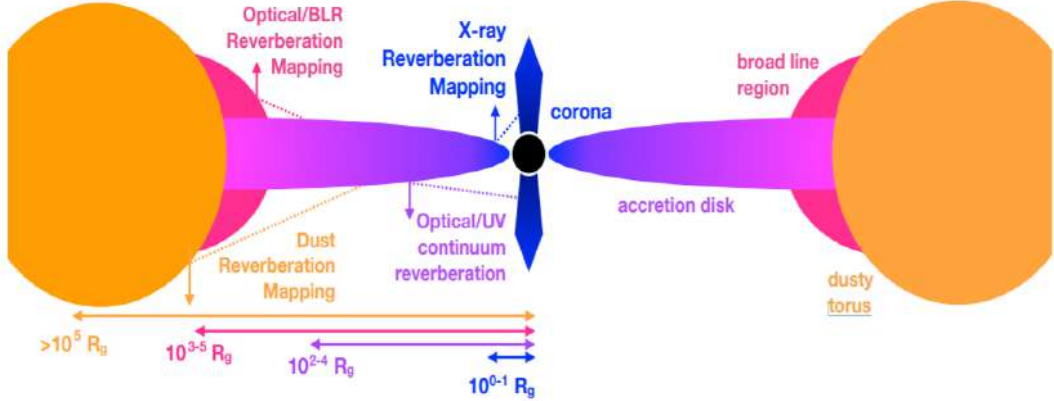

Fig. 1.6 A graphic showing the various scales involved in reverberation mapping. The components studied through RM are shown on the right side, while the length scales involved, in terms of the gravitational radius ($R_g$), are presented on the bottom right. Image obtained from Cackett et al. (2021).

**The radius luminosity relation:** By virtue of these campaigns, a relation between the size of the BLR (R) obtained through reverberation mapping of the $H\beta$ line and the luminosity of the AGN has been obtained. From the photoionization models, we know that,

$$U = \frac{Q(H)}{4\pi r^2 n_e c} \quad (1.8)$$

and,

$$Q(H) = \int \frac{L_\nu}{h\nu} d\nu \quad (1.9)$$

where U is the ionization parameter, Q(H) is the rate of the production of Hydrogen ions at a distance *r* and $n_e$ is the electron density. The assumption is that the conditions in AGN do not change much; hence, the values of U and $N_e$ can be taken as constant. From the above equations, we find that L ∝ r$^2$. This is expected from a theoretical point of view. Multiple reverberation mapping campaigns have yielded that it is true, and the correlation between L and r is very tight. The reverberation mapping studies have yielded the so-called radius luminosity relation (Kaspi et al., 2000a, Bentz et al., 2009, Du et al., 2016, see Figure 1.7). In (Bentz et al., 2009), the correlation has been proven as r ∝ $L^{0.563}$. This relation is significant as it provides a method to obtain the SMBH mass residing at the centre of AGN, using the virial relation (Bentz et al., 2013). The assumption is that the gravity of the central SMBH dominates the virial motion of gas in the BLR. If the velocity of the gas $\triangle V$ in that region is known, then:

$$M_{BH} = \frac{f \triangle V^2 R_{BLR}}{G} \quad (1.10)$$

where $R_{BLR}$ is the size of BLR obtained through reverberation mapping, G is the gravitational constant, and *f* is a scaling factor of the order unity. The mean value of this dimensionless correction factor depends on the region's geometry and inclination, and its exact values remain unknown. In the literature, different values for *f* have been used, ranging from 1 to 5.5, based on calibration of the mass obtained with estimates of the BH mass through other means (Yu et al., 2019).

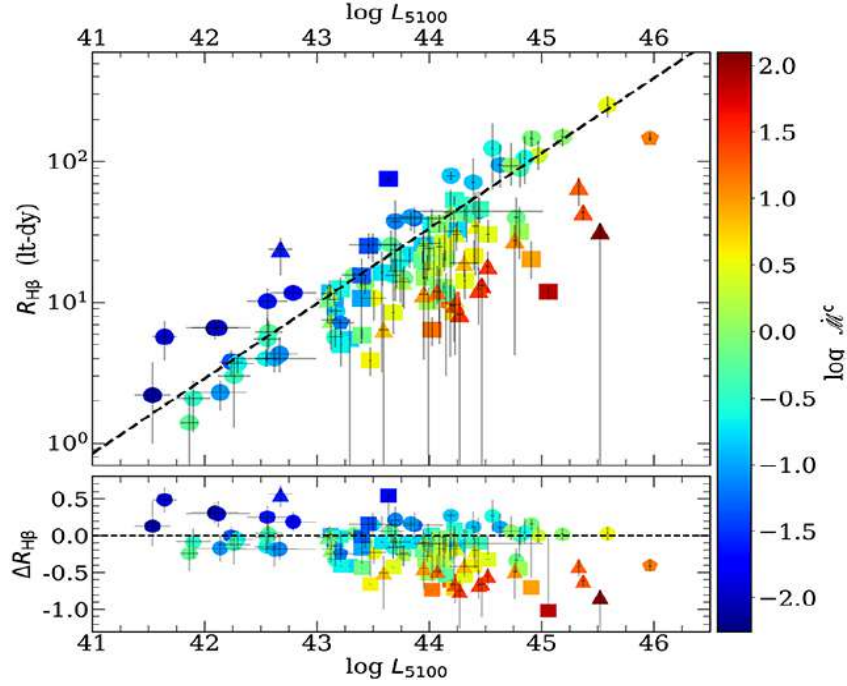

Fig. 1.7 The radius luminosity for the AGN obtained through the multiple reverberation mapping campaigns so far. The colour bar on the right shows the dimensionless accretion rate. The sources with high accretion rates generally lie below the best-fit curve. Image obtained from Panda et al. (2019a).

This relation has indeed been valid for most AGNS whose BLR size has been obtained through reverberation mapping but recently, with the reverberation mapping studies of the Super Eddington AGNs, it has been found that the lags do not follow the radius luminosity relation and the lags in some of the sources are significantly lower than expectations from this relation (Du et al., 2016) implying that the accretion rates could play a role in the RL relationship.

**Photometric reverberation mapping**

Reverberation mapping has been used to get the BLR sizes for around 120 AGN (Panda et al., 2019a). The spectrum obtained for RM is both resource and time-consuming, and thus an option is to obtain both the continuum and emission-line flux using a combination of broad and narrowband filters. This technique is known as Photometric Reverberation Mapping (PRM). PRM uses one band to trace the continuum and another narrowband to trace the emission line but needs much less telescope time than the spectral RM and can also be done with smaller 1-2m class telescopes. This technique has the potential to yield BLR sizes and SMBH masses for a large number of sources. PRM has been demonstrated as a successful alternative by Haas et al. (2011), Chelouche & Daniel (2012) etc. A few sources have been studied using PRM, and the lags obtained are in good agreement with the spectroscopic lags (Pozo Nuñez et al., 2012, Edri et al., 2012). Zhang et al. (2017) have published results using PRM for a few sources selected from the SDSS-Stripe 82 catalog. However, a systematic campaign involving a large number of sources using the PRM technique has not been executed so far. Hence, it is necessary to calibrate the technique in order to understand the difficulties and associated challenges.

Reverberation Mapping of the BLR faces difficult questions. The reverberation mapping studies have yielded the size of BLR for only a handful of AGN, yet the gas motion in these regions remains poorly understood. As for the sources with time-resolved lags, the velocity delay maps have not yet been available for the general AGN population. Sometimes double-peaked emission lines are also observed in some AGN, which is puzzling in understanding the dynamics of matter in the BLR.

**Accretion disk reverberation mapping**

Reverberation mapping of the accretion disk using multiple bands (Fausnaugh et al., 2017) has been a recent development in RM studies. The simple *lamppost* model implies a corona above the accretion disk, which irradiates in the X-ray wavelengths, and the photons are reprocessed in the form of UV and optical wavelength emissions. The emission from the accretion disk is anticipated to be of the black body type, with its peak wavelength varying based on the disk's temperature. To obtain information about the structure and temperature profile of the accretion disk, it is necessary to continuously monitor it in multiple wavelengths, covering both the hotter inner regions and the cooler outer regions. This technique has been used to estimate the accretion disk structure for a handful of AGN (e.g., see Jiang et al., 2017, Mudd et al., 2018, Cackett et al., 2020). These studies only used the optical wavelength lags, and the accretion disk profile from the X-ray (innermost regions) to optical wavelengths (outer accretion disk) could not be adequately constrained. The AGN 'Space Telescope and Optical Reverberation Mapping' (STORM) campaign was the first ever reverberation mapping campaign to study the innermost regions of the most studied AGN so far: NGC 5548. The study was performed using simultaneous observations from *Swift*, Hubble Space Telescope (HST), and several ground-based observatories (e.g., see De Rosa et al., 2015) with high cadence for a period of almost 6 months. Similar studies have been made for a handful of AGN (see Hernández Santisteban et al., 2020). Based on these recent multi-band monitoring campaigns, we now understand that:

- The continuum variations at large wavelengths lag in comparison to that at smaller wavelengths, which is expected from the SS disk model.

- The inter-band lag scales with wavelength as a power law of index 4/3, but the size of the disk is observed to be 3−4 times larger than expected from the standard SS disk (see Fausnaugh et al., 2016, and Figure 1.8).

- The UV wavelength flux variations lead the optical band flux variations in some cases (Edelson et al., 2019).

What can be the cause behind the accretion disk size discrepancy and whether this is true for the AGN of all types is not known clearly. The knowledge of interband lags provides us with an observational framework to check the accretion disk models. The observational *lags* can become an input parameter with which the various accretion disk models can be studied. Also, the scaling of the physical parameters, such as the black hole mass, luminosity, and accretion rate with the size of the accretion disk, has eluded us for the time being. However, this knowledge is fundamental in the context of SMBH growth and evolution.

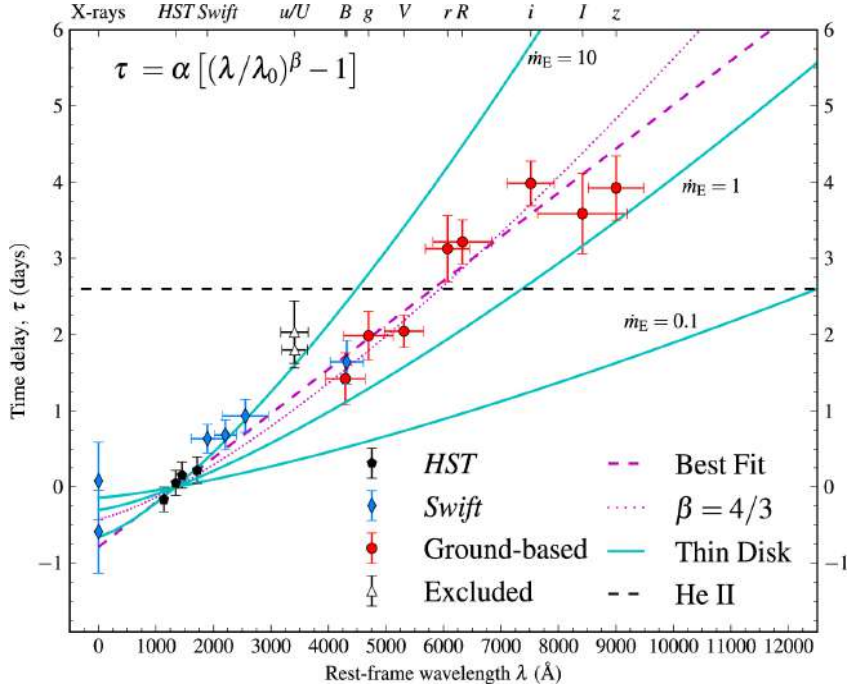

Fig. 1.8 The interband continuum lags for NGC 5548 from the AGN STORM campaign. The lags follow the 4/3 relation, but the magnitude of the lags is larger than the SS disk predictions. Image courtesy: Fausnaugh et al. (2016).

**Microvariability**

AGN has been known to be variable at all time scales. While the traditional RM uses the timescales of days, months, and even years in the case of luminous quasars, microvariability probes the timescales of minutes to hours. Microvariaility is also called Intra Night Optical Variability (INOV) or Intra Day Variability (IDV). Blazars have been found to show a higher probability of microvariability (Miller et al., 1989). Apart from blazars, microvariability has also been reported in radio-quiet quasars (Gopal-Krishna et al., 2003), and weak emission line quasars (Kumar et al., 2015). With a sample of NLSy1 galaxies, (Ojha et al., 2019) have also demonstrated significant INOV detection in the $\gamma-$ray detected NLSy1 galaxies. The cause of microvariability could be localized particle acceleration or non-thermal plasma flowing in a relativistic jet. It can be used to provide information about the size limits of the line-emitting regions. In recent works, microvariability has been applied to identify the presence or absence of jets and even deduce the properties of the various classes of AGN. Microvariabilty in the optical regime has also been applied to study the quasi-periodic oscillations in blazars (Hong et al., 2018). Through microvariability, indirect evidence of radio loudness is also found. Such short-time scales usually give insights into *pure* physical conditions occurring in jets of these sources because the corresponding spatial scales are too small to be characterized with different emission properties. Very short timescales probe conditions very close to the ionizing source. Moreover, microvariability has been used to detect polarization in blazars. In some sources, a significant polarisation value is associated with enhanced INOV detection.

### 1.4.2 Emission line diagnostics of AGN

Through the reverberation mapping experiments, and the dynamic modelling of RM data, the power of modelling the BLR has been demonstrated (see Pancoast et al., 2014, Williams et al., 2018, etc.), but for the dynamical modelling studies, only a handful of AGN with velocity-resolved lags obtained through RM campaigns are used. Recent spectroastrometry using the GRAVITY instrument has resolved the BLR for some nearby AGN (Gravity Collaboration et al., 2018, 2020, 2021). Even if the RM has the potential to yield the BLR sizes and SMBH masses for a significant sample of AGN, it is a resource-intensive task. These limits will always constrain it, and the ensemble study is quite challenging. The gas dynamics in the line emitting regions for a large sample remain poorly understood in this scenario if we depend on RM only.

Prominent emission lines are one of the defining features of the AGN spectra. In view of that, a slightly different approach to studying these regions is to exploit the power of single epoch spectroscopy. A remarkable feature of the AGN spectra is that they show similar features from the local Seyfert galaxies to the highly luminous quasars at $z \geq 7$. Since the onset of SDSS, multi-object spectroscopy has gained traction, and it is possible to obtain the spectra for hundreds of thousands of sources simultaneously with reasonable SNR. Ensemble studies of Type 1 AGN have provided vital insights into the inner regions of Type 1 AGN. The emission line diagnostics can be used to organize the Type 1 AGN neatly into a sequence known as the *quasar main sequence*, a pioneering work done by Boroson & Green (1992). The main sequence can be represented by a trend in the FWHM of H$\beta$ and the iron strength (see Figure 1.9). This correlation has been observed for a large number of AGN (Shen et al., 2011, Boroson & Green, 1992, Marziani et al., 2018a). The main sequence divides the sources into population A and population B with a break at 4000 km/s. The emission lines for the population A sources show Lorentzian profiles, while the emission lines for the population B sources can be modelled as Gaussian profiles.

Further, through single epoch spectroscopy, traces of outflow in the CIV emission line have been detected, for instance, Gaskell (2000). This has direct implications for the BLR geometry. The blue shifting observed in the C IV emission profiles rules out a Keplerian BLR as hypothesized in the models. A similar case for other AGNs exists too, or not is unclear as yet. However, single-epoch spectroscopy can be used for this. Whether or not asymmetric emission profiles exist for a large number of AGNs is not known very clearly. Also, correlations can be derived through the physical parameters to understand the dichotomy of different AGN types, even within the Type 1 AGN population. For example, the NLSy1 galaxies are classified as xA sources in the population A regime, and whether their physical parameters differ is not known very clearly. The multi-parameter correlation analysis can thus be used to characterize the AGNs based on their physical properties. The RL relation derived from the RM experiments can be used to get the SMBH mass estimates for a large sample of AGN through scaling relations. The main sequence is beneficial in developing physical models, and multi-parameter correlation analysis can be used to establish relations between the accretion disk and the line-emitting regions.

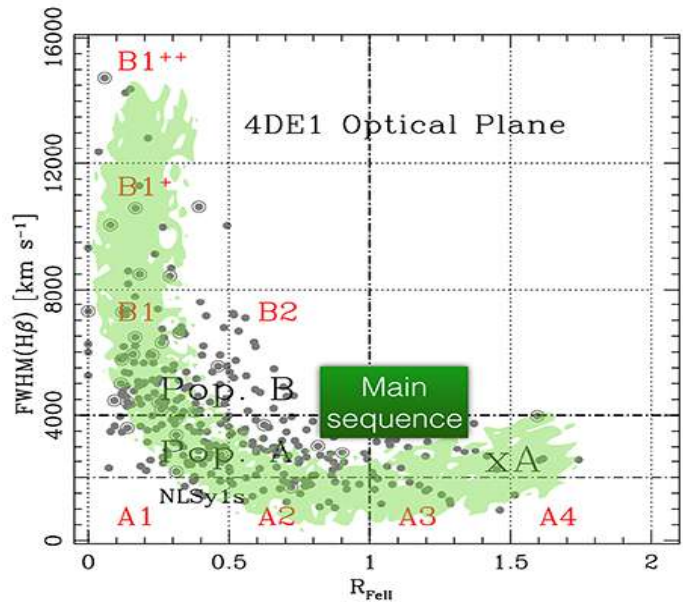

Fig. 1.9 The Quasar main sequence shows the anti-correlation between the FWHM of the H$\beta$ emission line and the iron strength in the BLR. Also shown are the divisions in Type 1 AGN in terms of the Population A and Population B Sources. Image obtained from Marziani et al. (2018a).

## 1.5 Motivation for this thesis

It is not clear currently what the accretion disk's structure is and how it correlates with the known physical parameters. This has been intriguing, as to date, only a handful of AGNs have proper accretion disk size measurements. Hence, it is necessary to measure accretion disk sizes for as many AGNs as possible to formulate the accretion disk models. The recent advancement in the reverberation mapping technique has yielded plenty of information related to the broad-line region, the accretion disk, and the Torus. Through the AGN-STORM campaign and subsequent works, for the first time, we can measure the accretion disk sizes and test the accretion mechanisms behind a handful of AGNs. However, the known number of AGN with accretion disk size estimates remains limited. Whether or not the disk sizes scale with the SMBH masses as expected from the SS disk reprocessing arguments remains to be seen. We want to address whether there are physical signatures that correlate with the accretion disk sizes or not. Knowledge of accretion disk sizes will help constrain the various accretion disk models. Coming to the BLR, since the emission line RM is a time-consuming and resource-intensive process, PRM has been proposed as a cost-effective alternative. It is necessary to calibrate the method with the traditional RM method to check for any inconsistencies and to determine how efficient it is at recovering lag estimates. PRM is a relatively new technique that has not been used extensively, and only a few objects have been studied using it. The method of photometric reverberation mapping, as compared to the traditional reverberation mapping experiments, needs to be calibrated to provide a prescription regarding the expected cadence, experiment length, variability amplitudes, and seeing conditions in order to design a robust photometric reverberation mapping campaign. Extensive simulations need to be developed in order to achieve this. In the upcoming era of large all-sky telescopes observing the sky in multiple bands, such as the Vera C. Rubin Observatory (VRO)/Large Synoptic Survey Telescope (LSST), photometric RM combined with single epoch spectroscopy can be used to measure BLR sizes for hundreds of AGNs.

Type 1 AGN with jets are a rare phenomenon, and only a few of them are known. The discovery of these AGN with jets evokes questions about whether they are related to the class of AGN, where jet emission dominates, and whether jets can play a part in their continuum variability. Whether these AGN relate to blazars or significant differences are present remains to be known. To seek the answers to these questions, Narrow Line Seyfert-1 (NLSy1) galaxies are excellent candidates. The NLSy1 galaxies occupy a peculiar class among the Type-1 AGN population with extremely high accretion rates and higher iron content in their BLR. NLSy1 galaxies are also assumed to be the younger version of the Type-1 AGN population. Some of the NLSy1 galaxies have been detected in $\gamma-$ray wavelengths, which contradicts the trend that relativistic jets are observed in AGN with larger SMBH masses. These peculiar NLSy1 galaxies can be used to understand the mechanism of jet formation and its connection to the innermost regions in relatively younger AGN.

As pointed out earlier, the exact nature of the gas in the BLR is not yet understood. Hence, if the NLSy1 galaxies show peculiar behaviour, whether their emission profile also show peculiarity? It can be tested with different emission line profiles. How these emission line profiles correlate with the physical parameters and whether these profiles themselves classify the NLSy1 galaxies as a subclass needs to be established clearly. In the general AGN population, the NLSy1 galaxies occupy the extreme end of the main sequence. What can be the possible reason behind this? Whether the physical parameters deriving the variation in parameters in the NLSy galaxies are similar to the general AGN population or different galaxies remains to be understood.

To address these questions from the primary motivations of this thesis, we have used data from multiple ground-based telescopes. Details about these telescopes and related instruments, along with the related web links, are included in the respective chapters of the thesis. A brief outline of the thesis is given below.

## 1.6  Structure of the thesis

In the following chapters, I will focus on addressing these crucial problems related to the inner regions in the AGNs. This thesis is structured into seven chapters, and we present a brief introduction to the problems addressed in the subsequent chapters as below:

**Chapter 2**: This chapter focuses on estimating the accretion disk sizes for a sample of AGN with previous reverberation mapping measurements. This work aims to estimate the accretion disk sizes for a large sample of AGNs and test whether the accretion disk sizes correlate with the SMBH masses. This work has been published in MNRAS as Jha et al. (2022b)

**Chapter 3**: This chapter describes a new accretion disk reverberation program titled: *Investigating the central parsec regions around supermassive black holes* (INTER-VAL) that we have started using the Growth India Telescope (GIT). We present the initial results, measuring the accretion disk sizes for one of the sources, namely IRAS04416+121. This work is in press as Jha et al. (2023).

**Chapter 4**: This chapter is related to the calibration of the photometric RM technique with the traditional RM technique using an archival dataset. Further, multiple simulations have been developed for the experiment design for a photometric reverberation mapping campaign to test the optimal cadence, experiment length, and variability to recover the lags using Photometric reverberation mapping.

**Chapter 5**: In this chapter, to explore the dichotomy of NLSy1 galaxies with jets, we have performed a micro variability experiment on a sample of NLSy1 with confirmed detected jets based on their radio observations. We have also studied the NLSy1 with $\gamma-$ray detection. This work has been published in MNRAS as Ojha et al. (2022).

**Chapter 6**: In this chapter, we have compared the physical properties of NLSy1 and BLSy1 galaxies using their optical spectra to test whether the physical parameters differ between the two types of galaxies. This work has been published in MNRAS as Jha et al. (2022a).

**Chapter 7**: In this chapter, we present the conclusions from the thesis and shed light on the upcoming and ongoing projects derived from the work presented in this thesis.

# Chapter 2

# Accretion Disk Size Measurements for AGN from the ZTF survey using Reverberation Mapping[1]

In this chapter, we present a study of the accretion disk sizes in a sample of 19 AGN using continuum reverberation mapping. We analyze the optical light curves in the $g$, $r$, and $i$ bands from the ZTF survey for these AGN, all of which have reliable SMBH mass estimates from previous reverberation mapping studies. By cross-correlating the light curves in different bands, we obtain the *lags* that are proportional to the accretion disk sizes. We use two methods for cross-correlation: the ICCF method and the Bayesian method implemented by the JAVELIN code. Our results show that the disk sizes derived from the lags are larger than those predicted by the standard SS analytical disk model. To further investigate this discrepancy, we fit the light curves directly with a thin disk model using the JAVELIN code, which allows us to estimate the disk sizes independently of the lags. Our goal is to explore the relation between the disk sizes and other physical parameters, such as the luminosity and the SMBH mass, and to test different scenarios for the accretion disks in AGN.

---

[1]**Based on the work published as:** *Accretion Disk Sizes from Continuum Reverberation Mapping of AGN Selected from the ZTF Survey.* Vivek Kumar Jha, Ravi Joshi, Hum Chand, Xue-Bing Wu, Luis C Ho, Shantanu Rastogi, Quinchun Ma, 2022, MNRAS, 511, 2.

## 2.1 Introduction

Accretion disks in AGN form as a result of inflowing matter onto the central Supermassive black hole [SMBH] (Salpeter, 1964, Lynden-Bell, 1969). Various accretion disk models have been proposed ranging from the standard SS disk (Shakura & Sunyaev, 1973), the slim disk (Abramowicz et al., 1988) to more complicated models like the advection dominated accretion disk presented by Narayan & Yi (1995) and the relativistic accretion disk around a Kerr black hole presented by Li et al. (2005). Based on these models, the accretion disk has a temperature profile, with higher temperatures in the inner regions and decreasing temperatures in the outer regions. The growing observational shreds of evidence suggest that the X-ray emission in the innermost regions of the AGN accretion disk are reprocessed as the UV/optical continuum flux. Further, the rapid X-ray variability and relative strength of the hard X-ray component point towards the presence of a corona in the proximity of the SMBH, which irradiates the accretion disk. However, the exact size and structure of this corona are not known precisely (e.g., see Meyer-Hofmeister et al., 2017, Arcodia et al., 2019, Sun et al., 2020).

According to the standard SS disk model, the effective temperature $T$ at a particular location in the accretion disk $R_{disk}$ is related to the disk size and the SMBH mass as follows:

$$T(R_{disk}) = \left( \frac{3GM_{BH}\dot{M}}{8\pi\sigma(R_{disk})^3} \right)^{1/4}, \tag{2.1}$$

where $\sigma$ denotes the Stefan-Boltzmann constant and $G$ and $\dot{M}$ represent the gravitational constant and the mass accretion rate respectively. As is evident from this equation, the accretion disk size (R) scales inversely with the temperature (T) by a power law of index 4/3.

The photon wavelength ($\lambda_0$) at a particular disk radius is related to the temperature by Wein's law. The size of the accretion disk at the wavelength $\lambda_0$ can be determined by two observable parameters, the SMBH mass, and the mass accretion rate. The radiation from the accretion disk is assumed to be composed of multiple black bodies, and the disk sizes at various wavelengths can be calculated using the equation below:

$$R_\lambda = R_{\lambda_0} \left[ \left( \frac{\lambda}{\lambda_0} \right)^\beta \right] \tag{2.2}$$

here, $R_{\lambda_0}$ is the disk size at a reference wavelength $\lambda_0$ and $\beta$ is 4/3=1.33 for the standard SS disk. The temperature profile of the disk generated using this formalism can be used to test the application of the SS disk to the AGN population.

Direct observation of these regions is difficult as they project to sub-micro arc second angular resolutions, which any current or near-future observational facilities cannot resolve. The disk sizes for a few AGN have been estimated using microlensing studies in (see Pooley et al., 2007, Dai et al., 2010, Morgan et al., 2010, Blackburne et al., 2011, Mosquera et al., 2013). Another technique that remains promising is Reverberation Mapping (RM). Utilizing the variable nature of AGN at most time scales, RM substitutes spatial resolution with time resolution to measure the lag between the flux emitted from different regions, thereby enabling us to reconstruct the echo images that encode the information about the structure and kinematics of AGN innermost region (see Bahcall et al., 1972, Blandford & McKee, 1982, Vestergaard & Peterson, 2006, Bentz et al., 2009). Traditionally the RM technique was used to estimate the size of the Broad Line Region (BLR) using the cross-correlation between the accretion disk continuum and the H$\beta$ emission line, which has yielded a significant correlation between the size of the BLR and the AGN luminosity (Kaspi et al., 2000a, Bentz et al., 2009, Du et al., 2016, 2018). Resolving the accretion disk structure using RM through multi-band observations is possible assuming the *lamppost* model, which implies a disk being irradiated by X-rays originating in the innermost central regions, e.g., corona above the disk, and the photons being reprocessed in the form of UV and optical wavelength emissions (Cackett et al., 2007, Fausnaugh et al., 2017). The accretion disk is expected to emit radiation that follows a black body spectrum, with the peak wavelength depending on the temperature of the disk. Therefore, by observing the disk simultaneously in multiple wavelength ranges, we can probe both the hotter inner regions and the cooler outer regions of the disk. This allows us to determine the structure and temperature profile of the accretion disk itself.

However, due to the complexity involved in executing such multi-band monitoring campaigns, accretion disk structure has been mapped for only a handful of AGN so far (see De Rosa et al., 2015, Fausnaugh et al., 2016, Starkey et al., 2016b, Pei et al., 2017, Edelson et al., 2019, Hernández Santisteban et al., 2020). Alternatively, with the availability of large all-sky surveys, it has been possible to constrain the accretion disk sizes based on interband lags for a sample of AGN using observations in the optical regime only. In Sergeev et al. (2005) the interband lags were detected for a sample of 14 AGN in the optical wavelengths. In Jiang et al. (2017), inter-band lags based on the PanStarrs light curves were presented while Mudd et al. (2018) presented results for 15 quasars from the OzDES survey and Yu et al. (2020a) performed the analysis with around 23 quasars from the same survey but a different field. In Homayouni et al. (2019), inter-band lags for around 33 quasars were calculated from the sample of SDSS RM quasars. These studies used the optical wavelength lags only, and the accretion disk profile from the X-ray (innermost regions) to optical wavelengths (outer accretion disk) could not be appropriately constrained. However, the presence of lags itself indicates evidence of the reprocessing of photons in the accretion disk. These results put an upper limit on the accretion disk sizes; nevertheless, the correlated reverberation disk signals are detected. Interestingly, although the interband lag scales with wavelength as a power law of index 4/3, the size of the disk has been observed to be 3-4 times larger than expected from the standard SS disk (Fausnaugh et al., 2018). Furthermore, Edelson et al. (2015) and Fausnaugh et al. (2016) have shown that the U band lags are significantly larger than the predictions from the SS disk models in some of the sources, and the Balmer emission from the BLR may likely affect the continuum flux.

The scaling of the physical parameters such as the SMBH mass, luminosity, and accretion rate with the size of the accretion disk has eluded us for the time being. Only a handful of AGN has both the accretion disk size measurements and SMBH mass estimation, making multi-band monitoring very important. In this work, using the publicly available Zwicky Transient Facility (ZTF)[2], we aim to constrain the accretion disk sizes and their correlations with the physical parameters such as the SMBH mass and bolometric luminosities of AGNs with well-constrained SMBH masses through prior RM studies. This chapter is structured as follows: In Section 3.2 we present the sample of AGN studied in this work along with the details of the observations and data. In Section 3.4 we present the analysis based on various lag estimation methods, and the results of our comparison of disk sizes with the standard SS disk, followed by the correlations with various physical parameters, are presented in Section 3.5. The discussion is presented in Section 3.6 and we conclude our results in Section 2.6.

## 2.2 The sample and data

To measure the accretion disk sizes and thereby test the accretion disk models, we assembled the reverberation mapped AGN selected from various RM campaigns, including 61 AGNs from the AGN Black hole mass database[3] (see Bentz & Katz, 2015), 44 AGNs from Sloan Digital Sky Survey - Reverberation Mapping (SDSS RM) project (Grier et al., 2019), and 25 AGNs from the Super Eddington Accreting Massive Black Holes (SEAMBH) sample available in Du et al. (2015, 2016, 2018). However, 8 AGNs from the SEAMBH list were already included in the AGN mass database. Thus, we were left with a total of 122 sources. Having an advantage of accurate BH mass estimation, an essential input for the disk size measurement, the sample spans a significant range of luminosities (42.57$\leq \log(L_{5100Å}) \leq$ 44.97) and redshifts (0.013 $\leq$ z $\leq$ 0.646) providing a homogeneous set to study the AGN population.

---

[2]https://www.ztf.caltech.edu/

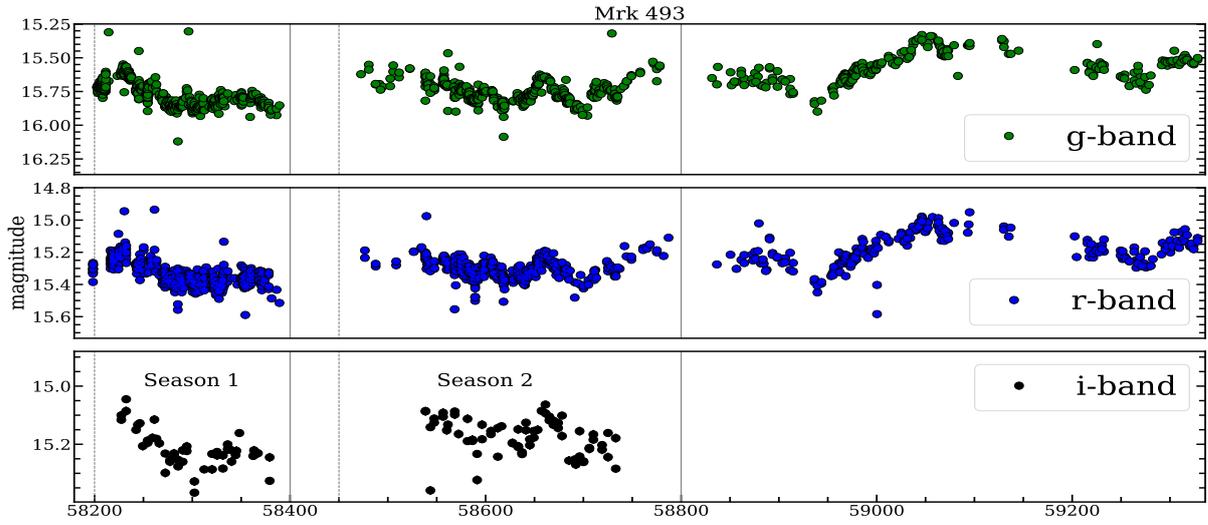

Fig. 2.1 The $g$, $r$, and $i$-band light curves for Mrk 493. There are seasonal gaps between the ZTF light curves after MJD 58400 and MJD 58800, based on which the light curves are divided in season 1 (MJD ranging from 58200 to 58400) and season 2 (MJD ranging from 58450 to MJD 58800), respectively. The dashed lines denote the beginning of a season, followed by a solid line denoting the end of a season. The $i$-band observations are not available beyond MJD 58850.

For the multi-band continuum light curves of these AGNs, we explore the Zwicky Transient Facility (ZTF) time-domain survey (Graham et al., 2019). The ZTF uses the 48 inch Samuel Oschin Schmidt telescope with a field of view of 47 deg$^2$ to map the sky in $g$, $r$, and $i$ optical bands with a typical exposure time of 30 seconds, reaching a magnitude $\sim$20.5 in $r-$band (Bellm et al., 2019). The processing of the data is done using the Infrared Processing and Analysis Centre (IPAC) pipeline (Masci et al., 2019). In addition, the average cadence of 3 days makes it a comparatively better survey for continuum RM observations which has otherwise been not possible in some recent works using all-sky surveys (see Jiang et al., 2017, Mudd et al., 2018, Yu et al., 2020a). We obtained the light curves using the ZTF-API [4] by providing the positions of individual objects in terms of their RA, DEC sky positions. Observations for all but 5 AGNs were available for at least one epoch in each band. We note that the $g$ and $r$ bands are well sampled in ZTF while the $i-$ band has the least number of observations (see Figure 2.1). Keeping this into account, we set a criterion of at least 15 observation epochs in each band. This limit was kept in order to obtain enough reverberation signal in order to get the interband lags. Since there are gaps between the observations in all the three bands, we divide the light curves in *two seasons*, the first season (season 1) running between MJD 58200 to 58400 broadly and the second season (season 2) running between 58450 and 58800. Since the light curves are sampled well, and the lags we expect are of the order of a few days, this division will not affect the lag estimates, and dividing the light curves has the additional advantage of getting two disk size measurements between these intervals.

---

[3]http://www.astro.gsu.edu/AGNmass/

This criterion resulted in the final sample of 57 reverberation-mapped AGNs for season 1, including 34 sources from the SDSS-RM project, 23 AGNs from the AGN-Mass sample, and 2 AGNs from the SEAMBH sample, which are part of the AGN-Mass sample. In season 2, our sample size was limited to 22 AGN primarily because of the poor sampling in the $i$- band. Light curves for some of the sources are presented in Figure 2.2. We moved forward with the final set of 57 AGNs from season 1 and 22 AGN from season 2 for inter-band lag estimation. All the 22 sources available in season 2 also satisfied the criterion in season 1 and thus, were part of the 57 sources selected from season 1. This provides us with the opportunity to measure the disk sizes for 22 AGN in two different time ranges.

## 2.3   Measurement of inter-band lags

Before measuring the lags between the multi-band continuum light curves, we first exclude any possible outliers in the light curves by applying a $3\sigma$ clipping. We employ two most commonly used lag estimation methods, namely the Interpolated Cross-Correlation Function (ICCF) method (Peterson et al., 1998) and the JAVELIN method (Zu et al., 2011) which employs a Bayesian approach to estimate the lags. Given that the expected lags between the $g$ versus $r$ band are expected to be shorter than the $g$ versus $i$ band, we used the $g$-band light curves as a reference for estimating the interband lags.

---

[4]https://irsa.ipac.caltech.edu/docs/program_interface/ztf_lightcurve_api.html

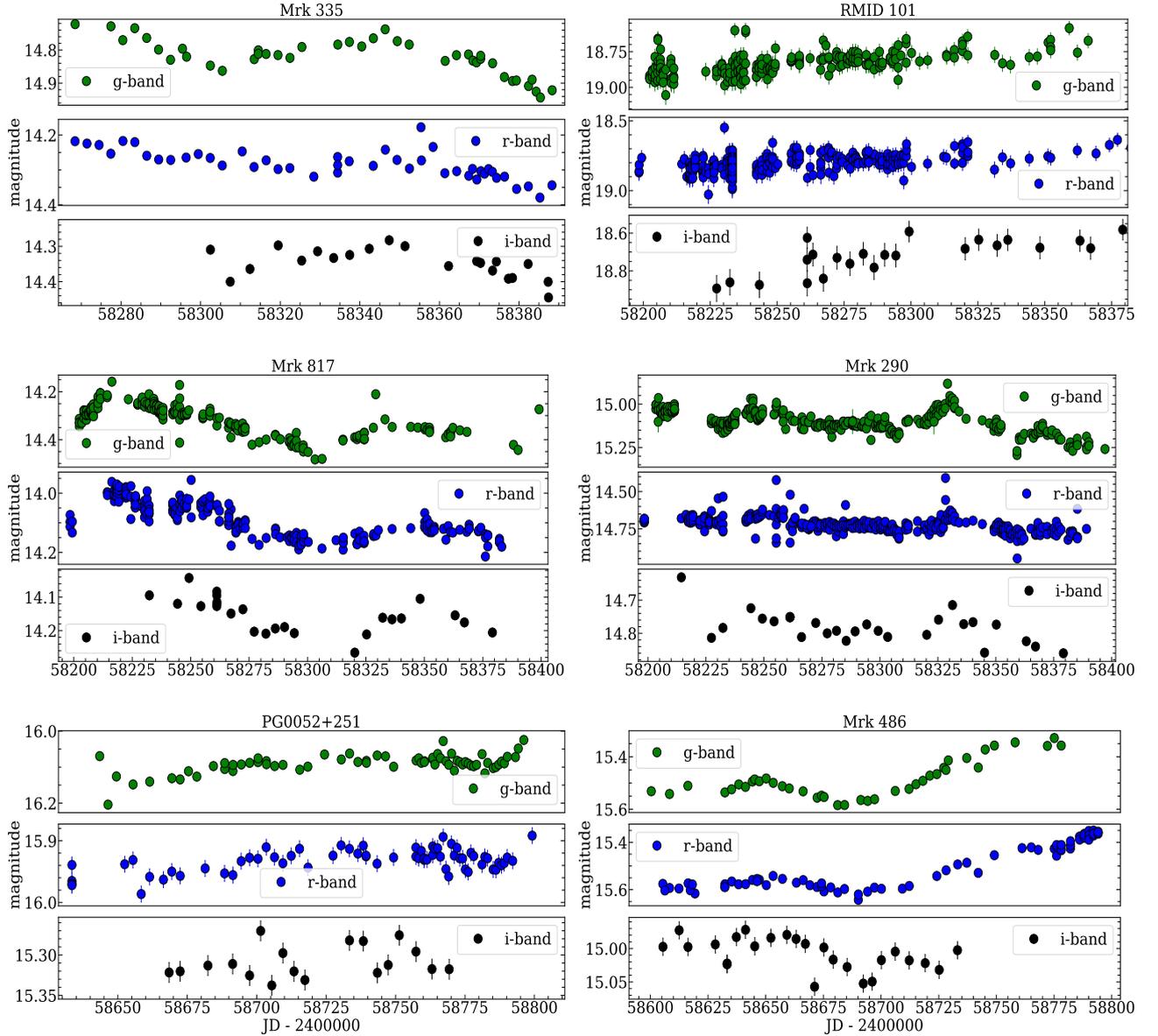

Fig. 2.2 The $g-$band (green), $r-$band (blue), and $i-$band (black) light curves for Mrk 335 and RMID101 are shown in the top panel, for Mrk 817 and Mrk 290 are shown in the middle panel, and for PG0052+251 and Mrk 486 are shown in the bottom panel. The first four light curves are from season 1, while for the sources in the bottom panel, the light curves are from season 2. The units for magnitude are calibrated magnitudes as available from the ZTF database. The light curves include the error bars; however, the error bars are smaller than the markers used here for some of the sources. The statistics for these light curves are available in Table 2.2.

### 2.3.1 JAVELIN based lags

We use the publicly available code JAVELIN [5] to estimate the interband reverberation lags. It models the variability of the AGN as a Damped Random Walk (DRW) or, in other terms, an Ornstein Uhlenbeck (OU) process for timescales longer than a few days (Zu et al., 2011, 2013). This approach has been demonstrated by Kelly et al. (2009) for modeling 100 quasar light curves from the OGLE database and further by MacLeod et al. (2010) for a sample of around 9000 quasars from the SDSS Stripe 82 region. The covariance function for the DRW process takes the following form:

$$S(\triangle t) = \sigma_d^2 e^{(-|\triangle t/\tau_d|)} \tag{2.3}$$

Where $\triangle t$ is the time interval between the two epochs and $\sigma_d$ and $\tau_d$ are the amplitude of variability and the damping time scale, respectively. Lag estimation in JAVELIN is based on the assumption that the responding light curve variability is a scaled, smoothed, and displaced version of the driving light curve. Here, we took the $g$-band light curve as the driving light curve, and the $r$ and $i$ band light curves were taken as the responding light curves while estimating the lags. Initially, using the driving light curves, we build the model to determine the parameters $\sigma_d$ and $\tau_d$. Then we fit the other two band light curves using this model to build the distribution for the time lag, the top hat smoothing factor, and the flux scaling factor.

We derive the lag, tophat width, and scale factor distribution by shifting, smoothing, and scaling the light curves. We use 20,000 Markov Chain Monte Carlo (MCMC) chains to get the best fit parameter distribution. We set the lag limits to [-50, 50] days for the initial run. Given that the observed inter-band lags for various AGNs in previous studies were typically smaller than 10 days (see, Jiang et al., 2017, Mudd et al., 2018, Yu et al., 2020a, Homayouni et al., 2019), we rejected the sources for which the lags were obtained outside the limits of $[-10, 10]$ days as these estimates could be unphysical.

Recall that in season 1, we had 57 sources with sufficient sampling (no. of observations in $i$ band $\geq$ 15), and in season 2, we had 22 sources with sufficient sampling. Out of 57 sources in season 1, we could constrain the lag within the range of $[-10, 10]$ days using JAVELIN for 25 sources, while in season 2, out of 22 sources, we could constrain the lags for 19 sources within the same limits. Comparing the lag estimates between the two seasons, we find that lags for 14 sources were available in both season 1 and season 2; hence we have two measurements for this sample of sources while we have 11 individual measurements from season 1 and 5 individual measurements from season 2.

---
[5] https://github.com/nye17/JAVELIN

### 2.3.2  ICCF based lags

The Interpolated Cross-Correlation Function (ICCF) method (Gaskell & Peterson, 1987, Peterson et al., 1998, Vestergaard & Peterson, 2006) is quite frequently used to estimate inter-band reverberation lags in reverberation mapping studies. We used the $g$-band light curve as the model light curve, which was assumed to drive the variations in the other two light curves. The observational gaps were interpolated linearly to create a uniformly sampled dataset, and the driving and responding light curves were sampled together, taking the effect of uncertainties into account. These interpolated light curves were cross-correlated to calculate the Cross-Correlation Function (CCF). This step was repeated multiple times to build the Cross-correlation Centroid distribution (CCCD) and Cross-correlation Peak distribution (CCPD). The CCCD has been used frequently as indicative of the interband lag (e.g., see, Vestergaard & Peterson, 2006, Homayouni et al., 2019). The interpolation frequency was set to 0.5 days as the ZTF light curves are quite well sampled in the $g$ and $r$ bands. The lag uncertainties were measured by employing the flux randomization (FR) and the random subset selection (RSS) sampling method over 5000 iterations with a significance of $r_{max} \leq 0.5$. We used a PYTHON implementation of this method, known as PyCCF developed by Sun et al. (2018).

The $g - r$ lags were shorter than the $g - i$ lags for most of the cases. For instance, in the case of Mrk 335, we obtained a lag of $1.5^{+1.2}_{-1.3}$ days between the $g$ and $r$ bands while we obtained a lag of $2.4^{+1.6}_{-2.0}$ days between the $g$ and $i$ bands. We note that ICCF lags suffer significant uncertainties compared to other methods (Fausnaugh et al., 2017, Li et al., 2019, Yu et al., 2020a, Homayouni et al., 2019) which is likely because the ICCF method linearly interpolates between the data points, rather than assuming an underlying model when there are observational gaps. We estimated the lags using the ICCF method for only the sources for which JAVELIN lags were reasonable. The reason was that since JAVELIN directly models the light curves as compared to the linear interpolation applied in the ICCF method, we expect the lag estimates to be more robust in the case of JAVELIN as has been observed using simulations (Li et al., 2019). However, the lags estimated using the ICCF method serve as a check for the lags estimated through JAVELIN. Figure 2.3 shows the distribution of lags obtained between the $g-r$ and $g-i$ bands for a few of the sources in our sample using both methods. The estimated lags are listed in columns B, C, D, and E of Table 2.3. Out of the sources with lag estimates in the reasonable range, we picked up the sources where at least one method showed positive lags that increase with the wavelength in any of the seasons. This criterion resulted in the final sample of 19 sources which are the sources we use for further analysis (information available in Tables 2.1, 2.2, 2.3). The lags obtained were larger than that predicted by the Standard SS disk model for 14 sources in this sample, while for 5 sources, the lags were within the expectations of the SS disk models.

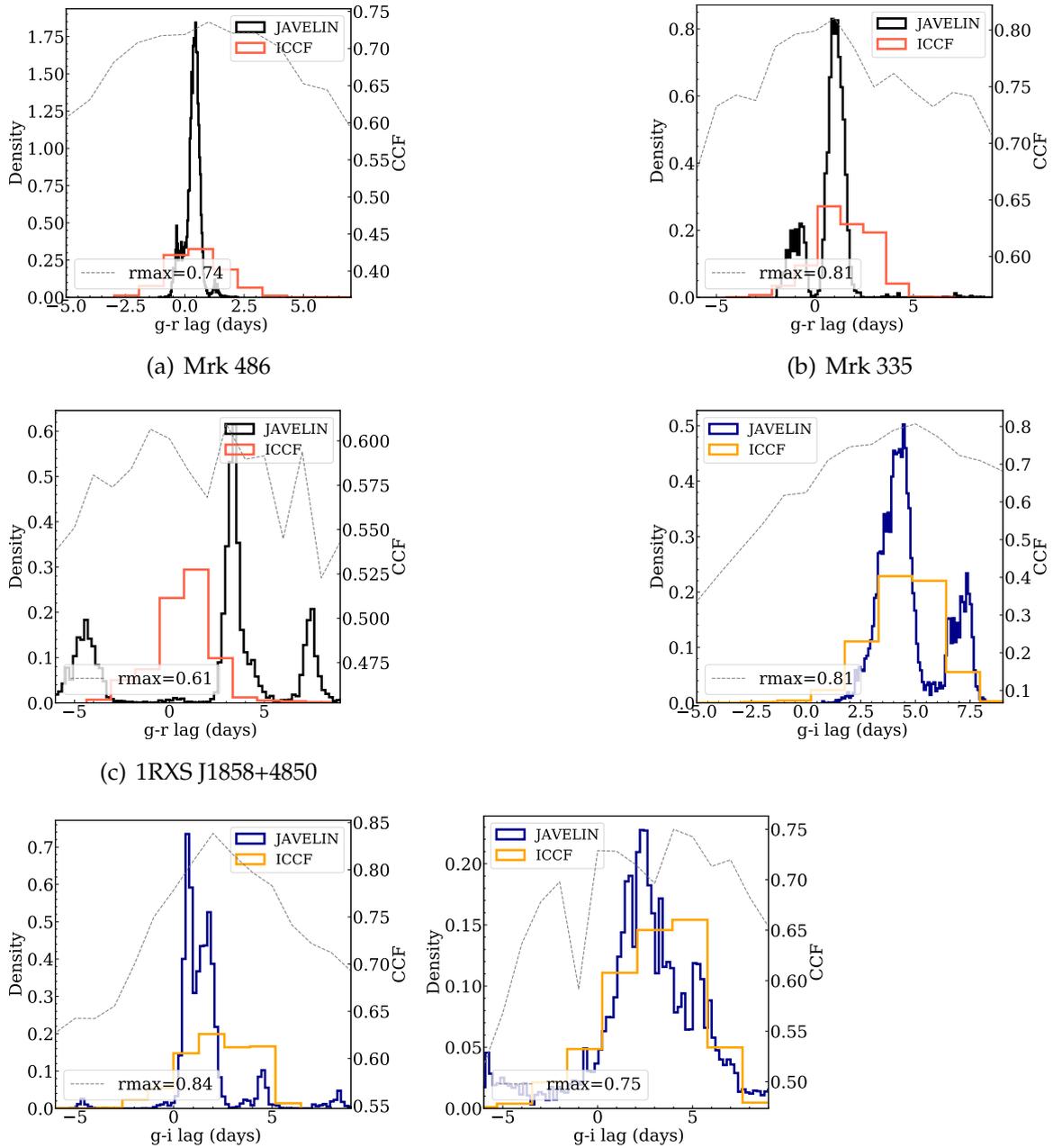

Fig. 2.3 The $g-r$ lag distribution obtained using JAVELIN (black) and ICCF (red) for 3 AGN: Mrk 486 (left), Mrk 335 (middle), and 1RXSJ1858+4850 (right) are shown in the top panel. The $g-i$ lag distribution is shown in the bottom panel with the JAVELIN distribution denoted as blue color and the ICCF distribution denoted as orange color. The CCF obtained using the ICCF method is shown in grey color on the plots. The maximum correlation coefficient obtained using the CCF method is denoted as rmax. Table 2.3 contains the results obtained from the ICCF and the JAVELIN methods.

We note that for 4 AGN in our sample, previous interband continuum lags are available in the literature, which can be directly compared with our lag measurements. For Mrk 335, in Sergeev et al. (2005), the B-R lag is reported to be $2.36^{+0.74}_{-0.85}$ days, and the B-I lag is reported to be $2.33^{+0.30}_{-1.86}$ days. We obtained a $g-r$ band lag of $1.5^{+1.2}_{-1.3}$ days and a $g-i$ band lag of $2.4^{+1.6}_{-2.0}$ days. Even though the filter set in ZTF is a bit different than the filters used by Sergeev et al. (2005), the interband lags between the two can be compared as the two sets of filters cover similar wavelength ranges. Mrk 817 is being monitored as part of the AGN STORM-II campaign (Kara et al., 2021) and based on the initial results, the $g-r$ band lag is $1.50^{+0.86}_{-0.76}$ days and the $g-i$ band lag is $2.31^{+0.70}_{-0.86}$ using the ICCF method, while we obtain a lag of $2.5^{+1.5}_{-1.5}$ days between the $g-r$ band and a lag of $3.6^{+1.1}_{-1.4}$ days between the $g-i$ band using the same methods. These results are in agreement within the limits of the uncertainties. Furthermore, the interband lags between the $g$ and $i$ bands were obtained for RMID 101 and RMID 300 in Homayouni et al. (2019) as $1.54^{+2.06}_{-3.08}$ and $2.93^{+1.06}_{-4.24}$ days respectively using the ICCF method. We obtained a lag of $1.5^{+5.5}_{-5}$ and $2.0^{+6.5}_{-4.5}$ days respectively for these sources using the ICCF method.

Our $g-r$ band lags have relatively smaller uncertainties than $g-i$ band lags, likely due to inadequate sampling in the $i$ band in the ZTF survey. To address the impact of limited cadence and the uncertainties in light curves, first, we took the g-band light curves for Mrk 335, which is at lower redshift and has smaller uncertainties in the light curve, and SDSS RMID101, which is at higher redshift and having relatively larger uncertainties in the light curve. We introduced lags ranging from 0.5 to 3 days with an interval of 0.5 days by shifting the g-band light curve accordingly. We could recover lags up to 1 day for Mrk 335 with typical uncertainties of $\pm$ 0.2. However, in the case of RMID101, where the uncertainty in the light curves is larger, we could recover lags larger than 1.5 days within the uncertainties of about $\pm 0.5$. Next, to check the effect of the length of light curves in the recovery of lags, in order to mimic the lag recovery using i-band light curves, we reduced the number of data points in the responding light curves of Mrk 335 to 15 data points (the minimum number of points used in this study) by randomly sampling the light curve and re-estimated the lags using JAVELIN. In this case, we could recover the lags beyond 1.5 days with minimal deviation, but faced greater uncertainties of $\pm$ 1 day. Based on the above exercise, we note that the uncertainties in the data points and the length of light curves introduces large uncertainties in recovering the inter-band lags.

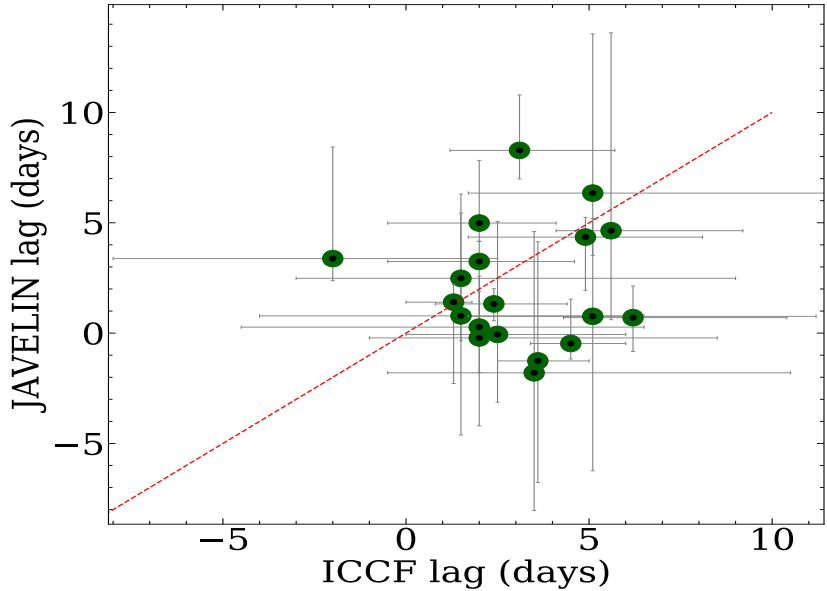

Fig. 2.4 Comparison of the $g - i$ band lags obtained using the JAVELIN and ICCF methods for the 19 sources used in this study. The red dashed line indicates a one-to-one correlation between the lags obtained using both methods.

Another concern is that emission from the BLR emission lines may contribute to the accretion disk continuum. However, the variability timescale of BLR is known to be higher than the timescale for disk emission based on the reverberation mapping measurements of the $H\beta$ emission line. Further, recent works (e.g. Jiang et al., 2017) have concluded that the accretion disk continuum contributes more than 90 % flux to the total flux, and thus emission line contamination may not be a significant contribution to broadband variations. However, the diffuse continuum may affect the lag estimates, especially for sources where the excess lags in the bands nearer to the UV wavelength have been reported (Edelson et al., 2015). In Korista & Goad (2019) the effect of the diffuse continuum (DC) has been found to affect the delays w.r.t the 1158 Å band. In our sample, 4 sources have redshifts where the DC can contribute in

Table 2.1 Physical properties of the sample of reverberation mapped AGNs used in this study.

| Name (A) | RA (B) | Dec. (C) | z (D) | $\log(M_{BH}/M_\odot)$ (E) | $\log(L_{5100})$ (F) |
| --- | --- | --- | --- | --- | --- |
| Mrk 335 | 00 06 19.44 | +20 12 07.2 | 0.026 | 7.23 | 43.76 |
| PG 0026+129 | 00 29 13.44 | +13 16 01.2 | 0.142 | 8.48 | 44.97 |
| PG 0052+251 | 00 54 52.08 | +25 25 37.2 | 0.154 | 8.46 | 44.81 |
| NGC 4253 | 12 18 26.40 | +29 48 43.2 | 0.013 | 6.82 | 42.57 |
| PG 1307+085 | 13 09 46.80 | +08 19 48.0 | 0.155 | 8.53 | 44.85 |
| RMID733* | 14 07 59.04 | +53 47 56.4 | 0.455 | 8.20 | 43.40 |
| RMID399* | 14 10 31.20 | +52 15 32.4 | 0.608 | 8.10 | 44.10 |
| RMID101* | 14 12 14.16 | +53 25 44.4 | 0.458 | 7.90 | 43.40 |
| PG 1411+442 | 14 13 48.24 | +44 00 10.8 | 0.089 | 8.54 | 44.56 |
| RMID779* | 14 19 23.28 | +54 22 01.2 | 0.152 | 7.40 | 42.60 |
| RMID300* | 14 19 41.04 | +53 36 46.8 | 0.646 | 8.20 | 44.00 |
| Mrk 817 | 14 36 22.08 | +58 47 38.4 | 0.031 | 7.58 | 43.74 |
| Mrk 290 | 15 35 52.08 | +57 54 07.2 | 0.029 | 7.27 | 43.17 |
| Mrk 486 | 15 36 38.16 | +54 33 32.4 | 0.039 | 7.63 | 43.69 |
| Mrk 493 | 15 59 09.36 | +35 01 44.4 | 0.031 | 6.19 | 43.11 |
| PG 1613+658 | 16 13 57.12 | +65 43 08.4 | 0.129 | 8.33 | 44.77 |
| PG 1617+175 | 16 20 11.28 | +17 24 25.2 | 0.112 | 8.66 | 44.39 |
| 1RXS J1858+4850 | 18 58 00.96 | +48 50 20.4 | 0.079 | 6.70 | 43.65 |
| PG 2130+099 | 21 32 27.60 | +10 08 16.8 | 0.063 | 7.43 | 44.20 |

**Note**: Column (A) denotes the common name of the sources, Columns (B) and (C) present the right ascension (RA) and declination (Dec.) of the sources as obtained from the literature. Column (D) is the redshift value. Column (E) denotes the SMBH mass in the units of $M_\odot$, and Column (F) is the luminosity at 5100Å. The information for all the parameters is obtained from the AGN Black Hole Mass database (Bentz & Katz, 2015) except for the sources marked with ∗ which are taken from Grier et al. (2017).

the g-band. However, in Fausnaugh et al. (2016), detailed simulations have yielded that the interband continuum lags can be biased by a factor of 0.6−1.2 days due to BLR contamination. Based on this, we conclude that the uncertainties encountered in the lag estimates for these sources may account for this offset.

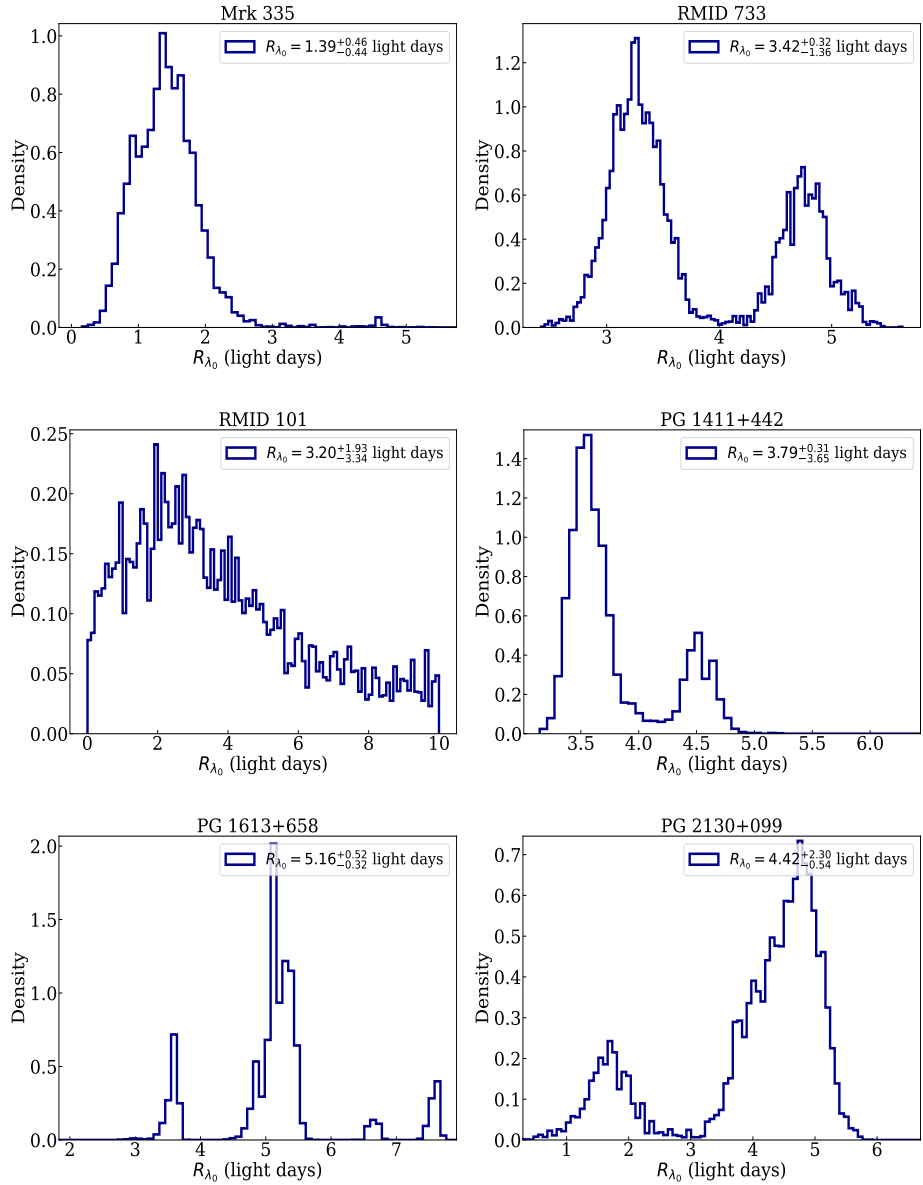

Fig. 2.5 Distribution for the disk size at $g$-band rest wavelength ($R_{\lambda_0}$) for 6 sources from our sample are shown here. This Distribution for $R_{\lambda_0}$ is obtained using the JAVELIN thin disk model assuming disk size scaling with temperature as a power-law index of 4/3. For most of the sources, the estimated $R_{\lambda_0}$ is larger than the predictions of the SS disk model (see columns F and G of Table 2.3).

Table 2.2 Light curve statistics for the sample of 19 AGN with reasonable lag estimates.

| Name | $g$-band | | $r$-band | | $i$-band | |
|---|---|---|---|---|---|---|
| | nobs | Mag | nobs | Mag | nobs | Mag |
| (A) | (B) | (C) | (D) | (E) | (F) | (G) |
| Mrk 335 (S1) | 48 | 14.82 ±0.01 | 52 | 14.28 ±0.01 | 27 | 14.34 ±0.01 |
| Mrk 335 (S2) | 72 | 14.80 ±0.01 | 67 | 14.25 ±0.01 | 24 | 14.27 ±0.01 |
| PG0026+129 (S2) | 70 | 15.31 ±0.01 | 120 | 15.11 ±0.01 | 19 | 14.68 ±0.01 |
| PG0052+251 (S1) | 43 | 16.21 ±0.01 | 52 | 16.02 ±0.01 | 24 | 15.36 ±0.01 |
| PG0052+251 (S2) | 68 | 16.09 ±0.01 | 67 | 15.93 ±0.01 | 18 | 15.31 ±0.01 |
| NGC4253 (S1) | 304 | 15.07 ±0.01 | 337 | 14.46 ±0.01 | 45 | 14.32 ±0.02 |
| NGC4253 (S2) | 168 | 15.17 ±0.01 | 143 | 14.51 ±0.01 | 29 | 14.35 ±0.02 |
| PG1307+085 (S1) | 49 | 16.00 ±0.01 | 102 | 15.83 ±0.01 | 20 | 15.35 ±0.01 |
| PG1307+085 (S2) | 63 | 15.70 ±0.01 | 72 | 15.59 ±0.01 | 19 | 15.21 ±0.01 |
| RMID733 (S1) | 231 | 18.34 ±0.04 | 331 | 17.85 ±0.03 | 25 | 17.46 ±0.03 |
| RMID399 (S1) | 113 | 20.77 ±0.19 | 237 | 20.37 ±0.15 | 19 | 19.76 ±0.13 |
| RMID101 (S1) | 227 | 18.84 ±0.06 | 325 | 18.81 ±0.06 | 23 | 18.72 ±0.06 |
| PG1411+442 (S1) | 452 | 14.98 ±0.01 | 447 | 14.75 ±0.01 | 41 | 14.66 ±0.01 |
| PG1411+442 (S2) | 196 | 14.97 ±0.02 | 211 | 14.75 ±0.01 | 32 | 14.65 ±0.02 |
| RMID779 (S1) | 340 | 19.69 ±0.14 | 321 | 19.26 ±0.09 | 24 | 18.83 ±0.07 |
| RMID300 (S1) | 392 | 19.24 ±0.09 | 318 | 19.20 ±0.07 | 21 | 19.23 ±0.09 |
| Mrk 817 (S1) | 227 | 14.31 ±0.01 | 246 | 14.08 ±0.01 | 25 | 14.15 ±0.01 |
| Mrk 817 (S2) | 93 | 14.36 ±0.01 | 102 | 14.11 ±0.01 | 16 | 14.16 ±0.01 |
| Mrk 290 (S1) | 375 | 15.10 ±0.01 | 412 | 14.72 ±0.01 | 28 | 14.78 ±0.01 |
| Mrk 486 (S1) | 377 | 14.78 ±0.01 | 432 | 14.28 ±0.01 | 27 | 14.32 ±0.01 |
| Mrk 486 (S2) | 144 | 14.80 ±0.01 | 188 | 14.28 ±0.01 | 21 | 14.31 ±0.01 |
| Mrk 493 (S1) | 681 | 15.77 ±0.01 | 623 | 15.33 ±0.01 | 47 | 15.22 ±0.01 |
| Mrk 493 (S2) | 297 | 15.77 ±0.01 | 236 | 15.31 ±0.01 | 49 | 15.18 ±0.01 |
| PG1613+658 (S1) | 644 | 15.08 ±0.02 | 615 | 15.07 ±0.01 | 48 | 14.51 ±0.01 |
| PG1613+658 (S2) | 335 | 15.45 ±0.02 | 318 | 15.38 ±0.01 | 53 | 14.74 ±0.01 |
| PG 1617+175 (S2) | 40 | 15.49 ±0.01 | 78 | 15.49 ±0.01 | 25 | 15.01 ±0.01 |
| 1RXS J1858+4850 (S1) | 452 | 16.57 ±0.01 | 728 | 16.22 ±0.01 | 28 | 16.27 ±0.01 |
| 1RXSJ1858+4850 (S2) | 288 | 16.66 ±0.01 | 276 | 16.32 ±0.01 | 28 | 16.36 ±0.01 |
| PG 2130+099 (S1) | 63 | 14.54 ±0.01 | 68 | 14.27 ±0.01 | 26 | 14.38 ±0.01 |
| PG2130+099 (S2) | 41 | 14.65 ±0.01 | 37 | 14.37 ±0.01 | 25 | 14.51 ±0.01 |

**Note**: Column (A) denotes the common name of the sources as available in Bentz & Katz (2015) and Grier et al. (2017). S1 and S2 denote the light curves for the two seasons used in this study. Columns (B), (D), and (F) are the number of observations (nobs) in the $g$, $r$, and $i$ filters, respectively, as available in ZTF-DR6. The mean magnitudes and their uncertainties in the $g$, $r$, and $i$ filters (Mag) throughout the observations are presented in columns (C), (E), and (G).

Table 2.3 The lags and the disk size estimates along with the SS disk prediction for 19 sources in our sample.

| Common name | $\tau_r$ (JAVELIN) (days) | $\tau_i$ (JAVELIN) (days) | $\tau_r$ (ICCF) (days) | $\tau_i$ (ICCF) (days) | $R_0$ ($g$-band) (light-days) | $R_0$ (SS disk) (light-days) |
|---|---|---|---|---|---|---|
| (A) | (B) | (C) | (D) | (E) | (F) | (G) |
| Mrk 335 (S1) | $0.94^{+0.45}_{-1.60}$ | $1.32^{+0.76}_{-0.68}$ | $1.5^{+1.2}_{-1.3}$ | $2.4^{+1.6}_{-2.0}$ | $1.39^{+0.46}_{-0.44}$ | 1.33 |
| PG 0026+129 (S2) | $3.30^{+3.70}_{-7.62}$ | $2.48^{+2.82}_{-2.95}$ | $0.6^{+0.9}_{-0.2}$ | $2.0^{+2.5}_{-2.6}$ | $6.29^{+1.24}_{-0.87}$ | 2.81 |
| PG 0052+251 (S2) | $2.88^{+1.75}_{-2.37}$ | $-1.79^{+6.24}_{-6.40}$ | $2.5^{+2.5}_{-0.0}$ | $3.5^{+4.0}_{-7.0}$ | $2.79^{+1.98}_{-3.26}$ | 2.44 |
| NGC 4253 (S1) | $-0.01^{+2.06}_{-1.98}$ | $-0.22^{+3.98}_{-2.79}$ | $0.5^{+3.9}_{-2.0}$ | $-2.0^{+3.0}_{-6.5}$ | $5.98^{+5.89}_{-2.09}$ | 0.55 |
| PG 1307+085 (S2) | $0.10^{+7.27}_{-0.46}$ | $-0.05^{+3.07}_{-5.12}$ | $3.0^{+2.6}_{-2.6}$ | $1.5^{+4.5}_{-7.5}$ | $9.16^{+4.96}_{-0.62}$ | 2.52 |
| RMID733 (S1) | $0.92^{+0.47}_{-2.47}$ | $3.37^{+1.00}_{-5.05}$ | $-0.5^{+1.9}_{-2.5}$ | $-2.0^{+6.0}_{-4.5}$ | $3.42^{+0.32}_{-1.36}$ | 0.56 |
| RMID399 (S1) | $2.62^{+5.12}_{-5.59}$ | $4.64^{+4.03}_{-8.97}$ | $6.8^{+2.2}_{-10.3}$ | $5.6^{+1.5}_{-3.6}$ | $5.91^{+3.51}_{-3.11}$ | 0.80 |
| RMID101 (S1) | $-2.95^{+5.13}_{-5.28}$ | $0.78^{+5.39}_{-5.52}$ | $3.3^{+4.1}_{-6.2}$ | $1.5^{+5.5}_{-5.0}$ | $3.20^{+1.93}_{-3.34}$ | 0.56 |
| PG 1411+442 (S2) | $-3.67^{+0.67}_{-0.82}$ | $-0.46^{+0.70}_{-2.00}$ | $1.5^{+7.0}_{-6.4}$ | $2.5^{+2.5}_{-3.5}$ | $3.79^{+0.31}_{-3.65}$ | 2.22 |
| RMID779 (S1) | $4.56^{+3.90}_{-3.04}$ | $6.34^{+2.83}_{-7.21}$ | $9.0^{+0.0}_{-2.8}$ | $5.1^{+3.4}_{-9.1}$ | $6.21^{+2.91}_{-2.82}$ | 0.46 |
| RMID300 (S1) | $3.77^{+1.62}_{-6.41}$ | $0.75^{+7.00}_{-4.43}$ | $2.0^{+2.6}_{-3.0}$ | $2.0^{+6.5}_{-4.5}$ | $7.84^{+3.39}_{-1.56}$ | 0.71 |
| Mrk 817 (S1) | $4.31^{+0.21}_{-1.99}$ | $4.98^{+0.82}_{-2.89}$ | $2.5^{+1.5}_{-1.5}$ | $3.6^{+1.1}_{-1.4}$ | $4.70^{+0.18}_{-3.83}$ | 1.30 |
| Mrk 290 (S1) | $0.84^{+3.28}_{-1.64}$ | $0.27^{+2.06}_{-1.62}$ | $4.1^{+1.9}_{-5.9}$ | $2.0^{+2.5}_{-2.1}$ | $5.14^{+2.05}_{-2.62}$ | 0.84 |
| Mrk 486 (S2) | $0.38^{+0.21}_{-0.36}$ | $4.35^{+2.41}_{-0.89}$ | $0.5^{+1.0}_{-1.0}$ | $4.5^{+1.1}_{-1.5}$ | $0.93^{+0.11}_{-2.05}$ | 1.23 |
| Mrk 493 (S2) | $0.68^{+1.68}_{-0.29}$ | $0.70^{+1.53}_{-1.43}$ | $2.0^{+0.6}_{-0.5}$ | $4.9^{+3.2}_{-3.2}$ | $5.52^{+1.10}_{-0.98}$ | 0.81 |
| PG 1613+658 (S1) | $-0.38^{+5.86}_{-0.46}$ | $-1.26^{+5.51}_{-5.39}$ | $2.0^{+1.9}_{-2.6}$ | $5.1^{+3.4}_{-6.1}$ | $5.16^{+0.52}_{-0.32}$ | 2.46 |
| PG 1617+175 (S2) | $3.41^{+2.04}_{-1.98}$ | $8.28^{+1.29}_{-2.51}$ | $5.2^{+2.9}_{-1.4}$ | $6.2^{+1.9}_{-4.2}$ | $8.14^{+1.75}_{-1.31}$ | 1.89 |
| 1RXS J1858+4850 (S2) | $0.37^{+0.13}_{-0.91}$ | $1.40^{+3.68}_{-0.81}$ | $1.0^{+1.1}_{-1.5}$ | $3.1^{+1.9}_{-2.6}$ | $2.74^{+1.90}_{-0.25}$ | 1.13 |
| PG 2130+099 (S2) | $2.67^{+0.73}_{-3.05}$ | $3.24^{+3.06}_{-0.90}$ | $1.2^{+0.6}_{-0.5}$ | $1.3^{+1.3}_{-0.5}$ | $4.42^{+2.30}_{-0.54}$ | 1.76 |

**Note**: Column (A) denotes the common name of the sources as available in Bentz & Katz (2015) and Grier et al. (2017). S1 and S2 denote the respective season in which the interband lags increasing with wavelength were obtained using either of the methods. The $g-r$ and $g-i$ reverberation lags obtained using JAVELIN are shown in Columns (B) and (C), while the lags obtained using the ICCF method for $g-r$ and $g-i$ are presented in columns (D) and (E) respectively. The disk sizes obtained using the JAVELIN thin disk model are presented in column (F) and the expected theoretical disk sizes at $g-$band effective wavelength assuming the SS disk model are presented in column (G).

## 2.4 Accretion disk sizes

### 2.4.1 Disk sizes using JAVELIN thin disk model

A recent extension in the JAVELIN code known as the *thin disk model* introduced by Mudd et al. (2018) allows us to directly model and estimate the size of the accretion disk at a particular wavelength based on the multi-band light curves. It fits the light curves assuming the SS disk approximation, and the accretion disk size is given according to equation 2.2.

This method was used by Mudd et al. (2018) to obtain the accretion disk size at 2500Å for a set of 15 quasars from the DES survey. Further, Kokubo (2018) and Yu et al. (2020a) have also used this model to derive the accretion disk sizes for different quasars. We used the thin disk model for the light curves to obtain the disk sizes at $g$-band rest wavelength and compare it to the predicted SS disk model based on individual SMBH mass and accretion rates. It is noticeable that the disk sizes predicted for JAVELIN assume the SS disk as we have fixed $\beta$ to 1.33 (4/3). Leaving the parameter $\beta$ free does not converge to any particular value, as noticed in the recent works (Kokubo, 2018, Yu et al., 2020a). However, a possible concern is that this model could be unable to recover the accurate parameters for AGN accreting at a faster rate which may require additional physical processes or special geometry to be explained (Castelló-Mor et al., 2017).

The distribution for $R_{\lambda_0}$ gives the size of the accretion disk at the respective wavelength $\lambda_0$. Since the sources in our sample were located at varying redshifts, the corresponding wavelength ($\lambda_0$) for the disk size varied from 2717 Å to 4677 Å according to the $g$-band rest-frame wavelength. The values for $R_{\lambda_0}$ ranged from 0.93 light days to 9.16 light days with a mean value of 4.88 light days. Since the sources for this analysis have large SMBH mass and luminosity ranges, it is plausible that the estimated disk sizes vary from source to source. The distribution for $R_{\lambda_0}$ for a few sources is shown in Figure 2.5.

### 2.4.2 Comparison with analytical SS disk model

The standard SS disk model predicts the disk sizes based on the SMBH mass, and the mass accretion rate. We compare the disk sizes obtained using this analytical model with the disk sizes obtained by fitting the light curves through the JAVELIN thin disk model. Following Jiang et al. (2017), the light travel time across two different radii where photons with wavelengths $\lambda_g$ and $\lambda_x$ are emitted, can be put as:

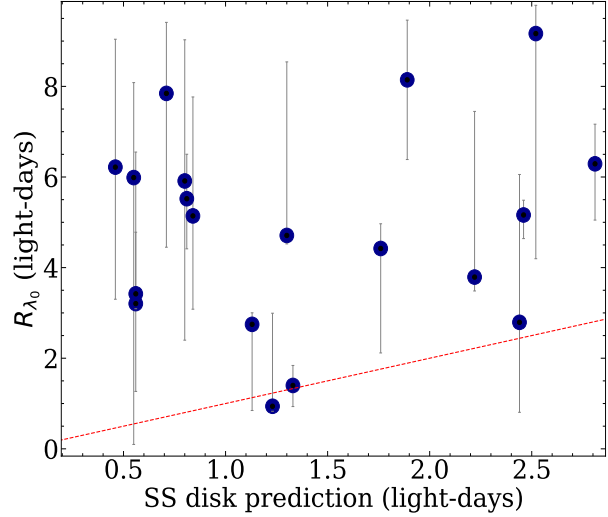

Fig. 2.6 Comparison of the disk sizes ($R_{\lambda_0}$) at $g$-band rest wavelength. The estimation for $R_{\lambda_0}$ has been made using the JAVELIN thin disk model, while the analytical disk size has been calculated using the SS disk model dependent on the SMBH mass and accretion rates. The red dashed line indicates a one-to-one correlation between the two estimates.

$$\Delta t_{g-x} = \left(X\frac{k_\text{B}\lambda_g}{hc}\right)^{4/3} \left(f_i\frac{3GM_{BH}\dot{M}}{8\pi\sigma}\right)^{1/3} \left[\left(\frac{\lambda_x}{\lambda_g}\right)^{4/3} - 1\right] \quad (2.4)$$

Where, $k_B$, h, c, G and $\sigma$ are the Boltzmann constant, Planck constant, the speed of light, the gravitational constant and Stefan's constant respectively, $M_{BH}$ is the SMBH mass, $\dot{M}$ is the mass accretion rate, and X is a scaling factor. The value of X has been chosen to be 2.49 as per the recent works (see Fausnaugh et al., 2016, Edelson et al., 2019).

The disk size estimated by modeling the light curves through the JAVELIN thin disk model should agree with the lags if the SS disk assumption is held valid. Figure 2.6 shows the comparison of the disk sizes obtained using the JAVELIN thin disk model and the prediction of the SS disk model. For 16 sources in our sample, the obtained disk sizes are larger than the theoretical disk size estimates, which has been observed in previous works as well (see Shappee et al., 2014, Fausnaugh et al., 2016, Starkey et al., 2016a, Edelson et al., 2019, and references therein). However, for 3 sources, the disk sizes agree with the SS disk predictions. For instance, our estimated disk size of $1.39^{+0.46}_{-0.44}$ days at 4400 Å for Mrk 335 is in close agreement with the disk size of 1.33 light days predicted based on the standard SS disk at the same wavelength.

We calculated the lag spectrum based on the predicted SS disk model, the obtained thin disk sizes by modeling the light curves using the JAVELIN thin disk model, and the $g-r$ and $g-i$ lags obtained using the JAVELIN and ICCF methods. We had only 2 interband lag measurements, while we used the combination of $R_{\lambda_0}$ obtained using the JAVELIN thin disk model and the equation 2.2 to compare the accretion disk sizes at different wavelengths. We find out that most sources have disk sizes and interband lags larger than the SS disk prediction. The results are shown in Figure 2.7.

### 2.4.3 Scaling of Disk size with physical parameters

We used the disk sizes obtained using the JAVELIN thin disk model to study the relation of disk sizes with the physical parameters, namely the SMBH mass and the luminosity at 5100 Å. According to predictions of the SS disk model, the disk size scales with the SMBH mass as $M^{2/3}$ (see Morgan et al., 2010). The RM-based SMBH mass estimates have been the most robust so far in the absence of any other reliable techniques for SMBH, covering a significant range of redshifts and luminosities. In Figure 2.8, we show the relation between the SMBH masses and accretion disk sizes obtained using the JAVELIN thin disk model. We include the previous microlensing-based results from Morgan et al. (2010). We converted the units of the accretion disk size to light days and converted the reference wavelength at 2500Å. We also included the results from Mudd et al. (2018) and Yu et al. (2020a) in this relation. We also plot the expected relation between the SMBH mass and the accretion disk size based on the dimensionless accretion rates of 0.01, 0.1, and 1. The dimensionless accretion rate is $\dot{M}/\dot{M}_{\text{EDD}}$, where $\dot{M}$ is the mass accretion rate defined as $L_{\text{BOL}}/\eta c^2$ and $\dot{M}_{\text{EDD}}$ is the Eddington mass accretion rate defined as $L_{\text{EDD}}/c^2$ and $L_{\text{EDD}} = 1.45 \times 10^{38} \times \frac{M_{BH}}{M\odot} erg/s$. $L_{\text{BOL}}$ is the bolometric luminosity calculated as roughly 9 times the luminosity at 5100Å (Kaspi et al., 2000a). The radiative efficiency is denoted as $\eta$, which we set as 0.1 based on recent works.

We find a weak correlation between the obtained disk sizes and the SMBH masses, with the Spearman rank correlation coefficient being 0.38. Although Morgan et al. (2010) found a relation linking the SMBH mass and disk sizes obtained using microlensing, their sample covered the higher end of SMBH masses only, while our sample covers a range of log ($M_{BH}/M_\odot$) ranging from 6.19 to 8.66 which is larger than the range covered in that work. Although the correlation is not very strong, the general trend of AGN with higher SMBH mass tends to have larger accretion disk sizes. Both the disk size and SMBH mass estimates suffer large uncertainties, and thus, proper monitoring and robust estimates of the accretion disk size will be very helpful in establishing this relation further.

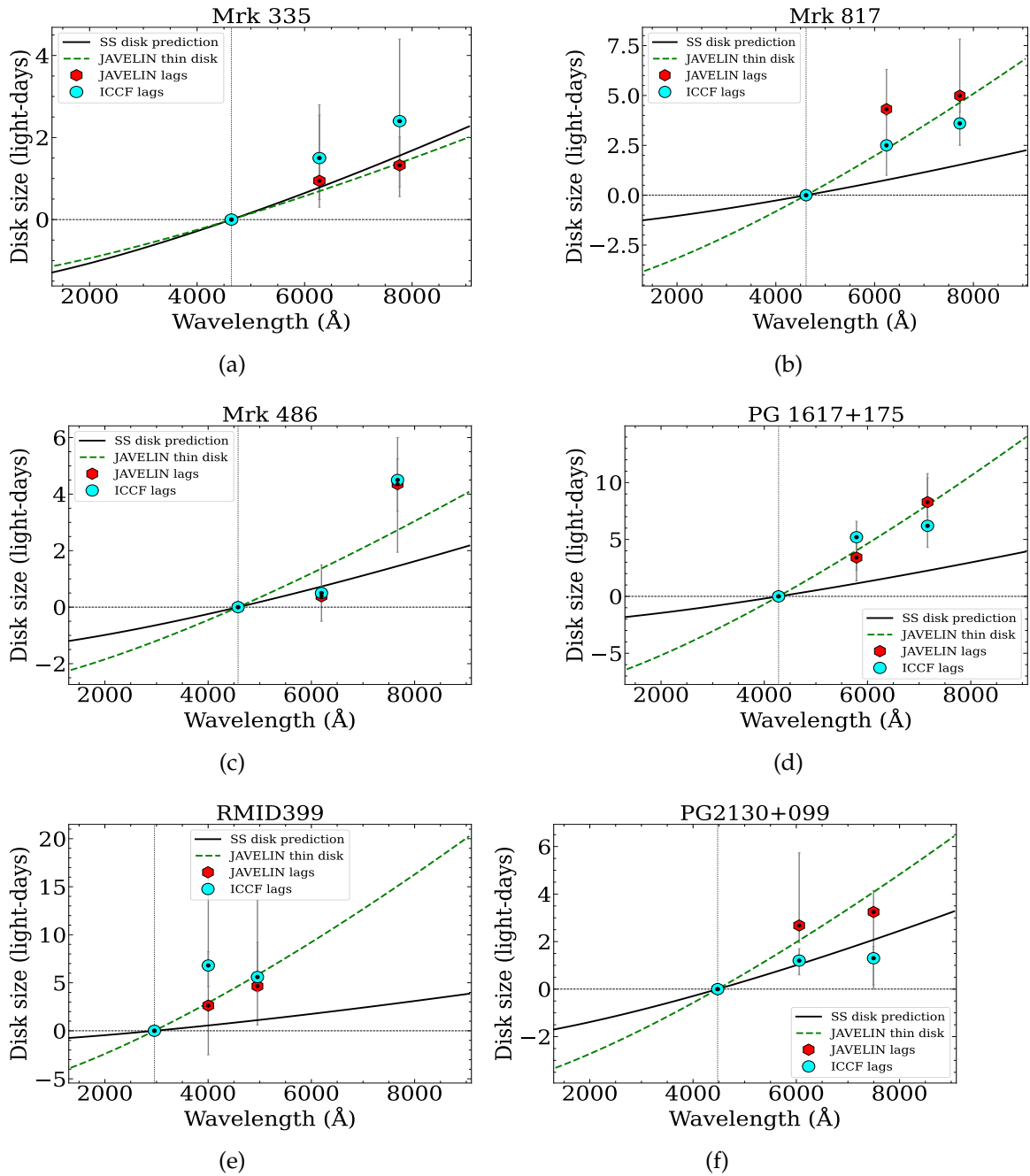

Fig. 2.7 The SS disk prediction (black line) and the disk sizes as obtained through the JAVELIN thin disk model (green dashed line) for 6 sources in our sample are shown here. Also, the $g-r$ and $g-i$ lags estimated using JAVELIN and ICCF are shown in red and cyan colors, respectively. The results for all the sources with lag estimates constrained between $[-10, 10]$ days are made available as supplementary material online.

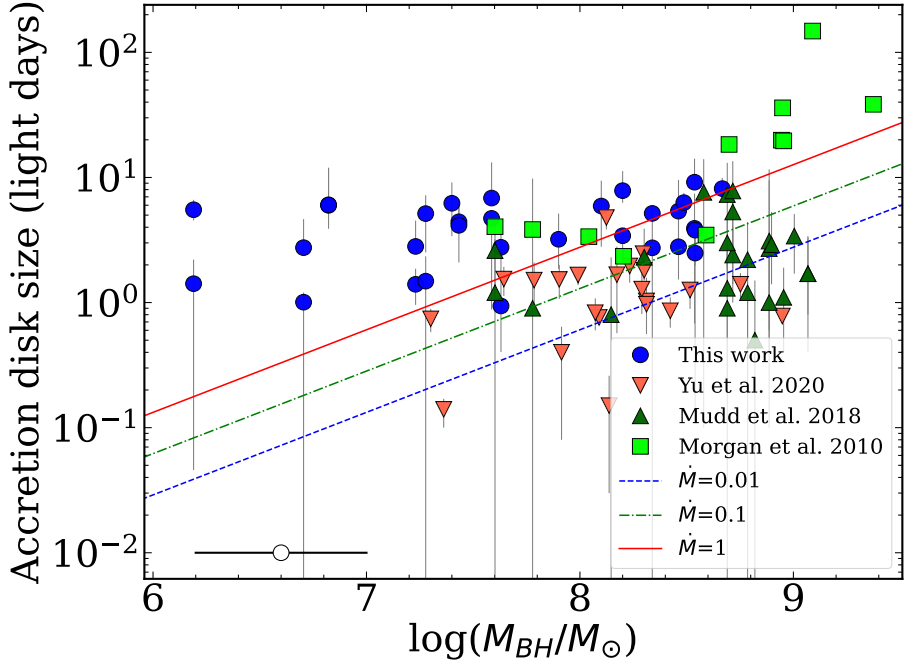

Fig. 2.8 The scaling of the disk sizes with the SMBH masses are shown here. The disk size used here is the effective size of the accretion disk at the wavelength of 2500 Å in the units of light days. The sample used in this study is denoted as blue circles, the sources from Yu et al. (2020a) are shown as red inverted triangles, sources from Mudd et al. (2018) are shown as green triangles, and the sources taken from Morgan et al. (2010) are shown as green squares. The uncertainty in SMBH mass estimate is approximately 0.4 dex, which is shown as a black line on the bottom left. The predictions for disk size based on the dimensionless mass accretion rates of 0.01 (blue dashed line), 0.1 (green dotted line), and 1 (solid red line) are also shown in the figure.

Further, in Figure 2.9, we show the relation between the accretion disk sizes and the luminosity at 5100Å. We expect that the disk sizes should be related to intrinsic luminosity as more luminous quasars are expected to have large accretion disks and hence larges inter-band lags, based on the light travel times (Jiang et al., 2017). In Sergeev et al. (2005), the interband lags are found to scale with the luminosity as $\tau = L^{0.4-0.5}$. However, we find a very weak correlation between the luminosities and the disk size, with a Spearman rank correlation coefficient of 0.18 and a null hypothesis ($p_{null}$) value of 0.31.

## 2.5   Discussion

To constrain the size of the accretion disk for a sample of quasars with known SMBH masses, we estimated the inter-band reverberation lags between the optical $g$, $r$, and $i$ bands observations from the ZTF survey. The reverberation lags represent the light travel time across the two different regions of the disk. We found that the interband light curves are correlated, and for 19 sources, the lags increasing with wavelength were recovered successfully using the JAVELIN and ICCF methods. The $g-r$ inter-band lags were shorter than the $g-i$ inter-band lags for these sources, which is consistent with the disk reprocessing *lamppost* model, according to which the photons arising from the innermost regions are reprocessed in the form of emission from the outer regions resulting in a lag. In multi-wavelength monitoring campaigns involving X-ray, UV, and optical wavebands, such as the AGN STORM campaign (Starkey et al., 2016a), inter-band lags have been reported in all the wavelengths, with longer wavelengths lagging the short wavelengths. Further, even using only optical band observations, inter-band lags have been successfully recovered for a large sample of AGN (see Jiang et al., 2017, Mudd et al., 2018, Homayouni et al., 2019, Yu et al., 2020a). Our results complement these important observations.

    We find out that the size of the accretion disk obtained at a reference wavelength is larger than predicted by the SS accretion disk model for a majority of the sources (see Figure 2.7). We note that the interband lags for 5 sources, namely PG0026+129, NGC 4253, PG1411+442, RMID300, and PG2130+099, follow the predictions of the SS disk model, while 14 sources have lags of about 3 to 4 times larger than the expectations from the SS disk model. However, we could not zero in on a parameter that distinguished these sources from the sources where the disk sizes are much larger than the SS disk assumption. In previous works, the disk size has been known to scale with the wavelength as a power law of index 4/3, but the sizes have been reported to be larger than predicted by the SS disk (Fausnaugh et al., 2016, Jiang et al., 2017, Fausnaugh et al., 2018, and references therein). Contrarily, Mudd et al. (2018) reported that the disk sizes for their sample of 15 sources did not differ much from the SS disk analytical model. Interestingly the disk sizes obtained through microlensing have also reached similar conclusions (Pooley et al., 2007, Blackburne et al., 2011, Mosquera et al., 2013). Larger accretion disk sizes obtained through both the continuum reverberation mapping and the microlensing methods for a majority of the sources may indicate that the standard SS disk assumption does not hold for the AGN in general and additional components may be needed while modeling the accretion disks in AGN. We note that some of the sources in our sample have

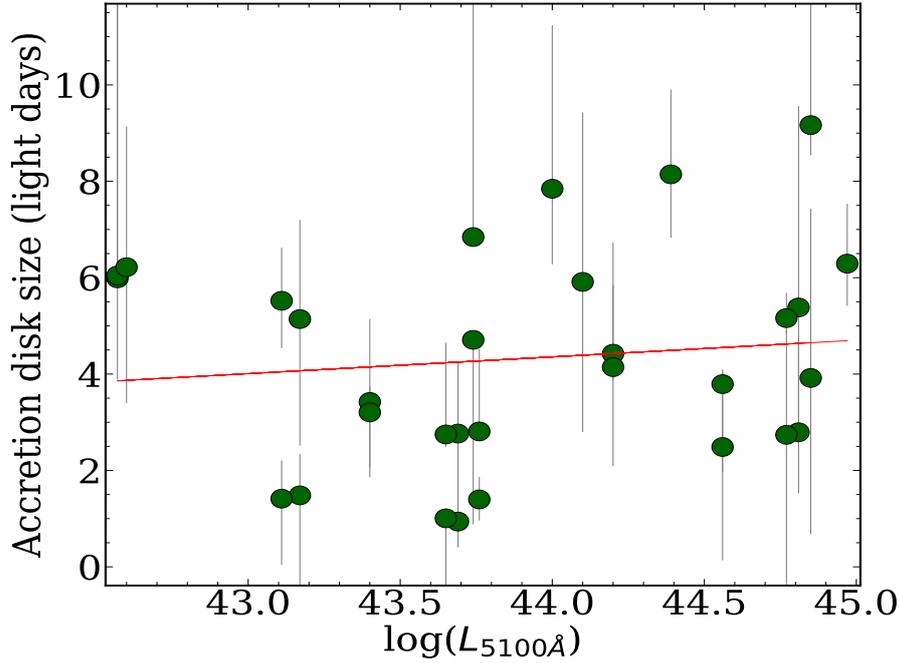

Fig. 2.9 The scaling of the disk sizes corresponding to the $g$-band rest wavelength with the luminosity at 5100 Å for 19 AGN from this sample. The red line denotes the fitted straight line.

higher accretion rates (Du et al., 2015). It is possible that the accretion mechanism in the AGN with higher accretion rates is different and thus, the disk sizes obtained assuming the SS disk model may not hold true for these kinds of AGN. Evidence of non-disk components in the optical continuum for Mrk 279 has been reported by Chelouche et al. (2019). This might be a possibility for the larger-than-expected lags for some of the sources in this sample too. However, the lag measurements suffer large uncertainties in the sources where this effect might take place. Also, the exact contribution from such a component can be achieved through either a combination of narrowband filters or simultaneous spectra, which has not been possible in our case. Another possible explanation for the larger disk sizes has been proposed by Gaskell (2017) where the reddening effects could lead to larger than expected accretion disk sizes. However, Nuñez et al. (2019) find out that for Mrk 509, this effect is not able to explain the larger disk sizes.

Continuum reverberation mapping gives more robust estimates with multi-wavelength campaigns, as we see in the case of NGC5548 from the AGN STORM campaign. Nevertheless, with optical data points only, we have been able to constrain the disk sizes for a large number of objects. While the exact structure of the AGN accretion disks remains unknown so far, there is overwhelming evidence in favor of disk reprocessing based on confirmed inter-band lags, increasing with the wavelength (McHardy et al., 2014, Fausnaugh et al., 2018). Our results for a set of reverberation-mapped AGN differing in SMBH mass and luminosities provide another set of observational evidence for testing the accretion mechanism in a wide range of such objects. While we find a weak correlation between the obtained disk size, the SMBH mass, and the luminosity, this might be due to the limit of observations as we cannot cover a significant wavelength range using the ZTF dataset. Another reason behind the weak correlation could be the wide range of SMBH masses and luminosities covered in the sample. In the future, with the availability of surveys such as the Vera C. Rubin Observatory Legacy Survey of Space and Time (LSST) (Ivezić et al., 2019), it will be possible to cover a larger sample with a much better cadence and hence will be helpful in better constraining these correlations.

## 2.6 Conclusions

We obtained the accretion disk size measurements of AGN with previous SMBH mass estimates through reverberation mapping, with a wide range of luminosities, redshifts, and SMBH masses, based on $g$, $r$, and $i$ band observations from the ZTF survey. The high cadence observations from ZTF provide an excellent dataset to constrain interband reverberation lags efficiently. The primary conclusions from this work are as follows:

1. The $g$, $r$, and $i$ band light curves are correlated, and the $g - r$ inter-band lags are shorter than $g - i$ inter-band lags for 19 sources in our sample, which provide strong evidence in favor of the disk reprocessing.

2. The interband lags and the disk sizes obtained using the JAVELIN thin disk model are larger than the ones predicted by the standard SS thin disk model for a majority of sources, which is consistent with the recent findings and raises the question of the usage of the simple SS disk model for AGN accretion disks.

3. For 5 sources, we obtained the interband lags in agreement with the SS disk predictions. However, no significant parameter distinguishes these AGN from the ones where we found larger disk sizes.

4. There is a weak correlation found in the SMBH mass versus disk size and the luminosity versus disk size, which may be due to the uncertainties in both the accretion disk size measurements and the SMBH mass measurements, which have been known to suffer uncertainties up to 0.4 dex.

Although mapping the entire disk profile is only possible with multi-wavelength campaigns such as the AGN-STORM campaign of NGC 5548 (Edelson et al., 2015), public surveys as the ZTF provide us the opportunity to use light curves for a large number of sources and obtain inter-band lags to understand the accretion mechanisms responsible for the interband lags, albeit with a smaller wavelength coverage. This work can be extended with larger AGN samples and derive the accretion mechanisms powering them to understand the AGN accretion disks better.

# Chapter 3

# Accretion Disk sizes for Super Eddington AGN using the Growth India Telescope (GIT)[1]

In this chapter, we present the initial results of our study of the structure of accretion disks around active galactic nuclei (AGN) using multiple-band reverberation mapping, a technique that measures the time lag between variations in the continuum emission at different wavelengths, which depends on the temperature profile of the disk. By observing the disk continuously and simultaneously in multiple wavelength ranges, we can probe both the hotter inner regions and the cooler outer regions of the disk and determine its structure and temperature profile. Our study focuses on a sample of AGN that have super-Eddington accretion rates and show smaller broad-line region (BLR) sizes than expected from the standard R-L relation. We use the Growth India telescope (GIT) to observe these AGN with Super High Accretion Rates and perform a broadband analysis on one of the sources, IRAS 04416+1215. Our analysis reveals that the size of the accretion disk for this source, calculated by cross-correlating the continuum light curves, is larger than expected from the theoretical model. To further investigate this discrepancy, we fit the light curves directly using the thin disk model available in JAVELIN and find that the disk sizes are approximately four times larger than expected from the Shakura Sunyaev (SS) disk model. We are continuing our observations for a sample of Super Eddington AGN to better understand their accretion disks.

---

## 3.1 Introduction

Active galactic nuclei (AGN) are among the most luminous and energetic objects in the universe. They are powered by the accretion of matter onto supermassive black holes at the centres of galaxies (Salpeter, 1964). The accretion process occurs in a disk-like structure known as an accretion disk. These disks are composed of gas and dust that orbit the black hole, gradually spiralling inward and releasing energy in the form of radiation. Understanding the properties of these disks, including their size, is crucial for understanding the behaviour of AGN and their role in the evolution of galaxies (Kormendy & Ho, 2013).

Reverberation mapping is a powerful technique for studying the inner structure of AGN (Bahcall et al., 1972, Blandford & McKee, 1982, Cackett et al., 2021). It involves measuring the time delay between variations in the continuum emission from different regions. This time delay provides a measure of the continuum emitting region. By analyzing these time delays for multiple AGNs, it is possible to measure the mass of the black hole residing at the centre (Dalla Bontà et al., 2020). Reverberation mapping measurements have also yielded a relation between the luminosity of the AGN and the size of the Broad Line Region (BLR), the so-called R-L relation (Kaspi et al., 2000a, Bentz et al., 2009, Du et al., 2016). Accretion disk reverberation mapping, which used the multi wavelength light curves to measure the lags between the various regions of the accretion disk, has been used to infer the disk sizes for a variety of objects, (Edelson et al., 2015, Starkey et al., 2016b, Hernández Santisteban et al., 2020, Kara et al., 2021, etc.). These studies measured the inter-band lags from X-ray to the optical IR wavelengths to generate a complete profile of the accretion disk. However, only a handful of objects have such intensive measurements. Ground-based studies have also proven successful in constraining the accretion disk sizes for about a hundred AGN, although covering a smaller wavelength range (Jiang et al., 2017, Mudd et al., 2018, Homayouni et al., 2019, Jha et al., 2022b, Guo et al., 2022b).

The theoretical SS disk model has been widely used to describe the accretion disks (Shakura & Sunyaev, 1973). However, results from observations have yielded that the size of the accretion disk in AGN is larger than the expectations from this model (Starkey et al., 2016b). This implies that either the SS disk model does not hold for these objects or additional complexities are involved. The disk sizes being larger than predicted by standard models also imply that some key ingredients in the classical thin disk theory may need to be added.

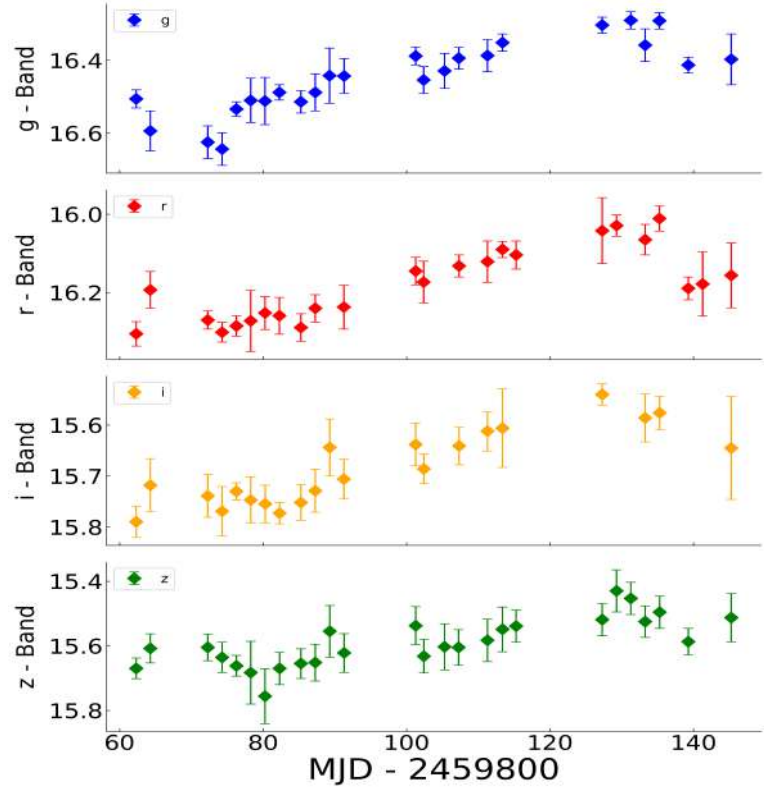

Fig. 3.1 The g, r, i, and z band light curves for IRAS 04416+1215 were obtained using the Growth India Telescope (GIT).

An interesting subset of AGN are the objects accreting at super Eddington rates (see Wang et al., 2013). These objects have been observed to be accreting at many times the Eddington Accretion rates. Through reverberation mapping campaigns, the BLR sizes for these objects are significantly smaller than the empirical RL relation observed in other AGN studies (Du et al., 2016). How their accretion disk sizes scale with respect to the other AGNs remains to be seen. In order to study the accretion disk structure in these objects, we are carrying out a disk reverberation mapping campaign in the optical wavelength.

In this work, we present initial results from the accretion disk reverberation mapping campaign of AGN, which we perform using data from the GROWTH India Telescope (GIT). This paper is structured as follows; Section 3.2 presents the sample being used for this study and the details of the observations. Section 3.3 presents the methods being used for our study, while Section 3.4 presents the initial results for one of the sources. We present the discussion in Section 3.5 followed by conclusions in Section 3.6.

## 3.2 The Sample and Observations

We have compiled a sample of 18 AGNs with high accretion rates obtained from Du et al. (2015). The SMBH masses for these AGNs are well constrained through the reverberation mapping studies. The BLR sizes for these sources are significantly lower than the radius luminosity relation, which has been observed to be very tight for the other AGNs with RM measurements. Whether this peculiar behaviour is seen in the accretion disk measurements of such AGN is unexplored at the moment. To map the accretion disk of these AGNs, we have started a large program titled: 'Investigating the central parsec regions around supermassive black holes' (INTERVAL). In this campaign, we are performing multi-band (u, g, r, i, and z) monitoring of AGNs using a 70cm GROWTH India telescope (GIT) and 50cm telescope at the Indian Astronomical Observatory (IAO) in Hanle, Ladakh.

The 70cm GIT operates in robotic mode through a queue-based observation schedule. Calibration frames are taken every night, and the automatic pipeline for photometry yields the aperture and the Point Spread Function (PSF) based magnitudes for the objects (Kumar et al., 2022). The observations for four of the AGNs in our sample have been completed with GIT while we continue the observations for the remaining sources from our sample. The data is reduced and calibrated using standard procedures to produce light curves for each object in each band using PSF photometry. We then perform cross-correlation analysis to measure the time delay between variations in the continuum emission arising from the accretion disk.

## 3.3 Methods

We aim to estimate the disk sizes in the AGN by measuring the interband lags between the continuum emission from different regions of the accretion disk represented by the u, g, r, i, and z band light curves. We employ two methods to derive the interband lags: JAVELIN and ICCF. We also use the JAVELIN thin disk model developed by Mudd et al. (2018) to fit the light curves directly to a disk model.

JAVELIN is a method that models the AGN variability as a Damped Random Walk (DRW) and employs a Bayesian approach to infer the posterior distribution of the lags and their uncertainties (Zu et al., 2011). This method has been demonstrated to be accurate and reliable as compared to the other methods being used (Li et al., 2019). ICCF is another method that computes the cross-correlation function of the light curves and identifies the peak of the function as the lag (Peterson et al., 1998).

Table 3.1 The light curve statistics for IRAS 04416+1215 being used for this study.

| Bands | Date range (MJD) | Points | Median PSF Magnitude | Error |
|---|---|---|---|---|
| g | 2459835 - 2459945 | 24 | 16.44 | 0.04 |
| r | 2459835 - 2459945 | 24 | 16.18 | 0.04 |
| i | 2459835 - 2459945 | 21 | 15.71 | 0.04 |
| z | 2459835 - 2459945 | 26 | 15.60 | 0.05 |

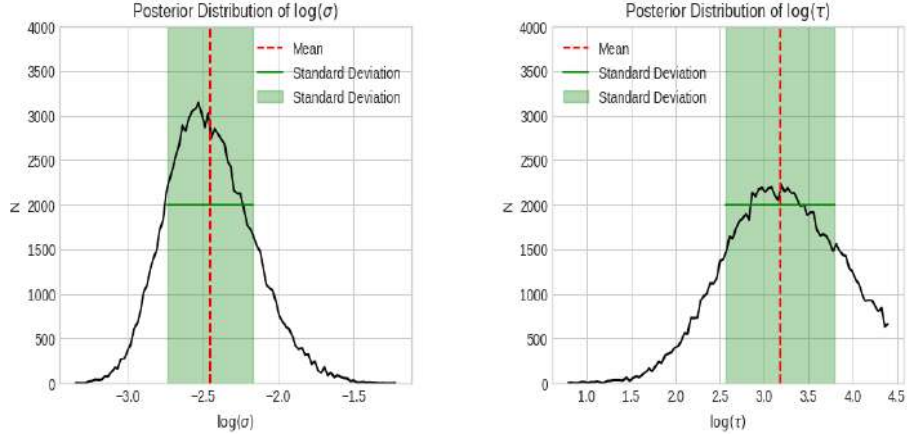

Fig. 3.2 Posterior distribution for the logarithms of the amplitude of variability ($\sigma$) and the damping timescale ($\tau$) obtained for the driving continuum light curve, which we assume to be the g-band light curve.

This method has been extensively applied in reverberation mapping studies and performs well when the data quality is high (Sun et al., 2018).

We compare the outcomes of both methods to assess their consistency and robustness. We also conduct various tests to evaluate the validity of our measurements and respective errors. We run Markov Chain Monte Carlo (MCMC) iterations in JAVELIN with the parameters: nwalkers 1000, nburn=500, and nchain=1000. These parameters are sufficient, and increasing them does not affect the results. Similarly, for the ICCF method, we run 5000 iterations of Random Subset Sampling and flux Randomization in ICCF. This enables us to obtain the lags between the light curves and their associated uncertainties.

## 3.4 Results:

The observations for the first 4 sources in our sample, namely IRAS 4416+1215, Mrk 382, Mrk 42 and Mrk 1044, are complete, and in this work, we report on the initial results obtained for IRAS 04416+1215. It is located at a redshift of 0.0889 and categorized as a hyper Eddington source (Tortosa et al., 2022). For this source, the observations are available for a total of 27-30 points in each band. The average photometric error is about 0.04 magnitudes in all the bands. We reject the outlier points whose magnitude is greater or less than 2 magnitudes from the mean magnitude. This reduces about 2-4 data points in each band. We use the PSF light curves for our analysis of this source. We note that the source is not prominent in the u band. Hence we use only the 4 band light curves −, namely the g, r, i, and z bands, for our analysis. The object appeared to be variable in all 4 bands (see Figure 3.1). The inter-band lags are measured with respect to the shortest wavelength available, which happens to be the g-band.

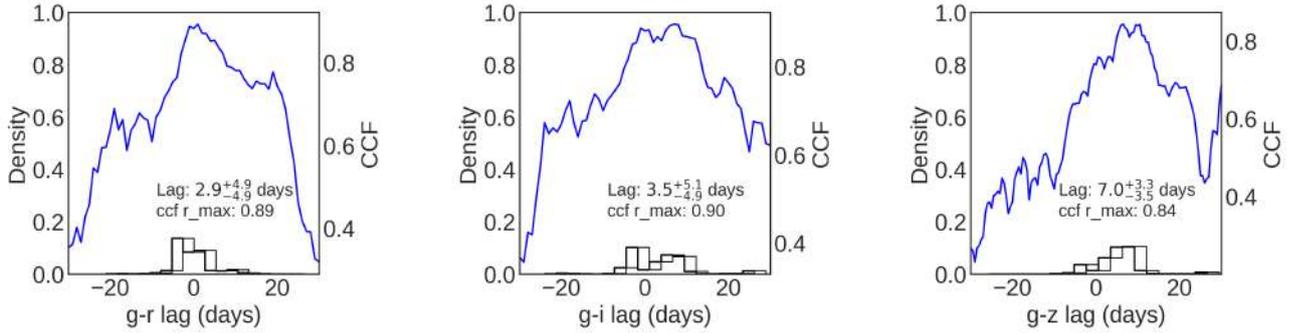

Fig. 3.3 The Cross Correlation Posterior Distribution (CCPD) along with the correlation coefficient for IRAS 04416+1215 for the g-r, g-i, and g-z bands are shown here.

Table 3.2 Estimated lags using JAVELIN and ICCF (in days) for IRAS 04416+1215.

| Bands | JAVELIN lags | ICCF Lags |
|---|---|---|
| g - r | $0.09^{+1.05}_{-1.07}$ | $2.9^{+4.9}_{-4.9}$ |
| g - i | $6.93^{+3.69}_{-9.11}$ | $3.5^{+5.1}_{-4.9}$ |
| g - z | $6.26^{+5.23}_{-5.78}$ | $7.0^{+3.3}_{-3.5}$ |

First, we estimate the lags using JAVELIN (Zu et al., 2011). It models the driving light curve as a DRW process, building a posterior distribution for the variability amplitude ($\sigma$) and the damping timescale ($\tau$) (see Figure 3.2). It then shifts, scales, and smooths the light curve to generate a responding light curve and builds a distribution of the lag between the two light curves. We obtained a lag of $0.09^{+1.05}_{-1.07}$ days between the g-r bands, $6.93^{+3.69}_{-9.11}$ days between the g-i bands and $6.26^{+5.23}_{-5.78}$ days between the g-z bands. Noticeably, the g-r band lag is larger than the g-i band lags, although the uncertainties are significantly higher at the 1 $\sigma$ level.

We also implemented the ICCF method, a widely used method for the lag measurements (Peterson et al., 1998). It does not assume an underlying model, but rather it interpolates between the gaps and then cross-correlates the individual points to generate a distribution for the cross-correlation function. The uncertainties are estimated using flux randomization (FR) and random subset selection (RSS). We use the centroid of the CCF to estimate the lag. We obtain a lag of $2.9^{+4.9}_{-4.9}$ days for g-r band, $3.5^{+5.1}_{-4.9}$ days for g-i band and $7.0^{+3.3}_{-3.5}$ days for g-z bands (see Figure 3.3). The increasing lags with the increase in wavelength are seen using the ICCF method. We find that the obtained lags through both methods are slightly different with higher uncertainties, especially with the measurement using JAVELIN.

We notice that the uncertainties in the lag estimates are quite large using both JAVELIN and the ICCF methods. One of the primary reasons could be the cadence of the light curves. The median cadence that we achieve is about 4-5 days, which makes it difficult to recover shorter lags. For other sources in our sample, we aim to get denser sampling so as to recover lags with better precision.

Taking the advantage of our multi-band data set, we fit the light curves using the JAVELIN thin disk model. The g-band light curve is assumed to be the driving light curve, while the r, i, and z-band light curves are responding light curves. The disk model gives us the disk size at the driving wavelength, which is the g-band rest wavelength. We fixed the $\beta$ parameter to 1.33 while calculating the disk size, which implies a SS disk. The disk size at the g-band rest wavelength is estimated to be $2.96^{+3.28}_{-2.03}$ light days.

We also calculate the disk size $R$ size at the g-band rest wavelength $\lambda_0$, using the equation in Mudd et al. (2018), based on the SS disk model as:

$$R_{\lambda 0} = 9.7 \times 10^{15} \left(\frac{\lambda_0}{\mu m}\right)^{\beta} \left(\frac{M}{10^9 M_\odot}\right)^{2/3} \left(\frac{L}{\eta L_E}\right)^{1/3} \text{ cm} \qquad (3.1)$$

Where $\lambda_0$ is the wavelength at which the disk size is estimated, $\beta = 4/3$, M is the SMBH mass in units of Solar mass ($M_\odot$), and L is the luminosity obtained from (Du et al., 2015), $L_E$ is the Eddington luminosity. We take the efficiency parameter $\eta$ as 0.1, as has been used in previous works.

We find that the disk size obtained using the JAVELIN thin disk model is 4 times larger than what we would obtain based on the theoretical prediction of the SS disk as given in equation 3.1 . If we plot the thin disk model, we find that the calculated lags lie near to the curve generated by this model, implying that the disk is following the 4/3 scaling relation, but the size of the disk is larger than the prediction for the SS disk (see Figure 3.4).

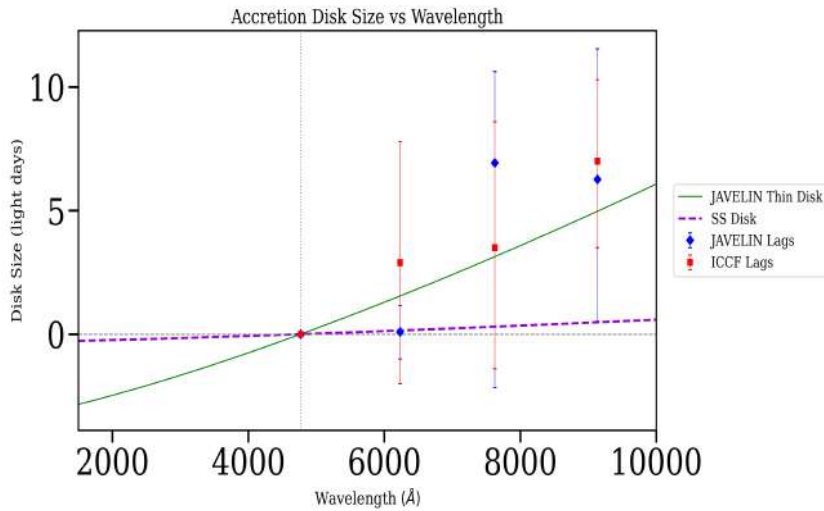

Fig. 3.4 The relation between the accretion disk size and the wavelength for IRAS 04416+1215. We use the g-band rest wavelength as the reference wavelength for estimating the disk sizes. The green solid line shows the extrapolation on the JAVELIN thin disk model. The blue points show the lags obtained using JAVELIN, while the red points show the lags obtained using ICCF. The dashed purple line shows the standard SS disk model prediction for this source.

## 3.5 Discussion

In this study, we present initial results from the 'INTERVAL' campaign measuring the accretion disk sizes for SEAMBH AGN using the GIT. For one of the sources, we measure the inter-band time delays using both the JAVELIN and ICCF methods. Our analysis reveals that the size of the accretion disks for this object is larger than expected from the theoretical SS disk model. Our findings are consistent with recent studies that have found that AGN accretion disk sizes are generally larger than predicted by standard models (Starkey et al., 2016b, Fausnaugh et al., 2018, Edelson et al., 2019). Recent studies using optical data from the Zwicky Transient Facility have also found that the accretion disk sizes for AGN are multiple times larger than predicted by the standard SS disk model (Jha et al., 2022b, Guo et al., 2022a,b).

To address this discrepancy, it has been proposed that the larger-than-expected accretion disk sizes found in AGN studies are primarily due to the neglect of reddening (Gaskell, 2017). They estimated that neglecting internal extinction leads to an underestimate of the luminosity at 1200Å by a factor of seven, and therefore the size scale of the accretion disk has been underestimated by a factor of about 2.6. This is similar to the accretion disk size discrepancy found in other studies and supports the proposal that internal reddening plays a significant role in explaining the discrepancy. We aim to estimate the reddening for our sources, to test whether this phenomenon affects the lag estimates. The contribution of the Diffuse Continuum (DC) to the continuum light curves has also been observed (see Chelouche et al., 2019), which can have a direct consequence of increasing the interband continuum lags on account of the contribution from the BLR.

Another possibility is that the accretion disks in AGN are not geometrically flat but instead have a slim (Narayan & Yi, 1995) or even a clumpy structure, which would not agree well with the SS disk model. Analytical modelling of the accretion disk will be very helpful in resolving this discrepancy. The results suggest that there are some key ingredients in the classical thin disk theory that need to be added. Further studies are needed to investigate this possibility and to better understand the structure and physics of AGN accretion disks.

While the BLR sizes for SEAMBH AGN are estimated to be lower than estimated from the R-L relation, we also want to explore whether this phenomenon is true for the accretion disk sizes. However, our result for one of the SEAMBH sources, namely IRAS 04416+1215, shows that the disk size is about 4 times larger than the SS prediction, a phenomenon observed for the other AGNs as well. We continue to monitor these AGNs in order to better understand their accretion disk structure.

## 3.6 Conclusions

We present results from our accretion disk reverberation mapping campaign targeting AGN with Super High Accretion Rates. Initial results for one of the sources, IRAS 04416+1215, identified as a hyper Eddington accreting source, are presented. Here are the conclusions from this work:

1. The interband continuum lags analysis for IRAS 04416+1215 reveals that the size of the accretion disk is larger than expected from the standard SS disk model.

2. We fit the light curves directly using the thin disk model available in the JAVELIN package, which reveals the disk sizes to be about 4 times larger than the prediction of the theoretical SS disk model.

3. Our findings are consistent with recent studies that have found that AGN accretion disk sizes are generally larger than predicted by the SS disk model. These results suggest that there can be some key ingredients in the classical thin disk theory that need to be added.

4. For further understanding, we continue to monitor the sample of SEAMBH AGN in order to get the accretion disk size estimates based on continuum reverberation mapping. Such studies are needed to understand better the structure and physics of AGN accretion disks and their role in the evolution of galaxies.

# Chapter 4

# Calibration of the Photometric Reverberation Mapping Technique[1]

Photometric reverberation mapping is a promising tool to understand the structure of BLR and obtain AGN Super Massive Black Hole (SMBH) masses with relatively small 1-2m class telescopes. Monitoring of low luminosity AGN with shorter expected lags can be completed within an observation cycle lasting 3-4 months. Thus PRM has the potential to yield BLR sizes for a large sample of AGN with relatively lesser telescope time. In this chapter, we present the calibration of the technique of PRM with regard to the spectroscopic method by using archival data obtained from the Lick AGN monitoring project. We re-estimate the reverberation lags using the photometric technique and compare them with the spectroscopic lags to see whether the results agree pretty well or not. We develop simulations to study the effect of campaign length, cadence, and intrinsic variability amplitude, and seeing conditions on the recovery of actual reverberation lags to find the optimal cadence and campaign duration to execute a successful PRM campaign. We also look for a minimum fractional variability amplitude to recover the lags with minimal deviation. The simulations also aim to demonstrate how the presence of a prominent host galaxy contribution can affect the variability trends in the long-term light curves, which will be used for cross-correlation and lag estimates.

---

[1]*under preparation*

## 4.1 Introduction

The details of the inner regions of AGN are not understood because they project to an angular resolution of only micro arcseconds across, which would be un-resolvable with any present or near future observational facilities. Reverberation Mapping (RM) (Bahcall et al., 1972, Blandford & McKee, 1982, Peterson & Horne, 2004) remains the primary method to probe these inner regions of the Active Galactic Nuclei( AGN). RM can be used to get the size of the accretion disk structure (see Chapter 2), the BLR, and the dusty torus. The campaigns to estimate the size and structure of the BLR are lengthy, targeted, and quite resource-consuming (Netzer & Peterson, 1997). Due to these factors, RM-based BLR sizes are available for only about 120 sources so far (Panda et al., 2019a). The fundamental goal of all reverberation mapping campaigns is to obtain the lag between the emission line and the continuum, which is the first moment of the transfer function. Basically, $R = c\tau$, where R is the size of the BLR and $\tau$ is the estimated *lag* usually of the order of a few days, although a reverberation lag as low as 3.5 hours for a low luminosity AGN, NGC 4395 has been obtained Chelouche & Daniel (2012). The size of the BLR can be used to estimate the mass of the SMBH using a virial relation (Du et al., 2015). An empirical relation linking the BLR's size to the AGN's luminosity, the so-called R-L relation, has been obtained (Kaspi et al., 2000b, Bentz et al., 2009). However, the validity of this relation in the low luminosity AGN population remains to be established.

The spectrum obtained for RM is both resource and time-consuming; thus, it is an option to obtain both the continuum and emission line flux using a combination of broad and narrow-band filters. This technique, known as Photometric Reverberation Mapping (PRM), uses one band to trace the continuum and another narrow band to trace the emission line but needs much less telescope time than the spectral RM and can also be achieved with smaller 1-2m class telescopes. PRM has been demonstrated as a successful alternative (Haas et al., 2011, Chelouche & Daniel, 2012). A handful of sources have been studied using PRM, and the lags obtained are in good agreement with the spectroscopic lags. A campaign for 3C120 has been reported by Pozo Nuñez et al. (2012), and NGC 4395 has been monitored using broadband filters by Edri et al. (2012). Nuñez et al. (2017) present an ongoing systematic PRM performed using narrow band filters at the WISE observatory. Zhang et al. (2017) have published results using PRM for quasars selected from the SDSS-Stripe 82 catalog. PRM is a relatively new technique, and a large systematic project which covers a large number of sources has not been undertaken. Monitoring sources through PRM has been restricted to single-object campaigns most of the time.

There have not been many efforts to carry out PRM experiments due to the unavailability of large telescopes in our country, and only recently with the installation of telescopes such as the 1.3m[2] and 3.6m[3] telescopes at the Devasthal Observatory in ARIES, such campaigns have been deemed feasible. The ARIES Devasthal Optical Telescope Photometric Reverberation Mapping (DOT-PRM) (see Chand et al., 2018) project aims to utilize the concept of PRM and intends to study the nearby low luminosity AGN having luminosities $L < 10^{43}$ erg/s. This regime has not been explored thoroughly in past campaigns. As part of this project, 8 narrow band filters with their central wavelengths falling at redshifted $H\alpha$ and $H\beta$ wavelengths have been procured. A sample of carefully selected AGN has been chosen with redshift such that they fall in the range of the filters (see Figure 4.1).

The upcoming VRO/LSST will monitor the large sky areas in 6 broad bands with a photometric cadence of 3-4 days (MacLeod et al., 2010), which brings us a unique opportunity to obtain the BLR sizes for hundreds of AGNs/quasars with PRM (Haas et al., 2011). Thus, it is imperative to understand, improve and modify the technique of PRM. The fundamental purpose of this project is to proceed in this direction and prepare the tools for photometric RM based on the tests with photometric and spectroscopic data of AGN. In the first step, we calibrate the lag measurements obtained using PRM with the traditional RM based on spectroscopic observations. Further, we develop simulations to study the effect of campaign length, observational cadence, variability amplitudes, and the effect of seeing conditions on the recovery of actual lags, which will help derive the optimal strategy for successful large-scale PRM campaigns in the near future.

This chapter is structured as follows: In section 4.2, we describe the data obtained for calibrating the PRM technique. In section 4.3 we calibrate the method of PRM by estimating the spectroscopic lags between $H\beta$ emission line and V band flux signifying the continuum and then by re-estimating the lags using the photometric technique. In section 4.4 we present the simulations developed for this project followed by a summary and the conclusions in 4.5.

---

[2]https://aries.res.in/facilities/astronomical-telescopes/130cm-telescope
[3]https://aries.res.in/facilities/astronomical-telescopes/360cm-telescope

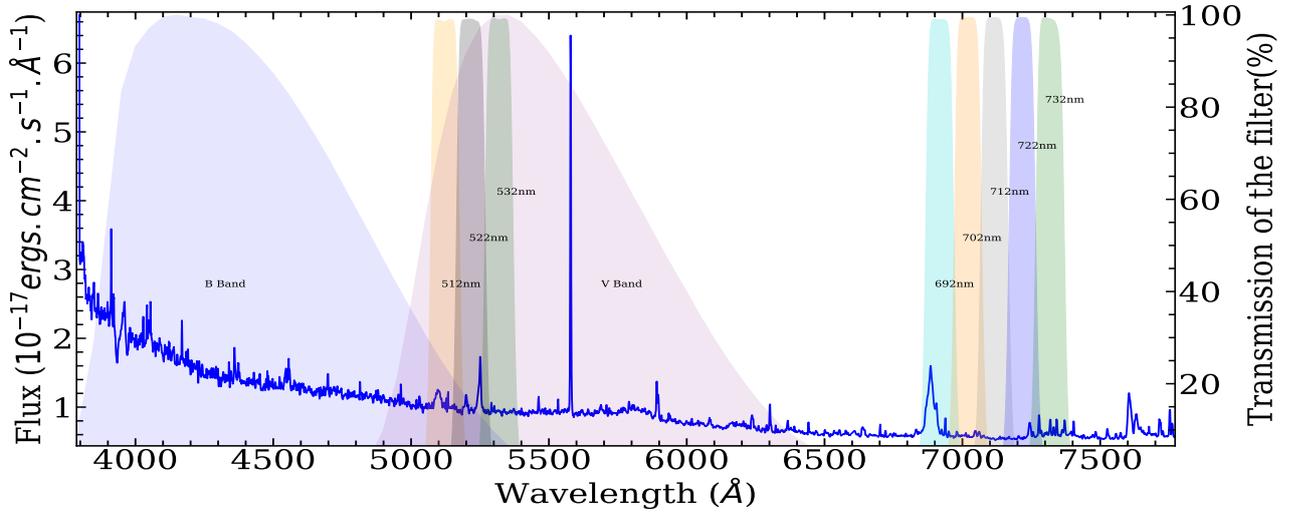

Fig. 4.1 The transmission curve of the 8 narrow band filters along with their central wavelengths and Johnson's B and V broadband filters overplotted on the spectrum of a type 1 AGN from our sample. The narrow band filters cover the redshifted H$\beta$ and the H$\alpha$ emission lines. (spectrum courtesy: SDSS)

## 4.2 Data

To calibrate the method of PRM, we obtained data from the LICK AGN monitoring project (LAMP) 2008 [4] in the form of multi-epoch spectra (see Table 4.1). The LAMP campaign targeted 13 low redshifted ($0.005 < z < 0.044$), low luminous Seyfert 1 galaxies. The expected SMBH masses were between $10^6$ and $10^7 M_\odot$, and the expected $H\beta$ reverberation lags were between 5-20 days. The campaign also included the well-studied AGN NGC 5548 by previous RM campaigns. A spectroscopic monitoring campaign of 64 nights was executed to achieve this goal using the 3m Shane telescope at the Lick Observatory. The broadband B and V imaging were supplemented using four smaller telescopes: the 0.76m robotic Katzman Automatic Imaging Telescope (KAIT), the 2m Multicolor Active Galactic Nuclei Monitoring telescope (MAGNUM), the Palomar 60-inch telescope (P60), and the 0.8m Tenagra II telescope. Further details about the dataset are available in Walsh et al. (2009). These B and V band light curves were used as the driving continuum for cross-correlation rather than the flux measurement at 5100 Å, which has commonly been used in spectroscopic RM campaigns. The reason was that higher SNR could be attained using images rather than spectra. Lags for 9 out of 13 were presented in Bentz et al. (2009). We obtained the archived spectra for these sources and extracted the $H\beta$ emission line light curves by decomposing the spectra into multiple components. The continuum light curves were obtained from the observations in the B and V bands. The light curves for these bands are in terms of Vega magnitudes, which were converted into flux units for our analysis.

---

[4] https://www.physics.uci.edu/~barth/lamp.html

Table 4.1 The sample of AGN selected for this analysis from LAMP Dataset.

| S No. | Name | RA (J2000) (hh mm ss) | Dec (J2000) (deg mm ss) | Redshift (z) | $\sigma_{H\beta}$ ($km/sec$) |
|---|---|---|---|---|---|
| 1 | Mrk 142 | 10 25 31.3 | +51 40 35 | 0.04494 | 1183.24 |
| 2 | SBS 1116+583A | 11 18 57.7 | +58 03 24 | 0.02787 | 4405.94 |
| 3 | Arp 151 | 11 25 36.2 | +54 22 57 | 0.02109 | 2098.24 |
| 4 | Mrk 1310 | 12 01 14.3 | -03 40 41 | 0.01941 | 2903.30 |
| 5 | Mrk 202 | 12 17 55.0 | +58 39 35 | 0.02102 | 7171.31 |
| 6 | NGC 4253 | 12 18 26.5 | +29 48 46 | 0.01293 | 1207.91 |
| 7 | NGC 4748 | 12 52 12.4 | -13 24 53 | 0.01463 | 1181.71 |
| 8 | NGC 5548 | 14 17 59.5 | +25 08 12 | 0.01718 | 4746.46 |
| 9 | NGC 6814 | 19 42 40.6 | -10 19 25 | 0.00521 | 2289.15 |

## 4.3 Calibration of the method

Since only a handful of sources have been studied using the photometric reverberation mapping technique, the critical difference between this method and the traditional RM is not yet established. Thus, it is necessary to calibrate this technique with the already available spectroscopic results. From the multi-epoch LAMP-2008 spectra, we chose the $H\beta$ emission line for our study as it is the most widely used emission line for reverberation mapping of the broad line region. For the continuum, we used the Johnsons B and V broadband light curves. We obtained the reverberation lags using the spectroscopic and photometric techniques and compared the two for the calibration of the PRM technique.

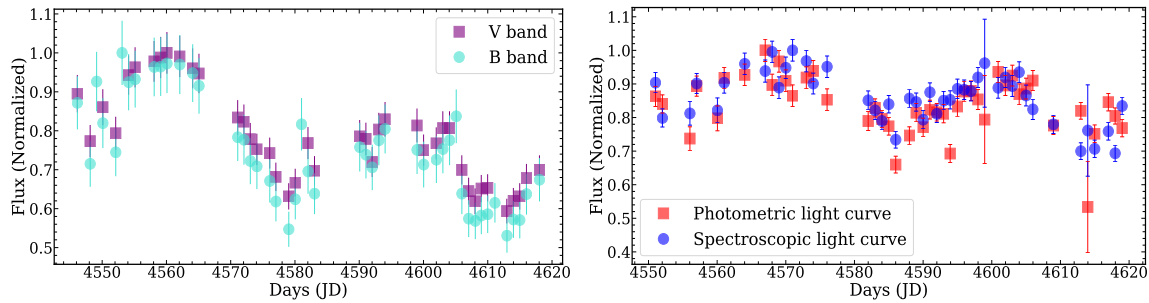

Fig. 4.2 The B and V band continuum (left) and the $H\beta$ emission line (right) light curves obtained for NGC 6814 obtained during the LAMP-2008 campaign. The emission line light curves are initially obtained through the spectroscopic method outlined in section 4.3.1 and the photometric method mentioned in section 4.3.2. The flux values are normalized with respect to the respective maximum flux value to bring two light curves to the same scale.

### 4.3.1 Estimation of spectroscopic lags

We obtained the spectra for all the sources and performed spectral decomposition of the spectra using PyQSOFit [5] (Guo et al., 2018). It simultaneously fits the emission lines as single or multiple Gaussians, the continuum as a power law, and the iron blends using given optical/UV templates (a detailed description of this method is available in Chapter 6, Section 6.3). Based on the galaxy templates, the host galaxy contribution was removed for a defined redshift range. We obtained the fit for $H\beta$ emission profiles and the flux by integrating over the line limits (4600Å- 4920Å). We prepared the $H\beta$ emission line light curves by repeating the process for the spectrum at every epoch for each source. We estimated the reverberation lags by cross-correlating the $H\beta$ line light curve with the B and V band continuum light curves. Among the various methods of lag estimation, we found out that JAVELIN and the ICCF methods are most widely used and that JAVELIN performs better than the ICCF method in the recent works (Yu et al., 2020b, Li et al., 2019, **?**). As a result, we used JAVELIN throughout this work. The lag estimates are in the observers' frame, and the actual values would be reduced due to time dilation effects. We compared the lags obtained using our method with the lags available in Bentz et al. (2009). For example, we obtained a BLR size of $3.43^{+1.11}_{-0.13}$ light days for ARP151, $7.35^{+0.10}_{-0.08}$ light days for NGC6814, $1.74^{+0.66}_{-0.34}$ light days for SBS1116+583A and $4.14^{+0.31}_{-0.53}$ for NGC5548, while published values in Bentz et al. (2009) for the same sources are $3.52^{+0.82}_{-0.72}$, $6.49^{+0.95}_{-0.96}$, $2.24^{+0.65}_{-0.61}$ and $4.24^{+0.1}_{-1.35}$ light days respectively. Note that the lag estimates in Bentz et al. (2009) were made using the ICCF method, while in this work, we have used JAVELIN and there is a good correlation between the two (see Figure 4.3).

---

[5] https://github.com/legolason/PyQSOFit

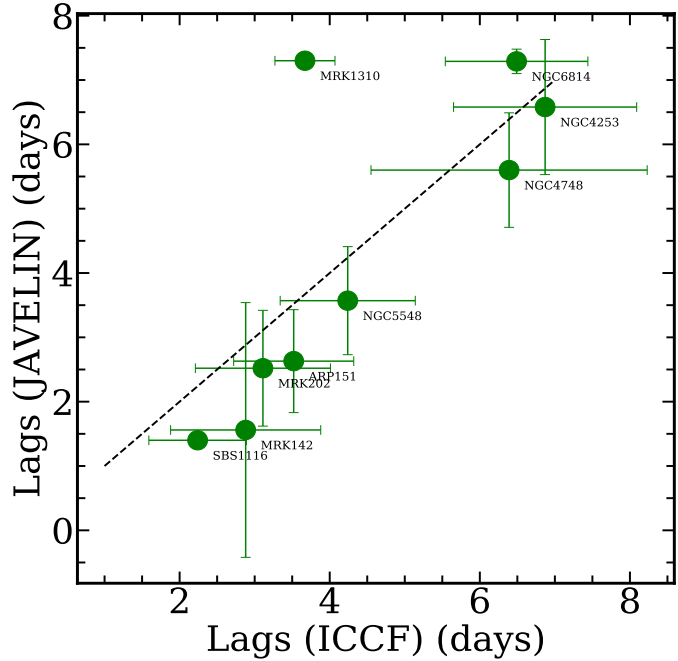

Fig. 4.3 The comparison between the lags obtained through the JAVELIN method and the ones published in Bentz et al. (2009) using the ICCF method. The black dashed line in the middle shows a one-to-one correlation.

### 4.3.2 Estimation of Photometric Lags

In order to understand the complications involved with the Photometric Reverberation mapping technique, we need narrow and broadband photometric light curves. We used the transfer function of the narrow band filters, and the photometric light curves were prepared by convolving this transfer function with the $H\beta$ emission line to get the flux as it would be obtained if we observed through these filters (see Figure 4.1). This resulted in the generation of $H\beta$ photometric light curves from multi-epoch spectra (see Figure 4.2). The continuum light curves were already available in the form of B and V broadband light curves, and we converted them from magnitude units to flux units. The cross-correlation was performed between the continuum and the emission line light curves.

The first complication arising from this method is that the underlying continuum cannot be removed from the narrow-band emission line flux; hence, this contamination can affect the lag estimates. The lags obtained without subtracting the continuum flux from the narrow band light curves represent the autocorrelation of the continuum flux. This is a problem related to PRM, and Haas et al. (2011) overcame it by subtracting the scaled V band light curve from the narrow band $H\beta$ line light curve. Edri et al. (2012) performed broadband photometric RM on NGC 4395. They simultaneously undertook observations in SDSS $g'$, $r'$, and $i'$ bands. This way, the $i'$ band consisted of the continuum flux only, while the $g'$ and $r'$ bands consisted of the continuum and $H\beta$ and $H\alpha$ flux, respectively. Later, they subtracted the $i'$ band flux to get pure $H\beta$ and $H\alpha$ line flux. Since we have multiple narrowband filters, a solution to this problem could be observing the continuum using the narrowband filters red and bluewards of the emission line profile to obtain the pure continuum flux and later subtract it from the emission line profile in order to get the emission line cross-correlation (see Figure 4.4).

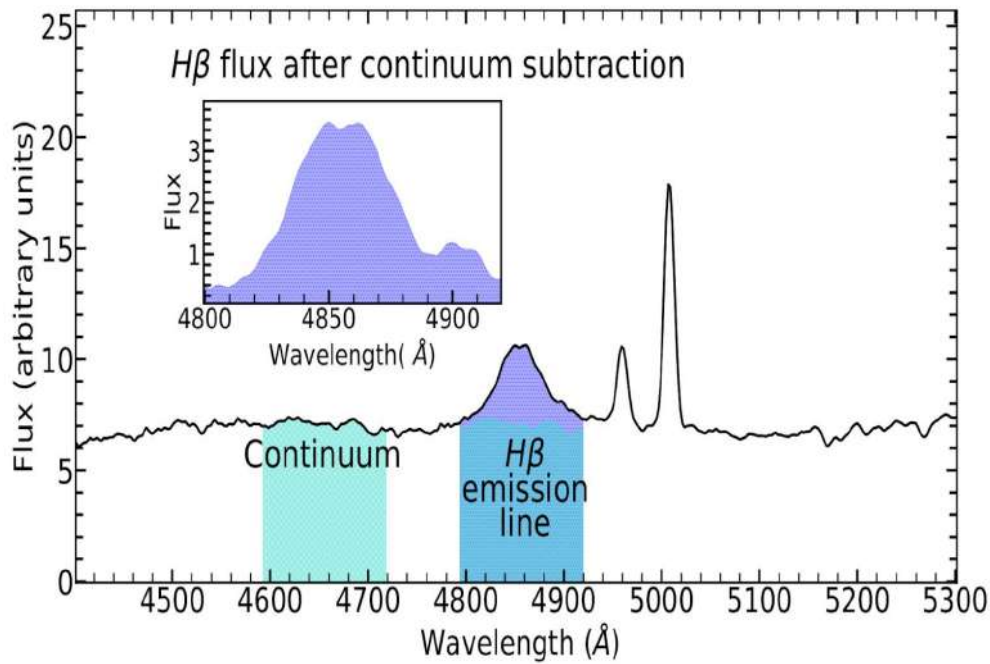

Fig. 4.4 Demonstration of the continuum contribution removal technique for the H$\beta$ emission line using 2 narrow band filters simultaneously. The figure in the inset shows the continuum removed H$\beta$ flux by placing a narrow band filter in the line-free region to catch the continuum flux.

### 4.3.3 Comparison of Photometric and Spectroscopic lags

Since the photometric light curves contain a contribution from the continuum region, there can be a significant deviation in the reverberation lags because the timescale of variability in the continuum region is usually smaller than the response of the BLR emission line. Hence, we found large deviations from the spectroscopic lags (see Figure 4.5). To check how continuum subtraction would help, we again convolved the transfer function of the filters with the continuum level on both the blue and red sides of the emission line. This way, we obtained the continuum flux and, after averaging the two, subtracted the flux from the $H\beta$ emission line. Thus, we obtained continuum free $H\beta$ flux. Through this technique, deviation in lag values was reduced significantly. We estimated lags for all the sources, and the photometric lags do not deviate much from the lags obtained through spectroscopic techniques (see Figure 4.6). We apply the same observation criteria for our monitoring campaign to reduce the continuum flux. Another problem with photometric monitoring is the host galaxy's contribution to flux. If the spectroscopic method is used, the galaxy contamination can be removed using spectral decomposition techniques, which is impossible in photometric monitoring. We have developed simulations to address this problem, and the results are highlighted in the subsequent section.

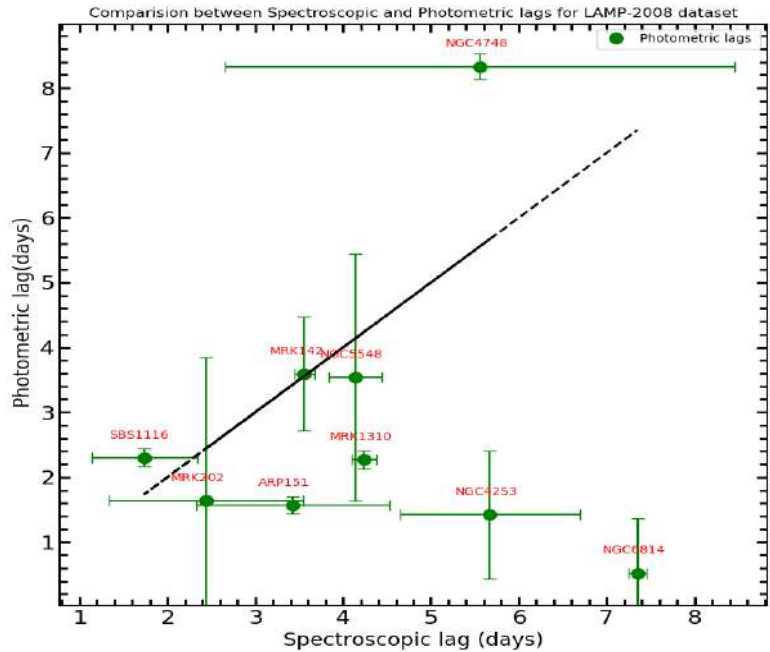

Fig. 4.5 The comparison between the spectroscopic and photometric lags. If the continuum contribution is not removed, it is evident that significant deviation in the H$\beta$ emission lags arises.

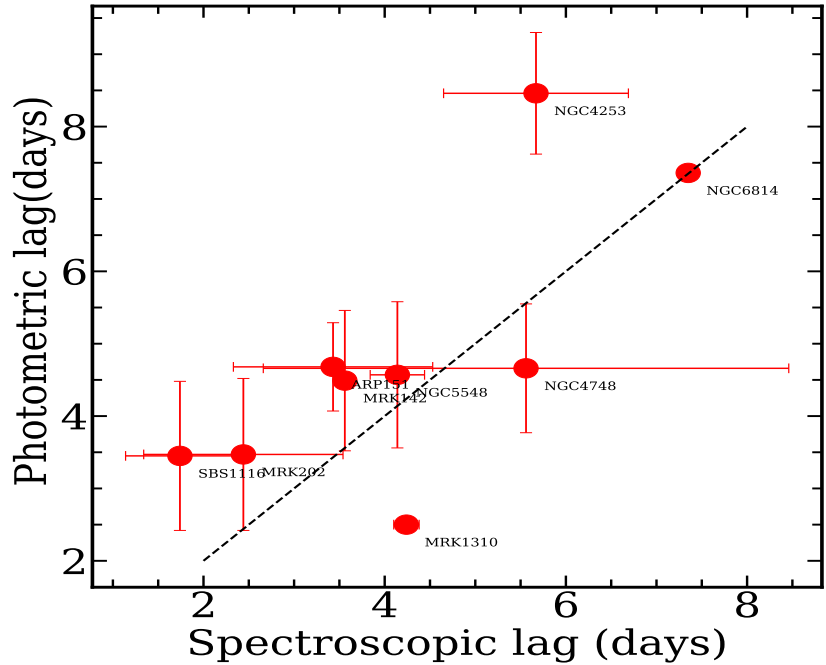

Fig. 4.6 The comparison between the spectroscopic and photometric lags after removing the continuum contribution from the H$\beta$ emission line light curves. The black dashed line in the middle shows a one-to-one correlation.

## 4.4 Simulations

We develop simulations to obtain the best observation strategy for our campaign and understand the complications and limitations of the PRM technique. We generated light curves using the damped random walk (DRW) model. This light curve acts as the continuum light curve. After simulating the light curves, we generate the responding light curves and vary the observation cadence, the campaign duration, and the variability amplitude. We recover the lags under these conditions and see the permissible limit for these parameters. We introduce 3, 5, and 10 days lag between the continuum and emission light curves. These lags lie in the range of expected lags we intend to obtain for most sources with lower luminosities and hence shorter BLR sizes as expected from the RL relation of Bentz et al. (2013). We recover the input lags and note the deviations with various cadences, campaign duration and variability amplitudes. Subsequently, with the simulation of the host galaxy profiles based on the luminosity functions, we study the effect of seeing conditions on the recovery of actual lags.

Table 4.2 A comparison of reverberation lags for $H\beta$ line obtained in this work along with the SMBH mass estimated through the spectra. (All lag estimates are in light days).

| S No. | Name | Lag in Bentz et al. (2009) | Spectroscopic Lag | Photometric Lag |
|---|---|---|---|---|
| 1 | Mrk142 | $2.88^{+1.00}_{-1.01}$ | $3.56^{+0.11}_{-1.03}$ | $4.48^{+0.97}_{-0.90}$ |
| 2 | SBS1116 | $2.24^{+0.65}_{-0.61}$ | $1.74^{+0.66}_{-0.34}$ | $3.46^{+4.11}_{-0.86}$ |
| 3 | Arp 151 | $3.52^{+0.82}_{-0.72}$ | $3.43^{+1.11}_{-0.13}$ | $4.68^{+0.61}_{-1.09}$ |
| 4 | Mrk 1310 | $3.67^{+0.46}_{-0.50}$ | $4.24^{+0.14}_{-0.68}$ | $2.39^{+0.97}_{-0.96}$ |
| 5 | Mrk 202 | $3.11^{+0.91}_{-1.12}$ | $2.44^{+1.10}_{-0.93}$ | $3.47^{+1.05}_{-0.98}$ |
| 6 | NGC 4253 | $6.87^{+1.22}_{-1.84}$ | $5.67^{+1.02}_{-0.19}$ | $4.72^{+0.83}_{-0.18}$ |
| 7 | NGC 4748 | $6.39^{+1.84}_{-1.46}$ | $5.56^{+2.90}_{-0.95}$ | $8.54^{+0.96}_{-0.15}$ |
| 8 | NGC 5548 | $4.24^{+0.91}_{-1.35}$ | $4.14^{+0.31}_{-0.53}$ | $4.57^{+1.00}_{-0.13}$ |
| 9 | NGC 6814 | $6.49^{+0.95}_{-0.96}$ | $7.35^{+0.10}_{-0.08}$ | $7.36^{+0.10}_{-0.80}$ |

### 4.4.1 Generation of mock light curves

Using the Damped Radom Walk model, we generate mock light curves for the continuum and emission line variability. The damped random walk is an Ornstein and Uhlenback process. It has been demonstrated that several AGN light curves can be modeled by this process (see Kelly et al., 2009, Kozłowski, 2017). For generating the light curves we use the DRW function available in the astroML[6] package.

Basically, the DRW model depends on two free parameters, the variability amplitude ($\sigma$) and the damping timescale ($\tau$). We generate mock light curves mimicking the observations for a particular source, denoting the continuum variability. We keep the value of $\tau$ fixed at 300 days based on the median value obtained for quasars from the SDSS survey (Li et al., 2019). This light curve works as the driving light curve originating due to the fluctuations in the accretion disk. It is assumed that the UV continuum ionizes the accretion disk giving rise to the variability, which is stochastic. The responding light curve is obtained by convolving the continuum light curve with a transfer function and shifting it by a few days. The transfer function is a top hat function that mimics the lamp post model for the BLR emission clouds. The exact form of the transfer function is still unknown. However, the top hat function has been quite commonly used in the literature (Zu et al., 2011), and changing the transfer function does not necessarily affect the lag estimates (Zu et al., 2016). The shift in the light curves is the *lags*, which we aim to recover using our techniques. We introduce a lag of 3, 5, and 10 days by shifting the continuum light curve by the same number of days. See Figure 4.7 for an example of the continuum and the emission light curves.

---

[6]https://www.astroml.org/modules/generated/astroML.time_series.generate_damped_RW.html

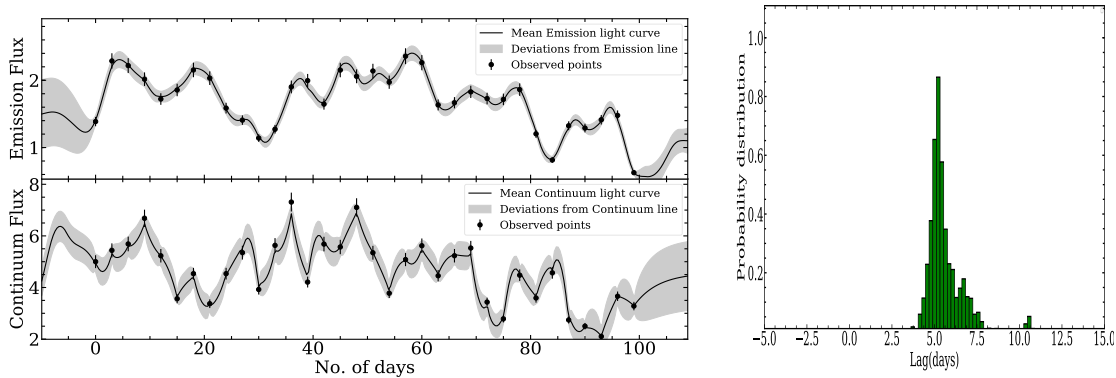

Fig. 4.7 Example of the continuum and emission line light curves generated using the DRW model is shown in the left panel. The probability distribution of the recovered lag for an input lag of 5 days is shown on the right.

### 4.4.2 Effect of cadence

We consider various observational cadences for different cases from every day to every 10 days. This is done in order to check at what minimal observing cadence the observations should be taken to recover actual lags with no significant deviation as it may not be possible to get daily cadence due to weather and technical issues. We set the deviation of $\pm 1$ day in the lag recovery as a tolerable limit. For an input lag of 3 days, we vary the observation cadence from every day to every 2 days and recover the lags to see the difference. For an input lag of 5 days, we sample the data points from every day incrementally to every 5 days, and for an input lag of 10 days, we sample the points from every day to every 10 days. We also look for the effect of an observational gap of up to the expected reverberation lag on the recovery. We ensure that at least 40 data points are available in both the light curves for the lag recovery. Since JAVELIN performs better than ICCF, as demonstrated in previous works, we use JAVELIN to recover the lags. We find that in the case of 5-day lags, we can recover the lags without deviation for almost all the cases. In the case of 10-day lags, we can recover the lags with only half the cadence. Once the cadence reaches 6 days and beyond, deviations start to appear. From this exercise, we can conclude that a sampling of approximately half the expected lag can recover lags within the error range. Going beyond that, the lag values deviate from the tolerable limits, as evident in Figure 4.9.

### 4.4.3 Effect of campaign duration

Based on the surveys for AGN undertaken in the past, we aim to observe a source within an observation cycle lasting 3 months. However, we need to decide the optimal duration within which we can recover the reverberation lags. To check whether this introduces some bias in the estimation of lags or not, we start with an observing duration of 50 days to 150 days and a lag of 5 and 10 days and recover the lags. We set the cadence as per the previous section. So, we have 4 measurements at 50 days, 60 days, 80 days, and 100 days in length with varying cadence for the 5-day input lag. For the 10-day input lag, we have the measurements at 60 days, 100 days, and 150 days campaign length with a varying cadence from 1 to 10 days. We notice that the increase in the observation length does not affect the lag estimates if a cadence of less than half the input lags is observed. However, if the cadence is increased by more than half, in the case of a 10-day input lag, there are deviations from the input lag. Horne et al. (2004) concluded that a campaign duration of at least 3 times the expected lags is necessary to obtain high-quality RM results. For a lag of 10 days, if we monitor the source with a 2-day median cadence for 30 days, we get a deviation of 3 days, while if we monitor for 50 days and even if the median cadence is reduced to 3 days, the deviation in lag is within 1 day only. Thus we conclude that JAVELIN can recover smaller lags, but the cadence and campaign duration should be high. A campaign duration of 10 times the lag values with good sampling can recover lags for larger lag values. (King et al., 2015) performed simulations for the OzDES RM project and concluded that not only cadence but the campaign length also play a role in the recovery of actual lags, which is evident from our results. Increasing the campaign duration does not necessarily guarantee lag recovery, especially if the cadence is lower.

### 4.4.4 Effect of variability amplitude

AGN with lower luminosities have been shown to have greater variability amplitudes (Simm et al., 2016). In order to pick the best candidates for PRM, it is necessary to quantify a lower limit to the variability amplitudes, which causes deviation in the recovery of actual lags. In the next step, we study the effect of the variability

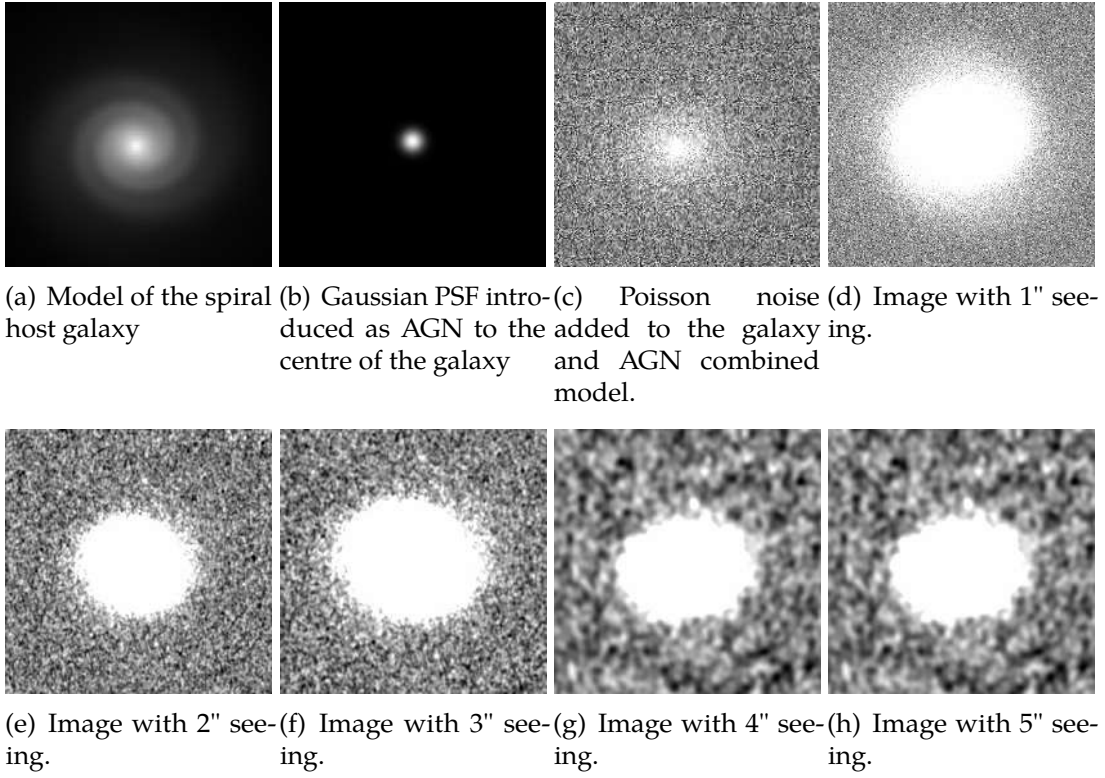

(a) Model of the spiral host galaxy (b) Gaussian PSF introduced as AGN to the centre of the galaxy (c) Poisson noise added to the galaxy and AGN combined model. (d) Image with 1" seeing.

(e) Image with 2" seeing. (f) Image with 3" seeing. (g) Image with 4" seeing. (h) Image with 5" seeing.

Fig. 4.8 The galaxy model along with the AGN profile developed using a Sersic function along with different seeing conditions generated by convolution with a Gaussian kernel are shown here.

amplitude. Following Vaughan et al. (2003), we calculate the fractional variability amplitude $F_{\text{var}}$, as:

$$F_{\text{var}} = \sqrt{\frac{X^2 - \overline{\sigma_{\text{err}}^2}}{\bar{x}^2}}. \tag{4.1}$$

where X is the sample variance calculated as:

$$X^2 = \frac{1}{N-1} \sum_{i=1}^{N} (x_i - \bar{x})^2, \tag{4.2}$$

N is the number of data points in the light curve, and $\bar{x}$ is the arithmetic mean of the flux value of the light curve. $\bar{\sigma}$ is the mean square error calculated as:

$$\overline{\sigma_{\text{err}}^2} = \frac{1}{N} \sum_{i=1}^{N} \sigma_{\text{err,i}}^2 \tag{4.3}$$

where $\sigma^2_{\text{err,i}}$ is the error in each data point. $F_{\text{var}}$, gives us a measure of how variable the light curve is with respect to the mean. We simulate various conditions with different variability amplitudes, with $F_{var}$ ranging from 0.02 to 0.5 in the continuum light curves. We provide an input lag of 5 days and recover it with different variability amplitudes. We find that a value of $F_{\text{var}} \geq 0.1$ in the continuum light curves can recover the lags with lesser uncertainties. As the emission line, light curve is assumed to be a shifted, scaled, and smoothed version of the continuum, the variability amplitude in the emission light curves is expected to be smaller than the continuum. The JAVELIN method deviates from the lag values for shorter variability amplitudes, while the ICCF method detects the lag even with small variability amplitudes, although the uncertainties are large. Since smaller variability amplitudes mean a feeble chance to catch the reverberation signals, especially in the emission line light curves, since RM is based on the cross-correlation of these two light curves, such minor variations may cause considerable deviations in the lag estimates. Hence, we may exclude sources whose variability amplitudes are lower for more extensive campaigns. In the era of all-sky surveys such as the ZTF survey, the light curves for plenty of sources are available publicly. It would be helpful to note the variability amplitudes of the sources and possibly exclude sources with a lower variability amplitude if genuine reverberation lags are to be estimated with higher accuracy.

### 4.4.5 Effect of seeing conditions

For the photometric monitoring campaign, the effect of seeing conditions and the prominence of the host galaxy in the total flux plays an important role. As seeing, conditions vary night by night, and within the night itself, it becomes imperative to understand the effect of seeing conditions on the variability of quasars. The seeing conditions also have some aperture effects, as is evident in (Cellone et al., 2000). Since seeing conditions can induce artificial variability in the light curves, we simulated images of the host galaxy and the AGN to understand and quantify the impact of seeing and the host galaxy's contribution to the AGN flux. This is obtained by modeling the host galaxy profile using GALFIT (Peng, 2003). The program is designed to fit galaxy profiles but allows various galaxy profiles to be simulated by varying the luminosity functions. For the galaxy, we use a Sersic profile, defined as:

$$S(r) = S_e \exp\left[-\kappa\left(\left(\frac{r}{r_e}\right)^{1/n} - 1\right)\right]. \tag{4.4}$$

where $S_e$ is the pixel brightness of the surface at effective half-light radius $r_e$. n is the concentration parameter of the profile, and $\kappa$ is a dependent variable on n to ensure half the flux of the galaxy is enclosed in the half-light radius $r_e$(Peng et al., 2010). In the center of the galaxy, we introduce the AGN profiles as a gaussian Point Spread Function (PSF). We introduced Poisson noise to the galaxy profile to mimic the actual observations. Once the galaxy and AGN composite profile is prepared, we introduced the continuum and emission variability to generate multi-epoch data points.

We performed aperture photometry on the images to extract the light curves under various apertures. The motive for simulating this profile is to study the effects caused by atmospheric seeing. We convolved the images with 2D Gaussian functions to mimic the seeing PSF, to choose various seeing profiles ranging from $1"$ to $5"$ to mimic the best and worst conditions at the telescope site. We studied the impact of seeing conditions on the variability trends for the AGN. We find that as the seeing conditions worsen, the variability amplitude decreases significantly. We chose different apertures to extract the fluxes and found that the effect of seeing reduces the variability amplitudes. This is evident most prominently if we use smaller apertures. Choosing a larger aperture does not cause a significant decrease in the variability amplitudes, but it is reduced even in the case of best-seeing conditions. The smaller aperture has a high amplitude of variability, while due to the dominant host galaxy flux in the case of the larger aperture, the variability amplitude is lesser. Thus an optimum aperture to observe the sources and extract pure photometric flux needs a balance between the host galaxy contribution and the worsening seeing conditions. Thus, the seeing-related effects must be considered while selecting the light curves for observations. In the worse seeing conditions, the effect of seeing is evident in sources with prominent host galaxies. The host galaxy contribution can be reduced by modeling the galaxy profile using 2D profiles, which can be subtracted to get the pure AGN flux in such cases.

## 4.5 Summary and Conclusions

In order to calibrate the method of PRM with the traditional emission line RM achieved through spectroscopic observations, we estimated the BLR sizes for AGN in the LAMP 2008 dataset. We convolved the spectra with the transfer function of our narrow band filters to convert spectra into the photometric light curve, and there were considerable deviations in the lag estimates. We tried to overcome these deviations by measuring the continuum flux alongside the $H\beta$ emission line and subtracting it, and through this technique reduced deviations in the recovered lags significantly. Since PRM is a technique that has not been applied to a large sample of AGN, we have demonstrated that PRM is a suitable and cost-effective alternative to RM. Based on the simulated light curves, we have studied the effect of observational cadence, the campaign duration, the variability amplitude, and the effect of seeing conditions in recovering the lags. The results from this work can be summarized as follows:

1. We calculated the $H\beta$ emission line lags using the JAVELIN code for the LAMP-2008 dataset, and the results agree with the lags obtained through the ICCF methods in (Bentz et al., 2009).

2. We convolved the spectra with the transmission curve of the narrow band filters available at the Devasthal Observatory to get the photometric data points. We find significant deviations in the lag estimates without properly removing the continuum from the emission line region. However, if the continuum contribution is removed, the lag deviations reduce significantly and agree with the spectroscopic lags.

3. Through simulated light curves, we found that a cadence of roughly more than half the expected lags causes a significant deviation in the lag estimates. Further, we find that a campaign length of roughly 10 times the expected lags is sufficient to recover actual lags. Increasing the campaign duration does not contribute significantly to reducing the deviations in lags.

4. We find that in the light curves, a variability amplitude of less than 10% is insufficient to recover the actual lags, with huge deviations through both the JAVELIN and ICCF methods.

5. We find out that prominent host galaxy contribution to the AGN flux reduces the variability amplitudes significantly if the seeing conditions worsen. With a smaller aperture for photometry, this effect is the most evident. In comparison, if a larger aperture is used, the variability patterns are less affected. However, the contribution of the host galaxy can be significant, thus causing a problem in recovering the actual AGN flux. A proper balance between the aperture size and the seeing conditions is needed to recover proper AGN flux for calculating the lags.

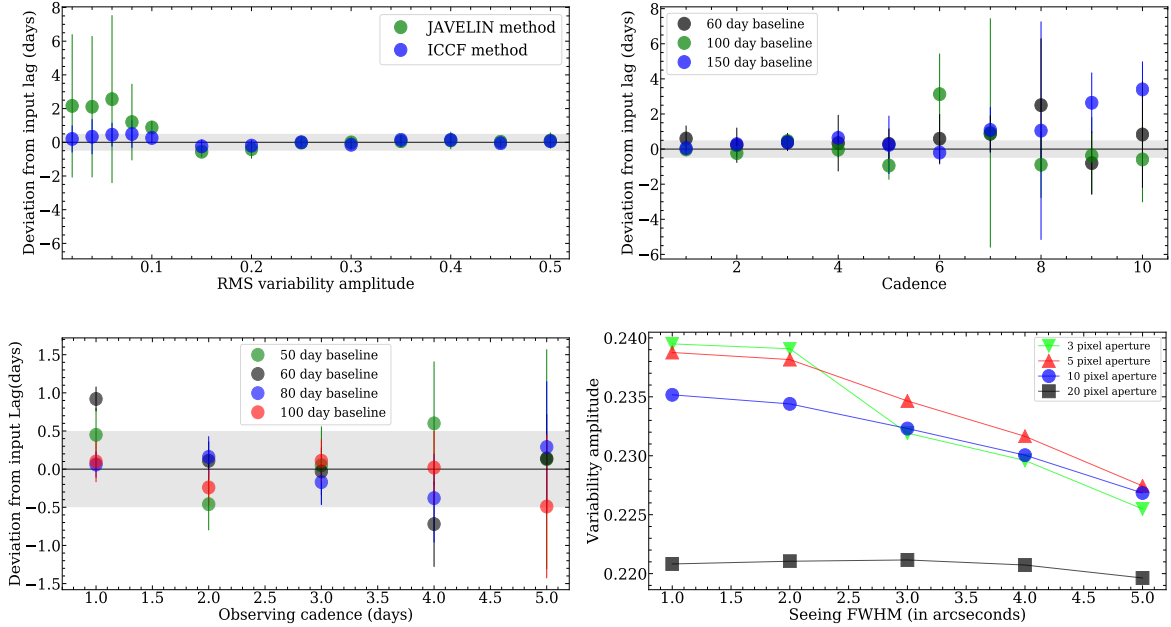

Fig. 4.9 The effect of variability amplitude (top left), cadence for an input lag of 10 days (top right), cadence for an input lag of 5 days (bottom left), and the seeing conditions (bottom right) on the estimation of lag are shown here.

Since for the PRM, we would be targeting nearby, low luminosity AGN, and based on RL relation, the lags for such sources should be in the range of 5-15 days. Hence based on the results obtained in this work, we may design the narrow-band PRM campaign efficiently. With the calibration results available, we may use it for a proper, efficient PRM campaign providing lags for a large sample of such AGN.

# Chapter 5

# Evidence of Jet induced Microvariability in the AGN Continuum[1]

Variability of AGN on timescales ranging from minutes to hours, known as INOV or Intra night Optical variability, can provide insights into their innermost regions. The relationship between the jets and the variability of the accretion disk can be determined by examining the INOV characteristics. NLSy1 galaxies beleived to be younger AGN, are radio loud and have relativistic jets. In this chapter, we present a systematic comparative study of the INOV properties of 23 RLNLSy1 galaxies, based on their jet detection from Very Long Baseline Array observations. We divide them into two groups: 15 jetted RLNLSy1s and 8 non-jetted RLNLSy1s. We observe these groups with multiple optical telescopes for 37 and 16 sessions, respectively, each lasting at least 3 hours. Our goal is to assess the impact of radio jets on the INOV phenomenon in Radio-Loud NLSy1 (RLNLSy1) galaxies. We also want to determine if the jetted RLNLSy1s that are detected in $\gamma$-rays ($\gamma$-ray) have a similar Duty Cycle (DC) to blazars. We are interested in finding out if relativistic beaming plays a significant role in INOV for low-luminous high accreting AGNs like NLSy1s. Our aim is to understand if the jetted $\gamma$-ray RNLSy1 galaxies exhibit blazar-like variability on INOV timescales among the NLSy1s.

---

[1]**Based on the work published as:** *Evidence of jet induced optical microvariability in radio-loud Narrow Line Seyfert 1 Galaxies* . Vineet Ojha, **Vivek Kumar Jha**, Hum Chand, Veeresh Singh, 2022. MNRAS, 514, 4.

## 5.1 Introduction

Narrow-line Seyfert 1 (NLSy1) galaxies are a subclass of active galactic nuclei (AGN), emitting electromagnetic radiations from radio to gamma-ray wavebands. Although in the optical wavelengths, both permitted and forbidden emission lines are present in their spectra, the width of their broad component of Balmer emission lines is narrower than the population of general type-1 Seyfert galaxies, with the full width at half maximum of the broad component of Balmer emission line (FWHM(H$\beta$)) being less than 2000 km s$^{-1}$ (Osterbrock & Pogge, 1985, Goodrich et al., 1989). Other optical characteristics such as flux ratio of [O$_{III}$]$_{\lambda 5007}/H\beta < 3$, and strong permitted Fe II emission lines, in addition to the criterion of FWHM(H$\beta$) are used to characteristically define NLSy1 galaxies (Shuder & Osterbrock, 1981). Besides, these galaxies also display peculiar characteristics in another wavelength, especially in X-ray wavelength, such as strong soft X-ray excess emission below 2 keV (e.g., Brandt et al., 1997, Vaughan et al., 1999), steep soft X-ray spectra (e.g., Grupe et al., 1998), rapid X-ray (sometimes optical) flux variability (e.g., Liu et al., 2010, Kshama et al., 2017), and blue-shifted emission line profiles (e.g., Boroson, 2005, Jha et al., 2022a). Furthermore, NLSy1s are believed to be relatively young AGNs, and they represent an early phase of their evolution (e.g., Mathur, 2000b). Observationally, it is suggested that the majority of NLSy1s have relatively lower Super Massive Black Hole (SMBH) masses of $10^6$ - $10^8$ M$_\odot$ (Grupe & Mathur, 2004, Wang et al., 2014, Rakshit et al., 2017), and higher accretion rates $\lambda_{Edd} \sim 0.05$ - $1.00$, in contrast to luminous class of AGN such as quasars (Boroson & Green, 1992, Peterson et al., 2000). However, relatively lower SMBH mass is not uncontested since a systematic underestimation of their SMBH has been suggested (Decarli et al., 2008, Viswanath et al., 2019). These highly accreting galaxies are generally hosted in spiral/disc galaxies (Ohta et al., 2007, Olguín-Iglesias et al., 2020), although in a few $\gamma$-ray detected NLSy1s, elliptical hosts have been suggested (hereafter $\gamma$-NLSy1s, D'Ammando et al., 2017, 2018).

Interestingly, although the NLSy1 exhibit both radio-quiet and radio-loud characteristics, defined by the radio parameter R$_{5GHz} \equiv f_{5GHz}/f_{4400\text{Å}}$ with R$\leq$ 10 and $> 10$ are being used to parameterized radio-quiet and radio-loud AGNs, respectively (e.g., see, Kellermann et al., 1994, 1989), the population is dominated by radio-quiet objects (Kellermann et al., 2016) and only a small fraction $\sim 7$% of NLSy1s are radio-loud (hereafter RLNLSy1s, Singh & Chand, 2018). This implied that in a few of these galaxies, jets may be present, making them radio-loud (Zhou et al., 2003, Yuan et al., 2008). Indeed, Very Long Baseline Array (VLBA) observations have discovered parsec-scale blazar-like radio jets in a few RLNLSy1s (Lister et al., 2013, Gu et al., 2015).

The existence of relativistic jets in such AGN (although in a few of the sources) that have relatively higher accretion rates and lower black hole masses contradicts the trend of the existence of relativistic jets in larger black hole masses and lower accretion rates (Urry et al., 2000, Boroson, 2002). Hence, studying NLSy1s from the standpoint of jets is essential to understanding the physical processes that can launch relativistic jets in this subclass of AGN.

Nonetheless, despite a double-humped spectral energy distribution (SED) like the blazars for a few RLNLSy1s (e.g., Abdo et al., 2009c, Paliya et al., 2013b), a small fraction of RLNLSy1s, especially very radio-loud (R > 100) RLNLSy1s exhibit interesting multi-wavelength characteristics such as compactness of the radio core, high brightness temperature, superluminal motion, flat radio and X-ray spectra, and rapid infrared and X-ray flux variability similar to blazar class of AGN (Boller et al., 1996, Grupe et al., 1998, Berton et al., 2018). All these characteristics provide indirect evidence for the presence of jets in them. However $\gamma$-ray detection by *Fermi*-Large Area Telescope (*Fermi*-LAT)[2] from about two dozen RLNLSy1s gives conclusive evidence that $\gamma$-ray detected NLSy1s are capable of ejecting relativistic jets (Abdo et al., 2009a, Foschini et al., 2010, D'Ammando et al., 2012, Yao et al., 2019).

Variability of an AGN's optical flux from a few minutes to a day time scales is variously known as microvariability (Miller et al., 1989), Intraday Variability (IDV, Wagner & Witzel, 1995) or Intra-Night Optical Variability (INOV, Gopal-Krishna et al., 1993). This alternative tool is also used to verify the presence or absence of jets in other sub-classes of AGN indirectly because of the well-established observational fact that for radio-loud jet-dominated sources such as blazars, both INOV amplitude ($\psi$) and the duty cycle (DC) are found to be distinctively high in comparison to non-blazars, including weakly polarised flat-radio-spectrum (i.e., radio-beamed) quasars (Goyal et al., 2013b, Gopal-Krishna & Wiita, 2018). Interestingly, such an indirect tool has been used for a decade in searching for the Doppler-boosted optical jets in low luminous AGNs such as NLSy1s and weak emission line QSOs (e.g., see Liu et al., 2010, Paliya et al., 2013a, Kumar et al., 2015, Ojha et al., 2021). However, this indirect evidence is based upon the observed high amplitude, and duty cycle of INOV as seen in blazars consisting of strongly Doppler-boosted jets. More importantly, an INOV study comparing a subclass of AGN with jets and without jets has not yet been explored so far. Therefore, to establish stronger INOV amplitude ($\psi$) with high DC as evidence of the existence of jet in an AGN, we have carried out an INOV study with two sub-samples of RLNLSy1s with and without radio VLBA jets.

---
[2]https://heasarc.gsfc.nasa.gov/docs/heasarc/missions/fermi.html

The general consensus regarding the radio structures of NLSy1s has been that they harbor sizes of less than 300 pc (Ulvestad et al., 1995, Lister, 2018) and are generally compact sources with a steep spectrum (Foschini, 2011, 2012, Berton et al., 2015). The appearance of radio structures in the radio observations of AGNs depends upon the resolution and sensitivity of the radio telescopes; therefore, the non-detection of the radio jets in the radio images of RLNLSy1s may not necessarily imply that they do not have jets. Here, we have selected our sources (see Sect. 5.2) based on their available observations with the radio telescopes, which mainly consist of VLBA observations.

Furthermore, as pointed out above, $\gamma$-ray detections in several RLNLSy1s suggest the presence of relativistic jets in them; therefore, a comprehensive INOV study of the jetted with $\gamma$-ray detected RLNLSy1s (hereafter J-$\gamma$-RLNLSy1s) and the jetted without $\gamma$-ray detected RLNLSy1s (hereafter J-RLNLSy1s) is essential to understand the nature of their variability and jets. Therefore, we have also discussed the INOV nature of J-$\gamma$-RLNLSy1s and J-RLNLSy1s sub-samples in the present work.

The layout of this chapter is as follows. In Sect. 5.2, we outline the sample selection procedure. Sect. 5.3 provides details of our intra-night optical monitoring and the data reduction procedure. The statistical analysis is presented in Sect. 5.4, and our main results, followed by a brief discussion, are given in Sect. 5.5. In Sect. 5.6, we summarize our main conclusions.

## 5.2 Sample selection

The bulk of our sample for intra-night monitoring is drawn from Gu et al. (2015) where they have reported good quality VLBA observations at 5 GHz for the 14 RLNLSy1 galaxies having a flux density above 10 mJy at 1.4 GHz and a radio-loudness parameter[3] $R_{1.4GHz}$ > 100. Out of 14 RLNLSy1s, they confirm the presence of a jet in 7 of the sources (hereafter, "jetted") based on the core-jet structure at 5 GHz, and the remaining 7 sources are termed as "non-jetted" based on only the compact core. However, in reference to the jetted nature, it may be noted that VLBA radio images at typical resolution could generally resolve parsec scale core jet structures (Lister, 2018) which may not be relativistically beamed. Hence some of the sources could be steep spec-trum sources as seen in mini-radio galaxies (see Gu et al., 2015). We also note that the "non-jetted" source J144318.56+472556.7 has a ∼ 15-mas long quasi-linear radio component resolved into seven components along the south-west direction, in addition to diffuse emission extending up to ∼ 30 mas (see figure 12 of Gu et al., 2015). Therefore, we have included this source in our jetted subsample. We next expanded this sample by including another 9 sources from the VLBA literature, which also satisfy the above twin criteria (i.e., $f_{\nu\ 1.4GHz} \geq 10$ mJy and $R_{1.4GHz} > 100$). The jetted (or non-jetted) classification of the 8 sources out of 9 is possible using their published VLBA observations. Thus, three of these 8 sources, viz., J163323.59+471859.0, J164442.53+261913.3 and J170330.38+454047.3 are taken from the Doi et al. (2011) and the remaining 5 sources, viz., J032441.20+341045.0, J084957.98+510829.0, J094857.32+002225.6, J122222.99+041315.9 and J150506.48+032630.8 are taken from Zhou et al. (2007), D'Ammando et al. (2012), Giroletti et al. (2011), Lister et al. (2016) and Orienti et al. (2012), respectively. The frequencies at which observations of 23 RLNLSy1s were carried out are tabulated in the last column of Table 5.1. However, for the NLSy1 J120014.08−004638.7, Doi et al. (2012) have confirmed its lobe-dominated nature based upon its Very Large Array (VLA) 1.4 GHz FIRST images. Note that this source is also not in the latest sample of Monitoring Of Jets in Active galactic nuclei with VLBA Experiments (MOJAVE) XVII program (see Lister et al., 2019). Therefore, we have included this source in our non-jetted set. Based on the published VLBA observations of 8 sources, 7 sources have a confirmed jet, and one source falls in the non-jetted category. Thus, overall, our sample consists of 23 RLNLSy1s, including 15 of which are jetted, and the remaining 8 RLNLSy1s are non-jetted. Table 5.1 summarizes the basic properties of our sample. The SDSS names of the sources with *Fermi*-LAT detection are suffixed with a "▲" sign and the references are given in the footnote "b" to Table 5.1.

---

[3]$R_{1.4GHz}$ is the ratio of the monochromatic rest-frame flux densities at 1.4 GHz and 4400Å (see Yuan et al., 2008).

Table 5.1 The present sample of 23 RLNLSy1s galaxies selected for INOV monitoring.

| SDSS Name,[a] | R-mag[b] | $z^c$ | $R_{1.4GHz}{}^d$ | Apparent jet speed[e] | Optical Polarization[f] | Radio Polarization[g] | log $(M_{BH})$[h] $(M_\odot)$ | Observing freq. (GHz) |
|---|---|---|---|---|---|---|---|---|
| jetted NLSy1s | | | | | | | | |
| J032441.20+341045.0▲ | 13.10 | 0.06 | 318* | $9.1c \pm 0.3c$ | 1-3%$^\alpha$, 0.7-0.8%$^\beta$, 1.2%$^\gamma$ | 4%$^\psi$, 0.2-1%$^\Lambda$ | 7.30$^\vee$ | 2.2/8.4 |
| J081432.12+560958.7 | 18.10 | 0.51 | 339† | - | - | * | 8.00$^\diamond$ | 4.9 |
| J084957.98+510829.0▲ | 17.79 | 0.58 | 4496* | $6.6c \pm 0.6c$ | 10%$^\delta$, 10%$^\gamma$ | 0.3-3%$^\Lambda$, 3.3%$^\tau$ | 7.59$^\perp$ | 5.0/8.4/15.3 |
| J090227.20+044309.0 | 18.20 | 0.53 | 1047† | - | - | * | 7.64$^\perp$ | 4.9 |
| J094857.32+002225.6▲ | 18.17 | 0.58 | 846* | $9.7c \pm 1.1c$ | 36%$^\alpha$, 18.8%$^\beta$, 2.4%$^\gamma$ | 0.2-3%$^\Lambda$ | 7.50$^\perp$ | 22.2 |
| J095317.10+283601.5 | 18.60 | 0.66 | 513† | - | - | - | 7.73$^\perp$ | 4.9 |
| J104732.78+472532.0 | 18.20 | 0.80 | 7413† | - | - | - | 8.10$^\diamond$ | 4.9 |
| J122222.99+041315.9▲ | 17.06 | 0.97 | 1534* | $0.9c \pm 0.3c$ | - | 0.2-3.3%$^\Lambda$ | 8.30$^\Upsilon$ | 15.4 |
| J130522.75+511640.2▲ | 15.80 | 0.79 | 219† | - | 1.0%$^\gamma$ | * | 8.15$^\perp$ | 4.9 |
| J142114.05+282452.8 | 17.10 | 0.78 | 205† | - | - | - | 7.72$^\perp$ | 4.9 |
| J144318.60+472557.0▲ | 17.70 | 0.70 | 1175† | - | - | * | 7.80$^\diamond$ | 4.9 |
| J150506.48+032630.8▲ | 17.72 | 0.41 | 3364* | $0.1c \pm 0.2c$ | 4%$^\gamma$ | 0.2-2.5%$^\Lambda$ | 7.26$^\perp$ | 15.3 |
| J154817.92+351128.0 | 18.40 | 0.48 | 692* | - | 2.1%$^\gamma$ | * | 7.84$^\perp$ | 4.9 |
| J164442.53+261913.3▲ | 16.60 | 0.14 | 447* | $> 1.0c$ | 2.2%$^\gamma$ | - | 7.21$^\perp$ | 1.7 |
| J170330.38+454047.3 | 12.80 | 0.06 | 102* | - | 3-5%$^\kappa$ | - | 6.77$^+$ | 1.7 |
| non-jetted NLSy1s | | | | | | | | |
| J085001.17+462600.5 | 18.40 | 0.52 | 170† | - | - | - | 7.34$^\perp$ | 4.9 |
| J103727.45+003635.6 | 19.10 | 0.60 | 457† | - | - | - | 7.48$^\perp$ | 4.9 |
| J111005.03+365336.2 | 19.00 | 0.63 | 933† | - | - | - | 7.43$^\perp$ | 4.9 |
| J113824.53+365327.2 | 18.30 | 0.36 | 219† | - | - | - | 7.29$^\perp$ | 4.9 |
| J120014.08−004638.7 | 17.70 | 0.18 | 169 | - | - | - | 7.40$^\zeta$ | 1.4 |
| J124634.65+023809.1 | 17.50 | 0.36 | 277† | - | - | - | 7.42$^\perp$ | 4.9 |
| J163323.59+471859.0 | 14.50 | 0.12 | 154* | - | 2.4%$^\gamma$ | - | 6.70$^\perp$ | 1.7 |
| J163401.94+480940.2 | 19.10 | 0.49 | 204† | - | - | - | 7.56$^\perp$ | 4.9 |

[a]The SDSS names of the sources with *Fermi*-LAT detection are suffixed with a "▲" sign and the references (Abdo et al., 2009a,b, Foschini et al., 2010); (Abdo et al., 2009c); (Foschini, 2011, D'Ammando et al., 2012); (D'Ammando et al., 2015); (Liao et al., 2015); (Yao et al., 2015); and (Ajello et al., 2020) are for the sources, J094857.32+002225.6; J032441.20+341045.0 and J150506.48+032630.8; J084957.98+510829.0; J164442.53+261913.3; J130522.75+511640.2, J122222.99+041315.9, and J144318.60+472557.0 respectively.

[b]Taken from Monet (1998).

[c]Emission-line redshifts are taken either from Gu et al. (2015) or from Paliya et al. (2019).

[d]$R_{1.4GHz} \equiv f_{1.4GHz}/f_{4400Å}$ values are taken from Gu et al. (2015) for the sources marked with a '†', and are taken from Ojha et al. (2020a) for the sources marked with an '*', except for J120014.08−004638.7 for which $R_{1.4GHz}$ is estimated using its total flux density of 27.1 mJy at 1.4 GHz and k-corrected B-band optical flux density of 0.16 mJy (Doi et al., 2012).

[e]The available jet speed of RLNLSy1s from literature which is as follows: for the jet speed of J032441.20+341045.0, J084957.98+510829.0, J094857.32+002225.6, J150506.48+032630.8 see Lister et al. (2019), and for J122222.99+041315.9, J164442.53+261913.3 see Lister et al., 2016, and Doi et al., 2012.

[f]The optical polarization values as reported in these papers: $^\alpha$Itoh et al. (2014); $^\beta$Ikejiri et al. (2011); $^\gamma$Angelakis et al. (2018); $^\delta$Maune et al. (2014); $^\kappa$Leighly (1999).

[g]The radio polarization values as reported in these papers: $^\psi$Neumann et al. (1994); $^\Lambda$Hodge et al. (2018); $^\tau$Homan et al. (2001). *fractional radio polarisation images presented in Fig-6 of Gu et al. (2015).

[h]Derived black hole masses of the present sample based upon single-epoch optical spectroscopy virial method were compiled from the literature. The references for the black hole mass are as follows: $^\vee$Zhou et al. (2007); $^\diamond$Yuan et al. (2008); $^\Upsilon$Yao et al. (2015); $^\perp$Rakshit et al. (2017); $^+$Wang & Lu (2001); $^\zeta$Greene & Ho (2007).

## 5.3 Observations and Data Reduction

### 5.3.1 Photometric monitoring observations

Intra-night monitoring of all 23 RLNLSy1s of our sample was performed in the broadband Johnson-Cousin filter R due to the optimum response of the used CCDs in this filter. Four telescopes namely, the 1.04 meter (m) Sampurnanand telescope (ST, Sagar, 1999), 1.30-m Devasthal Fast Optical Telescope (DFOT, Sagar et al., 2010), 3.60-m Devasthal Optical Telescope (DOT, Sagar et al., 2012) and 2.01-m Himalayan Chandra

Telescope (HCT, Prabhu & Anupama, 2010) were used for the intra-night monitoring of the present sample. Out of these four telescopes, the 1.04-m ST is located at Nainital, while the 1.30-m DFOT and the 3.60-m DOT are located at Devasthal near Nainital, and all three are managed by the Aryabhatta Research Institute of Observational Sciences (ARIES). The fourth telescope, the 2.01-m Himalayan Chandra Telescope (HCT), is located in Ladakh and operated by the Indian Institute of Astrophysics (IIA), Bangalore, India. All four telescopes are equipped with Ritchey-Chretien (RC) optics and were read out 1 MHz rate during our observations, except for the 3.60-m DOT, which was read out additionally at 500 kHz. The monitoring sessions lasted between $\sim$ 3.0 and $\sim$ 5.5 hours (median 3.75 hrs). Our sources were observed in 4×4 binning mode with 1.04-m ST and the same with the 3.6-m DOT on 2017.04.11. A 2×2 binning mode was adopted for the remaining sessions with the DOT. No binning was done for the DFOT and the HCT telescope observations. The basic parameters of the four telescopes and their CCDs used in the present observations are listed in Table 5.2. In order to improve the INOV statistics, at least two intra-night sessions were managed for each of our 23 RLNLSy1s. In this work, $\sim$ 1.04-3.60m class telescopes have been used, therefore depending upon the apparent magnitude of the target NLSy1, moon illumination, and sky condition, we had set a typical exposure time for each science frame between 4 and 15 minutes. The median seeing (FWHM of the point spread function (PSF)) for the sessions typically ranged between $\sim$ 1 - 3 arcsec, except for a single session dated 2019.03.25 when seeing became considerably poorer (see Fig. 5.5).

### 5.3.2 Data reduction

During each observation night, images of the sky-flat-field were captured at dusk and dawn, and a minimum of three bias frames were obtained. Dark frames were not necessary due to the low temperature of the CCD detectors, which were cooled using either liquid nitrogen (to approximately $-120°$C) or thermoelectric cooling (to approximately $-90°$C for the 1.3-m DFOT). The standard routines within the IRAF[4] software package were followed for preliminary processing of the observed frames. Aperture photometry (Stetson, 1987, 1992) was selected in this work for extracting the instrumental magnitudes of the targets and the comparison stars recorded in the CCD frames, using DAOPHOT II algorithm[5] due to less crowded fields of the monitored NLSy1s. The prime parameter in the aperture photometry is the size of the optimal aperture, which is used to estimate the instrumental magnitude and the corresponding signal-to-noise ratio (SNR) of the individual photometric data points recorded in each CCD frame. As emphasized in Howell (1989), the SNR of a target recorded in a CCD is maximized for the aperture radius $\sim$ PSF. However, as suggested by Cellone et al. (2000) when the underlying host galaxy significantly contributes to the total optical flux, its contribution to the aperture photometry can vary significantly due to PSF variation, mimicking INOV. The possibility of such spurious INOV can be significant for the lower redshift NLSy1s in our sample, particularly, J032441.20+341045.0 (z = 0.06), J164442.53+261913.3 (z = 0.14), J170330.38+454047.3 (z = 0.06), J120014.08−004638.7 (z = 0.18) and J163323.59+471859.0 (z = 0.12) (Table 5.1). This issue is further addressed in Sect. 5.5. Nonetheless, bearing the above in mind, we have chosen an aperture radius equal to 2×FWHM for our final analysis, as already elaborated in Ojha et al. (2021).

---

[4]Image Reduction and Analysis Facility (http://iraf.noao.edu/)
[5]Dominion Astrophysical Observatory Photometry (http://www.astro.wisc.edu/sirtf/daophot2.pdf)

Using the instrumental magnitudes extracted from the aperture photometry, DLCs of each NLSy1s were derived for each session, relative to a minimum of two (steady) comparison stars that were chosen based on their closeness to the monitored target NLSy1, both in position and brightness, as recorded in the CCD frames. The importance of these procedures for genuine INOV detection has been highlighted by Howell et al. (1988) and further focused in Cellone et al. (2007). In the case of 12 targets, we could identify at least a comparison star within $\sim$ 1 instrumental magnitude to the target NLSy1. The median magnitude offsets ($\Delta m_R$) for the remaining 10 targets were also not significant and within $\sim$ 1.5-mag, except for a source, viz., J103727.45+003635.6 for which $\Delta m_R$ was found to be 1.74 (see Figs 2-6). Table A.1 lists coordinates together with some other parameters of the steady comparison stars used for all the sessions. It has been shown in Ojha et al. (2021) that the color differences of such orders can be safely discounted while analyzing the variability of the DLCs.

## 5.4  Statistical analysis

In recent years, multi-testing has been widely used to ensure the reliability of microvariability event detection (e.g., Joshi et al., 2011, Goyal et al., 2012, de Diego, 2014, Ojha et al., 2021). Therefore, for unambiguous detection of INOV in a DLC, we have used in the current work two different versions of $F$-test, which are the standard $F$-test (hereafter $F^\eta$-test) and the power-enhanced $F$-test (hereafter $F_{enh}$-test). However, in the case of $F^\eta$-test, it is suggested that mismatching in the brightness levels among target AGN (in the current work, NLSy1 galaxy) and the two chosen steady comparison stars should be within $\sim$ 1-mag in order to avoid photon statistics and other random-noise terms (e.g., see Howell et al., 1988, Cellone et al., 2007, Goyal et al., 2012). Therefore, care was taken while selecting two (non-varying) comparison stars to be within 1-mag of the respective NLSy1s. Thus, for the present set of jetted-RLNLSy1s, the median magnitude mismatch of *0.82* between the reference star (i.e., the comparison star with the closest match with the target AGN'S instrumental magnitude) and target NLSy1. The corresponding median values for the non-jetted-RLNLSy1s and the entire set of 23 NLSy1s are *1.28* and *0.91*, respectively. Additionally, while implementing the $F^\eta$-test, it is also crucial to use the correct RMS errors on the photometric data points due to underestimated magnitude errors by a factor ranging between 1.3 and 1.75, returned by the routines in the data reduction software DAOPHOT and IRAF (Gopal-Krishna et al., 1995, Garcia et al., 1999, Sagar et al., 2004, Stalin et al., 2004, Bachev et al., 2005). Therefore, the '$\eta$' value is taken here to be 1.54±0.05, computed using the data of 262 intra-night monitoring sessions of AGNs by Goyal et al. (2013a).

Table 5.2 Details of system parameters of the telescopes and detectors used in the observations of 23 RLNLSy1s.

| Telescope (s) | Detector (s) | Field of view (arcmin$^2$) | Readout noise (e$^-$) | Gain (e$^-$/ADU) | Focal ratio of telescope | Pixel size of CCD ($\mu$m) | Plate scale of CCD ($''$/pixel) |
|---|---|---|---|---|---|---|---|
| 1.04-m ST$^a$ | 4k×4k | 15.70×15.70 | 3.0 | 10.0 | f/13 | 15.0 | 0.23 |
| 1.30-m DFOT$^b$ | 2k×2k | 18.27×18.27 | 7.5 | 2.0 | f/4 | 13.5 | 0.53 |
| 3.60-m DOT$^c$ | 4k×4k | 6.52×6.52 | 8.0$^\star$, 5.0$^\dagger$ | 1.0$^\star$, 2.0$^\dagger$ | f/9 | 15.0 | 0.10 |
| 2.01-m HCT$^d$ | 2k×2k | 10.24×10.24 | 4.1 | 2.20 | f/9 | 15.0 | 0.30 |

$^a$Sampoornand Telescope (ST), $^b$Devasthal Fast Optical Telescope (DFOT)
$^c$Devasthal Optical Telescope (DOT), $^d$Himalayan Chandra Telescope (HCT).
$\star$Readout noise and corresponding gain at readout speed of 1 MHz and,
'$\dagger$' represents the same at readout speed of 500 kHz.

Following Goyal et al. (2012), $F^\eta$-statistics defined as

$$F^\eta_{s1} = \frac{\sigma^2_{(q-cs1)}}{\eta^2 \langle \sigma^2_{q-cs1} \rangle}, \quad F^\eta_{s2} = \frac{\sigma^2_{(q-cs2)}}{\eta^2 \langle \sigma^2_{q-cs2} \rangle}, \quad F^\eta_{s1-s2} = \frac{\sigma^2_{(cs1-cs2)}}{\eta^2 \langle \sigma^2_{cs1-cs2} \rangle} \qquad (5.1)$$

where $\sigma^2_{(q-cs1)}, \sigma^2_{(q-cs2)}$, and $\sigma^2_{(cs1-cs2)}$ are the variances with $\langle \sigma^2_{q-cs1} \rangle = \sum_{i=1}^{N} \sigma^2_{i,\,err}(q-cs1)/N$, $\langle \sigma^2_{q-cs2} \rangle$, and $\langle \sigma^2_{cs1-cs2} \rangle$ being the mean square (formal) rms errors of the individual data points in the 'target NLSy1 - comparison star1', 'target NLSy1 - comparison star2', and 'comparison star1 - comparison star2' DLCs, respectively.

The $F$-values were computed in this way for individual DLC using Eq. 5.1, and compared with the critical $F$-value, set, viz., $F^{(\alpha)}_\nu$, where $\alpha$ is the level of significance set by us for the $F^\eta$-test, and $\nu$ ($= N_j - 1$) is the degree of freedom for the DLC ($N_j$ being the number of data points in the DLC). The chance of a false INOV detection signal becomes lower for a smaller value of $\alpha$. Therefore, in the present work, similar to our previous work, two critical significance levels, $\alpha = 0.01$ and $\alpha = 0.05$ are set by us (e.g., see Ojha et al., 2021). Following Ojha et al. (2021), a NLSy1 is designated as variable for a given session according to this test if the computed value of $F^\eta$ is found to be greater than its $F_c(0.99)$. Table A.2 summarizes the computed $F^\eta$-values and the correspondingly inferred status of INOV detection for all the 53 sessions (columns 6 and 7).

The second flavor of $F$-test for INOV employed in the present study is the $F_{enh}$-test (e.g., de Diego, 2014). The statistical criteria for the $F_{enh}$-test can be described as

$$F_{\text{enh}} = \frac{s^2_{\text{NLSy1}}}{s^2_{\text{comb}}}, \quad s^2_{\text{comb}} = \frac{1}{(\sum_{j=1}^{q} N_j) - q} \sum_{j=1}^{q} \sum_{i=1}^{N_j} B^2_{j,i}. \quad (5.2)$$

here $s^2_{\text{NLSy1}}$ is the variance of the 'target NLSy1-reference star' DLC, while $s^2_{\text{comb}}$ is the combined variance of 'comparison star-reference star' DLC having $N_j$ data points and $q$ comparison stars, computed using scaled square deviation $B^2_{j,i}$ as

$$B^2_{j,i} = \omega_j (c_{j,i} - \bar{c}_j)^2 \quad (5.3)$$

where, $c_{j,i}$'s is the 'j$^{th}$ comparison star-reference star' differential instrumental magnitudes value and $\bar{c}_j$ represent the corresponding average value of the DLC for its $N_j$ data points. The scaling factor $\omega_j$ is taken here as described in Ojha et al. (2021).

The principal feature of the $F_{enh}$-test is that it takes into account the brightness differences of the target AGN and the selected comparison stars, a frequently encountered problem with the $C$ and $F$-statistics (e.g., see Joshi et al., 2011, de Diego, 2014). Thus $F_{enh}$-values were computed using Eq. 5.2 for individual DLCs and compared with the set critical values for this study (see above). Based upon this test, a NLSy1 DLC is assigned a designation "variable (V)" when the computed value of $F_{enh}$ found for 'target NLSy1-reference star' DLC, is greater than its $F_c(0.99)$ (i.e., $F_{\text{enh}} > F_c(0.99)$), and "probable variable (PV)" if same is found to be greater than $F_c(0.95)$ but less or equal to $F_c(0.99)$ (i.e., $F_c(0.95) < F_{\text{enh}} \leq F_c(0.99)$). In Table A.2, we tabulate the computed $F_{enh}$-values and the correspondingly inferred INOV status for our entire 53 sessions in columns 8 and 9.

Table 5.3 The DC and $\overline{\psi}$ of INOV, for the sample of 23 RLNLSy1 galaxies studied in this work, based on the $F_{enh}$-test and $F^{\eta}$-test.

| RLNLSy1s | No. of Sources | $F_{enh}$-test | | $F^{\eta}$-test | | SMBH mass[†] |
|---|---|---|---|---|---|---|
| | | *DC (%) | *$\overline{\psi}$[†] (%) | *DC (%) | *$\overline{\psi}$[†] (%) | log $(M_{BH}/M_\odot)$ |
| jetted-RLNLSy1s | 15 | 18 (30)[⊥] | 09 (07)[⊥] | 12 (30)[⊥] | 11 (05)[⊥] | 7.72 |
| non-jetted-RLNLSy1s | 8 | 05 (16)[⊥] | 09 (01)[⊥] | 00 (16)[⊥] | – | 7.42 |
| J-$\gamma$-RLNLSy1s | 8 | 34 (16)[⊥] | 10 (07)[⊥] | 29 (16)[⊥] | 11 (05)[⊥] | 7.59 |
| J-RLNLSy1s | 7 | 00 (14)[⊥] | – | 00 (14)[⊥] | – | 7.73 |

*We used the 46 sessions for this estimation, as explained in Sect. 5.4.1. [†]The mean value for all the DLCs belonging to the type 'V'.
[⊥]Values inside parentheses are the number of observing sessions used to estimate the parameters DC or $\overline{\psi}$.
[†]The median value of the SMBH mass

## 5.4.1 Computation of INOV duty cycle and amplitude of variability

To compute the duty cycle (DC) of INOV for the present sets of RLNLSy1s, we have adopted, following the definition given by Romero et al. (1999) (see, also Stalin et al., 2004)

$$DC = 100 \frac{\sum_{j=1}^{n} R_j (1/\Delta t_j)}{\sum_{j=1}^{n} (1/\Delta t_j)} \text{ per cent} \quad (5.4)$$

where $\Delta t_j = \Delta t_{j,\ observed}(1+z)^{-1}$ ($z$ is the redshift of the target NLSy1 galaxy in the current study) is the target AGN's redshift corrected time duration of the $j^{th}$ monitoring session (see details in Ojha et al., 2020a). For $j^{th}$ session, $R_j$ is considered to be 1 in Eq. 5.4 only when INOV is detected, otherwise taken to be zero. Note that to avoid introducing bias, we have used only 2 sessions for each AGN. For sources observed in more than 2 sessions (e.g., see Table A.2), the computation of DC used only the longest two sessions, as pointed out in Ojha et al. (2020a, 2021). The computed INOV duty cycles (DCs) for the different sets of RLNLSy1s are listed in Table 5.3, based on two statistical tests.

To compute the peak-to-peak amplitude of INOV ($\psi$) detected in a given DLC, we followed the definition given by Heidt & Wagner (1996)

$$\psi = \sqrt{(H_{max} - H_{min})^2 - 2\sigma^2} \quad (5.5)$$

with $H_{min,\ max}$ = minimum (maximum) values in the DLC of target NLSy1 relative to steady comparison stars and $\sigma^2 = \eta^2 \langle \sigma^2_{NLSy1-s} \rangle$, where, $\langle \sigma^2_{NLSy1-s} \rangle$ is the mean square (formal) rms errors of individual data points. The mean value of ($\overline{\psi}$) for different sets (e.g., see Table 5.3) of RLNLSy1 galaxies is computed by taking average of the computed $\psi$ values for the DLCs belonging to the "V" category. In Table 5.3, we have summarized the computed $\overline{\psi}$ values based on the two statistical tests for the different sets of RLNLSy1s in our sample.

## 5.5 Results and discussion

The INOV characterization of RLNLSy1 galaxies presented here is likely to be more representative in comparison to the previous studies based on significantly smaller samples (Liu et al., 2010, Paliya et al., 2013a, Kshama et al., 2017). Also, we have paid particular attention to guarding against the possibility of spurious INOV claims arising from a varying flux contribution to the aperture photometry from the host galaxy of the AGN caused due to seeing disc variation during the session. As pointed out by Cellone et al. (2000), under such circumstances, false claims of INOV can result in $low-z$ AGNs. Based on recent deep imaging studies of NLSy1 galaxies by Olguín-Iglesias et al. (2020), it can be inferred that any variable contamination arising from the host galaxy is very unlikely to matter when studying the variability of AGNs at least at z > 0.5. From Table A.2, it is seen that the INOV detection ($\psi > 3\%$) has occurred for just 3 sources in our sample having z < 0.5. These are: J032441.20+341045.0 (at $z = 0.06$; 4 sessions), J164442.53+261913.3 (at $z = 0.14$; two sessions), and J163323.59+471859.0 ($z = 0.12$, one session). However, since the seeing disk remained steady in all these sessions (Fig. 5.1), except for the flickering in the PSF for the 3-4 points only in the case of J164442.53+261913.3 for its first session (Fig. 5.4) and a non-negligible systematic PSF variation in the case of J163323.59+471859.0 (Fig. 5.6). A closer checkup of these two intranight sessions shows that either PSF remained fairly steady during the time of AGN's flux variations (Fig. 5.4) or the gradients in the DLCs of the target AGN are seen to be anti-correlated with systematic variations in the PSF (Fig. 5.6).

This contradicts the expected outcome if the aperture photometric measurements were significantly affected by the underlying galaxy (see Cellone et al., 2000). Therefore, the possibility of a significant variation in the fractional contribution from the host galaxy can be safely discounted. We thus conclude that the present cases of INOV detection for $low-z$ RLNLSy1s are genuine and not artifacts of seeing disk variation through the monitoring sessions.

From Table 5.3, it is seen that the $F_{enh}$-test resulted in the DCs of 18% and 5% for the jetted and non-jetted-RLNLSy1s sets; however, the DCs of 12% and 0% are estimated based on the conservative $F^\eta$-test for the same sets. Thus, regardless of which of the two statistical tests is applied, the INOV DCs for the jetted sample is higher than the non-jetted sample. Since we are using only two steady comparison stars in this study, therefore power-enhanced F-test becomes similar to that of the scaled F-test. However, it has been suggested in Goyal et al. (2013b) that in such a condition power-enhanced F-test does not follow the standard F-distribution. Therefore, for further discussion and conclusion, we will be using our results based upon $F^\eta$-test. Additionally, in the case of powerful quasars/blazars, such DCs characterize non-blazars, including weakly polarised flat-radio-spectrum (i.e., radio-beamed) quasars (Goyal et al., 2013b, Gopal-Krishna & Wiita, 2018). It is further seen from Table 5.3 that a higher DC ( $\sim$30%, i.e., approaching blazar-like values, e.g., see Goyal et al., 2013b) is only exhibited by the $\gamma$-ray detected subset, which consists of 8 RLNLSy1s, all of which belong to the jetted category. This independently reinforces the premise that $\gamma$-ray detected NLSy1 galaxies emit blazar-like compact radio jets, which are likely conspicuous due to relativistic beaming. The absence of $\gamma$-ray detection among the non-jetted RLNLSy1s is interesting because it suggests the non-detection of jets in these sources. It is probably related to relativistic dimming (due to misalignment) rather than due to the jets being beamed but still too small physically to be resolved by VLBA. This inference is based on the currently popular notion that $\gamma$-rays in AGN primarily arise from the vicinity of the central engine, i.e., the base of the jets (e.g., see Neronov et al., 2015). It would then appear that the radio jets in the non-jetted RLNLSy1s either are not strongly beamed or weak. In this context, we also note from Table 5.3, that the DC of 8 J-$\gamma$-RLNLSy1s is 29% while none of the sources in the sample of 7 J-RLNLSy1s has shown INOV. As also noted in Sect. 5.2 that the VLBA-detected jets might not be relativistically beamed, which may result in the non-detection of INOV in these sources. The higher INOV DC in 8 J-$\gamma$-RLNLSy1s may be related to the relativistic beaming of jets in these sources, and the mere presence of jets does not guarantee an INOV detection in NLSy1 galaxies. For instance, J032441.20+341045.0 has the lowest polarization value (e.g., see Table 5.1) and radio loudness value (e.g., see Zhou et al., 2007) among 8 J-$\gamma$-RLNLSy1s but shows strong INOV activity which perhaps could be due to very high jets speed (e.g., see Table 5.1). This supports our above argument that relativistic beaming plays a significant role for INOV in the case of gamma-ray detected radio-loud NLSy1s.

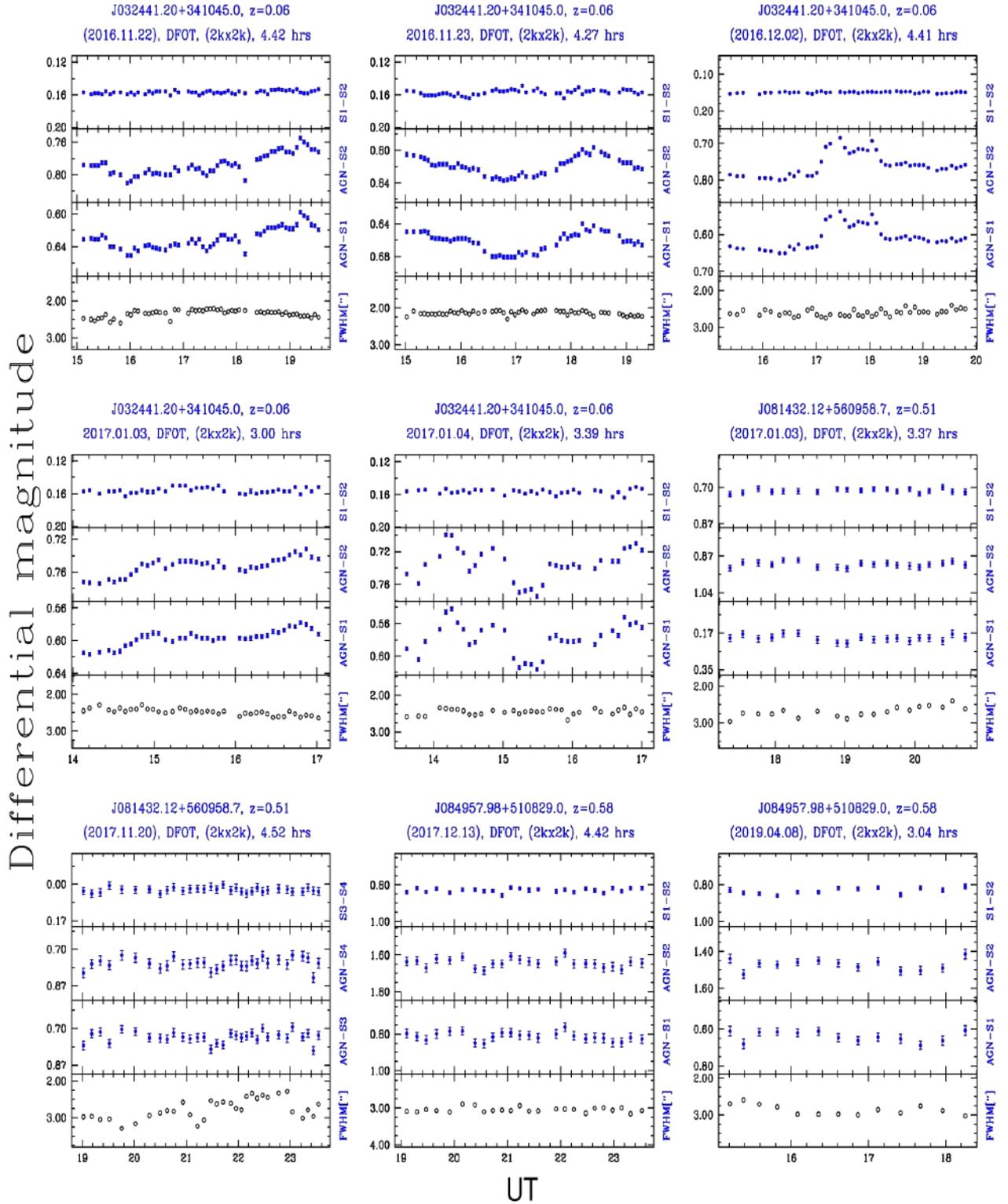

Fig. 5.1 The differential light curves (DLC) for the first 3 jetted-RLNLSy1s from our sample of 15 jetted-RLNLSy1s are shown here. The name of the RLNLSy1 galaxy, the redshift (z), the name of the telescope used, and the duration of the observations are shown on the top of each panel.

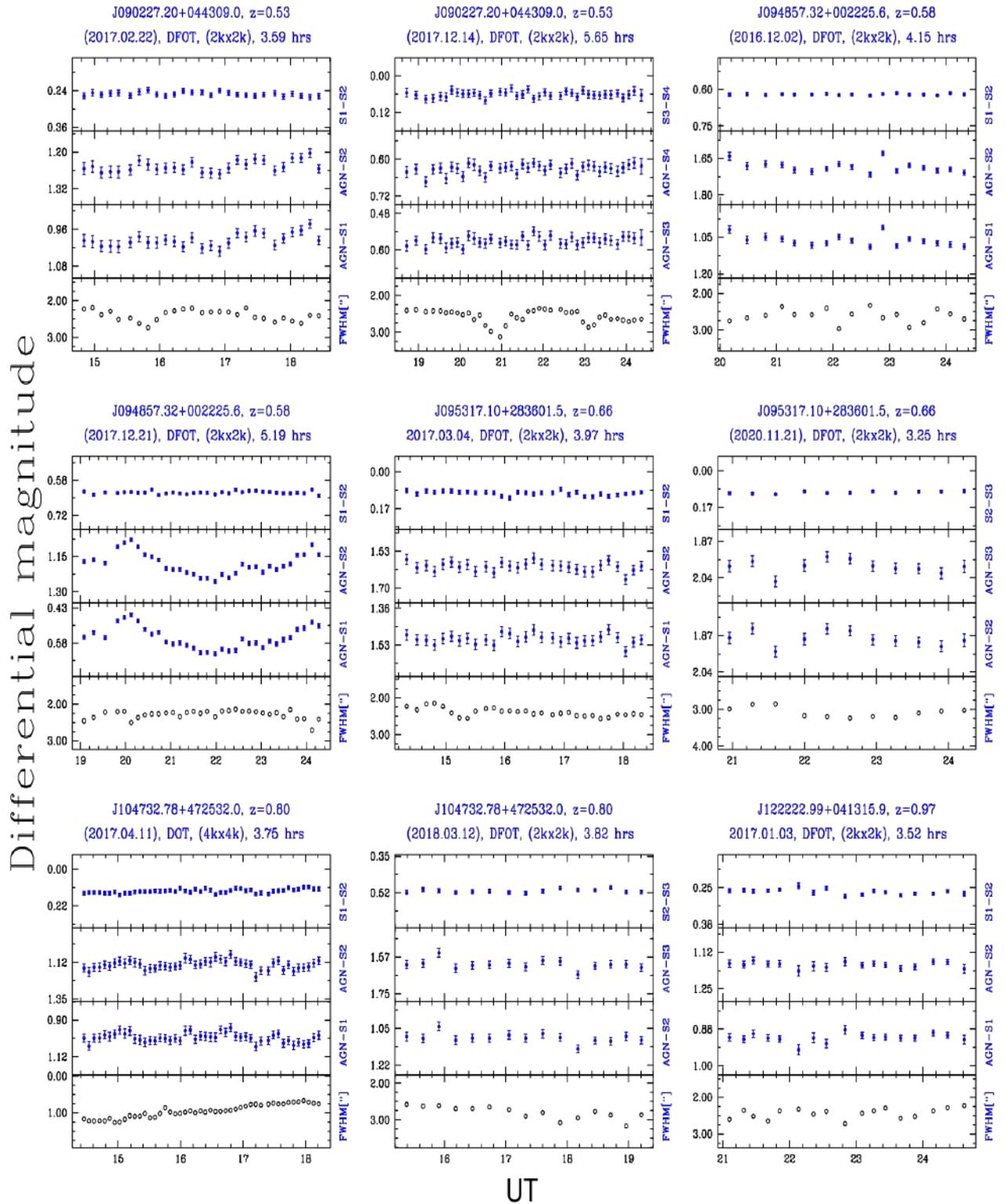

Fig. 5.2 Similar to Fig. 5.1, but for the subsequent 5 jetted-RLNLSy1s from the current sample of 15 jetted-RLNLSy1s.

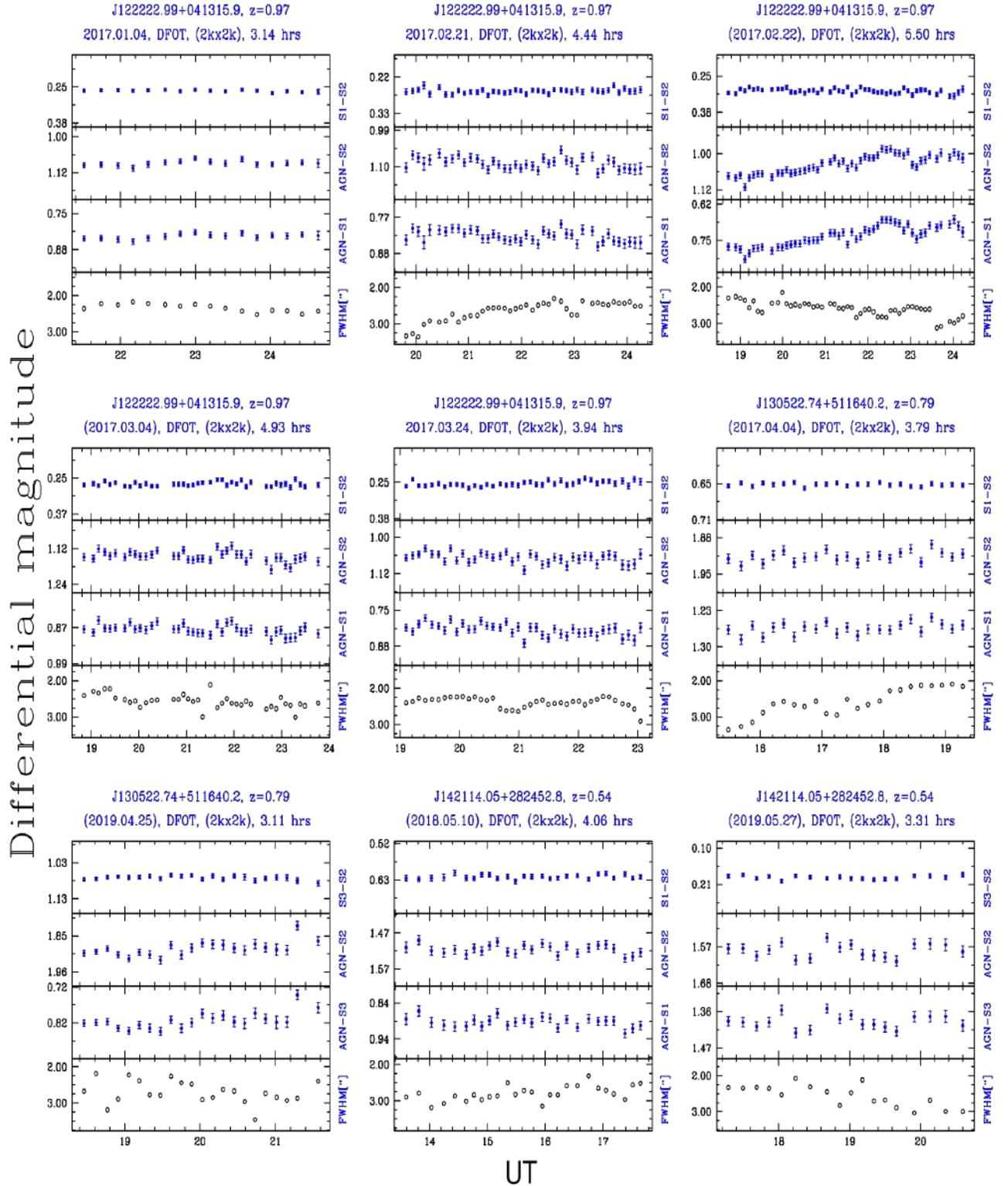

Fig. 5.3 Similar to Fig. 5.1, but for another 2 jetted-RLNLSy1s from the current sample of 15 jetted-RLNLSy1s.

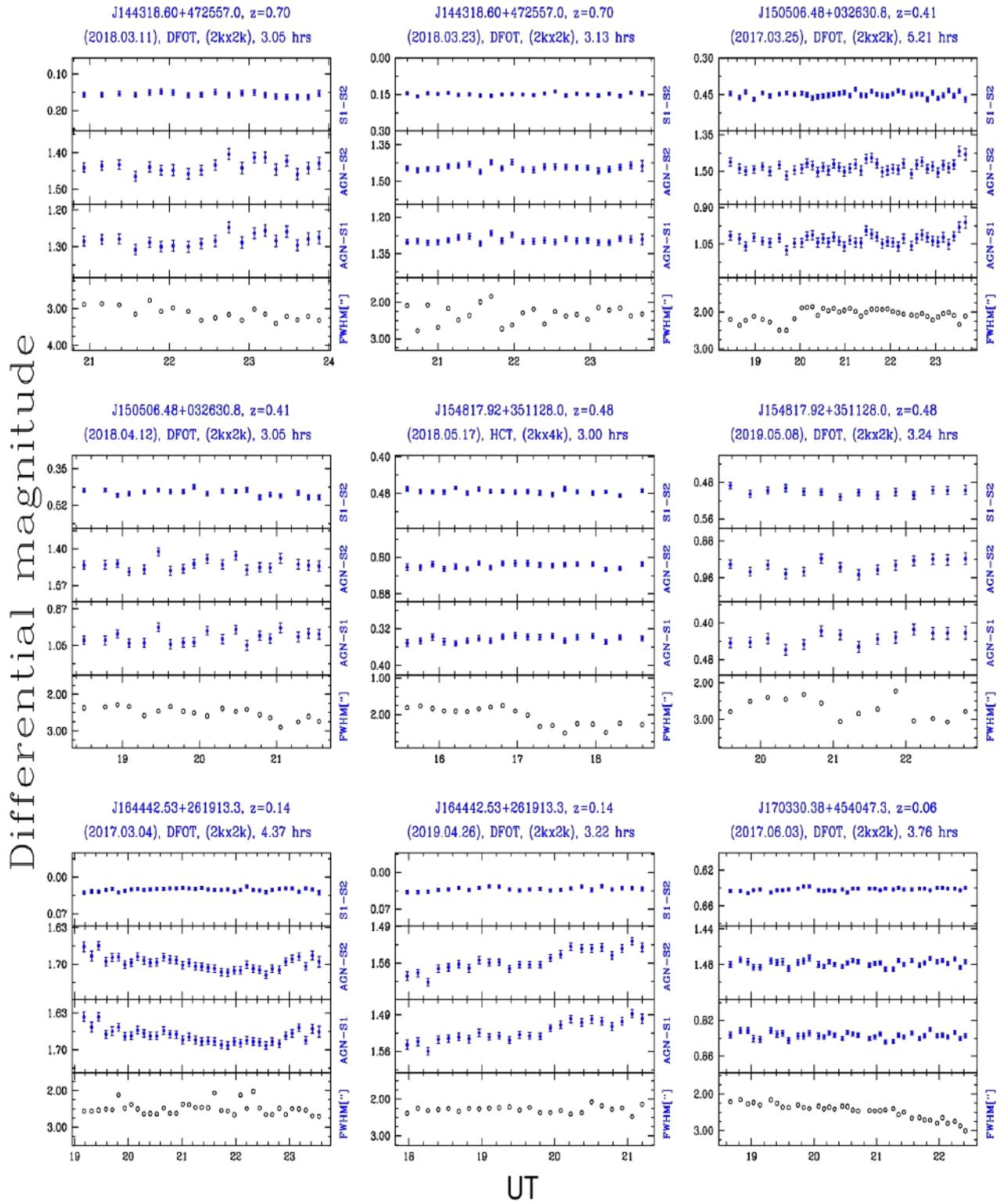

Fig. 5.4 Similar to Fig. 5.1, but for the final 5 jetted-RLNLSy1s from the current sample of 15 jetted-RLNLSy1s.

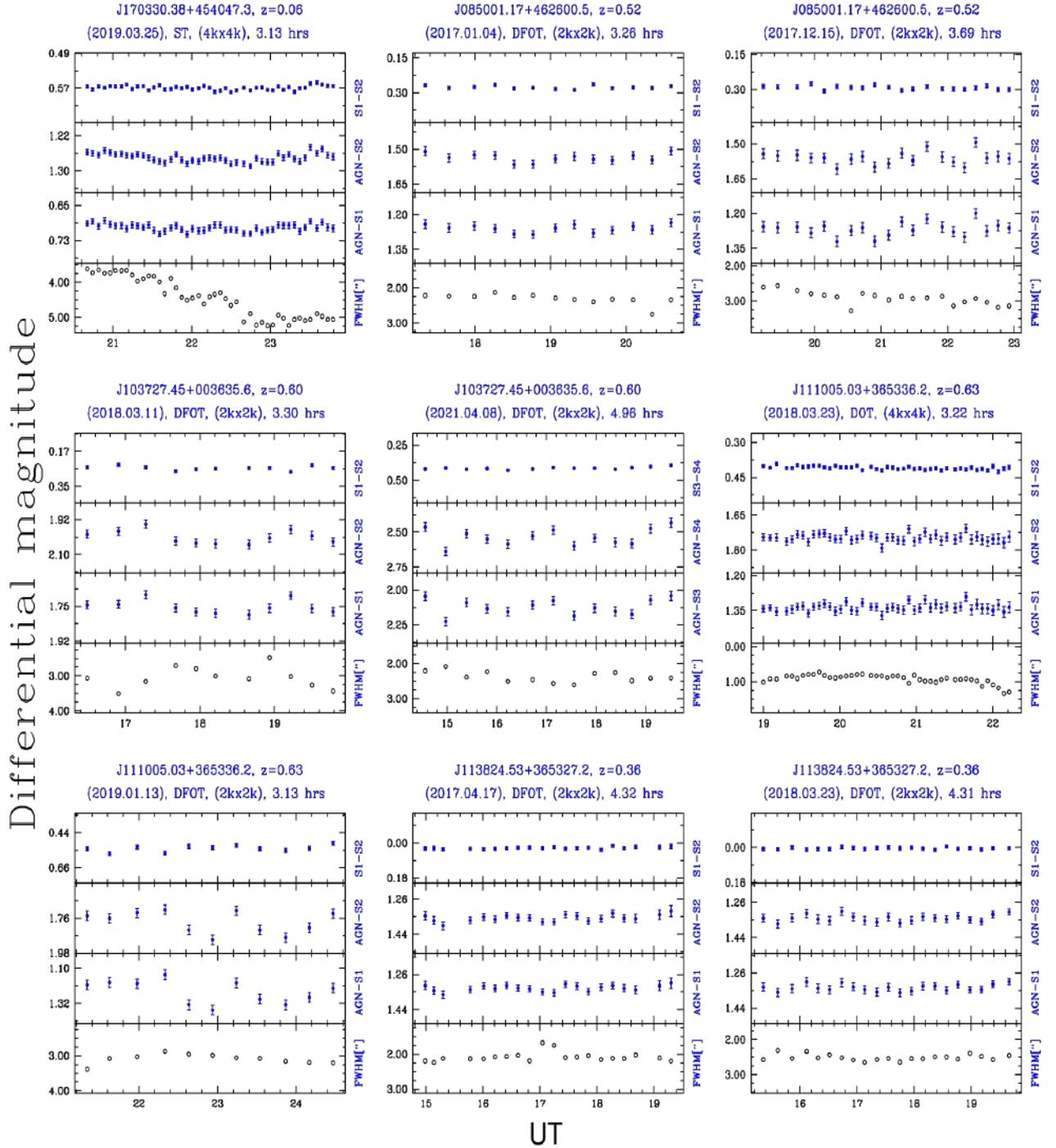

Fig. 5.5 Similar to Fig. 5.1, but for the first 4 non-jetted-RLNLSy1s from our sample of 8 non-jetted-RLNLSy1s.

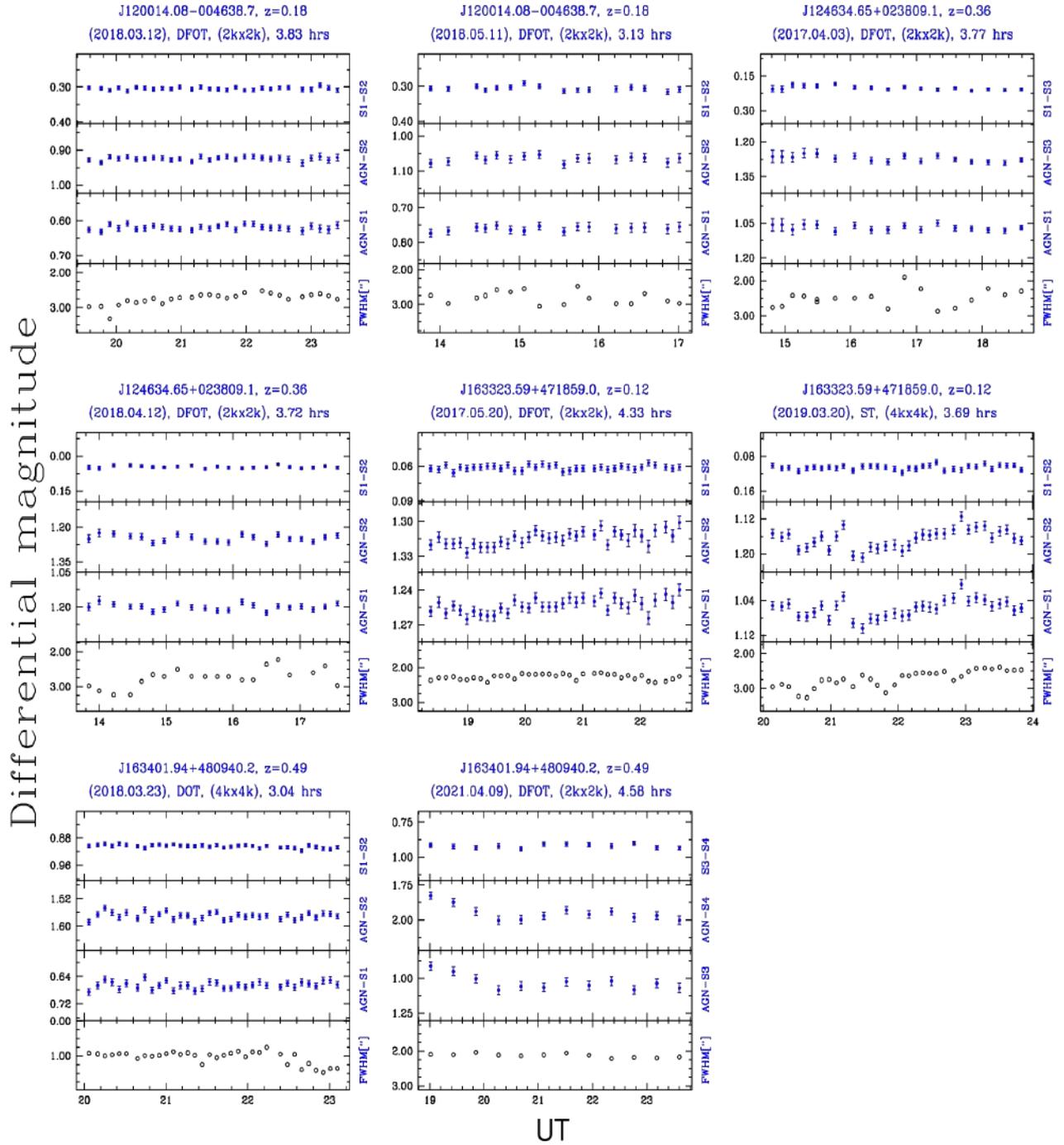

Fig. 5.6 (Continued) DLC for the last 4 non-jetted-RLNLSy1s from our sample of 8 non-jetted-RLNLSy1s.

From Table 5.3, the DC estimates, based upon $F^\eta$-test for the samples of jetted and non-jetted RLNLSy1s, have contrasting differences. Thus, our results suggest that the jetted-RLNLSy1s sample shows higher DC in comparison to the non-jetted-RLNLSy1s. However, as noted above that, instead of the mere presence of a jet, relativistic beaming seems to play a dominant role for INOV in the case of low-luminous high accreting AGNs such as NLSy1 galaxies. As has been emphasized in Sect. 5.1 that the central engine of NLSy1s operates in a regime of a higher accretion rate, and contributions from their host galaxies are prominent in the case of lower redshift sources (see Sect. 5.3.2). However, our method of analysis for INOV detections is very unlikely to be affected by the host galaxy's contributions (hence spurious INOV detection) from the present sources, but due to higher accretion rates of NLSy1s, a relative enhancement in the AGN's optical emission (i.e., thermal component) as compared to its synchrotron emission is expected (e.g., see Zhou et al., 2007, Paliya et al., 2014). Since AGN's optical emission is likely to be less amenable to being variable in comparison to synchrotron emission, resulting from the Doppler-boosted synchrotron jet; therefore the amplitude of INOV is expected to be suppressed by this thermal contamination.

The DC of the jetted and non-jetted RLNLSy1s samples would be more reliable if it were possible to remove thermal contamination from the disc due to their higher Eddington accretion rates in NLSy1s. A relatively lower DC of the jetted-RLNLSy1s sample could be attributed to either the sub-luminal speed of jets (e.g., see Ojha et al., 2019) or their primarily misaligned relativistic jets towards the observer line of sight (Berton et al., 2018). Therefore, we divided our sample of jetted-RLNLSy1s into two subsamples: J-$\gamma$-RLNLSy1s (8 sources) and J-RLNLSy1s (7 sources), based on their detection in $\gamma$-ray by *Fermi*-LAT. The detection of $\gamma$-ray emission supports the presence of Doppler boosted relativistic jets.

To further confirm the above scenario, we compiled the apparent jet speed of our J-$\gamma$-RLNLSy1s from the literature. We found that out of 8 members of J-$\gamma$-RLNLSy1s, six members have available apparent jet speeds. These six J-$\gamma$-RLNLSy1s are: J032441.20+341045.0, J084957.98+510829.0, J094857.32+002225.6, J150506.48+032630.8, with $v_{app}/c$ of $9.1 \pm 0.3$, $6.6 \pm 0.6$, $9.7 \pm 1.1$, $0.1 \pm 0.2$, respectively (e.g., see Lister et al., 2019), and J122222.99+041315.9, J164442.53+261913.3, with $v_{app}/c$ of $0.9 \pm 0.3$, $>1.0$, respectively (e.g., see Lister et al., 2016, and Doi et al., 2012). Thus with the current INOV study and available jet speed of J-$\gamma$-RLNLSy1s, a correlated INOV detection with the superluminal motion in the radio jet is inferred, except for J084957.98+510829.0. Although unfortunately, we could not find any INOV in both $> 3$ hrs long monitoring sessions of the source, J084957.98+510829.0 in the present study, this source has shown in the past a fading by $\sim 0.2$ mags in its INOV study within just $\sim 15$ minutes during its high $\gamma$-ray active phase (see figure 6 of Maune et al., 2014). Other than this instance, this source previously had also shown significant INOV in all its six intra-night sessions (e.g., see Paliya et al., 2016). Therefore, the non-detection of INOV in our current study may be due to currently undergoing the quiescent $\gamma$-ray phase of this source. Nonetheless, the above correlation could be firmly established once a more comprehensive INOV database and apparent jet speeds of J-$\gamma$-RLNLSy1s become available.

On the other hand, in the case of RLNLSy1s (with $R > 100$), where most of the electromagnetic emission (radio, optical, X-ray, and $\gamma$-ray) supposedly comes from their jets, it is suggested that more massive black holes are more amenable to launch powerful relativistic jets (Urry et al., 2000, Hyvönen et al., 2007, Chiaberge & Marconi, 2011, Olguín-Iglesias et al., 2016). Therefore, we compared the median black hole masses of jetted-RLNLSy1s and non-jetted-RLNLSy1s samples, derived based upon single-epoch optical spectroscopy virial method (e.g., see second last column of Table 5.1). This has resulted in a nominal difference with the median values of log ($M_{BH}/M_\odot$) of *7.72* and *7.42* for the jetted-RLNLSy1s and non-jetted-RLNLSy1s samples (see the last column of Table 5.3), respectively. A contrasting difference has not been found between the black hole masses of the two sets, which may be either due to smaller sample sizes or due to the use of the single-epoch optical spectroscopy virial method, which is suggested to have a systematic underestimation while estimating black hole masses (Decarli et al., 2008, Marconi et al., 2008, Calderone et al., 2013, Viswanath et al., 2019, Ojha et al., 2020b). Therefore, to firmly establish the above scenario, estimation of black hole masses of a large sample of jetted and non-jetted RLNLSy1s is needed with the method that is less likely to be affected by underestimation of black hole masses such as the standard Shakura-Sunyaev accretion-disc model method (Calderone et al., 2013, Viswanath et al., 2019).

## 5.6 Conclusions

To quantify the role of the absence/presence of radio jets for INOV in the case of RLNLSy1s, we have carried out a systematic INOV study based on an unbiased sample of 23 RLNLSy1s. Among them, 15 RLNLSy1s have confirmed detection of jets (jetted), and the remaining 8 RLNLSy1s have no detection of jets (non-jetted) with the VLBA observations. Our study spans 53 sessions of a minimum 3-hour duration each. The main conclusions from this work are as follows:

1. We estimated the INOV DC based upon $F^\eta$-test for the sample of jetted RLNLSy1s to be 12%, however, none of the sources showed INOV in the sample of non-jetted RLNLSy1s, at the 99% confidence level for a typical threshold $\psi > 3$%.

2. Among the jetted RLNLSy1s, the DC for jetted $\gamma$-ray detected RLNLSy1s is found to be 29% in contrast to null INOV detection in the case of non-$\gamma$-ray detected RLNLSy1s. It suggests that the INOV detection in RLNLSy1 galaxies does not solely depend on the presence of radio jets, but relativistic beaming plays a dominant role.

3. The predominance of beamed jet for INOV is also supported in our study based on the correlation of the INOV detection with the apparent jet-speed available for 6 jetted $\gamma$-ray detected RLNLSy1s.

4. The higher DC of $\sim 30\%$, approaching blazar-like DC, is only exhibited by the $\gamma$-ray detected subset, suggests that $\gamma$-ray detected NLSy1 galaxies emit blazar-like compact radio jets in which relativistic jet motion (speed) and/or small jet's angles to the observer's line of sight seems to be correlated with the presence of INOV.

For further improvement, it will be helpful to enlarge the sample and conduct similar systematic INOV studies for other subclasses of AGN with and without a confirmed jet, along with the proper estimate of apparent jet speeds.

# Chapter 6

# A comparison of physical parameters of NLSy1 and BLSy1 galaxies [1]

NLSy1 galaxies are believed to be younger versions of AGN that have different physical characteristics from the rest of the AGN population. In this chapter, we compare the physical characteristics of a uniform sample of 144 NLSy1 and 117 BLSy1 galaxies. Both samples have similar redshifts and luminosities and their optical and X-ray spectra are publicly accessible. We use ten parameters derived from fitting the optical spectra and the soft X-ray photon indices as another parameter to perform direct correlation analysis and Principal Component Analysis (PCA). Our goal is to determine if the NLSy1 galaxies follow the same correlations as the general quasar population, even though they have noticeable differences in their observational properties. We also use the line shape parameters for the H$\beta$ emission line, such as the asymmetry and kurtosis indices, to describe the sample and to see if the NLSy1 galaxies have any specific type of asymmetries in their emission profiles and if the asymmetry in the emission lines can indicate outflows in the inner regions of Active Galactic Nuclei. Lastly, we use PCA to examine if the NLSy1 galaxies are a subset of BLSy1 galaxies or if they have a different parameter space.

---

## 6.1 Introduction

Seyfert-1 type galaxies have a lower luminosity nucleus as compared to quasars. These Active Galactic Nuclei (AGN) exhibit both broad and forbidden narrow emission lines in their spectra. The Seyfert class of AGN has been broadly divided into various subcategories based on the strength or the Full Width at Half Maximum (FWHM) of the Balmer $H\beta$ emission line (e.g., see Netzer, 2015, for a review). Narrow-line Seyfert 1 galaxies (NLSy1 galaxies) are a special subclass of AGN having narrower broad Balmer line widths with FWHM of broad $H\beta$ emission line $< 2000$ km s$^{-1}$ (Goodrich, 1989), a small flux ratio of the [O III] $\lambda 5007$ to $H\beta$ line ($[OIII]/H\beta < 3$) (Osterbrock & Pogge, 1985), stronger optical Fe II emissions. These AGN show usually steeper soft X-ray spectra and rapid X-ray and sometimes optical flux variability (see Rakshit et al., 2017, Ojha et al., 2021, and references therein). It is assumed that these properties are due to the central supermassive black hole (SMBH) being less massive but accreting at a very high rate (Sulentic et al., 2000b, Boroson, 2004). It has been proposed that the NLSy1 galaxies are *younger* versions of Broad-line Seyfert 1 galaxies (BLSy1) galaxies only and can be assumed to be in different evolutionary stages (see Mathur, 2000a, Williams et al., 2018). They have been known to have higher accretion rates as compared to the BLSy1 galaxies, which makes it one of the defining parameters in the classification (Pounds et al., 1994, Collin & Kawaguchi, 2004, Sulentic et al., 2000b).

The prominent emission lines arising from the Broad Line Region (BLR) are broadened to line widths of thousands of km s$^{-1}$ due to gas in virial motion around the central SMBH (Gaskell, 2000, Peterson & Horne, 2004). Direct observation of the BLR is desired in order to understand the dynamics of gas in these regions, but it has been a difficult challenge, and only recently Gravity Collaboration et al. (2018, 2020, 2021) have spatially resolved the inner regions of a few AGN. Reverberation mapping (RM), which works in the time domain instead of the spatial domain, has been a method of choice to explore the dynamics of matter in the BLR (see Bahcall et al., 1972, Blandford & McKee, 1982, Peterson & Horne, 2004, Bentz et al., 2010, Grier et al., 2012, De Rosa et al., 2015, Du et al., 2016, 2018, Lira et al., 2018, Wang et al., 2019).

The RM measurements have yielded a strong relationship between the size of the BLR ($R_{BLR}$) and the optical luminosity measured at 5100 Å ($L_{5100}$) (Kaspi et al., 2000b, Bentz et al., 2009). The velocity resolved reverberation mapping has enabled a more comprehensive understanding of these regions with the construction of velocity delay maps (Grier et al., 2013, Xiao et al., 2018). The dynamical modeling of a few AGN suggests that the gas in the BLR is dominated by Keplerian motion with traces of outflow and inflow (Pancoast et al., 2014). However, the origin and the geometry of the BLR so far remain a mystery, and several models have been explored and debated for a very long time. For instance, the disk wind model (Gaskell, 2000, Gaskell & Goosmann, 2013, Baskin & Laor, 2018, Matthews et al., 2020) has been successful in explaining the outflowing profiles seen in the CIV emission line (Gaskell & Goosmann, 2016). The failed radiation-driven winds model proposed by Czerny et al. (2017) states that the material rising from the disk is expected to emit the low ionization lines such as Mg II and H$\beta$ emission lines. This material spends more time in the rising phase than in the falling phase, hence the effect might appear like an outflow, although no actual outflow exists. Another model has been proposed by Wang et al. (2017a) in which clumps of cloud from the dusty torus, tidally disrupted by the central SMBH, can give rise to the emission lines in the BLR.

While the known number of AGN extends into hundreds of thousands (?), the reverberation mapping information is available for only approximately 120 AGN (see Bentz & Katz, 2015, for a comprehensive database of reverberation mapped AGN). Much of our understanding has relied on the statistical analysis of a sample of AGN constrained by various limits; to infer the physical properties of the inner regions of these AGN. A remarkable work was done by Boroson & Green (1992), where they performed a Principal Component Analysis (PCA) on the properties derived from X-ray, optical, and radio wavelength data for a set of 87 AGN. With this dataset, they derived that the Eigenvector 1 (E1), which is driven by the anti-correlation of the ratio of equivalent width (EW) of Fe II lines in the optical band and the FWHM of $H\beta$ emission line ($R_{fe}$) is the primary cause of variability in the parameters. Sulentic et al. (2000b) and later subsequent works e.g., Zamfir et al. (2010) established the foundations of the four-dimensional EV1 (4DE1) formalism, which includes FWHM of $H\beta$ and $R_{fe}$ as two of the main components. These two quantities are related to the SMBH mass and the Eddington ratio and the ongoing results have yielded a so-called quasar main sequence (see Shen & Ho, 2014, Sulentic & Marziani, 2015, Marziani et al., 2018a, and references therein).

One of the important consequences of such statistical studies on a set of AGN has been that the prominent broad emission lines in a few Type-1 AGN show asymmetric profiles (see Sulentic, 1989, Corbin, 1995, Brotherton, 1996, Marziani et al., 1996, Zamfir et al., 2010, Negrete et al., 2018, Wolf et al., 2020). The asymmetric emission line profiles raise essential questions in our understanding of the gas in the BLR itself, like whether the gas is in virial motion around the SMBH or is it in the form of an outflowing/inflowing motion related to the accretion disk. A red asymmetric emission profile may indicate the presence of inflowing gas rotating nearer to the SMBH, while the blue asymmetric emission profiles may indicate outflow being induced by the presence of a disk-wind structure (Zamfir et al., 2010). The asymmetry in the Balmer emission lines yields crucial information about the gas motion in the vicinity of the SMBH, and hence proper understanding of the cause of asymmetric profiles is imperative. While in the high ionization line like the C IV emission line, the blue ward asymmetry is explained by the two-component BLR model (e.g., see Gaskell & Goosmann, 2016, Sulentic et al., 2017), a similar analogy is not used for low ionization lines such as the $H\beta$ emission line. The characterization of the AGN in terms of their emission line shapes and shifts and their correlation with observational and physical parameters such as the FWHM of $H\beta$ emission line and accretion rate has been demonstrated in Zamfir et al. (2010) where they conclude that the AGN with $H\beta$ FWHM $\geq$ 4000 km s$^{-1}$ show different characteristics than the ones with the lower value of FWHM.

In the type 1 AGN family, the NLSy1 galaxies are located at the extreme end of the population and appear as a class of AGN with high accretion rates, which may be a defining criterion as compared to the traditional FWHM based classification of these AGN (Marziani et al., 2018b). In the recent works, statistical study of the properties of NLSy1 galaxies with a control sample of BLSy1 galaxies is explored (e.g., see Nagao et al., 2001, Zhou et al., 2006, Zhou & Zhang, 2010, Jin et al., 2012, Xu et al., 2012, Cracco et al., 2016, Ojha et al., 2020b, Waddell & Gallo, 2020); however, the line shape parameters have not been covered for a large sample. It remains to be understood what fraction of the NLSy1 galaxies shows asymmetric emission line profiles and possible causes behind this phenomenon. It is essential to know how emission line asymmetries correlate with other observational and physical parameters and whether this behavior is peculiar compared to the general type 1 AGN population. While high accretion rates in NLSy1 galaxies make them stand out from the general population of Seyfert galaxies, it becomes imperative to understand the variation in these properties for a representative population of these galaxies. A comparison of the properties of NLSy1 galaxies and a control sample of BLSy1 galaxies in the context of their BLR geometry characterized by the emission line features such as the asymmetry and kurtosis helps understand the relationship between the two types of AGN. This is also important to understand whether the NLSy1 galaxies form a particular subclass of the BLSy1 galaxies.

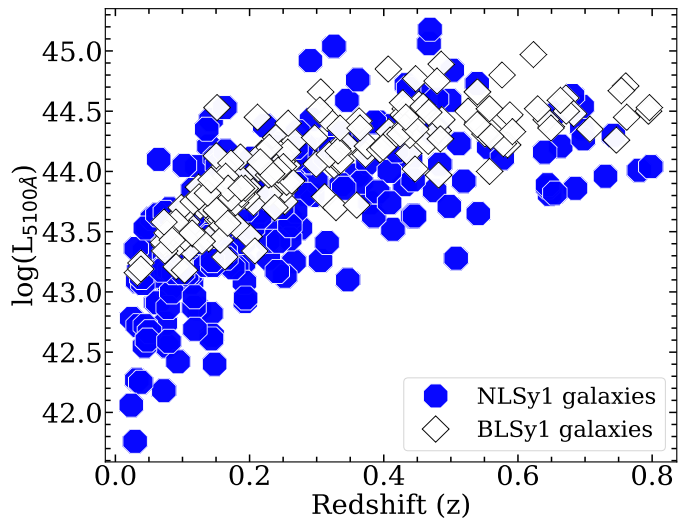

Fig. 6.1 The luminosity- redshift (L-z) distribution for the entire sample of 221 NLSy1 and 154 BLSy1 galaxies being used for this work. This sample is limited to a redshift of 0.8 due to the upper wavelength limit of $H\beta$ emission line in SDSS. The redshift values were obtained from the values available in the SDSS database while the $L_{5100}$ Å values were obtained using the spectral fitting procedure outlined in Sec. 6.3.

In this work, we have performed a statistical study to understand the observational and physical parameters responsible for driving the variations in the population of NLSy1 and BLSy1 galaxies. This chapter is organized as follows: we describe the sample in Sec. 6.2, which is followed by the analysis in Sec. 6.3 where we present the fitting procedure and the estimation of physical parameters. The results obtained in this work are presented in Sec. 6.4. The discussion and interpretation of the results are presented in Sec. 6.5, and we conclude with a summary in Sec. 6.6.

## 6.2 Sample

Our present sample comprising of both NLSy1 galaxies and BLSy1 galaxies, matching in redshift (see Figure 6.1) has been drawn from Ojha et al. (2020b) where they have presented a detailed comparative study of a redshift-matched sample of 221 NLSy1 galaxies and 154 BLSy1 galaxies having optical observation in Sloan Digital Sky Survey (SDSS) and X-ray observation either in Röntgensatellit (ROSAT[2]) or in X-ray Multi-Mirror Mission (XMM[3]). Firstly, we obtained the optical spectrum of all the sources from the sample within a position offset of 3 arcseconds from the recent data release 16 of SDSS (SDSS-DR16)[4]. More information about the SDSS database of spectra is available in York et al. (2000), Abazajian et al. (2009), Shen et al. (2011), Rakshit et al. (2017), Lyke et al. (2020). Here, we avoided the repetition of any source with multi-epoch spectra by retaining only the epoch with the highest signal-to-noise ratio (SNR). Furthermore, to get precise emission line detection which is important for fitting the spectra, we put a minimum limit of SNR $\geq$ 5. This criterion was satisfied by 206 NLSy1 galaxies and 153 BLSy1 galaxies. The single epoch optical spectrum of each source was analyzed using the publicly available code PyQSOFIT[5]. The spectra were dereddened using the dust reddening maps available in Schlegel et al. (1998)[6] and brought to the rest frame using the redshift values available in the headers of the individual spectrum.

---

[2] https://heasarc.gsfc.nasa.gov/docs/rosat/rosat.html
[3] https://www.cosmos.esa.int/web/xmm-newton

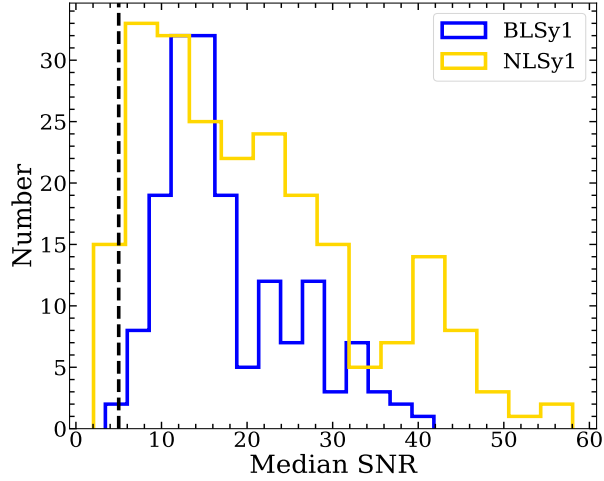

Fig. 6.2 Distribution of median Signal to Noise Ratio (SNR) for the parent sample of 221 NLSy1 and 154 BLSy1 galaxies obtained from (Ojha et al., 2020b). The vertical dashed red line is drawn at SNR = 5, rejecting 15 NLSy1 galaxies and 1 BLSy1 galaxy from the parent sample, with SNR lower than this limit.

## 6.3 Analysis

### 6.3.1 Spectral fitting

We used the publicly available code PyQSOFit (Guo et al., 2018) to decompose the spectra into multiple components. Initially, the continuum model was prepared using

---

[4]https://dr16.sdss.org/
[5]https://github.com/legolason/PyQSOFit
[6]https://github.com/kbarbary/sfdmap

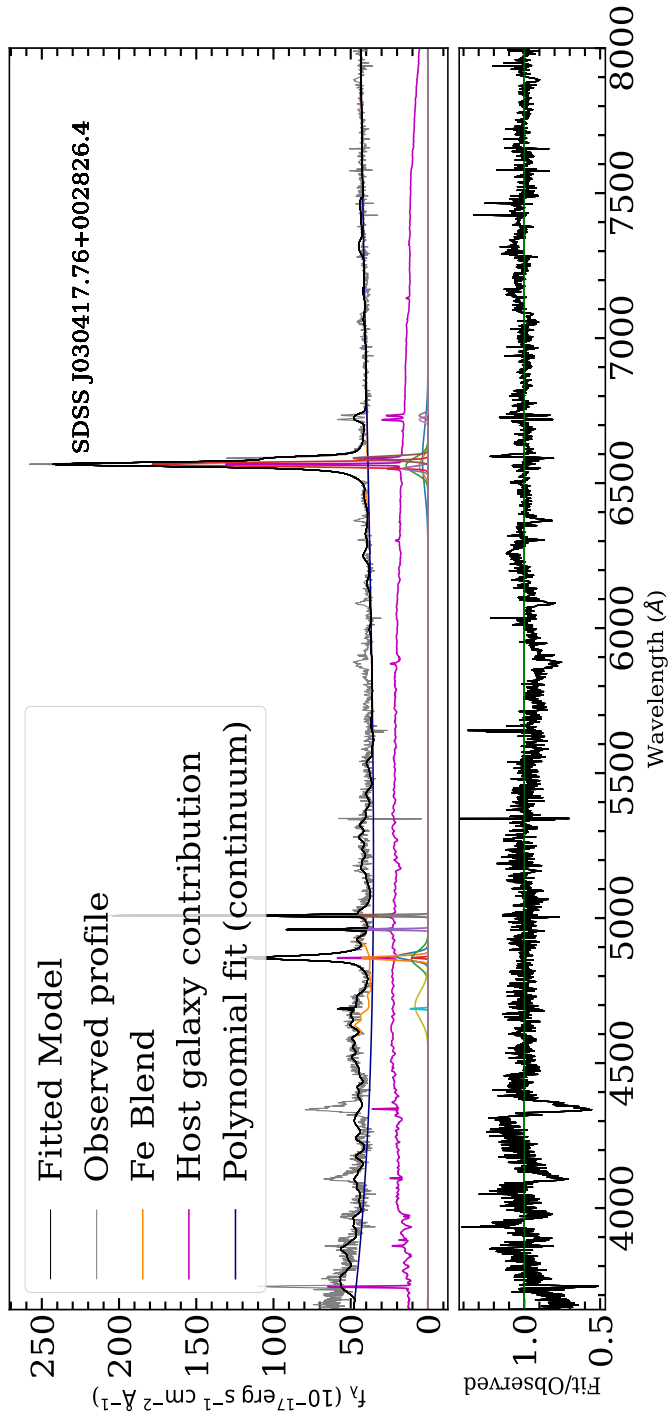

Fig. 6.3 Demonstration of the fitting procedure: The host galaxy has been decomposed using available templates in Yip et al. (2004), and the Fe contamination has been removed using the templates available in Boroson & Green (1992). The continuum has been fit using a polynomial and a power law component, in order to fit the emission lines, a combination of up to 3 Gaussian profiles was used: a narrow component with FWHM $\leq$ 1200 km s$^{-1}$, a broad component with FWHM $\leq$ 2200 km s$^{-1}$ for the NLSy1 galaxies and a very broad component $\geq$ 2200 km s$^{-1}$ in some cases. The limit of 2200 km s$^{-1}$ was removed in the case of BLSy1 galaxies. The [OIII] doublet is fitted using a single Gaussian component with its width tied to the $H\beta$ narrow component.

Table 6.1 The table of properties obtained for the sample of 144 NLSy1 and 117 BLSy1 galaxies.

| SDSS Name[a] | $z$[b] | FWHM($H\beta$)[c] | EW($H\beta$)[d] | [OIII]/$H\beta$[e] | $H\alpha/H\beta$[f] | $R_{Fe}$[g] | $\log(L_{5100\text{Å}})$[h] | $M_{BH}$[i] | $\log(R_{Edd})$[j] | AI[k] | KI[l] | $\Gamma_X$[m] |
|---|---|---|---|---|---|---|---|---|---|---|---|---|
| \multicolumn{13}{c}{NLSy1 galaxies} | | | | | | | | | | | | |
| J001137.20-144160.0 | 0.1319 | 705 ± 152 | 26.87 ± 7.09 | 0.36 ± 0.13 | 4.65 ± 0.31 | 1.35 ± 0.50 | 43.853 ± 0.002 | 6.64 ± 0.30 | 0.67 ± 0.29 | 0.404 ± 0.025 | 0.114 ± 0.025 | 2.52 ± 0.18 |
| J003859.28-005450.4 | 0.4137 | 1537 ± 183 | 70.08 ± 14.72 | 0.34 ± 0.10 | 1.79 ± 0.25 | 0.50 ± 0.15 | 43.915 ± 0.005 | 7.32 ± 0.17 | 0.03 ± 0.01 | 0.174 ± 0.001 | 0.321 ± 0.038 | 2.53 ± 0.21 |
| J011911.52-104532.4 | 0.1253 | 1478 ± 297 | 40.28 ± 13.16 | 0.00 ± 0.00 | 0.87 ± 0.22 | 0.63 ± 0.29 | 43.241 ± 0.006 | 7.26 ± 2.84 | -0.27 ± 1.10 | 0.129 ± 0.517 | 0.265 ± 0.534 | 2.97 ± 0.66 |
| J014644.88-004044.4 | 0.0824 | 1699 ± 184 | 44.76 ± 7.23 | 0.53 ± 0.12 | 2.85 ± 0.19 | 1.32 ± 0.30 | 43.545 ± 0.002 | 7.39 ± 0.15 | -0.014 ± 0.003 | 0.316 ± 0.034 | 2.92 ± 0.31 | |
| J014904.56-125746.8 | 0.7305 | 1454 ± 336 | 153.93 ± 44.27 | 0.00 ± 0.00 | 0.09 ± 0.04 | 0.15 ± 0.15 | 44.583 ± 0.015 | 7.43 ± 0.33 | 0.41 ± 0.20 | 0.229 ± 0.106 | 0.345 ± 0.080 | 2.12 ± 0.47 |
| J020337.20-051406.0 | 0.5193 | 1746 ± 242 | 74.09 ± 4.49 | 0.00 ± 0.00 | 2.03 ± 0.14 | 0.66 ± 0.06 | 44.160 ± 0.003 | 7.47 ± 0.20 | 0.04 ± 0.01 | 0.014 ± 0.004 | 0.228 ± 0.032 | 1.58 ± 0.58 |
| J020853.28-043354.0 | 0.5563 | 1513 ± 439 | 56.40 ± 8.48 | 0.28 ± 0.06 | 2.15 ± 0.32 | 0.96 ± 0.20 | 44.345 ± 0.002 | 7.38 ± 0.41 | 0.26 ± 0.15 | 0.040 ± 0.023 | 0.296 ± 0.086 | 2.84 ± 0.84 |
| J021329.28-051138.4 | 0.443 | 705 ± 254 | 8.39 ± 3.24 | 3.22 ± 1.75 | 6.93 ± 3.90 | 3.54 ± 1.93 | 44.080 ± 0.006 | 6.67 ± 0.57 | 0.79 ± 0.57 | 0.347 ± 0.017 | 0.347 ± 0.125 | 2.21 ± 0.80 |
| J021803.60-043337.2 | 0.3796 | 968 ± 506 | 68.57 ± 11.15 | 0.00 ± 0.00 | 1.61 ± 0.16 | 0.75 ± 0.17 | 44.061 ± 0.005 | 6.94 ± 0.74 | 0.50 ± 0.53 | -0.148 ± 0.155 | 0.325 ± 0.170 | 2.84 ± 0.19 |
| J022452.32-040520.4 | 0.6952 | 859 ± 100 | 56.08 ± 12.07 | 0.39 ± 0.12 | 0.00 ± 0.00 | 1.23 ± 0.38 | 44.589 ± 0.010 | 6.97 ± 0.16 | 0.87 ± 0.22 | -0.029 ± 0.007 | 0.299 ± 0.035 | 2.61 ± 0.27 |
| J022928.32-051124.0 | 0.3068 | 1338 ± 175 | 35.45 ± 2.62 | 0.42 ± 0.04 | 4.46 ± 0.45 | 1.63 ± 0.17 | 44.573 ± 0.001 | 7.35 ± 0.19 | 0.48 ± 0.13 | -0.121 ± 0.032 | 0.313 ± 0.041 | 2.97 ± 0.23 |
| ... | ... | ... | ... | ... | ... | ... | ... | ... | ... | ... | ... | ... |
| \multicolumn{13}{c}{BLSy1 galaxies} | | | | | | | | | | | | |
| J002113.20-020115.6 | 0.7621 | 9759 ± 1578 | 116.29 ± 19.26 | 0.39 ± 0.09 | 0.00 ± 0.00 | 0.30 ± 0.07 | 44.644 ± 0.010 | 9.11 ± 0.23 | -1.21 ± 0.41 | 0.548 ± 0.177 | 0.568 ± 0.092 | 1.83 ± 0.35 |
| J011254.96-000314.4 | 0.2389 | 6561 ± 347 | 76.43 ± 13.97 | 0.14 ± 0.04 | 3.29 ± 0.17 | 0.71 ± 0.18 | 44.397 ± 0.002 | 8.67 ± 0.07 | -0.99 ± 0.11 | 0.034 ± 0.004 | 0.290 ± 0.015 | 2.60 ± 0.24 |
| J012254.72-010108.4 | 0.1994 | 3506 ± 469 | 23.90 ± 5.20 | 0.36 ± 0.11 | 2.94 ± 0.27 | 0.73 ± 0.22 | 43.548 ± 0.003 | 8.02 ± 0.19 | -0.87 ± 0.24 | -0.048 ± 0.013 | 0.299 ± 0.040 | 1.79 ± 0.38 |
| J013517.52-001940.8 | 0.3119 | 7242 ± 471 | 36.79 ± 1.02 | 0.31 ± 0.01 | 4.04 ± 0.13 | 0.19 ± 0.01 | 43.931 ± 0.003 | 8.67 ± 0.09 | -1.31 ± 0.18 | 0.012 ± 0.001 | 0.340 ± 0.022 | 2.31 ± 0.30 |
| J014959.28-125656.4 | 0.432 | 4076 ± 275 | 87.32 ± 23.20 | 0.19 ± 0.07 | 0.80 ± 0.30 | 0.44 ± 0.06 | 44.670 ± 0.003 | 8.36 ± 0.10 | -0.44 ± 0.06 | 0.095 ± 0.013 | 0.311 ± 0.021 | 2.24 ± 0.19 |
| J020744.16-060957.6 | 0.6496 | 9877 ± 699 | 102.77 ± 9.19 | 0.26 ± 0.03 | 0.00 ± 0.00 | 0.44 ± 0.06 | 44.537 ± 0.003 | 9.07 ± 0.10 | -1.28 ± 0.19 | 0.477 ± 0.068 | 0.556 ± 0.039 | 2.07 ± 0.46 |
| J020840.56-062718.0 | 0.092 | 2509 ± 42 | 31.55 ± 4.61 | 0.00 ± 0.00 | 3.57 ± 0.13 | 1.10 ± 0.23 | 43.449 ± 0.001 | 7.72 ± 0.02 | -0.63 ± 0.02 | -0.039 ± 0.001 | 0.315 ± 0.005 | 2.21 ± 0.11 |
| J021139.12-042606.0 | 0.4856 | 3099 ± 193 | 62.47 ± 2.69 | 0.03 ± 0.00 | 2.51 ± 0.12 | 1.00 ± 0.06 | 45.025 ± 0.001 | 8.42 ± 0.09 | -0.02 ± 0.00 | 0.004 ± 0.001 | 0.284 ± 0.018 | 2.78 ± 0.19 |
| J021318.24-130643.2 | 0.4076 | 8659 ± 581 | 145.32 ± 17.39 | 0.17 ± 0.03 | 1.00 ± 0.06 | 0.29 ± 0.05 | 44.527 ± 0.004 | 8.95 ± 0.10 | -1.17 ± 0.16 | 0.030 ± 0.004 | 0.300 ± 0.020 | 2.07 ± 0.24 |
| J021434.80-042243.2 | 0.4369 | 11843 ± 582 | 159.30 ± 28.22 | 0.53 ± 0.13 | 0.00 ± 0.00 | 0.43 ± 0.11 | 44.210 ± 0.006 | 9.14 ± 0.07 | -1.60 ± 0.17 | -0.021 ± 0.002 | 0.306 ± 0.015 | 2.48 ± 0.14 |
| J021820.40-050426.4 | 0.6492 | 3440 ± 428 | 51.72 ± 10.87 | 0.30 ± 0.09 | 0.00 ± 0.00 | 0.00 ± 0.00 | 44.586 ± 0.009 | 8.18 ± 0.18 | -0.34 ± 0.09 | -0.036 ± 0.009 | 0.312 ± 0.039 | 1.92 ± 0.26 |
| ... | ... | ... | ... | ... | ... | ... | ... | ... | ... | ... | ... | ... |

Note: Only a portion of the table is available here to show the form and content. The entire table is available in the Appendix section (see Appendix B).

[a] Obtained from SDSS SAS.
[b] Source redshift based on single epoch spectra available in SDSS DR16.
[c] The broad component of $H\beta$ emission profile in the units of km s$^{-1}$.
[d] Equivalent width of $H\beta$ emission line.
[e] [OIII]/$H\beta$ emission line flux ratio.
[f] $H\alpha/H\beta$ broad component flux ratio.
[g] $R_{Fe}$, calculated as the flux ratio of area covered by the broad Fe line between 4433 Å and 4684 Å and the flux of the $H\beta$ emission line.
[h] Optical luminosity at 5100Å($\lambda L_{5100\text{Å}}$) in units of ergs/sec/Å
[i] SMBH mass in the units of $\log(M_{BH}/M\odot)$ obtained using the Radius-Luminosity relation and FWHM of $H\beta$ emission line.
[j] Eddington ratio, the ratio of bolometric to Eddington luminosity.
[k] Asymmetry index for the $H\beta$ emission line calculated using Equation 6.1(a).
[l] Kurtosis Index for the $H\beta$ emission line calculated using Equation 6.1(b).
[m] The soft X-ray photon index calculated between 0.2-2 KeV energy range taken from Ojha et al. (2020b).

the host galaxy components, the contribution from the iron line, and the accretion disk continuum, which reflects itself as a power-law component. A simple power-law component describes the accretion disk continuum satisfactorily (Sexton et al., 2020). We removed the host galaxy component using the Principal Component Analysis (PCA) method, obtained from the host galaxy templates available in Yip et al. (2004). We used five components of the PCA in order to subtract the host galaxy emission. However, for many of the sources in our sample, the host galaxy decomposition could not be applied. At higher redshifts, the contribution from the host galaxy does not have a large impact on the emission spectrum; hence we did not attempt to remove the host galaxy contribution from the spectra wherever it was not possible. The Fe blends were removed using the templates available in Boroson & Green (1992) which are available within the code itself. The final continuum model consisting of: the power-law component, the host galaxy template wherever applicable, and the Fe blends was subtracted from the original spectrum, which yielded the emission line components only. We were concerned with measuring the asymmetry in the $H\beta$ emission line; hence for the emission line fitting, we concentrated on the $H\beta$-[OIII] emission line complex. We used a combination of both narrow and broad Gaussian profiles to build the emission line model. We assumed the emission line complex to be composed of narrow and broad components representing the $H\beta$ emission from the Narrow line and Broad-line regions, respectively. The width of the narrow Gaussian components used for fitting the [OIII] doublet was tied with the narrow component of the $H\beta$ emission line, which physically indicates the emission coming from the same narrow-line region. This technique has been widely used while fitting the AGN spectra (e.g., see Chand et al., 2010, Rakshit et al., 2017, Sexton et al., 2020). For fitting the broad profile of the $H\beta$ emission line, we used a combination of up to 3 Gaussian components. The broad central component was needed in all the cases, while additional Gaussian components were needed to fit the wings of the line in a few of the cases. We also allowed a very broad Gaussian component for a few sources. The existence of a very broad component has been proposed for a few AGN in the recent works (see Gaskell & Goosmann, 2013, Sulentic & Marziani, 2015, Marziani et al., 2018a, Wolf et al., 2020) in which emission coming either from the innermost regions of the BLR or from the outer regions of the accretion disk results in a very broad component of the emission lines. Hao et al. (2005) set a criterion of FWHM larger than 1200 km s$^{-1}$ for defining broad line AGN. As a result, we set the limits for the width of the Gaussian profiles at 1200 km s$^{-1}$ for narrow [OIII] and $H\beta$ components, while 2200 km s$^{-1}$ was set for broad components and beyond 10000 km s$^{-1}$ for very broad components. The limit of 2200 km s$^{-1}$ was kept keeping in mind the previous works classifying the NLSy1 galaxies (e.g., see Rakshit et al., 2017). For fitting the BLSy1 galaxies, we removed the upper limit of 2200 km s$^{-1}$ on the broad component while still allowing up to three Gaussian components, including a very broad component. A demonstration of all the components used for fitting a particular AGN, namely: SDSS J030417.76+002826.4, is shown in Figure 6.3.

Since this work concerns the parameters related to the emission line shapes, precise profile measurements were required. We rejected the sources for which emission line detection was not possible significantly. Another issue we encountered was that either the H$\beta$ or the O[III] flux values reported zero values. In this case, the flux ratios become either 0 or infinite, which leads to the rejection of the respective source. It may be recalled that our parent sample of NLSy1 was chosen from Ojha et al. (2020b) based upon their X-ray detection where the strength of either the narrow or broad emission lines were not considered in the selection procedure. As a result, many sources with weak emission lines might have been included in the parent sample, which were rejected due to the criteria mentioned above. Finally, out of 206 NLSy1 galaxies, we could fit and get proper measurements of physical quantities for 144 objects, while out of 153 BLSy1 galaxies, we obtained proper measurements for 117 objects.

### 6.3.2 Emission line parameters

From the multiple Gaussians used in the emission line fit, first, we estimated the parameters characterizing the $H\beta$ emission line. In the case of NLSy1 galaxies, we picked up the single broad Gaussian component for the FWHM of $H\beta$, while we took the average of the broad Gaussian components in the case of BLSy1 galaxies. The broad Gaussian component represents the emission coming from the BLR, while the narrow component is representative of the Narrow Line Region (NLR) emission. The flux of the $H\beta$ emission line was calculated by integrating the flux between 4700Å and 4920Å. We calculated the equivalent width (EW) of the emission line using the same wavelength window, and the monochromatic luminosity at 5100Å ($L_{5100}$) was obtained from the fit. The iron strength ($R_{fe}$), a crucial parameter responsible for driving the variations in the properties of AGN, was calculated as the ratio of area covered by the broad Fe line between 4433Å and 4684Å and the flux of the $H\beta$ emission line. Further, to understand the influence of NLR emission in driving the variations in the properties of both the types of galaxies, we estimated the R5007 parameter, which is the flux ratio of the narrow [OIII] component at 5007Å and the broad $H\beta$ emission line (Gaur et al., 2019). Moreover, the ratio of broad components of $H\alpha$ and $H\beta$ was calculated using the flux of the broad components of the two emission lines. The Gaussian components used here were similar to those used for estimating the FWHM of the $H\beta$ emission line. We estimated the uncertainties in the respective parameters by fitting the individual spectra in 100 iterations using a Markov Chain Monte Carlo (MCMC) implementation in order to build the distribution of fitted parameters. We then took the 16th and the 84th percentile values as the uncertainty range in these parameters. The typical uncertainties in the FWMH were 10%, while in the EW, flux, and others, it was in the range of 10% to 30%. To get the uncertainties in the flux ratios, we propagated the uncertainties in the flux of the respective emission lines.

### 6.3.3 Asymmetry in the emission line

To characterize the line shapes in the $H\beta$ emission-line profiles, we estimated the asymmetry index (AI) and Kurtosis Index (KI) for all the AGN in our sample. AI has been calculated with different flux values in the recent past (Marziani et al., 1996, Brotherton, 1996, Du et al., 2018). We followed the technique used in the previous works (see Heckman et al., 1981, Marziani et al., 1996, Wolf et al., 2020) and chose a combination of 75% and 25% flux values in order to estimate AI. The wavelengths at which the emission line profile constructed by adding the broad Gaussian components reach the 75% and 25% of the flux values are recorded, and AI, as well as KI, are calculated a

$$AI = \frac{\lambda_R^L + \lambda_B^L - 2\lambda_0}{\lambda_R^L - \lambda_B^L}$$

(6.1)

$$KI = \frac{\lambda_R^H - \lambda_B^H}{\lambda_R^L - \lambda_B^L}$$

where $\lambda_B^H$ and $\lambda_B^L$ are the wavelength values at which the blue wing of the flux reaches 75% and 25% of the peak flux, respectively; while $\lambda_R^H$ and $\lambda_R^L$ the wavelength values at which the red wing of the emission profile reaches at 75% and 25% of the peak flux, respectively. $\lambda_0$ is the peak wavelength for the H$\beta$ emission line. The emission profile constructed by adding the broad Gaussian components of the emission line was used to estimate these parameters. The value of AI can range from $-1$ to 1, while the value of KI can range from 0 to 1. Negative AI means blue asymmetry, characteristic of the outflow component, while positive AI means red asymmetry, characteristic of the inflow component. Thus, AI can be used as an indicator of gas dynamics in the line emitting regions of AGN. We estimated the uncertainties in the AI and KI values by propagating the uncertainties in FWHM. The resultant typical uncertainties in AI and KI values were found out to be 0.05

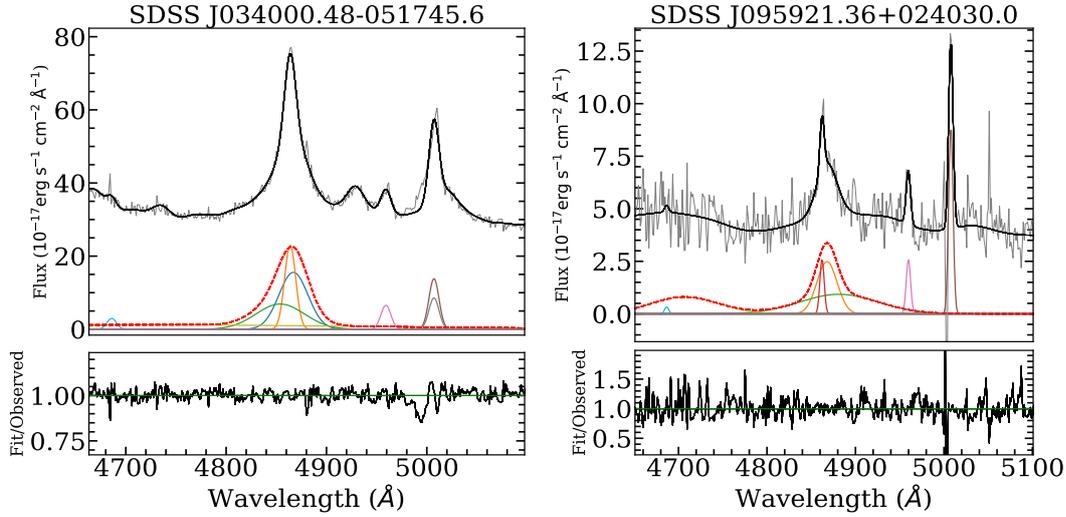

Fig. 6.4 Demonstration of asymmetric $H\beta$ emission profiles. The multiple Gaussian profiles used to fit the H$\beta$ emission line are also shown. The combination of broad components used to estimate AI is shown as a dashed red line. An example of a *blue* asymmetric profile having a negative asymmetry index of $-0.152$ is shown in panel (a), while the panel (b) shows an example of *red* asymmetric profile having a positive asymmetry index of $+0.227$. The AI values are calculated using Equation 6.1 by utilizing the flux values at 75% and 25%, respectively.

### 6.3.4　SMBH mass, Eddington ratio and X-ray photon indices

The SMBH mass has been estimated using various empirical relations in the recent past (see Shen, 2013, for a review). The SMBH mass has been measured using stellar dynamics for nearby quiescent galaxies. The single epoch SMBH mass estimation technique is based on the scaling relations obtained from local galaxy stellar velocity dispersion (Gebhardt et al., 2000). Reverberation mapping (see Bahcall et al., 1972, Blandford & McKee, 1982) based SMBH masses provide tighter constraints, and thus far, this technique has been the only reliable SMBH mass estimation method up to higher redshifts (Bentz et al., 2009). This technique assumes the virial motion of BLR clouds around the central SMBH and the Radius-Luminosity (R-L) relation holding true for the entire type-1 AGN population. The equation is:

$$\log\left[\frac{M_{BH}}{M\odot}\right] = 0.91 + 0.5\log\left[\frac{\lambda L_\lambda}{10^{44}ergs^{-1}}\right] + 2\log\left[\frac{FWHM}{kms^{-1}}\right] \quad (6.2)$$

This scaling equation has been obtained from Vestergaard & Peterson (2006), and 0.91 and 0.5 are the scaling constants to be used in the case of H$\beta$ emission line, the FWHM is in the units of km s$^{-1}$ and $\lambda L_\lambda$ is the luminosity at 5100 Å (L$_{5100}$). This equation is based on the assumption that the radius-luminosity (R-L) relation available for a set of approximately 120 reverberation mapped AGN so far (Bentz et al., 2009, Yu et al., 2020b) holds for the type 1 AGN in general. We estimated the SMBH mass using the broad component of the FWHM of the $H\beta$ emission line for all the AGN. The NLSy1 galaxies had smaller SMBH masses, owing to the small FWHM of the $H\beta$ emission line. The uncertainties in SMBH mass were estimated by propagating the errors in FWHM and the Luminosity values. We obtained a typical uncertainty of around 0.3 to 0.4 dex, which is consistent with the values obtained in the past works (Vestergaard & Peterson, 2006, Shen, 2013).

We estimated the bolometric luminosity ($L_{bol}$) for the AGN using the empirical relations available in the literature. The scaling for the bolometric luminosity is 9.1 times the one estimated at 5100Å (see Kaspi et al., 2000b). Using $L_{bol}$ and the obtained SMBH masses, we estimated the Eddington ratio $R_{\text{EDD}}$. The Eddington ratio is defined as the ratio of the bolometric to the Eddington luminosity ($L_{bol}/L_{\text{EDD}}$). The equations for $L_{bol}$ and $L_{\text{EDD}}$ are as follows:

$$L_{bol} = 9.1 \times L_{5100}; L_{\text{EDD}} = 1.45 \times 10^{38} \times \frac{M_{BH}}{M\odot} erg/s \quad (6.3)$$

In Ojha et al. (2020b), $R_{\text{EDD}}$ was calculated using X-ray observations which were consistent with the estimation obtained from optical parameters; hence we did not attempt to estimate $R_{\text{EDD}}$ using other methods and used the value obtained from optical spectra only. The soft X-ray photon indices ($\Gamma_X$) were obtained from (Ojha et al., 2020b). It has been reported that the NLSy1 galaxies have steeper X-ray spectra, thus a high value of $\Gamma_X$ (Waddell & Gallo, 2020). Being one of the fundamental components in the 4DE1 formalism of Boroson & Green (1992), the comparison of $\Gamma_X$ with the physical properties of both the types of galaxies can provide clues to the peculiar behavior of NLSy1 galaxies as we have done in this work. All the estimated parameters, along with the uncertainties, are presented in the form of a table (see Table 6.1).

## 6.4 Results

### 6.4.1 Intrinsic distribution of parameters

The properties of the line emitting regions, i.e., the BLR and NLR of AGN, can be characterized using a set of parameters derived from the optical spectra, while the X-ray photon index provides clues to the innermost regions of the accretion disk. We used 11 observational and physical parameters, namely the FWHM of the $H\beta$ emission line, Equivalent width of $H\beta$ (EW) emission line, the $[OIII]/H\beta$ flux ratio, the $H\alpha/H\beta$ flux ratio, $R_{fe}$, $L_{5100}$, SMBH mass, $R_{\rm EDD}$, AI, KI and $\Gamma_X$ to understand the diversity in the properties of BLSy1 and NLSy1 galaxies. The distribution for 6 of these parameters is shown in Figure 6.5, while the distribution for the FWHM of $H\beta$ emission line is shown in Figure 6.6. Orange and blue colors denote the BLSy1 galaxies and the NLS1 galaxies, and the combined sample is denoted by grey color in the histograms. The vertical lines in the middle denote the median value of the respective populations. We found the iron strength in NLSy1 galaxies to be higher than the BLSy1 galaxies for most of the AGN, which is consistent with the past works (see Gaskell, 2000). Results based on CLOUDY based simulations also indicate similar phenomenon (Panda et al., 2019b). In the current work, the median $R_{fe}$ was 1.02 for the NLSy1 galaxies while it was nearly half of that value, 0.53 for the BLSy1 galaxies. Eddington ratio is one of the important classifiers for NLSy1 galaxies. We found out that the Eddington ratio is higher for the NLSy1 galaxies, the median value of $\log(R_{\rm EDD})$ being 0.22 while the median $\log(R_{\rm EDD})$ was $-1.02$ in the case of BLSy1 galaxies. Furthermore, we found out that the SMBH masses are lower for the NlSy galaxies with the median value of $\log(\frac{M_{BH}}{M_\odot})$= 7.18, while in the case of BLSy1 galaxies, the median $\log(\frac{M_{BH}}{M_\odot})$ was higher at 8.47.

Based on the uncertainties observed in AI in Sect. 6.3.3, we set a criterion of AI $\leq -0.05$ and AI $\geq 0.05$ as significant blue and red asymmetries, respectively. We found out that more NLSy1 galaxies show blue asymmetries as compared to the BLSy1 galaxies. In the current sample, 46 NLSy1 galaxies show significant blue asymmetric profiles compared to 17 BLSy1 galaxies, while 20 NLSy1 galaxies show red asymmetric profiles compared to 52 BLSy1 galaxies. Blue asymmetries indicate that there is outflowing gas arising from that region. In recent works, the sources with high $R_{fe}$ values have been known to show blueward asymmetries (see Ganci et al., 2019, Wolf et al., 2020). The exact relation between the two is not clear, but high accretion rates could likely play a part as low accretors typically possess red asymmetries (Zamfir et al., 2010).

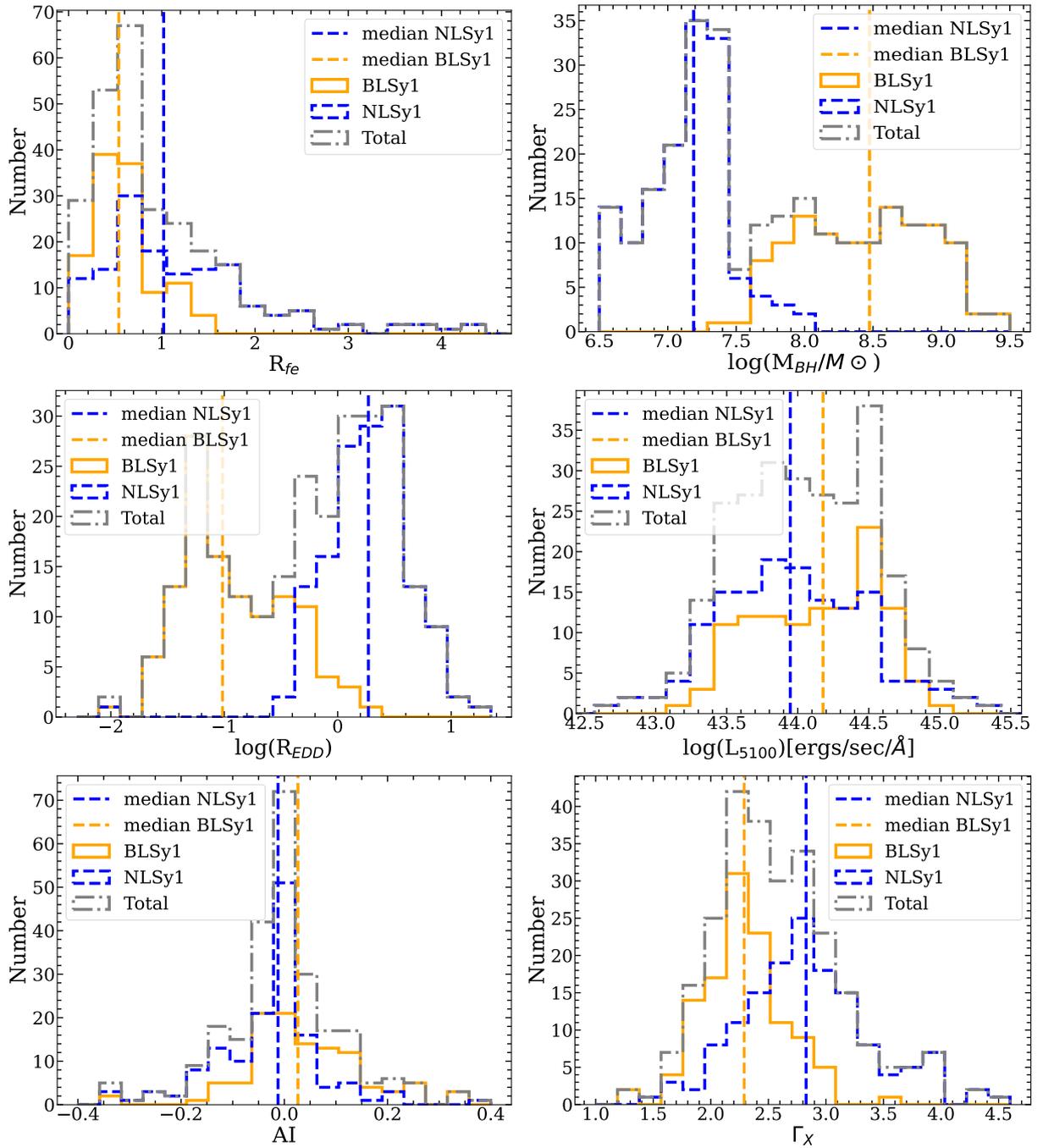

Fig. 6.5 Distribution of various physical parameters for the NLSy1 galaxies (blue dashed line), BLSy1 galaxies (solid orange line), and combined sample (dashed grey line). The top-left and right panels show the distribution for iron strength ($R_{fe}$), and the distribution for SMBH mass, respectively, the middle-left and middle-right panels show the distribution for Eddington ratio ($R_{\rm EDD}$), and the distribution for luminosity at 5100 Å, respectively. The bottom-left and bottom-right panels show the distribution of AI calculated using Equation 6.1 and the distribution of soft X-ray photon indices ($\Gamma_X$), respectively. The median values of the parameters for both types of galaxies are shown as vertical lines.

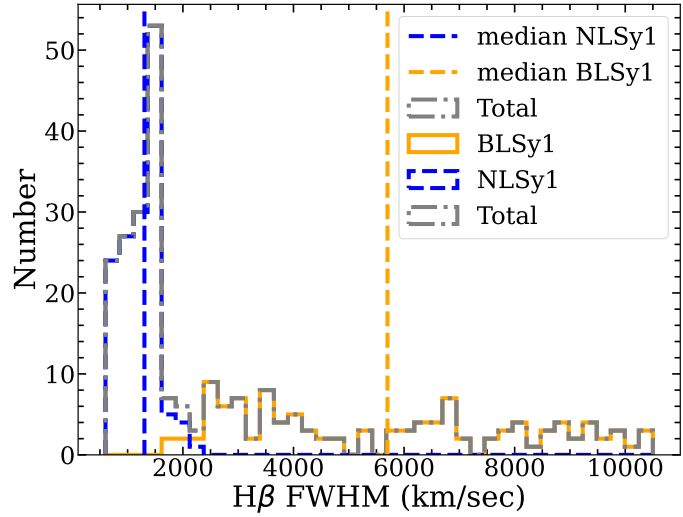

Fig. 6.6 The distribution for FWHM of the H$\beta$ emission line for the NLSy1 galaxies (blue dashed line), BLSy1 galaxies (solid orange line), and combined sample (solid grey line). The median values for the BLSy1 and NLSy1 galaxies are shown as blue dashed and orange solid vertical lines, respectively.

The broad $H\alpha/H\beta$ flux ratio has been used in the literature to understand the influence of the dust on the broad emission lines (Dong et al., 2007). Our sample shows slightly different behavior in both types of galaxies, with the median value for NLSy1 galaxies being 2.01, while the median value for BLSy1 galaxies is a bit higher at 2.47.

We performed a two-sample Kolmogorov Smirnov (KS) test (Jr., 1951) on the distributions of the parameters in order to check the similarity of both populations. If the p-value is small, then it implies that the two populations are significantly different. The p-value was $\leq 0.01$ for all but two of the parameters in the sample. This signifies that the distributions were not drawn from a common sample (see Table 6.2). Only $L_{5100}$ and the $[OIII]/H\beta$ flux ratio had a p-value greater than 0.01, which rejects the null hypothesis that the sources are drawn from the same parent sample. As our sources were in a similar luminosity range, we did not expect the luminosity characterized by $L_{5100}$ to be statistically different in the two populations. However, in this sample, the $[OIII]/H\beta$ flux ratio may not be a parameter responsible for the diversity between the two types of galaxies, as is evident from the very similar median values for both the types of galaxies as well.

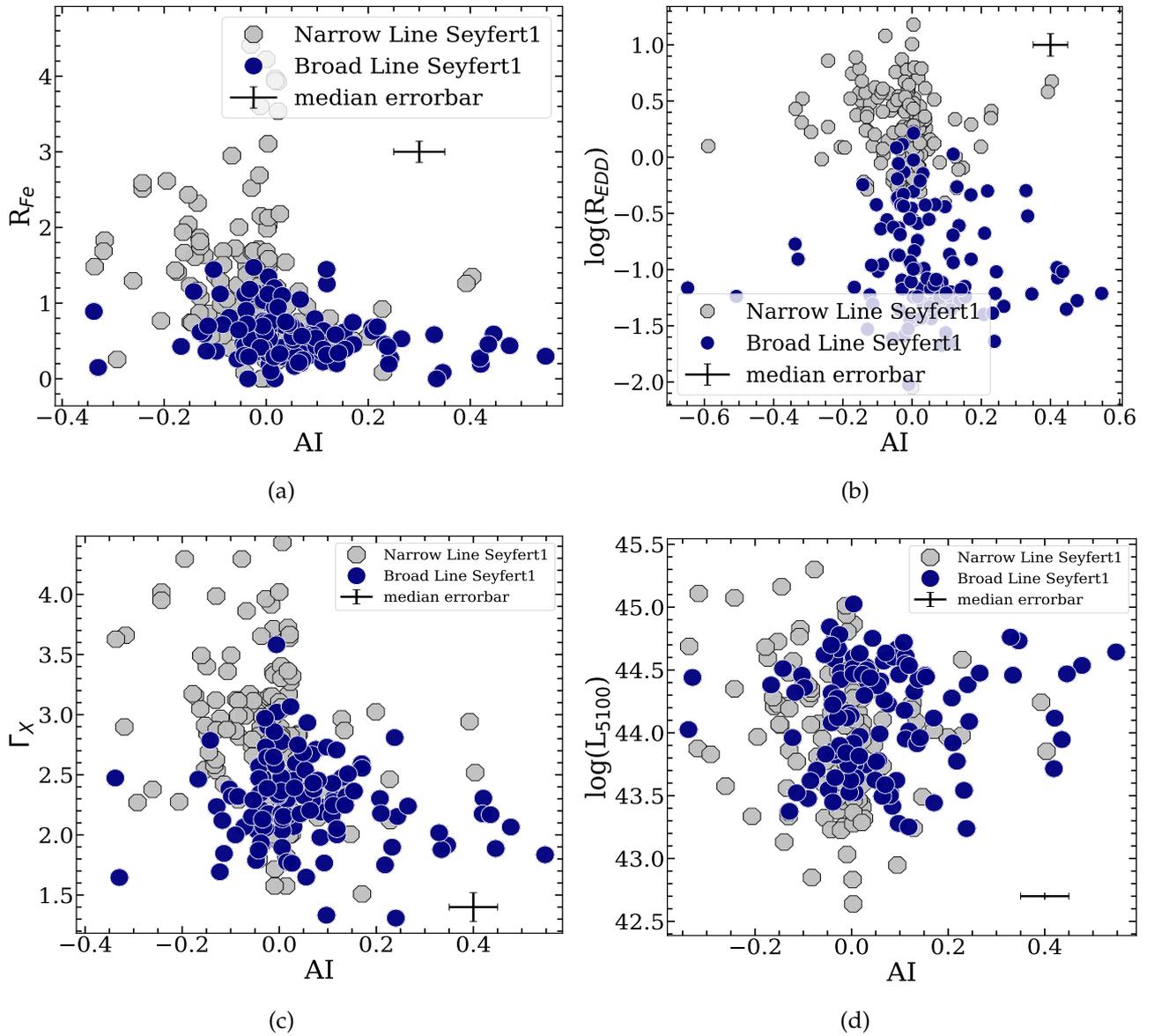

Fig. 6.7 Correlations between AI and 4 physical parameters for the NLSy1 galaxies (grey circles) and BLSy1 galaxies (blue circles). Relation of AI with $R_{fe}$ is shown in panel (a), with $log(R_{EDD})$ is shown in panel (b), with $\Gamma_X$ in panel (c), and with $L_{5100}$ in panel (d); representative of the physical processes in the central regions of AGN. The median errorbars on each parameter are also shown.

### 6.4.2 Spearman rank correlations

We calculated the Spearman rank correlation coefficients among all the parameters obtained for the entire sample. We performed the correlation analysis on NLSy1 and BLSy1 galaxies individually and on the combined population. The results are presented in Figures 6.8 and 6.9. We also included a cluster map that separates the parameters into two distinct groups, one dominated by FWHM, other by $R_{fe}$ in all three cases. The FWHM of $H\beta$ correlates negatively with $R_{fe}$ for both the classes, with the NLSy1 galaxies having a correlation coefficient of $-0.34$ while the BLSy1 galaxies have a correlation coefficient of $-0.52$. When the cross-correlation is run on the entire sample, the anti-correlation coefficient is $-0.57$. This is a well-known relation and forms the backbone of the quasar main sequence (see Marziani et al., 2018a, and references therein). $R_{\text{EDD}}$ is the most distinguishing parameter in our analysis. This parameter separates the population of NLSy1 galaxies and BLSy1 galaxies into two distinct classes, which is also apparent from the distribution in Figure 6.5. While $R_{\text{EDD}}$ anti-correlates weakly with the SMBH mass (correlation coefficient of $-0.30$) in the case of NLSy1 galaxies, a strong anti-correlation is seen between the two in the case of BLSy1 galaxies, with a correlation coefficient of $-0.8$. However, between the FWHM and $R_{\text{EDD}}$, we observe a similar anti-correlation for both types of galaxies, with the correlation coefficient being $-0.6$ for the NLSy1 galaxies and $-0.8$ for the BLSy1 galaxies. For the entire sample, a strong anti-correlation between $R_{\text{EDD}}$ and $H\beta$ FWHM is seen, with the anti-correlation coefficient of $-0.9$. Moreover, we also find a correlation between $R_{\text{EDD}}$ and $R_{fe}$ in the case of NLSy1 galaxies (correlation coefficient of 0.38) as well as the BLSy1 galaxies (correlation coefficient of 0.47).

The SMBH mass significantly correlates with the $H\beta$ FWHM in all the cases. Since the SMBH mass is a derived quantity based on the $H\beta$ FWHM among other parameters, it is expected to correlate to some degree. The Balmer decrement calculated using the ratio of the area of the $H\alpha$ and $H\beta$ emission lines, anti-correlates with the $H\beta$ FWHM in both the cases with the correlation coefficient of $-0.28$ in NLSy1 galaxies and a correlation coefficient of $-0.42$ in the case of BLSy1 galaxies. In Dong et al. (2007), it was observed that the ratio does not correlate much with the AGN physical properties. In their analysis for a sample of 446 low redshift AGN, the $H\alpha/H\beta$ flux ratio shows a very weak correlation of 0.078 with $H\beta$ FWHM. Our results are in deviation of this result.

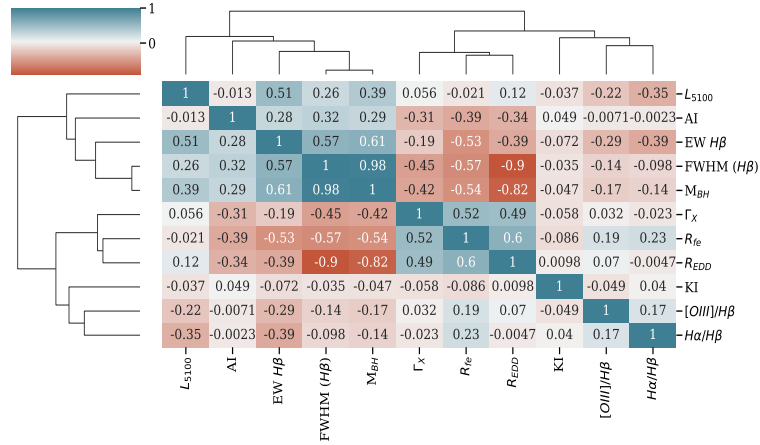

Fig. 6.8 The Spearman rank correlation matrix along with a cluster map for the entire sample of 144 NLSy1 galaxies and 117 BLSy1 galaxies. A combination of 11 parameters namely the $H\beta$ FWHM (broad component), Equivalent width of $H\beta$ (EW), the $[OIII]/H\beta$ flux ratio, the $H\alpha/H\beta$ flux ratio, iron strength ($R_{fe}$), $L_{5100}$, SMBH mass, $R_{EDD}$, asymmetry index (AI), kurtosis index (KI), and Soft X-ray photon index ($\Gamma_X$) have been used in this analysis. The cluster map denotes the linking of various parameters with each other. The orange rectangles denote the anti-correlations, while the blue rectangles denote the positive correlations. The Spearman rank correlation coefficient is written on the rectangular boxes.

$\Gamma_X$ is anti-correlated with the FWHM of $H\beta$ emission line if we take the entire sample into account, which is consistent with the results obtained from the sample analyzed by Grupe (2004). However, there is no significant correlation between the two if it is calculated separately. $\Gamma_X$ also correlates with $R_{EDD}$ and $R_{fe}$, which indicates that the inner regions of the accretion disk have a role in the diversity of the parameters in the two types of galaxies.

The FWHM of $H\beta$ correlates positively, although weakly, with AI. The correlation coefficient is 0.26 in the case of BLSy1 galaxies, while it is a meager 0.07 in the case of NLSy1 galaxies. Also, it anti-correlates with $R_{fe}$ in both the cases of NLSy1 galaxies and the BLSy1 galaxies. AI shows weak positive correlations with the SMBH mass in both cases, mimicking the behavior of $H\beta$ FWHM. Correlation between the $R_{EDD}$ and blueshift in the C IV emission line has been observed in Sulentic et al. (2017). However, in our study, we have found that AI is anti-correlated with the $R_{EDD}$ in the case of NLSy1 galaxies, while it is weakly correlated in the case of BLSy1 galaxies. There has been a postulation to use AI in the emission profile as a surrogate parameter in the 4DE1 formalism (Zamfir et al., 2010). The correlations between AI and other parameters mimic the behavior of FWHM, yet the correlation between these two parameters itself is not very strong. Thus, based on these results, we cannot conclusively say that AI can be used as a surrogate parameter in the 4DE1 formalism.

| Parameter | Median value | | p-null |
|---|---|---|---|
| | NLSy1 | BLSy1 | |
| $H\beta$ FWHM (km s$^{-1}$) | 1304.4 | 5699.8 | $1.67\times10^{-98}$ |
| $H\beta$ EW (Å) | 42.42 | 72.03 | $3.45\times10^{-07}$ |
| $[OIII]/H\beta$ | 0.20 | 0.16 | $3.2\times10^{-02}$ |
| $H\alpha/H\beta$ | 2.00 | 2.47 | $4.0\times10^{-03}$ |
| $R_{fe}$ | 1.02 | 0.53 | $3.94\times10^{-12}$ |
| $\log(L_{5100})$ | 43.94 | 44.17 | $1.17\times10^{-02}$ |
| $\log(M_{BH}/M\odot)$ | 7.18 | 8.47 | $4.48\times10^{-15}$ |
| $\log(R_{\rm EDD})$ | 0.27 | $-1.02$ | $6.91\times10^{-58}$ |
| AI | $-0.01$ | 0.02 | $2.56\times10^{-07}$ |
| KI | 0.30 | 0.28 | $3.37\times10^{-04}$ |
| $\Gamma_X$ | 2.83 | 2.29 | $1.99\times10^{-15}$ |

Table 6.2 The median values for all the 11 parameters for NLSy1 galaxies (column 2) and BLSy1 galaxies (column 3) are shown in this table. Column 4 shows the *p-null* values as a result of a two-sample Kolmogorov Smirnov (KS) test performed on the sample of NLSy1 galaxies vs BLSy1 galaxies. The low values of the *p-null* values signify that the distributions are not part of the same parent population.

### 6.4.3 Principal component analysis

Principal Component Analysis (PCA) is a technique used to reduce the dimensionality of a data set. PCA converts these observable parameters into principal components which are orthogonal to each other. These components are known as Eigenvectors, and these Eigenvectors reproduce the entire data set. The first Eigenvector (EV1) accounts for the maximum variance, followed by the subsequent orthogonal Eigenvectors. This technique is useful while looking for key parameters responsible for variability in a large sample like the one used here. PCA has often been used in the statistical study of the quasars (see Boroson & Green, 1992, Brotherton, 1996, Boroson, 2004, Grupe, 2004, Xu et al., 2012, Järvelä et al., 2015, Waddell & Gallo, 2020, Wolf et al., 2020).

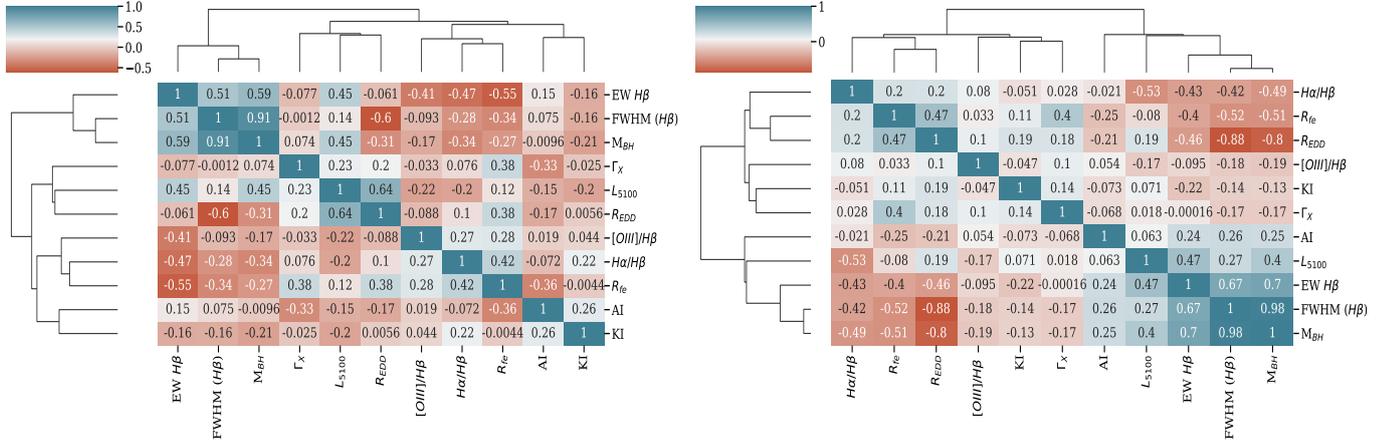

(a) Spearman correlations for NLSy1 galaxies.  (b) Spearman correlations for BLSy1 galaxies.

Fig. 6.9 The Spearmann rank correlation matrix along with a clustermap presented individually for the NLSy1 galaxies (left) and BLSy1 galaxies (right). The correlations and anti-correlations are denoted in the same way as Figure 6.8.

| Parameters | Narrow line Seyfert 1 (NLSy1) galaxies | | | | | Broad Line Seyfert 1 (BLSy1) galaxies | | | | |
|---|---|---|---|---|---|---|---|---|---|---|
| | EV1 | EV2 | EV3 | EV4 | EV5 | EV1 | EV2 | EV3 | EV4 | EV5 |
| FWHM $H\beta$ | 0.433 | −0.14 | −0.1 | −0.449 | −0.028 | 0.428 | −0.272 | −0.023 | −0.049 | 0.185 |
| EW $H\beta$ | 0.418 | 0.025 | 0.122 | −0.026 | −0.013 | 0.34 | −0.105 | −0.364 | 0.002 | 0.17 |
| $[OIII]/H\beta$ | −0.304 | −0.026 | −0.01 | −0.553 | −0.048 | −0.25 | −0.505 | −0.263 | −0.149 | −0.044 |
| $H\alpha/H\beta$ | −0.367 | −0.016 | 0.031 | −0.428 | 0.111 | −0.315 | −0.403 | 0.15 | −0.032 | −0.143 |
| $R_{fe}$ | −0.376 | 0.152 | 0.1 | −0.307 | 0.009 | −0.306 | 0.144 | −0.215 | −0.005 | 0.509 |
| $\log(L_{5100})$ | 0.185 | 0.248 | 0.621 | −0.147 | −0.054 | 0.184 | 0.179 | −0.7 | −0.108 | −0.205 |
| $log(M_{BH}/M_\odot)$ | 0.449 | −0.045 | 0.102 | −0.403 | −0.03 | 0.455 | −0.194 | −0.06 | −0.053 | 0.082 |
| $\log(R_{\rm EDD})$ | −0.121 | 0.319 | 0.619 | 0.156 | −0.035 | −0.361 | 0.323 | −0.385 | −0.002 | −0.212 |
| AI | −0.087 | −0.63 | 0.296 | 0.056 | −0.025 | 0.071 | −0.204 | −0.099 | 0.668 | −0.54 |
| KI | −0.083 | −0.627 | 0.305 | 0.052 | −0.029 | 0.117 | 0.082 | 0.142 | −0.687 | −0.516 |
| $\Gamma_X$ | −0.062 | 0.015 | −0.069 | 0.001 | −0.989 | −0.237 | −0.504 | −0.238 | −0.203 | 0.064 |

Table 6.3 The projections of the first 5 Eigenvectors on the physical parameters for both types of galaxies obtained using the PCA method.

We performed the PCA using Python's sklearn[7] package on the 11 calculated observational and physical parameters from the current sample to understand the correlation between these parameters in the Eigenvector space. We performed individual PCA on NLSy1 galaxies and BLSy1 galaxies and then jointly on the entire sample. We have included the line shape parameters, namely AI and KI, in the PCA analysis for the first time, keeping the separation between the NLSy1 and BLSy1 galaxies in mind.

The results of the PCA are available in Table 6.3. The first three Eigenvectors are responsible for almost 66% of the variation in the sample. The Eigenvector1 is responsible for 37% variation in the case of NLSy1 galaxies and 30% variation in the case of BLSy1 galaxies. The EV1 is dominated by the anti-correlation between the $H\beta$ FWHM and $R_{fe}$ for both the cases, clearly indicating the validity of the relationship obtained in Boroson & Green (1992). Figure 6.10 shows the projection of Eigenvector 1 on the various components for both cases. $R_{\text{EDD}}$ and $M_{BH}$ are the other parameters that dominate the EV1 projections. While $M_{BH}$ correlates highly with EV1, $R_{\text{EDD}}$ anti-correlates weakly with EV1 in the case of NLSy1 galaxies, it anti-correlates strongly with EV1 in the case of BLSy1 galaxies (see Table 6.4 for more information). AI shows a weak correlation with EV1 in the case of BLSy1 galaxies, while it shows a negligible correlation with EV1 in the case of NLSy1 galaxies. As only a few AGN show large asymmetry indices, the correlation may not significantly affect the parameter space when analyzing larger AGN populations. The EV2 is responsible for 19% (20%) variations in the case of NLSy1 (BLSy1) galaxies. The Eigenvector 2 of BLSy1 galaxies is driven by the anti-correlation of $[OIII]/H\beta$ flux ratio, $\Gamma_X$ and $H\alpha/H\beta$ flux ratio and the correlation of $R_{\text{EDD}}$. However, in the case of NLSy1 galaxies, the parameters driving the EV2 are AI, KI, and $R_{\text{EDD}}$ (see Table 6.3). The Spearman rank correlation coefficients between the EV2 and the parameters yield that $R_{\text{EDD}}$ highly correlates with EV2 in both cases, meaning that it is a significant driver of EV2. It is interesting to note that the projections of the EV1 and EV2 obtained in this study are different from the ones obtained in Boroson & Green (1992). While their EV1 shows a high correlation with $R_{\text{EDD}}$, in our study, $R_{\text{EDD}}$ shows anti-correlation with EV1 and a strong correlation with EV2. A similar correlation to our study was obtained in Järvelä et al. (2015). Since the Eigenvectors may yield different projections for different datasets (Grupe, 2004), it is thus imperative to perform the multi-parameter correlation analysis with different parameters in order to understand the underlying differences.

---

[7] https://scikit-learn.org/stable/

| Parameter | NLSy1 galaxies | | BLSy1 galaxies | | Combined | |
|---|---|---|---|---|---|---|
| | EV1 | EV2 | EV1 | EV2 | EV1 | EV2 |
| FWHM ($H\beta$) | 0.813 | −0.605 | 0.958 | −0.674 | 0.924 | −0.838 |
| EW $H\beta$ | 0.816 | −0.137 | 0.764 | −0.352 | 0.749 | −0.431 |
| $[OIII]/H\beta$ | −0.392 | −0.061 | −0.184 | −0.123 | −0.273 | 0.060 |
| $H\alpha/H\beta$ | −0.577 | 0.139 | −0.552 | −0.121 | −0.293 | −0.082 |
| $R_{fe}$ | −0.554 | 0.519 | −0.622 | 0.468 | −0.693 | 0.700 |
| $\log(L_{5100})$ | 0.376 | 0.578 | 0.397 | 0.234 | 0.366 | 0.066 |
| $\log(M_{BH}/M_\odot)$ | 0.871 | −0.324 | 0.972 | −0.601 | 0.936 | −0.777 |
| $\log(R_{\text{EDD}})$ | −0.313 | 0.948 | −0.780 | 0.806 | −0.798 | 0.917 |
| AI | 0.058 | −0.244 | 0.283 | −0.444 | 0.310 | −0.420 |
| KI | −0.226 | −0.032 | 0.187 | 0.132 | −0.027 | −0.083 |
| $\Gamma_X$ | −0.125 | 0.273 | −0.200 | 0.117 | −0.511 | 0.535 |

Table 6.4 The Spearman rank correlation coefficient between the different parameters with their corresponding projected values in EV1 and EV2 axis calculated for the sample of NLSy1 galaxies (left), BLSy1 galaxies (middle), and the combined sample (right).

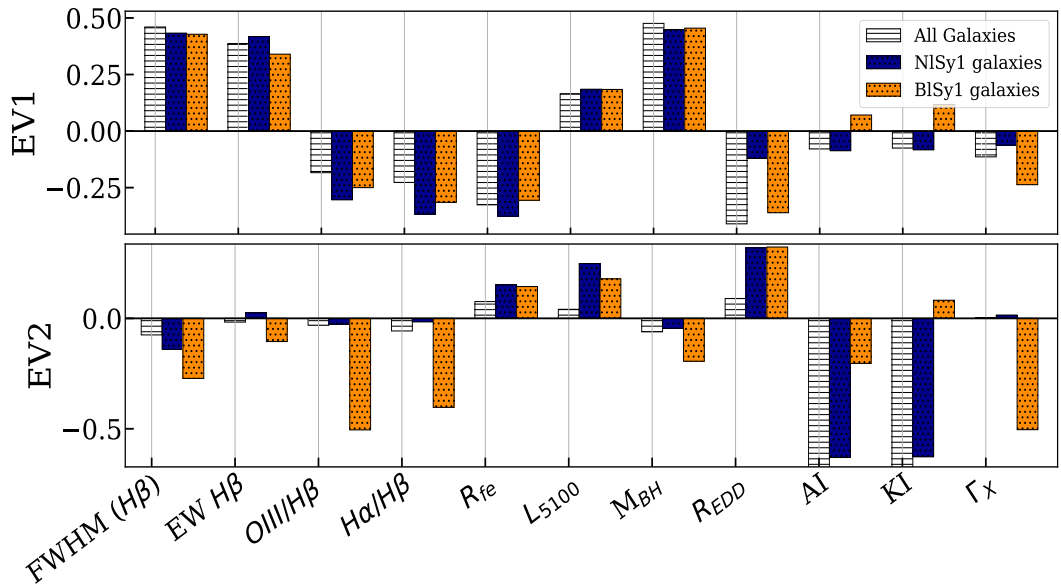

Fig. 6.10 The projection of the 11 physical parameters on the Eigenvector 1 (top) and Eigenvector 2 (bottom) as a result of the PCA performed on the sample of NLSy1 (blue color) and BLSy1 (orange color) galaxies individually and jointly (white color).

## 6.5 Discussion

To understand the differences in properties of Narrow line Seyfert galaxies compared to the general type-1 Seyfert population, we analyzed the single epoch optical spectrum for a large uniform X-ray selected sample of 144 NLSy1 galaxies and a comparison sample of 117 BLSy1 galaxies. This study is unique because the optical parameter correlation analysis involving emission line shapes for a large sample has been done, keeping the separation between the NLSy1 galaxies and BLSy1 galaxies in mind. Results based on our large sample of NLSy1 and BLSy1 galaxies are consistent with other studies with similar samples (see Grupe, 2004, Zamfir et al., 2010, Xu et al., 2012). The Spearman rank correlations point to higher anti-correlation of the $H\beta$ FWHM with $R_{fe}$ and $R_{EDD}$ in the case of NLSy1 galaxies. This provides a strong hint towards the younger age of NLSy1 galaxies. While there is a strong anti-correlation of FWHM with $R_{fe}$, it is slightly weaker in the case of BLSy1 galaxies. This may be because $R_{fe}$ is strongly dependent on the flux of the iron emission line, and thus the emission region of NLSy1 galaxies is richer in iron content as compared to the BLSy1 galaxies (Collin & Joly, 2000). We find out that the SMBH mass is correlated with FWHM in both the cases and that the NLSy1 galaxies have a higher $R_{EDD}$ than the BLSy1 galaxies. The lower SMBH mass coupled with high accretion rates in the NLSy1 galaxies may indicate that they are in evolving phase as compared to the BLSy1 galaxies. We also note here that the anti-correlation seen between SMBH mass and $R_{EDD}$ is influenced by the FWHM, which by definition is systematically lower in NLSy1 as compared to BLSy1 galaxies.

We find out that SMBH mass correlates highly with the FWHM of $H\beta$ emission line, which indicates that the more massive the central SMBH is, the faster the gas in the line emitting region rotates, supporting the virialized motion formalism, despite the significant difference in accretion rates characterized by $R_{EDD}$ in both types of galaxies. The $H\alpha$ to $H\beta$ emission line flux ratio is slightly higher in the case of BLSy1 galaxies as compared to the NLSy1 galaxies, which is also evident by its anti-correlations with the FWHM of $H\beta$ emission line. The possible reason could be that the $H\alpha$ and $H\beta$ emission come from different BLR locations. Another possibility could be that the $H\alpha$ profile is blended with contribution from the NLR emission lines, which are difficult to remove while fitting the $H\alpha$ emission complex. Hence the flux from the NLR might have a contribution.

We have used the line shape parameters, namely the Asymmetry and Kurtosis indices for the H$\beta$ emission line, to understand their influence in the peculiar behavior of the NLSy1 galaxies as compared to the BLSy1 galaxies. The higher number of NLSy1 galaxies showing blue asymmetric profiles adds another peculiar behavior to their nature among the general type-1 Seyfert population. The outflow signatures indicated by blue asymmetric H$\beta$ emission profile has been reported in the past works as well (see Ganci et al., 2019, Wolf et al., 2020) and this can be possible due to the accretion disk winds in these kinds of galaxies driven by high accretion rates (Grupe, 2004, Xu et al., 2012). It is worth noting that there is anti-correlation between $R_{fe}$ and AI in both the types of galaxies, implying that the sources with high $R_{fe}$ tend to show blue asymmetries. However, since the line emitting regions in AGN are still poorly understood, the exact connection between the outflowing components and the ratio of iron flux to the H$\beta$ emission line flux, signified by $R_{fe}$ is not clear.

The higher value of $R_{fe}$ in NLSy1 is a defining parameter, and this coupled with the high accretion rates may have an influence on these galaxies showing blue asymmetric profiles. As discussed in Grupe (2004), the NLSy1 galaxies with higher $R_{\text{EDD}}$ values may induce strong outflows, which might be a possible reason for this phenomenon. It will be interesting to know whether the outflow signatures are limited to only the H$\beta$ emission line in these galaxies or are seen in other emission lines as well.

We did not find significant correlations of $\Gamma_X$ with the observational and physical parameters in either NLSy1 or BLSy1 galaxies. However, when calculated on the combined population, the correlations of $\Gamma_X$ with $R_{fe}$ and $R_{\text{EDD}}$ emerged. This may be due to different ranges of the FWHM of H$\beta$ emission line covered in both the populations as has been suggested in Grupe (2004). The lack of significant correlations of $\Gamma_X$ with the line shape parameters, namely AI and KI, indicates that the emission from the inner regions of the accretion disk does not have a role in driving the emission line asymmetries at least.

Decarli et al. (2008) suggest that the NLSy1 galaxies do not differ from the BLSy1 galaxies generally, and their difference in observed properties can be accounted for by taking the orientation into account. Our results suggest that there are key significant properties that identify the NLSy galaxies as a distinct class in the type-1 AGN population. However, it is also possible that the difference in correlations seen in the NLSy1 and BLSy1 galaxies might be amplified due to the lower range of FWHM values in the former as compared to the latter.

### 6.5.1 NLSy1 galaxies in the context of Population A and B

It has been argued that the Population A AGN (FWHM of $H\beta \leq 4000$ km s$^{-1}$) are well fitted by Lorentzian profiles while the Population B AGN (FWHM of $H\beta >$ 4000 km s$^{-1}$) are well fitted using multiple Gaussian profiles (see Sulentic et al., 2009, Marziani et al., 2018a). Moreover, high and low accreting sources in this population show significant differences in the parameters. Further in Kovačević et al. (2010) a difference in properties is seen at even 3000 km/sec, which is challenging to our understanding of the AGN population and defining the limit of emission line FWHM at which the ensemble properties change. While most of the studies (e.g., Zamfir et al., 2010, Marziani et al., 2018a) have proposed a difference in the physical properties at $H\beta$ FWHM = 4000 km s$^{-1}$, we tried to see if the differences in physical properties arise even in the case of NLSy1 galaxies when compared to the general Seyfert galaxies population. Naturally, with the NLSy1 galaxies occupying extreme ends in some parameter space, it becomes imperative to understand the properties of NLSy1 galaxies compared to the general type-1 Seyfert population. In Grupe (2004), and many other studies (e.g., see Xu et al., 2012, Ojha et al., 2020b, Waddell & Gallo, 2020) the properties of NLSy1 galaxies have been studied comparatively with BLSy1 galaxies. However, these studies have used a different set of observational and physical parameters from those studied here. Also, it can be noted that in Grupe (2004) and Xu et al. (2012) the sample size was much smaller than the one used here. However, the results of these studies and ours are largely consistent. We also observe that the negative AI values which are prominent in NLSy1 galaxies are not present significantly in the BLSy1 galaxies with H$\beta$ FWHM values lower than 4000 km/sec. Thus, negative AI values may be another parameter that can distinguish the NLSy1 galaxies as a class from the general type 1 Seyfert population. In Figure 6.11, we show the 4DE1 plot based on this sample. The NLSy galaxies occupy the bottom right space (with low FWHM and larger $R_{fe}$ values). However, we do not find any BLSy1 galaxies with higher $R_{fe}$ values indicating that the NLSy1 galaxies may be occupying a parameter space that separates them from the population of BLSy1 galaxies.

### 6.5.2 Radio loud NLSy1 galaxies

Blue asymmetries have been observed in the narrow [OIII] emission line (see Berton et al., 2016, Gaur et al., 2019) for both the BLSy1 and NLSy1 galaxies, and in the case of NLSy1 galaxies, this phenomenon has been attributed to the NLR outflows generated due to higher accretion rates. Here, we have observed blue asymmetric H$\beta$ emission profiles predominantly in the NLSy1 galaxies. Marziani et al. (2003) explored the blue outliers in the context of the 4DE1 classification of quasars. In Berton et al. (2016) the influence of radio loudness on the gas dynamics is explored. They find that the likelihood of asymmetric [OIII] emission profiles is higher in the radio-loud AGN than the radio-quiet AGN. They find that the interaction with the relativistic jets in the radio-loud NLSy1 galaxies may give rise to the blue asymmetric profile in the [OIII] emission line. It is indeed possible that a similar mechanism could be responsible for the asymmetry in the H$\beta$ emission line in these kinds of galaxies. Singh & Chand (2018) presented the analysis for a sample of 11001 NLSy1 from (Rakshit et al., 2017) out of which only 498 are detected in the radio wavelength. In our sample of 144 NLSy1 galaxies, we find out that 23 are radio-loud, which may harbor a jet. However, we could not find out the physical properties which differentiate these two classes. We performed a KS test on the AI and KI values for the radio-loud and radio-quiet galaxies. The KS statistic for AI was 0.28 while the p$_{null}$ value was 0.08, which means that the two populations are not statistically different. Based on this, we suggest that the jets in these galaxies may not be drivers behind the asymmetries observed in the H$\beta$ emission profile. In (Marziani et al., 2003), the influence of radio loudness on the broad component of H$\beta$ emission profile is not detected, so our results are in line with this result.

### 6.5.3 Interpretation of PCA Eigenvectors

From the PCA performed on both the types of galaxies, we find out that the first three Eigenvectors are responsible for approximately 66% variance in the sample, which is similar to the results obtained in the recent works (e.g., see Grupe, 2004, Xu et al., 2012, Järvelä et al., 2015, Wolf et al., 2020). We have included AI and KI, signifying the emission line shapes measured for the $H\beta$ emission line in the PCA for a large sample of NLSy1 and BLSy1 galaxies for the first time. The problem with interpreting the Eigenvectors from the PCA is that every sample may have its own Eigenvector as the physical parameters chosen may be different for different studies (Grupe, 2004). Hence, different parameters may arise as the dominant parameters for individual samples. Through this work, we found out that the EV1 correlates with the FWHM of H$\beta$, SMBH mass, $L_{5100}$ while it anti-correlates with $R_{fe}$ and $R_{\rm EDD}$ and the [OIII]/H$\beta$ and $H\alpha/H\beta$ flux ratios in both the types of galaxies. However, a strong correlation of $R_{\rm EDD}$ with EV2 is observed. The PCA results are different if the two populations are analyzed separately. For instance, the behavior of AI, KI, and $[OIII]/H\beta$ flux ratio with the second Eigenvector in the sample of NLSy1 galaxies differ significantly from the BLSy1 galaxies (see Figure 6.10). This implies that the observational and physical parameters responsible for driving the variation in the NLSy1 and BLSy1 galaxies may differ significantly, and the NLSy1 galaxies occupy

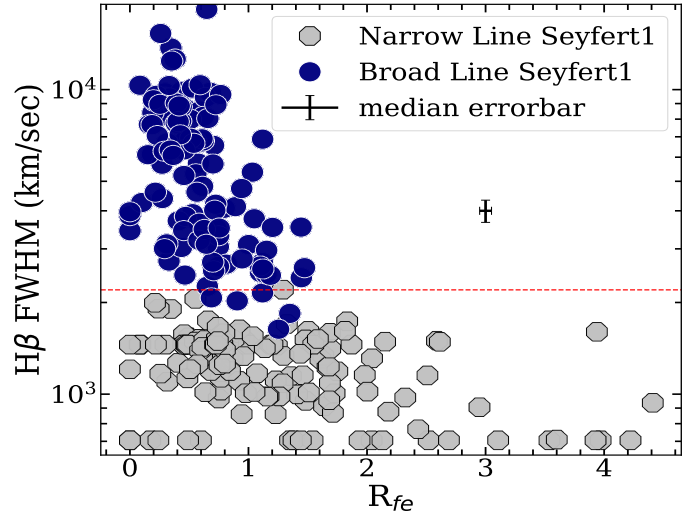

Fig. 6.11 The 4DE1 plot for the sample is shown here. The NLSy1 galaxies are shown in grey, while the BLSy1 galaxies are shown in blue. The limit of 2200 km sec$^{-1}$ chosen for differentiating the NLSy1 galaxies from the BLSy1 galaxies is shown as horizontal red dashed line.

their own parameter space, which is different from being a subclass of the BLSy1 galaxies.

## 6.6 Summary and Conclusions

In this work, we performed a statistical analysis on a large sample of NLSy1 and BLSy1 galaxies to understand the diversity in their parameters. This is a unique sample selected through X-ray observations and having almost similar luminosity and redshift values, which is intended to be a representative sample in parameter space for both types of galaxies. We analyzed the entire sample with direct correlation analysis by calculating the Spearman rank correlation coefficients and then performed a PCA on the entire sample. The main results are summarized as follows:

1. Spearman rank correlations yield that the 4DE1 formalism where $R_{fe}$ anti-correlates with the FWHM of H$\beta$ holds true for both the NLSy1 and BLSy1 galaxies despite the obvious differences in their spectral properties and Eddington ratio.

2. Higher fraction of NLSy1 galaxies show blue asymmetries (i.e., traces of outflowing gas) compared to the BLSy1 galaxies. This phenomenon may arise due to the presence of higher iron content, characterized by higher $R_{fe}$ values in the line emitting regions of NLSy1 galaxies.

3. The asymmetry indices correlate weakly with the FWHM of $H\beta$ and anti-correlate with the $R_{fe}$ values indicating that the sources with a lower value of $H\beta$ FWHM and higher $R_{fe}$ values tend to show blue asymmetries. This result also indicates that the asymmetry indices may be a surrogate parameter in 4DE1 formalism; however, the correlation of asymmetry indices and FWHM of $H\beta$ itself is not very strong in this sample.

4. Higher $R_{\text{EDD}}$ values for NLSy1 galaxies as compared to the BLSy1 galaxies confirms their peculiar behavior of low mass black holes accreting at higher rates as compared to their Broad-line counterparts. Moreover, we found strong anti-correlation between $R_{\text{EDD}}$ and FWHM of $H\beta$ in both the cases of NLSy1 and the BLSy1 galaxies, which can be interpreted as the NLSy1 galaxies being in the early phases of AGN development and accreting at a faster rate.

5. The PCA results signify that the first three Eigenvectors can describe around 66% of variance in both types of galaxies. The EV1 correlates with $H\beta$ FWHM, $H\beta$ EW, SMBH mass, $L_{5100}$ and AI while it anti correlates with $R_{\text{EDD}}$, $R_{fe}$, and $H\alpha/H\beta$ flux ratio for the combined population. However, PCA run separately yields different correlations indicating that the NLSy1 occupy a parameter space of their own which is different from that of the BLSy1 galaxies.

While this study is based on around 260 NLSy1 and BLSy1 galaxies, the NLSy1 galaxies occupy only a small fraction of the Seyfert 1 population, but we used a matching sample, which may not be representative of the entire population. Studies with even larger samples will be helpful in the understanding of the general behavior of NLSy1 galaxies compared with the type-1 Seyfert population and provide insights towards understanding the SMBH growth, dynamics of matter in the vicinity of the SMBH as well as the outflows and feeding of such galaxies.

# Chapter 7

# Conclusions and Future work

The all-sky surveys, such as the SDSS, have led to an explosion in the population of known AGN. The latest SDSS catalogue contains more than 700,000 confirmed AGNs (Lyke et al., 2020), and the MILLIQUAS catalogue has more than a million AGN candidates (Flesch, 2021). However, most of these AGNs are located at redshifts, where the inner regions can not be resolved with current or near-future observational techniques. Instruments such as GRAVITY can perform spectroastrometry for a handful of AGN currently (Gravity Collaboration et al., 2018, 2020, 2021), but for most of the AGN, we only have to rely on indirect techniques. Much physics about these regions can be derived using indirect methods only. The variability at all wavelengths and time scales remains a property of AGN, which can be exploited for this purpose. Using the RM technique, measuring the extent of the BLR, the size of the accretion disk, and the mass of the central SMBH for a handful of AGN has been possible. However, many questions pertaining to the central regions in these objects remain to be answered.

## 7.1 Conclusions

In this thesis, we have used indirect techniques to address problems related to the AGNs' innermost (sub-pc) regions. In the following paragraphs, I will briefly summarize the findings of this study.

The goal of chapter 2 was to estimate the accretion disk sizes for a sample of AGNs and find correlations between the accretion disk sizes measured and the physical properties, namely the SMBH mass and the optical luminosities. Reverberation mapping campaigns are challenging, time-consuming, and dependent on the amplitude of variability of the AGN during that particular time duration. Due to this limitation, the BLR sizes are known for almost 120 AGN (Panda et al., 2019a), and the accretion disk size measurements are available for even fewer AGNs (Hernández Santisteban et al., 2020). All sky surveys, such as the ZTF, have been able to provide light curves for a large number of sources (Bellm et al., 2019). Using the three optical band light curves (g, r, and i) available from the ZTF survey, we obtained the disk sizes based on the interband reverberation lags for 19 AGNs (Jha et al., 2022b). The SMBH masses for these AGNs are known through previous reverberation mapping campaigns. We found out that the disk sizes are, on average, 3.9 times larger than the predictions of the SS disk model. This result was also obtained in the recent literature based on microlensing studies. This challenges the assumption of using the SS disk model for AGN, and alternative models must be explored. We also did not find a strong correlation between the accretion disk sizes and the SMBH mass and between the disk sizes and luminosity. The results presented in this chapter have implications for the accretion disk models. As the SS disk model does not describe most of the AGN accretion disks, modelling the AGN interband lags in the context of different accretion disk models will be very helpful in understanding the accretion disk. Moreover, the potential of this technique is immense in measuring the extent of the accretion disk, especially in light of upcoming surveys such as the VRO/LSST, which will scan the sky using multiple filters with exceptional cadence.

In Chapter 3, we introduced the new accretion disk RM campaign titled 'Investigating the central parsec regions around supermassive black holes' (INTERVAL), which is aimed at obtaining the accretion disk sizes for AGN accreting at super Eddington rates. We are monitoring this sample using the GIT in ugriz bands. We present the initial results from one of the sources in our sample − IRAS 04416+121. By cross-correlating the interband light curves, we find evidence of lags increasing with the wavelength. However, we find that the accretion disk size obtained by using the thin disk model is approximately 4 times larger than the SS disk prediction, which has been observed in most of the sources studied so far. As the BLR sizes for these sources have been observed to be smaller than the empirical R-L relation (e.g., see Du et al., 2015), we aim to study whether the disk sizes for these sources are reduced too. However, for one of the sources in the sample, we have not found any reduction in the disk size. We aim to model the accretion disk further based on the results above to understand the disks in AGN better.

In Chapter 4, we calibrated the technique of Narrow Band PRM with respect to the traditional Spectroscopic RM. Narrowband Photometric Reverberation Mapping is a suitable alternative to the conventional spectroscopic RM (Haas et al., 2011). However, the sources with both spectroscopic and photometric RM are limited to only a few. Using the publicly available data set from the Lick AGN Monitoring Project-2008, we calibrated the technique of PRM. First, we converted the spectroscopic data points to photometric data points. These photometric data points were then used to estimate the reverberation lags. We found out that the lags estimated through the photometric technique agree with the spectroscopic lags within the error bars and that the continuum contributes significantly to deviation in the BLR lags. We developed simulations based on which a photometric RM campaign can be executed. We tested various cadences, campaign duration, variability amplitudes, and the effect of seeing conditions on the simulated light curves to derive an optimal observation strategy for executing a successful PRM campaign. Combined with single epoch spectroscopy, the narrow band PRM can be an excellent, cost-effective alternative to obtain the BLR sizes and the SMBH masses for a large number of AGN.

The following two chapters of this thesis deal with the peculiar class of AGN- the NLSy1 galaxies. While the general aim of understanding the innermost regions remains central throughout this work, in Chapter 5, we explored the connection between the jets and the accretion disk variability in Seyfert galaxies. To understand whether the NlSy1 galaxies with confirmed jets behave like blazars or not, we observed a sample of 23 AGN on 53 epochs using multiple telescopes. We studied the optical flux variability at extremely small time scales. As the blazars are known to show rapid optical flux variability, we tested if this phenomenon is true even in the case of NlSy1 galaxies with the presence of a jet. Further, a handful of AGN in our subset was detected in $\gamma-$ray wavelength. While the blazar jets are due to synchrotron, we demonstrated that these $\gamma-$ray NLSy1s approach blazar-like variability at shorter time scales. Surprisingly, we found that the jetted AGN without $\gamma-$ray detection does not necessarily show the presence of microvariability. This indicates that not only jet but relativistic beaming plays a dominant role in causing the microvariability in these AGNs. Our finding is extremely significant in this regard as the NLSy1 galaxies with confirmed jets have very high duty cycles that approach blazars' duty cycles. This may serve as a proxy to identify the jets in AGNs without the radio observations, which can then be followed up using dedicated radio observations (Ojha et al., 2022).

In chapter 6, we studied the peculiarity of the emission lines in NLSy1 galaxies using multi-epoch spectra from the SDSS database. The Broad Line regions of the AGNs are poorly understood. There has been evidence of viral motion, while disk winds (Matthews et al., 2020) and outflows have also been proposed(Waters & Li, 2019). The emission line profiles can be a tracer of the dynamics of matter in the BLR. Using a sample of 144 NLSy1 galaxies and 109 BLSy1 galaxies, we characterized the NLSy1 galaxies from their broad line counterparts through physical properties obtained using optical spectra. We calculated 11 physical parameters and found that the 4DE1 sequence holds for the NlSy1 galaxies despite significantly higher Eddington rations and iron content in their BLR. Further, we found out the NLSy1 galaxies are 3 times more likely to show blue asymmetric H$\beta$ emission line profiles. These blue asymmetric profiles indicate outflowing gas from the BLR. We also performed a PCA on the sample and found that the PCA Eigenvectors for NLSy1 galaxies differ from the BLSy1 galaxies, meaning they are not merely a subsample of BLSy1 galaxies but relatively young AGNs in the early stages of their evolution (Jha et al., 2022a). Through this work, we add a significant number of NLSy1 galaxies with outflowing emission profiles, which has a direct influence on understanding their Broad Line Regions. These can also be used to test whether similar outflowing profiles are observed in other emission profiles. If such asymmetry exists in other emission profiles, it would challenge the virial hypothesis used for the SMBH mass measurements for these AGNs.

To summarize, the main findings of the thesis are as follows:

1. We obtained the accretion disk sizes for 19 AGNs selected from the ZTF survey, and the disk sizes are, on average, 3.9 times larger than the predictions of the SS disk model. There is a weak correlation between the accretion disk sizes and the physical parameters: the SMBH mass and the luminosities for these AGNs.

2. We have started a new accretion disk RM campaign to understand the accretion disks in Super Eddington AGN. The initial results from one of the sources in our sample suggest the disk size to be about 4 times larger than the SS disk model.

3. The narrow band PRM can be a successful alternative to get the BLR sizes and the SMBH masses if combined with single epoch spectroscopy. However, the accretion disk continuum contribution in the narrow band filters must be considered. Cadence, campaign duration, variability amplitudes and the seeing conditions affect the lag estimates; hence they should be taken into account while planning the PRM campaigns.

4. Our study demonstrates that the NLSy1 galaxies with $\gamma-$ray detection display blazar-like variability on short time scales. At the same time, the mere presence of radio jets does not ensure variability in the AGN continuum. The relativistic beaming of the jets is a significant cause for the high-duty cycle observed in the $\gamma-$ray detected NLSy1 galaxies.

5. We find out that the NLSy1 galaxies are 3 times more likely to show the blue asymmetric H$\beta$ emission profiles compared to the BLSy1 galaxies, which means outflows may be present in the innermost regions of these AGN. Further, the PCA eigenvectors playing a role in the variation of parameters are different for the NLSy1 galaxies.

## 7.2 Future work

The results mentioned in the sections above can be used further to understand the relationship between the innermost regions of AGN in the context of their cosmological significance. As an extension to the abovementioned work, I plan to study the exciting phenomenon in AGNs further. Following are some of the projects and ideas which I aim to work on in the upcoming days:

- **Structure of the Accretion disk:** The accretion disk sizes measured in Chapter 2 were restricted in the wavelength coverage as only 3 filters are available in ZTF and in Chapter 3 we could obtain the results for one of the sources in 4 optical bands only. In contrast, dedicated surveys that include filters in X-ray, UV, optical, and IR wavelength ranges can be used to resolve the innermost regions in a significant number of AGNs with more precision. Such campaigns have covered only a handful of AGN at the moment (e.g., see Hernández Santisteban et al., 2020). Using the space-based observatory ASTROSAT, we aim to observe AGN in multiple wavelengths. Such a campaign will be equivalent to AGN STORM on NGC 5548 (Starkey et al., 2016a) and STROM 2 on MRK 817 (Kara et al., 2021) and will help understand the inner structure of the accretion disk in detail. Further, we continue this project with the ground-based telescope to cover optical bands. With the help of the GROWTH India telescope and the Devasthal optical telescopes, we aim to find time lags for a significant sample of AGN. We also intend to model the accretion disk using analytical models in the future. The premise is that since the predictions of the SS disk model fail for the AGNs, what other explanation could be there? If there is a significant deviation from the SS disk, can we model it with the advection-dominated or more complicated processes? With the help of space-based telescopes, we also aim to perform simultaneous accretion disk and corona monitoring to understand the connection between the accretion disk and the corona.

- **Modeling the variability in AGN light curves:** We aim to use data from the ZTF survey to model AGN light curves for a large sample. The benefit of using ZTF light curves is that it provides 3-day average cadence light curves in g, r, and i bands, which is higher than the previous surveys and can serve as a template for such science cases using the upcoming LSST/VRO data. These light curves can be used to assess the accuracy of the DRW approximation for variability in the AGN light curves. DRW modelling still has many questions, and deviation from DRW has been observed using dense light curves from the surveys like the Kepler mission (Edelson et al., 2014). Other models have also been explored to explain the AGN variability (Yu et al., 2022). We plan to investigate how physical parameters such as SMBH mass, luminosity and accretion rates can impact the variability in the light curves and its variability on diverse timescales.

- **The connection between optical variability and physical properties of the central engine:** We are investigating the correlation between the variability of UV/optical light curves and the physical parameters for a large sample of AGN selected from the Spectroscopic IDentification of eROSITA Sources (SPIDERS) AGN catalogue.To study the UV/optical variability, we use well-sampled g and r-band light curves obtained from the ZTF survey. We calculate the fractional excess variance in the light curves and model the light curves using a Damped Random Walk (DRW) process to extract the variability amplitudes and the damping timescales. Recently a tight correlation has been found between the damping timescale and the SMBH mass (Burke et al., 2021). We aim to find a correlation between these parameters for a large sample of AGN and see whether this relation holds true for most of the AGN population. We also plan to generate the Power Spectrum Density (PSD) of these light curves to extract the relation between the PSD slopes and the physical parameters.

- **Narrow band Photometric RM to get the BLR sizes and SMBH masses:** We aim to observe a sample of nearby ($0.01 \leq z \leq 0.1$), low luminosity ($L \leq 10^{43}$ ergs/s) AGN which are expected to have BLR size of fewer than 10 light-days using narrowband filters. We propose to monitor each source for 3 months with a median cadence of at least 3 days. Combining broadband UVBRI observations with simultaneous narrowband photometry, we hope to determine the BLR size and the lag between the continuum and emission lines. We will use the lag between the broadband continuum light curves to understand the structure of the accretion disk. For the first time, a systematic photometric reverberation mapping campaign covering a large sample of low-luminosity AGN can be carried out using multiple telescopes, providing BLR sizes and SMBH masses for these sources.

- **The emission line region of the Super Eddington accreting AGN:** Using the 3.6m Devasthal Optical Telescope (DOT), we are pursuing IR monitoring of a sample of select AGN, the SEAMBH. These AGN have been chosen based on their very high accretion rates compared to the general AGN population. Our IR spectroscopy will explore physical scales from the innermost region to the kpc scale, providing insights into AGN host galaxy feedback in SEAMBH. We can determine the excitation states in the core and extended NLR and the coronal line region of Seyfert galaxies with Super Eddington accretion rates. Many of our sources have been reverberation mapped, and the Paschen emission lines we observe will allow us to estimate the SMBH scaling relation between the H$\beta$ and Paschen lines. Photoionization modelling of these SEAMBH AGN will help determine potential photoionization mechanisms. Additionally, a comparison with a control sample of evolved Seyfert systems will help us understand the power sources and formation of AGN with Super Eddington accretion rates.

# Appendix A

# Details of Observations for the 23 NLSy1 galaxies

This appendix contains two tables:Table A1 consists of the observation details, the comparison stars used for calculating the variability for the NLSy1 galaxies studied in Chapter 4. Table A2 consists of the details of the INOV status and variability amplitudes for the 23 NLSy1 galaxies.

Table A.1 Basic parameters of the comparison stars along with their observation dates used in this study for the 23 RLNLSy1 galaxies.

| Target RLNLSy1s and the comparison stars (1) | Date(s) of monitoring (2) | R.A.(J2000) (h m s) (3) | Dec.(J2000) (° ′ ″) (4) | $g$ (mag) (5) | $r$ (mag) (6) | $g-r$ (mag) (7) |
|---|---|---|---|---|---|---|
| jetted NLSy1s | | | | | | |
| J032441.20+341045.0 | 2016 Nov. 22, 23; Dec. 02; 2017 Jan. 03, 04 | 03 24 41.20 | +34 10 45.00 | 14.50 | 13.70 | 0.80* |
| S1 | 2016 Nov. 22, Dec. 02 | 03 24 53.68 | +34 12 45.62 | 15.60 | 14.40 | 1.20* |
| S2 | 2016 Nov. 22, Dec. 02 | 03 24 53.55 | +34 11 16.58 | 16.20 | 14.40 | 1.80* |
| S3 | 2016 Nov. 23 | 03 24 10.92 | +34 15 01.90 | 16.30 | 15.10 | 1.20* |
| S4 | 2016 Nov. 23 | 03 24 14.04 | +34 18 20.10 | 15.80 | 15.00 | 1.00* |
| S5 | 2017 Jan. 03 | 03 24 14.92 | +34 15 21.20 | 15.90 | 15.10 | 0.80* |
| S6 | 2017 Jan. 03 | 03 24 08.44 | +34 08 15.80 | 15.80 | 14.20 | 1.60* |
| S7 | 2017 Jan. 04 | 03 24 38.14 | +34 13 53.40 | 15.90 | 15.20 | 0.70* |
| S8 | 2017 Jan. 04 | 03 24 14.08 | +34 16 48.60 | 15.40 | 14.80 | 0.60* |
| J081432.12+560958.7 | 2017 Jan. 03; Nov. 20 | 08 14 32.12 | +56 09 58.69 | 18.06 | 18.11 | −0.05 |
| S1 | 2017 Jan. 03 | 08 14 02.78 | +56 11 12.07 | 19.14 | 18.12 | 1.02 |
| S2 | 2017 Jan. 03 | 08 14 53.54 | +56 10 14.14 | 17.96 | 17.35 | −0.61 |
| S3 | 2017 Nov. 20 | 08 13 39.62 | +56 17 55.59 | 19.05 | 17.71 | 1.34 |
| S4 | 2017 Nov. 20 | 08 14 19.58 | +56 06 24.04 | 19.33 | 17.88 | 1.45 |
| J084957.98+510829.0 | 2017 Dec. 13, 2019 April 08 | 08 49 57.98 | +51 08 29.04 | 18.92 | 18.28 | 0.64 |
| S1 | | 08 50 12.62 | +51 08 08.03 | 19.45 | 18.06 | 1.39 |
| S2 | | 08 50 03.07 | +51 09 12.23 | 17.82 | 17.09 | 0.73 |
| J090227.20+044309.0 | 2017 Feb. 22; Dec. 14 | 09 02 27.20 | +04 43 09.00 | 18.96 | 18.63 | 0.33 |
| S1 | 2017 Feb. 22 | 09 02 01.94 | +04 37 32.90 | 19.10 | 17.74 | 1.36 |
| S2 | 2017 Feb. 22 | 09 02 23.40 | +04 35 44.57 | 18.70 | 17.31 | 1.39 |
| S3 | 2017 Dec. 14 | 09 03 04.11 | +04 48 19.65 | 18.85 | 17.95 | 0.90 |
| S4 | 2017 Dec. 14 | 09 03 07.29 | +04 38 57.34 | 18.92 | 17.93 | 0.99 |
| J094857.32+002225.6 | 2016 Dec. 02; 2017 Dec. 21 | 09 48 57.32 | +00 22 25.56 | 18.59 | 18.43 | 0.16 |
| S1 | | 09 48 36.95 | +00 24 22.55 | 17.69 | 17.28 | 0.41 |
| S2 | | 09 48 37.47 | +00 20 37.02 | 17.79 | 16.70 | 1.09 |
| J095317.10+283601.5 | 2017 March 04; 2018 March 23 | 09 53 17.10 | +28 36 01.48 | 18.99 | 18.97 | 0.02 |
| S1 | 2017 March 04 | 09 52 48.09 | +28 29 53.69 | 18.31 | 17.45 | 0.86 |
| S2 | 2017 March 04; 2020 November 21 | 09 53 07.49 | +28 37 17.10 | 18.46 | 17.32 | 1.14 |
| S3 | 2020 November 21 | 09 53 21.03 | +28 34 57.36 | 20.41 | 18.90 | 1.51 |
| J104732.78+472532.0 | 2017 April 11; 2018 March 12 | 10 47 32.78 | +47 25 32.02 | 18.97 | 18.76 | 0.21 |
| S1 | 2017 April 11 | 10 47 16.50 | +47 24 47.24 | 18.68 | 17.98 | 0.70 |
| S2 | 2017 April 11; 2018 March 12 | 10 47 27.51 | +47 27 58.94 | 18.79 | 17.87 | 0.92 |
| S3 | 2018 March 12 | 10 48 16.44 | +47 22 42.22 | 18.78 | 17.45 | 1.33 |
| J122222.99+041315.9 | 2017 Jan. 03, 04; Feb. 21, 22; March 04, 24 | 12 22 22.99 | +04 13 15.95 | 17.02 | 16.80 | 0.22 |
| S1 | | 12 22 34.02 | +04 13 21.57 | 18.63 | 17.19 | 1.44 |
| S2 | | 12 21 56.12 | +04 15 15.19 | 17.22 | 16.78 | 0.44 |
| J130522.75+511640.2 | 2017 April 04; 2019 April 25 | 13 05 22.74 | +51 16 40.26 | 17.29 | 17.10 | 0.19 |
| S1 | 2017 April 04 | 13 06 16.16 | +51 19 03.67 | 16.96 | 15.92 | 1.04 |
| S2 | 2017 April 04; 2019 April 25 | 13 05 57.57 | +51 11 00.97 | 16.35 | 15.26 | 1.09 |
| S3 | 2019 April 25 | 13 05 44.25 | +51 07 35.85 | 17.88 | 16.42 | 1.46 |
| J142114.05+282452.8 | 2018 May 10; 2019 May 27 | 14 21 14.05 | +28 24 52.78 | 17.73 | 17.74 | −0.01 |
| S1 | 2018 May 10 | 14 20 33.73 | +28 31 10.45 | 18.50 | 17.11 | 1.39 |
| S2 | 2018 May 10; 2019 May 27 | 14 21 08.78 | +28 24 04.99 | 16.16 | 16.21 | −0.05 |
| S3 | 2019 May 27 | 14 21 24.36 | +28 27 16.52 | 16.82 | 16.45 | 0.37 |
| J144318.56+472556.7 | 2018 March 11, 23 | 14 43 18.56 | +47 25 56.74 | 18.14 | 18.17 | −0.03 |
| S1 | | 14 43 37.14 | +47 23 03.03 | 17.51 | 16.82 | 0.69 |
| S2 | | 14 43 19.05 | +47 19 00.98 | 18.03 | 16.75 | 1.28 |
| J150506.48+032630.8 | 2017 March 25; 2018 April 12 | 15 05 06.48 | +03 26 30.84 | 18.64 | 18.22 | 0.42 |
| S1 | | 15 05 32.05 | +03 28 36.13 | 18.13 | 17.64 | 0.49 |
| S2 | | 15 05 14.52 | +03 24 56.17 | 17.51 | 17.14 | 0.37 |
| J154817.92+351128.0 | 2018 May 17; 2019 May 08 | 15 48 17.92 | +35 11 28.00 | 18.03 | 18.03 | 0.00 |
| S1 | | 15 47 57.43 | +35 14 05.24 | 18.31 | 17.50 | 0.81 |

Table A.1 – continued..

| Target AGN and the comparison stars | Date(s) of monitoring | R.A.(J2000) (h m s) | Dec.(J2000) (° ′ ″) | g (mag) | r (mag) | g-r (mag) |
|---|---|---|---|---|---|---|
| (1) | (2) | (3) | (4) | (5) | (6) | (7) |
| S2 | | 15 48 02.20 | +35 13 56.16 | 17.37 | 16.98 | 0.39 |
| J164442.53+261913.3 | 2017 April 03; 2019 April 26 | 16 44 42.53 | +26 19 13.31 | 18.03 | 17.61 | 0.42 |
| S1 | | 16 45 20.03 | +26 20 54.55 | 16.56 | 15.89 | 0.67 |
| S2 | | 16 44 34.40 | +26 15 30.27 | 16.28 | 15.80 | 0.48 |
| J170330.38+454047.3 | 2017 June 03; 2019 March 25 | 17 03 30.38 | +45 40 47.27 | 16.12 | 15.41 | 0.71 |
| S1 | | 17 04 02.02 | +45 42 16.56 | 15.02 | 14.39 | 0.63 |
| S2 | | 17 04 34.88 | +45 40 08.65 | 15.00 | 13.91 | 1.09 |
| non-jetted NLSy1s | | | | | | |
| J085001.17+462600.5 | 2017 Jan. 01; Dec. 15 | 08 50 01.17 | +46 26 00.50 | 19.12 | 18.82 | 0.30 |
| S1 | | 08 50 17.69 | +46 20 42.71 | 18.11 | 17.85 | 0.26 |
| S2 | | 08 49 48.29 | +46 21 11.81 | 18.72 | 17.63 | 1.09 |
| J103727.45+003635.6 | 2018 March 11, 22 | 10 37 27.45 | +00 36 35.60 | 19.57 | 19.21 | 0.36 |
| S1 | 2018 March 11 | 10 37 39.63 | +00 38 26.16 | 18.90 | 17.69 | 1.21 |
| S2 | 2018 March 11 | 10 37 28.03 | +00 37 59.88 | 18.97 | 17.52 | 1.45 |
| S3 | 2021 April 08 | 10 36 50.96 | +00 41 25.26 | 18.76 | 17.35 | 1.41 |
| S4 | 2021 April 08 | 10 37 38.72 | +00 40 28.00 | 17.26 | 16.82 | 0.44 |
| J111005.03+365336.2 | 2018 March 23; 2019 Jan. 13 | 11 10 05.03 | +36 53 36.22 | 20.60 | 20.49 | 0.11 |
| S1 | | 11 10 08.50 | +36 50 59.03 | 19.14 | 18.86 | 0.28 |
| S2 | | 11 10 10.76 | +36 55 26.97 | 19.74 | 18.45 | 1.29 |
| J113824.53+365327.2 | 2017 April 17; 2018 March 23 | 11 38 24.53 | +36 53 27.18 | 19.55 | 18.79 | 0.76 |
| S1 | | 11 38 25.03 | +36 54 44.02 | 18.90 | 17.52 | 1.38 |
| S2 | | 11 37 56.81 | +36 52 35.56 | 17.91 | 17.41 | 0.50 |
| J120014.08−004638.7 | 2018 March 12, May 11 | 12 00 14.08 | −00 46 38.74 | 18.51 | 17.81 | 0.70 |
| S1 | | 12 00 12.63 | −00 46 07.14 | 17.19 | 16.68 | 0.51 |
| S2 | | 12 00 25.99 | −00 51 45.21 | 16.73 | 16.37 | 0.36 |
| J124634.65+023809.1 | 2017 April 03; 2018 April 12 | 12 46 34.65 | +02 38 09.06 | 18.35 | 18.18 | 0.17 |
| S1 | 2017 April 03; 2018 April 12 | 12 47 00.55 | +02 37 31.37 | 17.91 | 16.89 | 1.02 |
| S2 | 2017 April 03 | 12 47 05.32 | +02 39 06.75 | 16.94 | 16.62 | 0.32 |
| S3 | 2018 April 12 | 12 46 49.50 | +02 37 11.64 | 17.24 | 16.77 | 0.47 |
| J163323.59+471859.0 | 2017 May 20; 2019 March 20 | 16 33 23.59 | +47 18 59.04 | 17.25 | 16.95 | 0.30 |
| S1 | | 16 32 59.26 | +47 26 05.45 | 15.57 | 15.18 | 0.39 |
| S2 | | 16 32 56.00 | +47 21 01.26 | 15.55 | 15.11 | 0.44 |
| J163401.94+480940.2 | 2018 March 22, 26 | 16 34 01.94 | +48 09 40.20 | 19.54 | 19.21 | 0.33 |
| S1 | 2018 March 22 | 16 34 04.24 | +48 11 32.47 | 19.65 | 18.77 | 0.88 |
| S2 | 2018 March 22 | 16 33 50.78 | +48 10 09.78 | 19.04 | 17.86 | 1.18 |
| S3 | 2021 April 09 | 16 33 31.81 | +48 04 31.89 | 19.73 | 18.34 | 1.39 |
| S4 | 2021 April 09 | 16 34 01.24 | +48 08 36.92 | 18.25 | 17.30 | 0.95 |

The SDSS DR14 catalog (Abolfathi et al., 2018) has been used for getting optical positions and apparent magnitudes of the sources and their comparison stars.

*The USNO-A2.0 catalog (Monet, 1998) has been used in case of non-availability of the SDSS 'g-r' color. The 'B-R' color has been used in such cases.

Table A.2 Details of the observations and the status of the INOV for the sample of 23 RLNLSy1 galaxies studied in this work (aperture radius used = 2×FWHM).

| RLNLSy1s (SDSS name) | Date(s)[a] yyyy.mm.dd | $T^b$ (hrs) | $N^c$ | Median[d] FWHM (arcsec) | $F^\eta$-test $F^\eta_{s1}, F^\eta_{s2}$ 99% | INOV status[e] 99% | $F^\eta$-test $F^\eta_{s1-s2}$ 99% | Variability status of s1−s2 | $F_{enh}$-test $F_{enh}$ | INOV status[f] 99% | $\sqrt{\langle \sigma^2_{i,err}\rangle}$ (AGN-s)[g] | $\overline{\psi}^g_{s1,s2}$ (%) |
|---|---|---|---|---|---|---|---|---|---|---|---|---|
| (1) | (2) | (3) | (4) | (5) | (6) | (7) | (8) | (9) | (10) | (11) | (12) | (13) |
| jetted NLSy1s ||||||||||||||
| J032441.20+341045.0 | (2016.11.22) | 4.42 | 56 | 2.32 | 14.34, 16.13 | V, V | 00.82 | NV | 17.39 | V | 0.003 | 5.38 |
|  | 2016.11.23 | 4.27 | 54 | 2.13 | 03.77, 02.46 | V, V | 00.70 | NV | 05.42 | V | 0.004 | 2.79 |
|  | (2016.12.02) | 4.41 | 44 | 2.60 | 85.80, 88.73 | V, V | 00.87 | NV | 98.29 | V | 0.003 | 11.44 |
|  | 2017.01.03 | 3.00 | 39 | 2.47 | 04.53, 08.53 | V, V | 00.34 | NV | 13.27 | V | 0.004 | 3.96 |
|  | 2017.01.04 | 3.39 | 33 | 2.45 | 21.60, 23.34 | V, V | 00.44 | NV | 49.35 | V | 0.004 | 7.99 |
| J081432.12+560958.7 | (2017.01.03) | 3.37 | 19 | 2.69 | 00.33, 00.34 | NV, NV | 00.16 | NV | 01.34 | NV | 0.022 | – |
|  | (2017.11.20) | 4.52 | 32 | 2.79 | 00.86, 00.55 | NV, NV | 00.55 | NV | 05.28 | NV | 0.031 | – |
| J084957.98+510829.0 | (2017.12.13) | 4.42 | 24 | 2.83 | 00.42, 00.51 | NV, NV | 01.06 | NV | 00.77 | NV | 0.033 | – |
|  | (2019.04.08) | 3.04 | 13 | 2.88 | 00.66, 00.89 | NV, NV | 00.62 | NV | 00.62 | NV | 0.032 | – |
| J090227.20+044309.0 | (2017.02.22) | 3.59 | 27 | 2.38 | 00.71, 00.57 | NV, NV | 00.25 | NV | 02.81 | NV | 0.025 | – |
|  | (2017.12.14) | 5.65 | 39 | 2.51 | 00.33, 00.37 | NV, NV | 00.24 | NV | 01.38 | NV | 0.024 | – |
| J094857.32+002225.6 | (2016.12.02) | 4.15 | 17 | 2.58 | 01.71, 01.88 | NV, NV | 00.16 | NV | 10.52 | V | 0.017 | 7.95 |
|  | (2017.12.21) | 5.19 | 33 | 2.24 | 13.95, 16.53 | V , V | 00.55 | NV | 25.26 | V | 0.012 | 16.42 |
| J095317.10+283601.5 | (2017.03.04) | 3.97 | 29 | 2.41 | 00.36, 00.35 | NV, NV | 00.48 | NV | 00.76 | NV | 0.035 | – |
|  | (2020.11.21) | 3.25 | 11 | 3.10 | 00.77, 00.68 | NV, NV | 00.33 | NV | 02.31 | NV | 0.035 | – |
| J104732.78+472532.0 | (2017.04.11) | 3.75 | 47 | 0.98 | 00.54, 00.53 | NV, NV | 00.53 | NV | 01.03 | NV | 0.035 | – |
|  | (2018.03.12) | 3.82 | 15 | 2.77 | 00.60, 00.63 | NV, NV | 00.38 | NV | 01.58 | NV | 0.028 | – |
| J122222.99+041315.9 | 2017.01.03 | 3.52 | 17 | 2.38 | 00.62, 00.30 | NV, NV | 00.91 | NV | 00.68 | NV | 0.018 | – |
|  | 2017.01.04 | 3.14 | 16 | 2.36 | 00.32, 00.37 | NV, NV | 00.16 | NV | 01.99 | NV | 0.014 | – |
|  | 2017.02.21 | 4.44 | 41 | 2.65 | 00.74, 00.76 | NV, NV | 00.35 | NV | 02.13 | V | 0.020 | 6.36 |
|  | (2017.02.22) | 5.50 | 50 | 2.59 | 03.98, 03.60 | V , V | 00.61 | NV | 06.51 | V | 0.017 | 13.33 |
|  | (2017.03.04) | 4.93 | 39 | 2.61 | 00.72, 00.86 | NV, NV | 00.53 | NV | 01.36 | NV | 0.019 | – |
|  | 2017.03.24 | 3.94 | 39 | 2.37 | 00.93, 00.75 | NV, NV | 00.56 | NV | 01.66 | NV | 0.020 | – |
| J130522.75+511640.2 | (2017.04.04) | 3.79 | 23 | 2.57 | 00.70, 00.66 | NV, NV | 00.24 | NV | 02.94 | PV | 0.012 | – |
|  | (2019.04.25) | 3.11 | 22 | 2.77 | 01.56, 01.42 | NV, NV | 00.39 | NV | 04.12 | PV | 0.018 | – |
| J142114.05+282452.8 | (2018.05.10) | 4.06 | 26 | 2.83 | 00.55, 00.56 | NV, NV | 00.35 | NV | 01.60 | NV | 0.019 | – |
|  | (2019.05.27) | 3.31 | 18 | 2.68 | 00.86, 00.86 | NV, NV | 00.36 | NV | 02.47 | PV | 0.021 | – |
| J144318.56+472556.7 | (2018.03.11) | 3.05 | 19 | 3.15 | 00.54, 00.56 | NV, NV | 00.21 | NV | 02.59 | PV | 0.022 | – |
|  | (2018.03.23) | 3.13 | 23 | 2.33 | 00.35, 00.36 | NV, NV | 00.35 | NV | 01.00 | NV | 0.018 | – |
| J150506.48+032630.8 | (2017.03.25) | 5.21 | 41 | 2.08 | 00.60, 00.59 | NV, NV | 00.58 | NV | 01.04 | NV | 0.028 | – |
|  | (2018.04.12) | 3.05 | 19 | 2.55 | 00.67, 00.63 | NV, NV | 00.80 | NV | 00.84 | NV | 0.032 | – |
| J154817.92+351128.0 | (2018.05.17) | 3.00 | 19 | 3.08 | 00.40, 00.38 | NV, NV | 00.38 | NV | 01.05 | NV | 0.008 | – |
|  | (2019.05.08) | 3.24 | 14 | 2.79 | 00.60, 00.65 | NV, NV | 00.30 | NV | 01.98 | NV | 0.017 | – |
| J164442.53+261913.3 | (2017.04.03) | 4.37 | 37 | 2.50 | 01.44, 01.28 | NV, NV | 00.41 | NV | 03.53 | V | 0.011 | 5.41 |
|  | (2019.04.26) | 3.22 | 24 | 2.27 | 03.06, 03.74 | V , V | 00.48 | NV | 06.40 | V | 0.011 | 7.50 |
| J170330.38+454047.3 | (2017.06.03) | 3.76 | 37 | 2.41 | 00.75, 00.67 | NV, NV | 00.64 | NV | 01.17 | NV | 0.004 | – |
|  | (2019.03.25) | 3.13 | 45 | 4.45 | 00.63, 01.08 | NV, NV | 00.90 | NV | 00.70 | NV | 0.010 | – |
| non-jetted NLSy1s ||||||||||||||
| J085001.17+462600.5 | (2017.01.04) | 3.26 | 13 | 2.28 | 00.39, 00.50 | NV, NV | 00.50 | NV | 00.78 | NV | 0.026 | – |
|  | (2017.12.15) | 3.69 | 20 | 2.89 | 00.80, 00.74 | NV, NV | 00.36 | NV | 02.22 | NV | 0.032 | – |
| J103727.45+003635.6 | (2018.03.11) | 3.30 | 11 | 3.12 | 01.18, 01.22 | NV, NV | 01.08 | NV | 01.09 | NV | 0.030 | – |
|  | (2021.04.08) | 4.96 | 13 | 2.42 | 01.54, 01.77 | NV, NV | 01.00 | NV | 01.53 | NV | 0.043 | – |
| J111005.03+365336.2 | (2018.03.23) | 3.22 | 44 | 0.90 | 00.34, 00.35 | NV, NV | 00.47 | NV | 00.72 | NV | 0.027 | – |
|  | (2019.01.13) | 3.13 | 11 | 3.07 | 02.61, 02.54 | NV, NV | 01.18 | NV | 02.11 | NV | 0.042 | – |
| J113824.53+365327.2 | (2017.04.17) | 4.32 | 20 | 2.11 | 00.32, 00.40 | NV, NV | 00.21 | NV | 01.58 | NV | 0.028 | – |
|  | (2018.03.23) | 4.31 | 21 | 2.53 | 00.36, 00.38 | NV, NV | 00.16 | NV | 02.26 | NV | 0.029 | – |
| J120014.08−004638.7 | (2018.03.12) | 3.83 | 28 | 2.72 | 00.31, 00.22 | NV, NV | 00.19 | NV | 01.63 | NV | 0.011 | – |
|  | (2018.05.11) | 3.13 | 16 | 2.97 | 00.13, 00.25 | NV, NV | 00.33 | NV | 00.39 | NV | 0.018 | – |
| J124634.65+023809.1 | (2017.04.03) | 3.77 | 18 | 2.50 | 00.21, 00.36 | NV, NV | 00.46 | NV | 00.46 | NV | 0.024 | – |
|  | (2018.04.12) | 3.72 | 21 | 2.71 | 00.59, 00.66 | NV, NV | 00.31 | NV | 01.89 | NV | 0.018 | – |
| J163323.59+471859.0 | (2017.05.20) | 4.33 | 36 | 2.26 | 00.85, 00.76 | NV, NV | 00.23 | NV | 03.76 | V | 0.007 | 2.56 |
|  | (2019.03.20) | 3.69 | 33 | 2.66 | 01.61, 01.79 | NV, NV | 00.36 | NV | 04.43 | V | 0.016 | 9.52 |
| J163401.94+480940.2 | (2018.03.23) | 3.04 | 34 | 0.98 | 00.62, 00.71 | NV, NV | 00.39 | NV | 01.61 | NV | 0.012 | – |
|  | (2021.04.09) | 4.58 | 12 | 2.12 | 01.27, 01.63 | NV, NV | 00.34 | NV | 03.76 | PV | 0.043 | – |

[a]Date(s) of the monitoring session(s). The dates given inside parentheses refer to the sessions we have used here for estimating the INOV duty cycle (e.g., see text in Sect. 5.4.1). [b]Duration of the monitoring session in the observed frame. [c]Number of data points in the DLCs of the monitoring session. [d]Median seeing (FWHM in arcsec) for the session. [e, f]INOV status inferred from $F^\eta$ and $F_{enh}$ tests, with V = variable , i.e. confidence level ≥ 99%; PV = probable variable, i.e. 95 − 99% confidence level; NV = non-variable, i.e. confidence level < 95%.
[g]Mean amplitude of variability in the two DLCs of the target NLSy1 (i.e., relative to the two chosen comparison stars).

# Appendix B

# The Complete Sample of NLSy1 and BLSy1 Galaxies.

This appendix contains the entire table of the physical properties of 144 NLSy1 galaxies (Table B1) and 117 BLSy1 galaxies (Table B2) studied in Chapter 5.

Table B.1 The table of properties obtained for the sample of 144 NLSy1 galaxies.

| SDSS Name[a] | z[b] | FWHM(Hβ)[c] | EW(Hβ)[d] | [OIII]/Hβ[e] | Hα/Hβ[f] | $R_{Fe}$[g] | $\log L_{5100\text{\AA}}$[h] | $M_{BH}$[i] | $\log(R_{Edd})$[j] | AI[k] | KI[l] | $\Gamma_X$[m] |
|---|---|---|---|---|---|---|---|---|---|---|---|---|
| J001137.20-144160.0 | 0.1319 | 705 ± 152 | 26.87 ± 7.09 | 0.36 ± 0.13 | 4.65 ± 0.31 | 1.35 ± 0.30 | 43.853 ± 0.002 | 6.64 ± 0.30 | 0.404 ± 0.174 | 0.114 ± 0.025 | 2.52 ± 0.18 |
| J003859.28-005450.4 | 0.4137 | 1537 ± 183 | 70.08 ± 14.72 | 0.34 ± 0.10 | 1.79 ± 0.25 | 0.50 ± 0.15 | 43.915 ± 0.005 | 7.32 ± 0.17 | 0.03 ± 0.001 | 0.321 ± 0.038 | 2.53 ± 0.21 |
| J011911.52-104532.4 | 0.1253 | 1478 ± 297 | 40.28 ± 13.16 | 0.00 ± 0.00 | 0.87 ± 0.20 | 0.63 ± 0.29 | 43.241 ± 0.006 | 7.26 ± 2.84 | 0.129 ± 0.517 | 0.265 ± 0.534 | 2.97 ± 0.66 |
| J016448.88-004044.4 | 0.0824 | 1699 ± 184 | 44.76 ± 7.23 | 0.53 ± 0.12 | 2.85 ± 0.19 | 1.32 ± 0.30 | 43.545 ± 0.002 | 7.39 ± 0.15 | -0.24 ± 0.05 | 0.316 ± 0.034 | 2.92 ± 0.31 |
| J014904.56-125746.8 | 0.7305 | 1454 ± 336 | 153.93 ± 44.27 | 0.00 ± 0.00 | 0.00 ± 0.00 | 7.43 ± 0.33 | 44.583 ± 0.015 | 7.43 ± 0.33 | 0.229 ± 0.106 | 0.345 ± 0.080 | 2.12 ± 0.47 |
| J020337.20-051406.0 | 0.5193 | 1746 ± 242 | 74.09 ± 4.49 | 0.00 ± 0.00 | 0.06 ± 0.04 | 0.04 ± 0.01 | 44.160 ± 0.003 | 7.47 ± 0.20 | 0.014 ± 0.004 | 0.228 ± 0.032 | 1.58 ± 0.58 |
| J020853.28-043354.0 | 0.5563 | 1513 ± 439 | 56.40 ± 8.48 | 0.28 ± 0.06 | 2.03 ± 0.14 | 0.96 ± 0.20 | 44.345 ± 0.003 | 7.38 ± 0.41 | 0.040 ± 0.023 | 0.296 ± 0.086 | 2.84 ± 0.84 |
| J021329.28-051138.4 | 0.443 | 705 ± 254 | 8.39 ± 3.24 | 3.22 ± 1.75 | 2.15 ± 0.32 | 3.54 ± 1.93 | 44.080 ± 0.006 | 6.67 ± 0.51 | 0.023 ± 0.017 | 0.347 ± 0.125 | 2.21 ± 0.80 |
| J021803.60-004337.2 | 0.3796 | 968 ± 506 | 68.57 ± 11.15 | 0.00 ± 0.00 | 1.61 ± 0.16 | 0.75 ± 0.17 | 44.061 ± 0.005 | 6.94 ± 0.74 | 0.50 ± 0.53 | 0.325 ± 0.170 | 2.84 ± 0.19 |
| J022452.32-040520.4 | 0.6952 | 859 ± 100 | 56.08 ± 12.07 | 0.39 ± 0.12 | 1.23 ± 0.38 | 0.00 ± 0.00 | 44.589 ± 0.010 | 6.97 ± 0.22 | -0.029 ± 0.007 | 0.299 ± 0.035 | 2.61 ± 0.27 |
| J022928.32-051124.0 | 0.3068 | 1338 ± 175 | 35.45 ± 2.62 | 0.42 ± 0.04 | 4.46 ± 0.45 | 1.63 ± 0.17 | 44.573 ± 0.001 | 7.35 ± 0.19 | -0.121 ± 0.032 | 0.313 ± 0.041 | 2.97 ± 0.23 |
| J023039.84-051557.6 | 0.1424 | 1027 ± 818 | 25.84 ± 2.08 | 0.32 ± 0.04 | 2.48 ± 0.19 | 0.77 ± 0.09 | 43.336 ± 0.004 | 6.94 ± 1.13 | -0.206 ± 0.329 | 0.277 ± 0.220 | 2.27 ± 0.60 |
| J030417.76-002826.4 | 0.0444 | 1304 ± 351 | 23.39 ± 3.71 | 0.46 ± 0.10 | 1.84 ± 0.26 | 1.61 ± 0.36 | 43.131 ± 0.003 | 7.15 ± 0.38 | -0.140 ± 0.075 | 0.268 ± 0.072 | 2.57 ± 0.54 |
| J030639.60-000343.2 | 0.1074 | 1454 ± 43 | 73.63 ± 4.59 | 0.35 ± 0.03 | 1.38 ± 0.14 | 0.36 ± 0.03 | 43.987 ± 0.001 | 7.28 ± 0.04 | 0.020 ± 0.001 | 0.305 ± 0.009 | 3.64 ± 1.04 |
| J030705.76-000010.8 | 0.2662 | 950 ± 972 | 16.08 ± 15.16 | 1.88 ± 2.51 | 4.46 ± 1.78 | 1.48 ± 1.45 | 43.612 ± 0.012 | 6.89 ± 1.45 | 0.30 ± 0.61 | 0.333 ± 0.341 | 3.06 ± 0.42 |
| J034000.48-051745.6 | 0.3451 | 1311 ± 212 | 36.63 ± 4.34 | 0.17 ± 0.03 | 2.57 ± 0.25 | 2.04 ± 0.34 | 44.726 ± 0.005 | 7.41 ± 0.23 | -0.153 ± 0.049 | 0.297 ± 0.048 | 2.54 ± 0.48 |
| J074207.68-251727.6 | 0.2941 | 1454 ± 86 | 111.02 ± 11.82 | 0.00 ± 0.00 | 1.44 ± 0.13 | 0.45 ± 0.07 | 44.026 ± 0.004 | 7.29 ± 0.08 | 0.13 ± 0.02 | 0.310 ± 0.018 | 2.03 ± 0.23 |
| J075525.20-391109.6 | 0.0332 | 1444 ± 431 | 30.88 ± 2.48 | 0.38 ± 0.04 | 1.51 ± 0.14 | 1.24 ± 0.14 | 43.325 ± 0.001 | 7.24 ± 0.42 | -0.21 ± 0.13 | 0.255 ± 0.076 | 2.74 ± 0.05 |
| J080101.44-184839.6 | 0.1395 | 1613 ± 118 | 68.17 ± 4.08 | 0.00 ± 0.00 | 2.00 ± 0.28 | 1.50 ± 0.13 | 44.270 ± 0.005 | 7.42 ± 0.10 | 0.16 ± 0.03 | 0.271 ± 0.020 | 3.12 ± 0.12 |
| J080608.16-244420.4 | 0.3579 | 1454 ± 107 | 135.19 ± 16.19 | 0.17 ± 0.03 | 3.03 ± 0.41 | 0.63 ± 0.17 | 44.396 ± 0.006 | 7.36 ± 0.16 | 0.33 ± 0.05 | 0.279 ± 0.021 | 2.44 ± 0.12 |
| J081442.00-212916.8 | 0.1626 | 1581 ± 179 | 70.21 ± 5.44 | 0.00 ± 0.00 | 1.49 ± 0.13 | 0.79 ± 0.09 | 43.980 ± 0.001 | 7.36 ± 0.16 | 0.04 ± 0.01 | 0.302 ± 0.004 | 2.78 ± 0.06 |
| J084719.79+380043.6 | 0.1364 | 1332 ± 1539 | 43.95 ± 12.91 | 0.12 ± 0.05 | 1.00 ± 0.19 | 0.73 ± 0.30 | 43.807 ± 0.004 | 7.19 ± 1.63 | -0.589 ± 1.360 | 0.139 ± 0.161 | 3.18 ± 0.98 |
| J085946.32-274536.0 | 0.2439 | 1146 ± 193 | 56.10 ± 4.93 | 0.15 ± 0.02 | 2.09 ± 0.22 | 1.36 ± 0.17 | 44.410 ± 0.001 | 7.16 ± 0.24 | -0.039 ± 0.006 | 0.286 ± 0.048 | 3.49 ± 0.70 |
| J090247.04+601048.0 | 0.4981 | 705 ± 39 | 32.87 ± 4.40 | 0.40 ± 0.08 | 0.00 ± 0.00 | 2.04 ± 0.38 | 44.863 ± 0.003 | 6.97 ± 0.08 | 0.004 ± 0.000 | 0.319 ± 0.018 | 3.04 ± 1.35 |
| J090741.52+500813.2 | 0.2088 | 1110 ± 175 | 28.28 ± 3.14 | 0.27 ± 0.04 | 2.26 ± 0.28 | 1.23 ± 0.19 | 44.067 ± 0.002 | 7.06 ± 0.22 | 0.39 ± 0.12 | 0.288 ± 0.046 | 3.40 ± 0.54 |
| J091448.96+085320.4 | 0.1399 | 705 ± 50 | 38.93 ± 3.39 | 0.00 ± 0.00 | 0.11 ± 0.01 | 0.17 ± 0.02 | 43.549 ± 0.002 | 6.62 ± 0.10 | 0.52 ± 0.08 | 0.337 ± 0.024 | 4.43 ± 0.16 |
| J091848.72+211716.8 | 0.149 | 906 ± 133 | 19.08 ± 6.05 | 2.90 ± 1.30 | 9.37 ± 1.17 | 2.95 ± 1.32 | 44.076 ± 0.002 | 6.88 ± 0.21 | 0.57 ± 0.17 | 0.302 ± 0.044 | 3.86 ± 0.61 |
| J092628.08+304543.2 | 0.2603 | 1171 ± 1135 | 30.55 ± 8.68 | 0.43 ± 0.17 | 1.58 ± 0.30 | 1.07 ± 0.43 | 43.860 ± 0.003 | 7.08 ± 1.37 | 0.24 ± 0.46 | 0.264 ± 0.256 | 2.54 ± 0.35 |
| J094106.24+033925.2 | 0.4318 | 709 ± 81 | 23.89 ± 3.98 | 0.15 ± 0.04 | 3.05 ± 0.34 | 3.60 ± 0.85 | 44.066 ± 0.004 | 6.67 ± 0.16 | -0.012 ± 0.003 | 0.338 ± 0.039 | 3.71 ± 1.51 |
| J094404.32+480646.8 | 0.3923 | 912 ± 166 | 70.82 ± 11.41 | 0.21 ± 0.05 | 3.54 ± 0.46 | 1.64 ± 0.37 | 44.467 ± 0.002 | 6.98 ± 0.26 | -0.047 ± 0.017 | 0.271 ± 0.049 | 3.03 ± 0.17 |
| J094610.80-095226.4 | 0.6974 | 998 ± 470 | 64.32 ± 10.63 | 0.00 ± 0.00 | 1.41 ± 0.33 | 0.00 ± 0.00 | 44.595 ± 0.024 | 7.11 ± 0.67 | -0.174 ± 0.163 | 0.255 ± 0.120 | 3.16 ± 0.10 |
| J094704.56+472143.2 | 0.5393 | 798 ± 157 | 24.56 ± 6.53 | 0.17 ± 0.06 | 5.87 ± 0.88 | 3.96 ± 1.47 | 45.042 ± 0.001 | 7.26 ± 0.28 | 1.16 ± 0.46 | 0.333 ± 0.066 | 3.66 ± 1.30 |
| J094903.60+474655.2 | 0.2145 | 1131 ± 247 | 65.85 ± 8.07 | 0.14 ± 0.03 | 2.04 ± 0.24 | 0.77 ± 0.13 | 43.879 ± 0.004 | 7.05 ± 0.31 | 0.28 ± 0.12 | 0.292 ± 0.064 | 3.17 ± 0.66 |
| J095833.84+560224.0 | 0.2155 | 1007 ± 689 | 30.14 ± 6.72 | 0.90 ± 0.28 | 2.28 ± 0.26 | 1.73 ± 0.54 | 43.954 ± 0.002 | 6.96 ± 0.97 | 0.42 ± 0.57 | 0.183 ± 0.126 | 3.17 ± 0.33 |
| J095921.36+024030.0 | 0.2598 | 1105 ± 1068 | 19.80 ± 2.90 | 0.30 ± 0.06 | 1.52 ± 0.25 | 0.92 ± 0.19 | 43.979 ± 0.004 | 7.04 ± 1.37 | 0.35 ± 0.67 | 0.223 ± 0.216 | 2.46 ± 1.12 |
| J100321.16-553632.4 | 0.215 | 1266 ± 1376 | 45.61 ± 12.30 | 1.14 ± 0.43 | 1.86 ± 0.28 | 0.41 ± 0.15 | 43.813 ± 0.003 | 7.15 ± 1.54 | -0.071 ± 0.154 | 0.284 ± 0.308 | 2.99 ± 0.40 |

[a] Obtained from SDSS SAS.
[b] Source redshift based on single epoch spectra available in SDSS DR16.
[c] The broad component of Hβ emission profile in the units of km s$^{-1}$.
[d] Equivalent width of Hβ emission line.
[e] [OIII]/Hβ emission line narrow component flux ratio in logarithmic units.
[f] Hα/Hβ broad component flux ratio.
[g] $R_{Fe}$, calculated as the flux ratio of area covered by the broad Fe line between 4433 Å and 4684 Å and the area covered by the Hβ emission line.
[h] Optical luminosity at 5100Å($\lambda L_{5100\text{\AA}}$) in units of ergs/sec/Å
[i] SMBH mass in the units of log($M_{BH}/M\odot$) obtained using the Radius-Luminosity relation and FWHM of Hβ emission line.
[j] Eddington ratio, the ratio of bolometric to Eddington luminosity
[k] Asymmetry index for the Hβ emission line calculated using Equation 1(a)
[l] Kurtosis Index for the Hβ emission line calculated using Equation 1(b)
[m] The soft X-ray photon index calculated between 0.2-2 KeV energy range taken from Ojha et al. 2020.

*(continued)*

| SDSS Name | z | FWHM($H\beta$) | EW($H\beta$) | [OIII]/$H\beta$ | $H\alpha/H\beta$ | $R_{Fe}$ | $\log(L_{5100\text{Å}})$ | $\log(M_{BH}/M\odot)$ | $\log(R_{EDD})$ | AI | KI | $\Gamma_X$ |
|---|---|---|---|---|---|---|---|---|---|---|---|---|
| J100055.68+314001.2 | 0.1946 | 975 ± 449 | 22.97 ± 3.90 | 0.11 ± 0.03 | 2.54 ± 0.38 | 2.32 ± 0.56 | 44.216 ± 0.002 | 6.97 ± 0.65 | 0.58 ± 0.53 | -0.134 ± 0.124 | 0.235 ± 0.109 | 2.88 ± 0.15 |
| J100232.16+023538.4 | 0.6573 | 1315 ± 398 | 65.17 ± 7.60 | 0.16 ± 0.02 | 0.00 ± 0.00 | 0.83 ± 0.14 | 44.827 ± 0.006 | 7.48 ± 0.43 | 0.62 ± 0.38 | -0.106 ± 0.064 | 0.235 ± 0.071 | 3.36 ± 0.98 |
| J101743.68+212006.0 | 0.2439 | 1454 ± 227 | 73.64 ± 14.05 | 0.06 ± 0.02 | 1.65 ± 0.21 | 0.93 ± 0.25 | 44.221 ± 0.002 | 7.32 ± 0.22 | 0.23 ± 0.07 | -0.012 ± 0.004 | 0.302 ± 0.047 | 3.12 ± 0.44 |
| J101837.68+532540.8 | 0.1751 | 1454 ± 190 | 74.59 ± 21.48 | 0.47 ± 0.19 | 1.18 ± 0.19 | 0.60 ± 0.24 | 43.498 ± 0.003 | 7.25 ± 0.19 | -0.13 ± 0.03 | -0.025 ± 0.006 | 0.300 ± 0.039 | 2.38 ± 0.54 |
| J101837.68+532540.8 | 0.1751 | 1335 ± 240 | 40.12 ± 16.47 | 0.72 ± 0.42 | 2.43 ± 0.42 | 1.30 ± 0.75 | 43.576 ± 0.005 | 7.18 ± 0.25 | -0.02 ± 0.01 | -0.261 ± 0.094 | 0.236 ± 0.042 | 2.38 ± 0.54 |
| J102531.20+514033.6 | 0.0446 | 1086 ± 285 | 54.94 ± 6.90 | 0.15 ± 0.03 | 1.49 ± 0.21 | 1.29 ± 0.23 | 43.558 ± 0.001 | 7.00 ± 0.37 | 0.15 ± 0.08 | -0.021 ± 0.011 | 0.333 ± 0.088 | 3.15 ± 0.10 |
| J102551.12+384008.4 | 0.1479 | 1454 ± 442 | 42.79 ± 7.72 | 0.00 ± 0.00 | 2.41 ± 0.20 | 0.45 ± 0.12 | 42.948 ± 0.009 | 7.24 ± 0.43 | -0.41 ± 0.25 | 0.095 ± 0.058 | 0.288 ± 0.087 | 2.36 ± 0.77 |
| J103438.64+393827.6 | 0.0431 | 912 ± 86 | 4.83 ± 2.57 | 3.37 ± 2.53 | 10.78 ± 1.56 | 7.16 ± 5.39 | 43.255 ± 0.001 | 6.84 ± 0.13 | 0.15 ± 0.03 | -0.017 ± 0.003 | 0.333 ± 0.031 | 3.91 ± 0.06 |
| J104336.96+453621.6 | 0.1399 | 868 ± 90 | 16.56 ± 2.68 | 0.21 ± 0.05 | 4.51 ± 0.46 | 1.71 ± 0.39 | 43.813 ± 0.002 | 6.82 ± 0.15 | 0.47 ± 0.10 | -0.015 ± 0.003 | 0.327 ± 0.034 | 3.15 ± 0.58 |
| J104613.68+525555.2 | 0.5021 | 1152 ± 368 | 28.33 ± 3.81 | 0.07 ± 0.01 | 0.00 ± 0.00 | 2.51 ± 0.48 | 45.074 ± 0.002 | 7.63 ± 0.45 | 0.86 ± 0.55 | -0.243 ± 0.155 | 0.263 ± 0.084 | 4.02 ± 0.12 |
| J104856.88+592826.4 | 0.0933 | 1454 ± 441 | 32.54 ± 3.37 | 0.52 ± 0.08 | 1.77 ± 0.12 | 0.08 ± 0.01 | 43.231 ± 0.003 | 7.24 ± 0.43 | -0.26 ± 0.16 | -0.042 ± 0.026 | 0.292 ± 0.089 | 1.80 ± 0.93 |
| J110016.08+393524.0 | 0.3129 | 1224 ± 220 | 49.64 ± 4.27 | 0.00 ± 0.00 | 2.17 ± 0.22 | 1.67 ± 0.20 | 44.209 ± 0.003 | 7.17 ± 0.25 | 0.37 ± 0.14 | -0.161 ± 0.058 | 0.309 ± 0.056 | 3.49 ± 0.62 |
| J110016.08+393524.0 | 0.3129 | 1328 ± 150 | 43.99 ± 4.15 | 0.27 ± 0.04 | 2.56 ± 0.24 | 1.63 ± 0.22 | 44.330 ± 0.003 | 7.26 ± 0.16 | 0.36 ± 0.08 | -0.139 ± 0.031 | 0.302 ± 0.034 | 3.49 ± 0.62 |
| J111006.96+612522.8 | 0.2625 | 1454 ± 31 | 121.62 ± 9.21 | 0.19 ± 0.02 | 1.17 ± 0.15 | 0.75 ± 0.08 | 44.269 ± 0.002 | 7.33 ± 0.03 | 0.25 ± 0.01 | 0.021 ± 0.001 | 0.316 ± 0.007 | 2.71 ± 0.11 |
| J111520.88+404326.4 | 0.0786 | 1454 ± 34 | 46.07 ± 1.20 | 0.00 ± 0.00 | 2.07 ± 0.07 | 0.47 ± 0.02 | 43.440 ± 0.003 | 7.25 ± 0.03 | -0.16 ± 0.01 | 0.020 ± 0.001 | 0.295 ± 0.007 | 2.70 ± 0.18 |
| J111650.16+424724.0 | 0.1754 | 705 ± 161 | 16.27 ± 3.78 | 0.00 ± 0.00 | 4.87 ± 0.57 | 1.38 ± 0.45 | 43.590 ± 0.002 | 6.63 ± 0.32 | 0.54 ± 0.25 | -0.038 ± 0.017 | 0.310 ± 0.071 | 2.89 ± 0.87 |
| J111702.88+421239.5 | 0.2479 | 705 ± 116 | 12.93 ± 2.73 | 0.44 ± 0.13 | 4.06 ± 0.60 | 3.97 ± 1.18 | 43.917 ± 0.004 | 6.65 ± 0.23 | 0.71 ± 0.24 | 0.018 ± 0.006 | 0.342 ± 0.056 | 3.73 ± 0.49 |
| J111757.84+640202.3 | 0.1934 | 1472 ± 150 | 87.48 ± 7.60 | 0.00 ± 0.00 | 2.18 ± 0.24 | 0.55 ± 0.07 | 43.967 ± 0.003 | 7.29 ± 0.14 | 0.09 ± 0.02 | 0.199 ± 0.041 | 0.272 ± 0.028 | 3.02 ± 0.53 |
| J112306.24+013748.0 | 0.6956 | 1512 ± 235 | 49.71 ± 6.35 | 0.09 ± 0.02 | 0.00 ± 0.00 | 1.48 ± 0.27 | 44.688 ± 0.016 | 7.51 ± 0.22 | 0.43 ± 0.14 | -0.336 ± 0.104 | 0.228 ± 0.035 | 3.63 ± 0.19 |
| J112328.08+052822.8 | 0.1011 | 1586 ± 264 | 49.07 ± 3.58 | 0.36 ± 0.04 | 2.22 ± 0.24 | 0.93 ± 0.10 | 44.513 ± 0.003 | 7.33 ± 0.24 | -0.20 ± 0.07 | -0.035 ± 0.011 | 0.260 ± 0.043 | 2.98 ± 0.08 |
| J112405.04+061250.4 | 0.2718 | 767 ± 139 | 17.06 ± 2.51 | 0.96 ± 0.20 | 3.07 ± 0.29 | 2.43 ± 0.51 | 44.227 ± 0.002 | 6.76 ± 0.26 | 0.79 ± 0.29 | -0.153 ± 0.055 | 0.219 ± 0.040 | 2.91 ± 0.51 |
| J113003.12+655627.6 | 0.1326 | 705 ± 68 | 15.49 ± 5.64 | 1.71 ± 0.88 | 3.25 ± 0.47 | 1.54 ± 0.79 | 43.732 ± 0.002 | 6.63 ± 0.14 | 0.61 ± 0.12 | 0.038 ± 0.007 | 0.293 ± 0.028 | 2.63 ± 0.79 |
| J113233.60+273957.6 | 0.6812 | 781 ± 95 | 49.53 ± 5.51 | 0.37 ± 0.06 | 0.00 ± 0.00 | 1.53 ± 0.24 | 44.840 ± 0.004 | 7.04 ± 0.17 | 1.08 ± 0.27 | 0.018 ± 0.005 | 0.333 ± 0.041 | 2.91 ± 0.32 |
| J113233.60+273957.6 | 0.6812 | 1454 ± 298 | 74.26 ± 10.08 | 0.25 ± 0.05 | 2.43 ± 0.24 | 1.00 ± 0.19 | 44.939 ± 0.004 | 7.67 ± 0.29 | 0.59 ± 0.25 | -0.011 ± 0.004 | 0.333 ± 0.068 | 2.91 ± 0.32 |
| J113447.28+490133.6 | 0.2527 | 705 ± 67 | 18.44 ± 2.42 | 0.47 ± 0.09 | 2.43 ± 0.24 | 2.12 ± 0.39 | 43.595 ± 0.003 | 6.63 ± 0.13 | 0.55 ± 0.11 | -0.031 ± 0.006 | 0.310 ± 0.029 | 2.05 ± 0.46 |
| J113450.16+490328.8 | 0.6443 | 983 ± 816 | 31.99 ± 5.68 | 0.53 ± 0.13 | 0.00 ± 0.00 | 1.26 ± 0.32 | 44.218 ± 0.008 | 6.98 ± 1.17 | 0.58 ± 0.97 | 0.392 ± 0.652 | 0.166 ± 0.138 | 2.94 ± 0.68 |
| J114008.64+030712.0 | 0.081 | 792 ± 115 | 6.64 ± 3.08 | 0.93 ± 0.61 | 5.29 ± 0.80 | 6.74 ± 4.42 | 43.225 ± 0.002 | 6.72 ± 0.21 | 0.26 ± 0.08 | -0.022 ± 0.006 | 0.341 ± 0.050 | 2.96 ± 0.09 |
| J114045.60+014424.0 | 0.6441 | 1903 ± 134 | 109.32 ± 9.88 | 0.39 ± 0.05 | 0.00 ± 0.00 | 0.35 ± 0.04 | 44.238 ± 0.005 | 7.56 ± 0.10 | 0.01 ± 0.00 | 0.126 ± 0.018 | 0.316 ± 0.022 | 2.25 ± 1.22 |
| J114116.08+215620.4 | 0.0633 | 1454 ± 65 | 102.28 ± 0.84 | 0.01 ± 0.00 | 1.86 ± 0.11 | 0.30 ± 0.01 | 43.902 ± 0.006 | 7.28 ± 0.06 | 0.07 ± 0.01 | 0.046 ± 0.004 | 0.268 ± 0.012 | 2.54 ± 0.12 |
| J114214.88+053939.6 | 0.1018 | 985 ± 191 | 84.14 ± 5.74 | 0.16 ± 0.02 | 1.30 ± 0.14 | 1.17 ± 0.11 | 43.560 ± 0.002 | 6.91 ± 0.28 | 0.24 ± 0.09 | -0.134 ± 0.052 | 0.296 ± 0.057 | 2.58 ± 0.34 |
| J114549.44+290643.2 | 0.1426 | 1519 ± 131 | 100.45 ± 18.80 | 0.37 ± 0.10 | 1.84 ± 0.27 | 0.58 ± 0.15 | 44.504 ± 0.001 | 7.43 ± 0.12 | 0.33 ± 0.06 | 0.038 ± 0.007 | 0.310 ± 0.027 | 2.83 ± 0.19 |
| J114815.36+013838.4 | 0.2686 | 981 ± 128 | 54.11 ± 8.70 | 0.01 ± 0.00 | 1.31 ± 0.20 | 1.16 ± 0.26 | 44.103 ± 0.003 | 6.96 ± 0.18 | 0.51 ± 0.14 | 0.064 ± 0.017 | 0.286 ± 0.037 | 2.39 ± 0.07 |
| J115138.16+561333.6 | 0.0508 | 1156 ± 283 | 11.98 ± 2.86 | 1.07 ± 0.36 | 3.65 ± 0.25 | 1.98 ± 0.67 | 42.832 ± 0.003 | 7.04 ± 0.35 | -0.26 ± 0.13 | -0.014 ± 0.002 | 0.361 ± 0.089 | 2.89 ± 0.39 |
| J115341.76+461243.2 | 0.0243 | 1305 ± 821 | 27.81 ± 7.19 | 0.00 ± 0.00 | 1.93 ± 0.41 | 0.87 ± 0.31 | 43.772 ± 0.005 | 7.17 ± 0.89 | 0.10 ± 0.13 | -0.114 ± 0.143 | 0.250 ± 0.157 | 2.42 ± 0.49 |
| J120051.12+341702.4 | 0.3764 | 1454 ± 88 | 95.89 ± 13.88 | 0.25 ± 0.26 | 5.20 ± 0.76 | 5.93 ± 6.20 | 44.563 ± 0.003 | 7.42 ± 0.09 | 0.40 ± 0.05 | 0.003 ± 0.000 | 0.321 ± 0.020 | 2.56 ± 0.66 |
| J120102.64+082439.6 | 0.2836 | 1189 ± 320 | 45.74 ± 10.23 | 0.60 ± 0.19 | 3.38 ± 0.65 | 1.75 ± 0.55 | 44.193 ± 0.004 | 7.14 ± 0.38 | 0.39 ± 0.21 | -0.129 ± 0.069 | 0.283 ± 0.076 | 2.87 ± 0.21 |
| J120123.28+295606.0 | 0.2279 | 705 ± 45 | 40.41 ± 3.45 | 0.51 ± 0.06 | 2.61 ± 0.29 | 1.44 ± 0.17 | 43.407 ± 0.002 | 6.66 ± 0.09 | 0.78 ± 0.10 | -0.014 ± 0.002 | 0.351 ± 0.022 | 2.97 ± 0.23 |
| J123436.48+163151.6 | 0.0733 | 705 ± 40 | 9.94 ± 1.09 | 0.19 ± 0.03 | 2.46 ± 0.33 | 3.93 ± 0.60 | 43.407 ± 0.002 | 6.62 ± 0.08 | 0.45 ± 0.05 | 0.023 ± 0.003 | 0.353 ± 0.020 | 3.27 ± 0.63 |
| J123748.48+092324.0 | 0.1248 | 1089 ± 722 | 20.65 ± 4.29 | 0.52 ± 0.15 | 1.79 ± 0.29 | 1.69 ± 0.50 | 43.876 ± 0.002 | 7.02 ± 0.94 | 0.31 ± 0.41 | -0.319 ± 0.423 | 0.221 ± 0.146 | 2.90 ± 0.17 |
| J123800.96+621337.2 | 0.44 | 1454 ± 321 | 42.04 ± 6.06 | 0.04 ± 0.01 | 0.00 ± 0.00 | 1.87 ± 0.38 | 44.478 ± 0.004 | 7.39 ± 0.31 | 0.36 ± 0.16 | -0.131 ± 0.058 | 0.284 ± 0.063 | 2.63 ± 0.19 |
| J129932.64+342222.8 | 0.6399 | 1907 ± 2475 | 18.17 ± 2.10 | 0.03 ± 0.01 | 1.37 ± 0.13 | 0.24 ± 0.04 | 43.615 ± 0.005 | 7.49 ± 1.84 | -0.31 ± 0.80 | 0.023 ± 0.060 | 0.347 ± 0.450 | 2.83 ± 0.39 |
| J124013.92+473354.0 | 0.1172 | 1098 ± 443 | 50.08 ± 4.00 | 0.20 ± 0.02 | 1.15 ± 0.12 | 0.84 ± 0.09 | 43.320 ± 0.004 | 7.00 ± 0.57 | 0.02 ± 0.02 | -0.037 ± 0.007 | 0.192 ± 0.017 | 3.65 ± 1.03 |
| J124124.00+321203.6 | 0.3857 | 1232 ± 159 | 41.59 ± 3.07 | 0.12 ± 0.01 | 1.73 ± 0.15 | 0.68 ± 0.03 | 43.807 ± 0.003 | 7.09 ± 0.10 | 0.20 ± 0.03 | 0.003 ± 0.000 | 0.302 ± 0.022 | 2.54 ± 0.09 |
| J124210.56+331702.4 | 0.0435 | 1454 ± 188 | 38.55 ± 1.21 | 0.11 ± 0.02 | 1.53 ± 0.21 | 0.55 ± 0.09 | 45.011 ± 0.001 | 7.77 ± 0.37 | 0.61 ± 0.32 | -0.012 ± 0.006 | 0.302 ± 0.080 | 3.36 ± 0.35 |
| J124341.76+310104.8 | 0.2907 | 1488 ± 392 | 55.62 ± 8.46 | 0.71 ± 0.14 | 2.12 ± 0.33 | 1.20 ± 0.23 | 43.587 ± 0.001 | 7.25 ± 0.18 | -0.09 ± 0.02 | 0.021 ± 0.006 | 0.327 ± 0.042 | 3.96 ± 0.70 |
| J124635.28+022208.4 | 0.0481 | 705 ± 54 | 16.85 ± 1.25 | 0.79 ± 0.08 | 2.40 ± 0.25 | 2.69 ± 0.28 | 43.588 ± 0.001 | 6.63 ± 0.11 | 0.54 ± 0.08 | -0.014 ± 0.002 | 0.356 ± 0.027 | - ± - |
| J124828.56+083112.0 | 0.1186 | 1207 ± 78 | 34.31 ± 1.93 | 0.47 ± 0.04 | 2.21 ± 0.09 | 0.00 ± 0.00 | 43.467 ± 0.004 | 7.09 ± 0.09 | 0.02 ± 0.00 | 0.014 ± 0.002 | 0.319 ± 0.021 | 1.82 ± 0.34 |

*(continued)*

| SDSS Name | z | FWHM(Hβ) | EW(Hβ) | [OIII]/Hβ | Hα/Hβ | $R_{Fe}$ | $\log(L_{5100\text{Å}})$ | $\log(M_{BH}/M_\odot)$ | $\log(R_{Edd})$ | AI | KI | $\Gamma_X$ |
|---|---|---|---|---|---|---|---|---|---|---|---|---|
| J125848.00+013603.6 | 0.3365 | 1410 ± 131 | 32.71 ± 6.28 | 0.00 ± 0.00 | 4.08 ± 0.55 | 1.69 ± 0.46 | 44.250 ± 0.004 | 7.30 ± 0.13 | 0.27 ± 0.05 | -0.025 ± 0.005 | 0.314 ± 0.029 | 2.70 ± 0.73 |
| J130052.08+564106.0 | 0.0718 | 1454 ± 12 | 85.25 ± 2.05 | 0.37 ± 0.01 | 1.32 ± 0.13 | 0.49 ± 0.02 | 43.515 ± 0.002 | 7.25 ± 0.01 | -0.112 ± 0.000 | 0.003 ± 0.000 | 0.317 ± 0.003 | 2.70 ± 0.10 |
| J130250.40+111828.8 | 0.2034 | 1498 ± 194 | 48.23 ± 5.23 | 0.09 ± 0.01 | 2.35 ± 0.20 | 2.59 ± 0.40 | 44.350 ± 0.002 | 7.37 ± 0.18 | 0.27 ± 0.07 | -0.243 ± 0.063 | 0.296 ± 0.038 | 3.95 ± 0.51 |
| J130749.46+080318.0 | 0.1476 | 1415 ± 208 | 27.73 ± 2.44 | 0.04 ± 0.00 | 2.30 ± 0.16 | 1.14 ± 0.14 | 44.006 ± 0.002 | 7.26 ± 0.21 | 0.15 ± 0.04 | 0.006 ± 0.006 | 0.295 ± 0.043 | 2.34 ± 0.73 |
| J130845.60+013054.0 | 0.1113 | 1462 ± 267 | 44.60 ± 6.05 | 0.36 ± 0.07 | 1.09 ± 0.10 | 0.75 ± 0.14 | 43.454 ± 0.003 | 7.25 ± 0.26 | -0.16 ± 0.06 | -0.014 ± 0.005 | 0.288 ± 0.053 | 2.20 ± 0.54 |
| J130955.92+530636.0 | 0.3206 | 705 ± 34 | 25.36 ± 1.02 | 0.00 ± 0.00 | 2.98 ± 0.32 | 1.94 ± 0.11 | 43.454 ± 0.003 | 6.70 ± 0.07 | 0.89 ± 0.09 | -0.163 ± 0.016 | 0.348 ± 0.017 | 3.05 ± 0.67 |
| J131232.64+372345.6 | 0.1179 | 1507 ± 486 | 43.42 ± 10.25 | 0.40 ± 0.13 | 0.77 ± 0.11 | 0.87 ± 0.29 | 43.334 ± 0.004 | 7.28 ± 0.46 | -0.24 ± 0.16 | -0.131 ± 0.084 | 0.269 ± 0.087 | 3.12 ± 0.69 |
| J131422.80+342938.4 | 0.0745 | 1487 ± 226 | 31.41 ± 6.43 | 0.41 ± 0.12 | 2.84 ± 0.41 | 2.15 ± 0.62 | 43.773 ± 0.006 | 7.28 ± 0.22 | -0.05 ± 0.02 | -0.011 ± 0.003 | 0.348 ± 0.053 | 3.33 ± 0.25 |
| J131906.00+310852.8 | 0.032 | 705 ± 145 | 24.80 ± 1.99 | 0.24 ± 0.03 | 0.00 ± 0.00 | 0.00 ± 0.00 | 43.690 ± 0.001 | 6.61 ± 0.29 | 0.26 ± 0.11 | -0.009 ± 0.004 | 0.342 ± 0.070 | 1.58 ± 0.47 |
| J132820.40+240928.8 | 0.2226 | 1506 ± 145 | 100.51 ± 4.10 | 0.00 ± 0.00 | 1.86 ± 0.15 | 0.98 ± 0.26 | 44.273 ± 0.003 | 7.42 ± 0.05 | -0.088 ± 0.107 | 0.021 ± 0.002 | 0.327 ± 0.012 | 3.67 ± 0.55 |
| J134300.72+360957.6 | 0.0237 | 1454 ± 423 | 22.74 ± 1.87 | 0.19 ± 0.02 | 1.92 ± 0.11 | 0.63 ± 0.04 | 42.639 ± 0.002 | 7.24 ± 0.41 | -0.56 ± 0.33 | 0.033 ± 0.002 | 0.352 ± 0.102 | 2.34 ± 0.76 |
| J134432.64+600101.1 | 0.0806 | 1388 ± 83 | 50.78 ± 1.44 | 0.00 ± 0.00 | 2.64 ± 0.05 | 0.00 ± 0.00 | 44.483 ± 0.002 | 7.42 ± 0.08 | 0.33 ± 0.03 | 0.147 ± 0.017 | 0.257 ± 0.015 | 2.00 ± 0.43 |
| J134452.80+000520.4 | 0.0747 | 705 ± 69 | 67.59 ± 3.80 | 0.00 ± 0.00 | 2.64 ± 0.13 | 0.60 ± 0.05 | 44.102 ± 0.010 | 6.67 ± 0.14 | 0.80 ± 0.17 | 0.007 ± 0.001 | 0.341 ± 0.033 | 2.67 ± 0.19 |
| J133137.92+013151.6 | 0.087 | 1463 ± 122 | 90.91 ± 18.09 | 0.21 ± 0.06 | 0.00 ± 0.00 | 0.78 ± 0.22 | 44.571 ± 0.004 | 7.42 ± 0.12 | 0.41 ± 0.07 | -0.020 ± 0.003 | 0.313 ± 0.026 | 2.49 ± 0.23 |
| J133141.04+015212.0 | 0.1451 | 1429 ± 847 | 56.58 ± 12.96 | 0.27 ± 0.09 | 1.57 ± 0.16 | 0.61 ± 0.20 | 43.648 ± 0.003 | 7.24 ± 0.84 | -0.04 ± 0.05 | 0.003 ± 0.004 | 0.300 ± 0.178 | 2.83 ± 0.42 |
| J133900.24+262609.6 | 0.2412 | 860 ± 201 | 39.14 ± 7.14 | 0.34 ± 0.09 | 2.94 ± 0.23 | 0.94 ± 0.24 | 43.773 ± 0.006 | 6.81 ± 0.33 | 0.46 ± 0.22 | -0.064 ± 0.025 | 0.323 ± 0.076 | 2.66 ± 0.46 |
| J133929.04+522725.2 | 0.269 | 1008 ± 612 | 61.51 ± 11.49 | 0.05 ± 0.01 | 1.86 ± 0.29 | 0.98 ± 0.26 | 44.273 ± 0.003 | 7.01 ± 0.86 | 0.57 ± 0.70 | -0.058 ± 0.107 | 0.342 ± 0.166 | 3.12 ± 0.86 |
| J135516.56+561243.2 | 0.1215 | 1019 ± 529 | 10.26 ± 2.38 | 0.05 ± 0.02 | 2.72 ± 0.44 | 2.00 ± 0.66 | 43.581 ± 0.001 | 6.95 ± 0.73 | 0.22 ± 0.23 | -0.054 ± 0.056 | 0.361 ± 0.188 | 3.11 ± 0.08 |
| J135516.56+561243.2 | 0.1215 | 1015 ± 798 | 30.26 ± 7.21 | 0.00 ± 0.00 | 2.31 ± 0.18 | 0.64 ± 0.21 | 44.156 ± 0.012 | 6.99 ± 1.11 | 0.51 ± 0.81 | -0.042 ± 0.067 | 0.323 ± 0.254 | 2.82 ± 0.27 |
| J135724.48+652504.8 | 0.1063 | 1590 ± 167 | 35.95 ± 3.53 | 1.93 ± 0.27 | 2.03 ± 0.27 | 1.69 ± 0.23 | 43.276 ± 0.001 | 7.37 ± 0.15 | 0.07 ± 0.02 | -0.084 ± 0.018 | 0.316 ± 0.033 | 2.81 ± 0.16 |
| J135810.32+653300.0 | 0.2269 | 1523 ± 1591 | 46.11 ± 11.83 | 0.86 ± 0.31 | 1.10 ± 0.18 | 0.73 ± 0.27 | 44.065 ± 0.003 | 7.28 ± 1.48 | -0.28 ± 0.59 | 0.007 ± 0.007 | 0.310 ± 0.324 | 3.11 ± 0.08 |
| J135810.32+653300.0 | 0.2269 | 705 ± 242 | 16.22 ± 4.18 | 0.00 ± 0.00 | 2.36 ± 0.56 | 0.49 ± 0.18 | 43.849 ± 0.003 | 6.64 ± 0.49 | 0.67 ± 0.47 | -0.021 ± 0.015 | 0.281 ± 0.119 | 2.89 ± 0.28 |
| J134945.64+632604.8 | 0.5409 | 1334 ± 198 | 55.12 ± 7.58 | 0.00 ± 0.00 | 1.31 ± 0.10 | 0.96 ± 0.19 | 43.690 ± 0.001 | 7.18 ± 0.21 | -0.05 ± 0.01 | 0.080 ± 0.024 | 0.308 ± 0.046 | 2.14 ± 0.45 |
| J141449.68+361239.5 | 0.181 | 1127 ± 1527 | 27.95 ± 4.47 | 0.17 ± 0.04 | 1.64 ± 0.14 | 0.48 ± 0.11 | 43.629 ± 0.006 | 7.04 ± 1.92 | 0.16 ± 0.42 | 0.052 ± 0.142 | 0.325 ± 0.440 | 2.42 ± 0.24 |
| J141451.36+281546.8 | 0.5079 | 705 ± 541 | 6.23 ± 1.76 | 0.04 ± 0.01 | 0.00 ± 0.00 | 0.57 ± 0.05 | 44.314 ± 0.005 | 7.21 ± 0.30 | 0.41 ± 0.17 | -0.084 ± 0.036 | 0.281 ± 0.060 | 2.29 ± 0.17 |
| J141519.44+003021.6 | 0.1345 | 705 ± 98 | 34.69 ± 5.34 | 0.14 ± 0.03 | 2.49 ± 0.38 | 2.13 ± 0.46 | 43.466 ± 0.003 | 6.64 ± 0.20 | 0.48 ± 0.14 | 0.005 ± 0.002 | 0.351 ± 0.083 | 3.06 ± 0.05 |
| J145108.88+270925.2 | 0.0645 | 1517 ± 370 | 26.40 ± 3.73 | 0.43 ± 0.09 | 2.74 ± 0.40 | 1.59 ± 0.32 | 44.279 ± 0.001 | 7.37 ± 0.34 | 0.22 ± 0.11 | 0.005 ± 0.002 | 0.341 ± 0.083 | 3.11 ± 0.08 |
| J145426.64+182956.4 | 0.2474 | 1154 ± 373 | 40.71 ± 7.83 | 0.25 ± 0.07 | 2.85 ± 0.43 | 1.46 ± 0.40 | 44.432 ± 0.002 | 7.17 ± 0.46 | 0.54 ± 0.35 | -0.020 ± 0.013 | 0.305 ± 0.099 | 2.78 ± 0.14 |
| J145512.24+584825.2 | 0.5766 | 1454 ± 126 | 107.98 ± 16.27 | 0.17 ± 0.04 | 4.93 ± 2.15 | 0.72 ± 0.15 | 44.484 ± 0.004 | 7.39 ± 0.12 | 0.36 ± 0.06 | 0.037 ± 0.006 | 0.300 ± 0.026 | 2.63 ± 0.42 |
| J144209.36+621026.4 | 0.3298 | 1254 ± 245 | 48.31 ± 5.24 | 0.24 ± 0.04 | 2.33 ± 0.19 | 1.55 ± 0.24 | 44.323 ± 0.004 | 7.34 ± 0.24 | 0.28 ± 0.10 | 0.004 ± 0.001 | 0.343 ± 0.058 | 3.41 ± 0.75 |
| J150148.96+014406.0 | 0.4835 | 1110.37 ± 7.03 | 2.58 ± 0.26 | 2.97 ± 0.22 | 0.92 ± 0.12 | | 43.763 ± 0.002 | 7.36 ± 0.16 | -0.10 ± 0.02 | -0.013 ± 0.003 | 0.295 ± 0.033 | 2.99 ± 0.21 |
| J150506.48+032631.2 | 0.4083 | 2203 ± 116 | 101.24 ± 11.52 | 0.03 ± 0.00 | 1.97 ± 0.22 | 1.29 ± 0.21 | 44.377 ± 0.001 | 7.72 ± 0.07 | -0.090 ± 0.009 | -0.084 ± 0.036 | 0.281 ± 0.105 | 2.09 ± 0.16 |
| J152114.16+222743.2 | 0.1361 | 1632 ± 184 | 61.35 ± 5.68 | 0.00 ± 0.00 | 1.97 ± 0.22 | 1.29 ± 0.21 | 44.377 ± 0.001 | 7.72 ± 0.07 | 1.01 ± 1.17 | 0.536 ± 0.822 | 0.138 ± 0.105 | 2.09 ± 0.16 |
| J151956.64+001615.6 | 0.1144 | 1632 ± 184 | 61.35 ± 5.68 | 0.00 ± 0.00 | 1.97 ± 0.22 | 1.29 ± 0.21 | 43.763 ± 0.002 | 7.36 ± 0.16 | -0.10 ± 0.02 | -0.013 ± 0.003 | 0.295 ± 0.033 | 2.99 ± 0.21 |
| J154818.00+351127.6 | 0.4786 | 1550 ± 248 | 72.19 ± 14.23 | 0.34 ± 0.09 | 1.04 ± 0.29 | | 44.805 ± 0.002 | 7.61 ± 0.23 | 0.47 ± 0.15 | -0.009 ± 0.003 | 0.325 ± 0.052 | 2.47 ± 0.15 |
| J155909.60+350148.0 | 0.031 | 705 ± 49 | 24.25 ± 1.76 | 0.26 ± 0.03 | 3.11 ± 0.32 | 2.18 ± 0.15 | 43.789 ± 0.011 | 6.77 ± 1.08 | 0.79 ± 0.32 | 0.027 ± 0.011 | 0.293 ± 0.058 | 3.32 ± 0.93 |
| J150948.72+336625.1 | 0.5119 | 876 ± 175 | 37.69 ± 12.76 | 0.29 ± 0.14 | 4.22 ± 1.67 | | 44.458 ± 0.004 | 6.94 ± 0.28 | 0.64 ± 0.26 | 0.027 ± 0.011 | 0.293 ± 0.058 | 3.32 ± 0.93 |
| J151115.12+380531.2 | 0.2475 | 1454 ± 118 | 111.99 ± 18.97 | 0.02 ± 0.01 | 1.62 ± 0.14 | 0.22 ± 0.05 | 43.913 ± 0.006 | 7.28 ± 0.11 | 0.08 ± 0.01 | 0.003 ± 0.001 | 0.321 ± 0.026 | 2.76 ± 0.57 |
| J151153.28+561658.8 | 0.391 | 1994 ± 180 | 128.38 ± 30.98 | 1.20 ± 0.41 | 1.60 ± 0.23 | 0.21 ± 0.07 | 44.087 ± 0.008 | 7.57 ± 0.13 | -0.11 ± 0.02 | 0.134 ± 0.024 | 0.290 ± 0.026 | 2.36 ± 0.44 |
| J151312.48+001937.2 | 0.1588 | 1245 ± 83 | 43.70 ± 5.24 | 0.44 ± 0.08 | 1.91 ± 0.22 | 0.76 ± 0.13 | 43.276 ± 0.007 | 7.13 ± 0.09 | 0.12 ± 0.02 | 0.064 ± 0.009 | 0.326 ± 0.022 | 2.24 ± 0.15 |
| J151743.68+070122.8 | 0.2825 | 1013 ± 484 | 41.01 ± 8.04 | 0.23 ± 0.05 | 2.68 ± 0.34 | 1.51 ± 0.32 | 44.350 ± 0.002 | 7.03 ± 0.68 | 0.61 ± 0.58 | -0.084 ± 0.081 | 0.228 ± 0.109 | 2.99 ± 0.21 |
| J160706.48+075710.8 | 0.2316 | 1398 ± 236 | 33.13 ± 5.82 | 0.39 ± 0.10 | 2.06 ± 0.20 | 0.80 ± 0.20 | 43.366 ± 0.001 | 7.24 ± 0.24 | 0.12 ± 0.06 | 0.023 ± 0.008 | 0.333 ± 0.056 | 2.36 ± 0.44 |
| J160806.72+424058.8 | 0.0844 | 1585 ± 267 | 33.46 ± 5.10 | 0.20 ± 0.04 | 1.84 ± 0.28 | 0.80 ± 0.17 | 43.586 ± 0.003 | 7.33 ± 0.24 | -0.16 ± 0.06 | -0.055 ± 0.019 | 0.229 ± 0.039 | 2.79 ± 0.24 |
| J161739.84+123607.2 | 0.1923 | 936 ± 700 | 6.44 ± 6.48 | 0.35 ± 0.49 | 8.04 ± 2.11 | 4.41 ± 6.27 | 43.477 ± 0.004 | 6.87 ± 1.06 | -0.24 ± 0.36 | -0.030 ± 0.045 | 0.330 ± 0.246 | 2.24 ± 0.99 |
| J162409.36+260430.0 | 0.0392 | 1454 ± 69 | 54.43 ± 3.57 | 0.00 ± 0.00 | 2.35 ± 0.07 | 0.72 ± 0.07 | 43.288 ± 0.001 | 7.42 ± 0.14 | -0.24 ± 0.02 | 0.327 ± 0.016 | 0.327 ± 0.016 | 2.33 ± 0.57 |
| J163502.88+412952.8 | 0.4723 | 1356 ± 132 | 59.58 ± 2.31 | 0.18 ± 0.01 | 1.44 ± 0.08 | | 44.683 ± 0.003 | 7.42 ± 0.14 | 0.52 ± 0.10 | -0.178 ± 0.035 | 0.258 ± 0.025 | 4.02 ± 0.75 |
| J163817.52+382248.0 | 0.3612 | 2054 ± 168 | 119.72 ± 6.75 | 0.02 ± 0.00 | 1.69 ± 0.11 | 0.55 ± 0.04 | 44.766 ± 0.001 | 7.83 ± 0.12 | 0.20 ± 0.03 | -0.108 ± 0.018 | 0.315 ± 0.026 | 3.17 ± 0.25 |
| J163817.52+382248.0 | 0.3612 | 1553 ± 106 | 119.47 ± 10.81 | 0.12 ± 0.02 | 0.83 ± 0.10 | 0.57 ± 0.07 | 44.804 ± 0.002 | 7.61 ± 0.10 | 0.46 ± 0.06 | 0.021 ± 0.003 | 0.316 ± 0.022 | 3.17 ± 0.25 |
| J164426.16+392931.2 | 0.1001 | 1663 ± 250 | 45.88 ± 6.12 | 0.00 ± 0.00 | 2.02 ± 0.20 | 0.74 ± 0.14 | 43.614 ± 0.002 | 7.37 ± 0.21 | -0.19 ± 0.06 | 0.002 ± 0.001 | 0.325 ± 0.049 | 2.88 ± 0.29 |
| J164508.16+392533.6 | 0.0693 | 1614 ± 406 | 34.45 ± 10.65 | 0.27 ± 0.12 | 2.16 ± 0.39 | 1.81 ± 0.79 | 43.369 ± 0.002 | 7.34 ± 0.36 | -0.29 ± 0.14 | -0.130 ± 0.065 | 0.300 ± 0.075 | 2.75 ± 0.26 |
| J165731.68+342558.8 | 0.1911 | 1485 ± 375 | 18.03 ± 3.69 | 0.53 ± 0.15 | 3.79 ± 0.66 | 2.61 ± 0.76 | 43.968 ± 0.002 | 7.30 ± 0.36 | 0.09 ± 0.04 | -0.195 ± 0.098 | 0.254 ± 0.064 | 3.99 ± 0.45 |
| J171131.20+333542.0 | 0.469 | 1089 ± 489 | 59.23 ± 9.73 | 0.07 ± 0.02 | 0.00 ± 0.00 | 1.05 ± 0.24 | 45.345 ± 0.001 | 8.09 ± 0.64 | 1.04 ± 0.94 | -0.058 ± 0.052 | 0.245 ± 0.110 | 4.29 ± 0.59 |
| J172322.56+321044.4 | 0.655 | 1260 ± 767 | 79.18 ± 17.32 | 0.10 ± 0.03 | 0.00 ± 0.00 | 0.80 ± 0.25 | 44.187 ± 0.008 | 7.19 ± 0.86 | 0.34 ± 0.42 | 0.124 ± 0.151 | 0.268 ± 0.163 | 2.38 ± 0.64 |
| J211853.04+073227.6 | 0.2601 | 1208 ± 490 | 56.96 ± 13.64 | 0.53 ± 0.18 | 2.27 ± 0.30 | 0.51 ± 0.17 | 44.006 ± 0.007 | 7.13 ± 0.57 | 0.171 ± 0.139 | 0.180 ± 0.073 | 0.180 ± 0.073 | 2.81 ± 0.36 |
| J211852.32-002132.4 | 0.3155 | 1167 ± 1998 | 59.29 ± 12.48 | 0.00 ± 0.00 | 0.73 ± 0.08 | 7.08 ± 2.42 | 43.829 ± 0.010 | 7.13 ± 0.57 | 0.23 ± 0.77 | -0.292 ± 1.000 | 1.542 ± 2.640 | 2.27 ± 0.49 |
| J221918.48+120751.6 | 0.0813 | 1602 ± 499 | 9.52 ± 2.95 | 0.81 ± 0.36 | 4.39 ± 0.50 | 1.96 ± 0.18 | 43.762 ± 0.001 | 7.35 ± 0.44 | 0.018 ± 0.011 | 0.342 ± 0.107 | 0.342 ± 0.107 | 3.36 ± 0.05 |
| J222607.68+134355.2 | 0.3257 | 1494 ± 213 | 78.47 ± 14.12 | 0.11 ± 0.03 | 1.96 ± 0.21 | 0.74 ± 0.19 | 45.162 ± 0.001 | 7.99 ± 0.20 | 0.68 ± 0.19 | -0.145 ± 0.041 | 0.222 ± 0.032 | 2.69 ± 0.08 |

Table B.2 The table of properties obtained for the sample of 117 BLSy1 galaxies.

| SDSS Name[a] | z[b] | FWHM(Hβ)[c] | EW(Hβ)[d] | [OIII]/Hβ[e] | Hα/Hβ[f] | $R_{fe}$[g] | $\log(L_{5100\text{Å}})$[h] | $M_{BH}$[i] | $\log(R_{\text{EDD}})$[j] | AI[k] | KI[l] | $\Gamma_X$[m] |
|---|---|---|---|---|---|---|---|---|---|---|---|---|
| J002113.20-020115.6 | 0.7621 | 9759 ± 1578 | 116.29 ± 19.26 | 0.39 ± 0.09 | 0.00 ± 0.00 | 0.30 ± 0.07 | 44.644 ± 0.010 | 9.11 ± 0.23 | -1.21 ± 0.41 | 0.548 ± 0.177 | 0.568 ± 0.092 | 1.83 ± 0.35 |
| J011254.96+000314.4 | 0.2389 | 6561 ± 347 | 76.43 ± 13.97 | 0.14 ± 0.04 | 3.29 ± 0.17 | 0.71 ± 0.18 | 44.397 ± 0.002 | 8.67 ± 0.07 | -0.99 ± 0.11 | 0.034 ± 0.004 | 0.290 ± 0.015 | 2.60 ± 0.24 |
| J012254.72-010108.4 | 0.1994 | 3506 ± 469 | 23.90 ± 5.20 | 0.36 ± 0.11 | 2.94 ± 0.27 | 0.73 ± 0.22 | 43.548 ± 0.003 | 8.02 ± 0.19 | -0.87 ± 0.24 | -0.048 ± 0.013 | 0.299 ± 0.040 | 1.79 ± 0.38 |
| J013517.52-001940.8 | 0.3119 | 7242 ± 471 | 36.79 ± 1.02 | 0.31 ± 0.01 | 4.04 ± 0.13 | 0.19 ± 0.01 | 43.931 ± 0.003 | 8.67 ± 0.09 | -1.31 ± 0.18 | 0.012 ± 0.001 | 0.340 ± 0.022 | 2.31 ± 0.30 |
| J014959.28-125656.4 | 0.432 | 4076 ± 275 | 87.32 ± 23.20 | 0.19 ± 0.07 | 0.00 ± 0.00 | 0.80 ± 0.30 | 44.670 ± 0.003 | 8.36 ± 0.10 | -0.44 ± 0.06 | 0.095 ± 0.013 | 0.311 ± 0.021 | 2.24 ± 0.19 |
| J020744.16-060957.6 | 0.6496 | 9877 ± 699 | 102.77 ± 9.19 | 0.26 ± 0.03 | 0.00 ± 0.00 | 0.44 ± 0.06 | 44.537 ± 0.003 | 9.07 ± 0.10 | -1.28 ± 0.19 | 0.477 ± 0.068 | 0.556 ± 0.039 | 2.07 ± 0.46 |
| J020840.56-062718.0 | 0.092 | 2509 ± 42 | 31.55 ± 4.61 | 0.00 ± 0.00 | 3.57 ± 0.13 | 1.10 ± 0.23 | 43.449 ± 0.001 | 7.72 ± 0.02 | -0.63 ± 0.02 | -0.039 ± 0.001 | 0.315 ± 0.005 | 2.21 ± 0.11 |
| J021139.12-042606.0 | 0.4856 | 3099 ± 193 | 62.47 ± 2.69 | 0.03 ± 0.00 | 2.51 ± 0.12 | 1.00 ± 0.06 | 45.025 ± 0.001 | 8.42 ± 0.19 | -0.02 ± 0.00 | 0.004 ± 0.001 | 0.284 ± 0.018 | 2.78 ± 0.19 |
| J021318.24+130643.2 | 0.4076 | 8659 ± 581 | 145.32 ± 17.39 | 0.17 ± 0.03 | 0.00 ± 0.00 | 0.29 ± 0.05 | 44.527 ± 0.004 | 8.95 ± 0.10 | -1.17 ± 0.16 | 0.030 ± 0.004 | 0.300 ± 0.020 | 2.07 ± 0.24 |
| J021434.80-004243.2 | 0.4369 | 11843 ± 582 | 159.30 ± 28.22 | 0.53 ± 0.13 | 0.00 ± 0.00 | 0.43 ± 0.11 | 44.210 ± 0.006 | 9.14 ± 0.07 | -1.60 ± 0.17 | -0.021 ± 0.002 | 0.306 ± 0.015 | 2.48 ± 0.14 |
| J021820.40-050426.4 | 0.6492 | 3440 ± 428 | 51.72 ± 10.87 | 0.30 ± 0.09 | 0.00 ± 0.00 | 0.35 ± 0.08 | 44.586 ± 0.009 | 8.18 ± 0.18 | -0.34 ± 0.09 | -0.036 ± 0.009 | 0.312 ± 0.039 | 1.92 ± 0.26 |
| J021910.56-055114.4 | 0.5579 | 13710 ± 1043 | 118.32 ± 19.55 | 0.06 ± 0.01 | 2.91 ± 0.61 | 0.35 ± 0.08 | 44.480 ± 0.004 | 9.33 ± 0.11 | -1.59 ± 0.25 | -0.022 ± 0.003 | 0.319 ± 0.024 | 2.18 ± 0.32 |
| J021923.28-045150.4 | 0.6295 | 2150 ± 132 | 57.04 ± 23.31 | 0.32 ± 0.19 | 0.00 ± 0.00 | 1.12 ± 0.64 | 44.668 ± 0.004 | 7.81 ± 0.09 | 0.11 ± 0.01 | -0.028 ± 0.003 | 0.308 ± 0.019 | 2.73 ± 0.36 |
| J022014.64-072858.8 | 0.2134 | 5689 ± 144 | 102.96 ± 1.41 | 0.18 ± 0.00 | 2.97 ± 0.04 | 0.27 ± 0.01 | 44.090 ± 0.006 | 8.48 ± 0.04 | -1.02 ± 0.06 | 0.243 ± 0.012 | 0.528 ± 0.013 | 2.15 ± 0.18 |
| J022105.76-044102.4 | 0.1987 | 6696 ± 666 | 43.15 ± 2.44 | 0.00 ± 0.00 | 2.84 ± 0.10 | 0.59 ± 0.05 | 43.961 ± 0.002 | 8.61 ± 0.14 | -1.23 ± 0.25 | -0.122 ± 0.024 | 0.293 ± 0.029 | 1.69 ± 0.60 |
| J022244.40-043346.8 | 0.7606 | 12600 ± 309 | 410.13 ± 60.63 | 0.12 ± 0.03 | 0.00 ± 0.00 | 0.38 ± 0.08 | 44.782 ± 0.008 | 9.41 ± 0.03 | -1.36 ± 0.08 | -0.034 ± 0.002 | 0.143 ± 0.004 | 2.48 ± 0.72 |
| J022258.80-045851.6 | 0.4662 | 2482 ± 201 | 34.68 ± 4.89 | 0.21 ± 0.04 | 4.17 ± 0.22 | 1.15 ± 0.22 | 44.423 ± 0.002 | 7.83 ± 0.11 | -0.13 ± 0.02 | -0.023 ± 0.004 | 0.320 ± 0.026 | 2.90 ± 0.23 |
| J022438.88-042707.2 | 0.2525 | 3857 ± 991 | 16.22 ± 3.59 | 0.43 ± 0.13 | 3.98 ± 0.43 | 0.00 ± 0.00 | 43.973 ± 0.004 | 8.13 ± 0.36 | -0.74 ± 0.39 | 0.017 ± 0.009 | 0.349 ± 0.090 | 2.38 ± 0.38 |
| J023012.48-041445.6 | 0.3258 | 5357 ± 254 | 72.03 ± 5.04 | 0.10 ± 0.01 | 2.53 ± 0.37 | 1.03 ± 0.19 | 44.153 ± 0.002 | 8.44 ± 0.07 | -0.94 ± 0.09 | -0.005 ± 0.001 | 0.333 ± 0.016 | 2.57 ± 0.33 |
| J023015.36+003556.4 | 0.4467 | 4127 ± 234 | 104.27 ± 8.84 | 0.11 ± 0.01 | 0.00 ± 0.00 | 0.89 ± 0.11 | 44.026 ± 0.014 | 8.19 ± 0.08 | -0.77 ± 0.10 | -0.338 ± 0.038 | 0.152 ± 0.009 | 2.47 ± 0.36 |
| J023040.80-040755.2 | 0.3307 | 9802 ± 363 | 78.29 ± 11.57 | 0.00 ± 0.00 | 2.79 ± 0.16 | 0.68 ± 0.14 | 44.277 ± 0.002 | 8.99 ± 0.05 | -1.40 ± 0.11 | 0.208 ± 0.015 | 0.225 ± 0.008 | 2.30 ± 0.47 |
| J023137.44+003704.8 | 0.5578 | 10393 ± 565 | 108.44 ± 20.51 | 0.16 ± 0.04 | 1.71 ± 0.26 | 0.56 ± 0.15 | 44.173 ± 0.003 | 9.02 ± 0.08 | -1.50 ± 0.17 | 0.029 ± 0.003 | 0.280 ± 0.015 | 2.57 ± 0.37 |
| J044759.52-043231.2 | 0.2078 | 8196 ± 606 | 86.48 ± 17.95 | 0.54 ± 0.16 | 2.42 ± 0.17 | 0.31 ± 0.09 | 43.415 ± 0.004 | 8.75 ± 0.09 | -1.67 ± 0.24 | 0.085 ± 0.013 | 0.152 ± 0.011 | 1.98 ± 0.28 |
| J074858.32+332822.8 | 0.5845 | 6114 ± 149 | 109.85 ± 2.41 | 0.59 ± 0.02 | 0.00 ± 0.00 | 0.15 ± 0.00 | 44.441 ± 0.004 | 8.62 ± 0.03 | -0.91 ± 0.05 | -0.330 ± 0.016 | 0.718 ± 0.017 | 1.65 ± 0.12 |
| J080421.84+235526.4 | 0.4817 | 8402 ± 848 | 109.15 ± 5.27 | 0.22 ± 0.02 | 2.96 ± 0.15 | 0.19 ± 0.01 | 44.383 ± 0.015 | 8.88 ± 0.14 | -1.21 ± 0.27 | 0.240 ± 0.049 | 0.361 ± 0.036 | 1.31 ± 0.19 |
| J080737.92+395334.8 | 0.4289 | 1841 ± 70 | 54.42 ± 25.90 | 0.57 ± 0.39 | 0.00 ± 0.00 | 1.35 ± 0.91 | 44.598 ± 0.005 | 7.64 ± 0.05 | 0.21 ± 0.02 | 0.004 ± 0.000 | 0.316 ± 0.012 | 2.18 ± 0.64 |
| J082658.56+190921.5 | 0.4282 | 10336 ± 466 | 126.99 ± 16.42 | 0.18 ± 0.03 | 0.00 ± 0.00 | 0.08 ± 0.02 | 44.732 ± 0.002 | 9.21 ± 0.06 | -1.22 ± 0.11 | 0.346 ± 0.031 | 0.483 ± 0.022 | 1.92 ± 0.61 |
| J082942.72+415436.0 | 0.1263 | 7688 ± 150 | 119.74 ± 4.18 | 0.13 ± 0.01 | 4.93 ± 0.06 | 0.16 ± 0.01 | 43.780 ± 0.006 | 8.71 ± 0.03 | -1.44 ± 0.07 | 0.055 ± 0.002 | 0.567 ± 0.011 | 1.65 ± 0.16 |
| J084230.48+495801.2 | 0.3046 | 3766 ± 109 | 72.99 ± 17.99 | 0.02 ± 0.01 | 3.79 ± 0.29 | 1.05 ± 0.36 | 44.571 ± 0.002 | 8.25 ± 0.04 | -0.42 ± 0.03 | 0.067 ± 0.004 | 0.287 ± 0.008 | 2.67 ± 0.27 |
| J085300.48+132446.8 | 0.1934 | 2261 ± 200 | 88.94 ± 22.57 | 0.00 ± 0.00 | 3.28 ± 0.40 | 0.65 ± 0.24 | 43.918 ± 0.003 | 7.66 ± 0.12 | -0.30 ± 0.06 | 0.004 ± 0.001 | 0.333 ± 0.029 | 2.45 ± 0.26 |
| J085615.84+014637.2 | 0.7072 | 10114 ± 1012 | 150.27 ± 20.21 | 0.38 ± 0.07 | 0.00 ± 0.00 | 0.53 ± 0.10 | 44.476 ± 0.007 | 9.07 ± 0.14 | -1.33 ± 0.28 | 0.265 ± 0.053 | 0.345 ± 0.035 | 2.24 ± 0.09 |

*(continued)*

[a] Obtained from SDSS SAS.
[b] Source redshift based on single epoch spectra available in SDSS DR16.
[c] The broad component of Hβ emission profile in the units of km s$^{-1}$.
[d] Equivalent width of Hβ emission line.
[e] [OIII]/Hβ emission line narrow component flux ratio in logarithmic units.
[f] Hα/Hβ broad component flux ratio.
[g] $R_{fe}$, calculated as the flux ratio of area covered by the broad Fe line between 4433 Å and 4684 Å and the area covered by the Hβ emission line.
[h] Optical luminosity at 5100Å($\lambda L_{5100\text{Å}}$) in units of ergs/sec/Å
[i] SMBH mass in the units of $\log(M_{BH}/M_\odot)$ obtained using the Radius-Luminosity relation and FWHM of Hβ emission line.
[j] Eddington ratio, the ratio of bolometric to Eddington luminosity
[k] Asymmetry index for the Hβ emission line calculated using Equation 1(a)
[l] Kurtosis Index for the Hβ emission line calculated using Equation 1(b)
[m] The soft X-ray photon index calculated between 0.2-2 KeV energy range taken from Ojha et al. 2020.

| SDSS Name | z | FWHM($H\beta$) | EW($H\beta$) | [OIII]/$H\beta$ | $H\alpha/H\beta$ | $R_{fe}$ | $\log(L_{5,100\text{Å}})$ | $\log(M_{BH}/M\odot)$ | $\log(R_{upo})$ | AI | KI | $\Gamma_x$ |
|---|---|---|---|---|---|---|---|---|---|---|---|---|
| J090247.52+600326.4 | 0.5396 | 2461 ± 153 | 47.82 ± 20.54 | 0.25 ± 0.15 | 0.00 ± 0.00 | 0.46 ± 0.28 | 44.843 ± 0.009 | 8.04 ± 0.09 | 0.08 ± 0.01 | -0.045 ± 0.006 | 0.293 ± 0.018 | 2.04 ± 0.55 |
| J090446.08+020842.0 | 0.7945 | 8509 ± 308 | 221.54 ± 22.68 | 0.18 ± 0.03 | 0.00 ± 0.00 | 0.59 ± 0.09 | 44.602 ± 0.009 | 8.97 ± 0.05 | -1.11 ± 0.10 | 0.086 ± 0.006 | 0.577 ± 0.021 | 2.19 ± 0.65 |
| J090810.32+509021.5 | 0.0981 | 6401 ± 330 | 70.59 ± 8.24 | 0.33 ± 0.05 | 2.30 ± 0.29 | 0.42 ± 0.07 | 43.745 ± 0.001 | 8.55 ± 0.07 | -1.29 ± 0.14 | 0.030 ± 0.003 | 0.287 ± 0.015 | 2.45 ± 0.22 |
| J090826.64+002136.0 | 0.1626 | 7675 ± 943 | 31.05 ± 3.26 | 0.20 ± 0.03 | 2.06 ± 0.16 | 0.19 ± 0.03 | 43.747 ± 0.003 | 8.71 ± 0.17 | -1.45 ± 0.36 | 0.336 ± 0.001 | 0.336 ± 0.041 | 2.04 ± 0.76 |
| J091029.04+542718.0 | 0.5257 | 9427 ± 753 | 104.94 ± 25.99 | 0.00 ± 0.00 | 0.00 ± 0.00 | 0.64 ± 0.22 | 44.608 ± 0.008 | 9.06 ± 0.11 | -1.20 ± 0.20 | 0.112 ± 0.018 | 0.283 ± 0.023 | 2.30 ± 0.18 |
| J091205.28+543140.8 | 0.4476 | 8166 ± 221 | 110.48 ± 2.35 | 0.16 ± 0.00 | 0.00 ± 0.00 | 0.62 ± 0.02 | 44.459 ± 0.012 | 8.88 ± 0.04 | -1.15 ± 0.08 | 0.153 ± 0.008 | 0.307 ± 0.008 | 2.36 ± 0.24 |
| J091417.28+503426.4 | 0.1864 | 9638 ± 2001 | 99.95 ± 18.95 | 0.34 ± 0.09 | 1.85 ± 0.27 | 0.77 ± 0.20 | 44.160 ± 0.002 | 8.95 ± 0.29 | -1.44 ± 0.60 | -0.006 ± 0.002 | 0.319 ± 0.066 | 3.58 ± 0.50 |
| J091737.28+284048.0 | 0.4388 | 9443 ± 834 | 119.02 ± 15.30 | 0.17 ± 0.03 | 0.00 ± 0.00 | 0.30 ± 0.05 | 44.413 ± 0.010 | 8.99 ± 0.13 | -1.30 ± 0.25 | 0.059 ± 0.010 | 0.198 ± 0.017 | 2.38 ± 0.18 |
| J092421.12+365338.4 | 0.1065 | 5732 ± 378 | 36.56 ± 3.25 | 0.08 ± 0.01 | 2.29 ± 0.11 | 0.58 ± 0.07 | 43.622 ± 0.002 | 8.45 ± 0.09 | -1.26 ± 0.17 | 0.093 ± 0.012 | 0.284 ± 0.019 | 1.77 ± 0.99 |
| J093347.76+211438.4 | 0.1722 | 3047 ± 101 | 55.70 ± 10.25 | 0.59 ± 0.15 | 3.90 ± 0.22 | 0.75 ± 0.19 | 43.437 ± 0.002 | 7.95 ± 0.05 | -0.45 ± 0.03 | 0.325 ± 0.000 | 0.325 ± 0.013 | 2.32 ± 0.37 |
| J095532.75+460400.3 | 0.1561 | 3914 ± 268 | 24.45 ± 1.48 | 0.23 ± 0.02 | 1.81 ± 0.08 | 0.53 ± 0.05 | 43.803 ± 0.002 | 8.13 ± 0.10 | -0.84 ± 0.12 | 0.006 ± 0.001 | 0.340 ± 0.023 | 1.90 ± 0.69 |
| J100025.20+015851.6 | 0.3728 | 9146 ± 379 | 139.57 ± 7.08 | 0.18 ± 0.01 | 3.02 ± 0.16 | 0.21 ± 0.01 | 44.261 ± 0.005 | 8.92 ± 0.06 | -1.35 ± 0.12 | 0.068 ± 0.006 | 0.174 ± 0.007 | 2.29 ± 0.07 |
| J100744.64+507748.0 | 0.2118 | 9391 ± 1131 | 117.64 ± 20.40 | 0.09 ± 0.02 | 1.12 ± 0.22 | 0.33 ± 0.08 | 44.513 ± 0.001 | 9.02 ± 0.17 | -1.24 ± 0.30 | 0.004 ± 0.001 | 0.294 ± 0.035 | 2.48 ± 0.15 |
| J101326.16-000137.2 | 0.2552 | 1633 ± 257 | 20.83 ± 46.54 | 12.59 ± 40.32 | 14.93 ± 2.52 | 1.25 ± 4.02 | 44.016 ± 0.003 | 7.39 ± 2.39 | 0.03 ± 0.09 | 0.119 ± 0.402 | 0.134 ± 0.227 | — ± — |
| J101953.76+451922.8 | 0.4328 | 9578 ± 187 | 148.14 ± 6.23 | 0.00 ± 0.00 | 0.00 ± 0.00 | 0.22 ± 0.01 | 44.551 ± 0.009 | 9.05 ± 0.03 | -1.24 ± 0.07 | 0.111 ± 0.004 | 0.269 ± 0.005 | 2.28 ± 0.34 |
| J102035.28+445938.4 | 0.588 | 4734 ± 628 | 33.70 ± 16.54 | 0.10 ± 0.03 | 0.94 ± 0.65 | 0.65 ± 0.17 | 44.632 ± 0.006 | 8.47 ± 0.19 | -0.59 ± 0.16 | 0.017 ± 0.004 | 0.333 ± 0.044 | 2.16 ± 0.19 |
| J102054.96+195450.4 | 0.2502 | 18349 ± 807 | 77.20 ± 14.22 | 0.10 ± 0.03 | 1.44 ± 0.19 | 0.65 ± 0.17 | 44.118 ± 0.003 | 9.50 ± 0.06 | -2.02 ± 0.19 | 0.338 ± 0.001 | 0.338 ± 0.015 | 2.86 ± 0.36 |
| J102152.56+131145.6 | 0.5508 | 3536 ± 203 | 53.47 ± 20.01 | 0.11 ± 0.06 | 4.86 ± 0.72 | 1.44 ± 0.77 | 44.459 ± 0.003 | 8.15 ± 0.08 | -0.42 ± 0.05 | -0.103 ± 0.012 | 0.276 ± 0.016 | 2.38 ± 0.18 |
| J102216.80+415544.4 | 0.1152 | 3313 ± 167 | 31.31 ± 1.64 | 0.25 ± 0.02 | 2.84 ± 0.10 | 0.75 ± 0.06 | 43.524 ± 0.003 | 7.97 ± 0.07 | -0.83 ± 0.09 | -0.007 ± 0.001 | 0.345 ± 0.017 | 2.13 ± 0.42 |
| J104048.48+061818.0 | 0.1631 | 6784 ± 1099 | 54.02 ± 6.44 | 0.00 ± 0.00 | 2.15 ± 0.12 | 0.62 ± 0.10 | 43.375 ± 0.004 | 8.58 ± 0.23 | -1.53 ± 0.50 | -0.042 ± 0.042 | 0.195 ± 0.032 | 2.24 ± 0.44 |
| J104901.68+535425.2 | 0.4844 | 6612 ± 708 | 99.81 ± 11.54 | 0.00 ± 0.00 | 0.00 ± 0.00 | 0.36 ± 0.06 | 44.361 ± 0.007 | 8.67 ± 0.15 | -1.02 ± 0.23 | -0.098 ± 0.021 | 0.510 ± 0.055 | 2.32 ± 0.48 |
| J110223.28+223921.6 | 0.4522 | 6839 ± 647 | 72.36 ± 8.76 | 0.00 ± 0.00 | 0.00 ± 0.00 | 0.47 ± 0.08 | 44.720 ± 0.003 | 8.84 ± 0.13 | -0.86 ± 0.17 | 0.108 ± 0.020 | 0.298 ± 0.028 | 2.16 ± 0.51 |
| J111330.72+092537.2 | 0.4094 | 7992 ± 407 | 142.40 ± 19.69 | 0.00 ± 0.00 | 0.00 ± 0.00 | 0.51 ± 0.19 | 44.374 ± 0.005 | 8.83 ± 0.07 | -1.17 ± 0.13 | -0.002 ± 0.005 | 0.317 ± 0.016 | 2.54 ± 0.36 |
| J112348.80+465649.1 | 0.1567 | 2672 ± 1036 | 14.43 ± 6.26 | 0.93 ± 0.57 | 0.81 ± 0.50 | 0.81 ± 0.50 | 43.704 ± 0.006 | 7.79 ± 0.55 | -0.56 ± 0.44 | 0.058 ± 0.008 | 0.262 ± 0.102 | 2.07 ± 0.25 |
| J112728.80+634319.2 | 0.1643 | 8037 ± 528 | 84.35 ± 15.47 | 0.41 ± 0.11 | 2.49 ± 0.31 | 0.66 ± 0.17 | 43.992 ± 0.002 | 8.77 ± 0.09 | -1.37 ± 0.18 | 0.073 ± 0.005 | 0.295 ± 0.019 | 2.93 ± 0.24 |
| J113129.28+310943.2 | 0.1802 | 4258 ± 213 | 46.23 ± 2.31 | 0.10 ± 0.01 | 3.62 ± 0.10 | 0.10 ± 0.01 | 43.643 ± 0.003 | 8.19 ± 0.07 | -0.99 ± 0.10 | 0.009 ± 0.001 | 0.342 ± 0.017 | 2.15 ± 0.22 |
| J113515.84+213730.0 | 0.2224 | 2627 ± 219 | 75.14 ± 13.24 | 0.25 ± 0.06 | 3.15 ± 0.37 | 0.75 ± 0.19 | 44.117 ± 0.003 | 7.81 ± 0.12 | -0.34 ± 0.06 | 0.170 ± 0.028 | 0.230 ± 0.017 | 2.59 ± 0.31 |
| J113801.92+490506.0 | 0.4794 | 9678 ± 178 | 121.50 ± 5.43 | 0.09 ± 0.01 | 0.00 ± 0.00 | 0.45 ± 0.03 | 44.859 ± 0.003 | 9.24 ± 0.03 | -1.10 ± 0.05 | 0.051 ± 0.005 | 0.234 ± 0.011 | 2.36 ± 0.77 |
| J113849.68+574243.2 | 0.1161 | 6758 ± 305 | 84.59 ± 10.44 | 0.48 ± 0.08 | 2.29 ± 0.27 | 0.56 ± 0.10 | 43.771 ± 0.002 | 8.60 ± 0.06 | -1.33 ± 0.12 | -0.005 ± 0.001 | 0.316 ± 0.044 | 2.32 ± 0.60 |
| J114116.56+102826.4 | 0.1034 | 3708 ± 518 | 20.23 ± 3.60 | 0.86 ± 0.21 | 2.31 ± 0.26 | 0.41 ± 0.10 | 43.525 ± 0.003 | 8.07 ± 0.20 | -0.93 ± 0.26 | 0.110 ± 0.022 | 0.298 ± 0.010 | 3.02 ± 0.78 |
| J114356.88+195649.2 | 0.3358 | 12416 ± 1259 | 149.60 ± 27.66 | 0.06 ± 0.02 | 0.00 ± 0.00 | 0.35 ± 0.09 | 44.179 ± 0.005 | 9.17 ± 0.14 | -1.65 ± 0.35 | 0.071 ± 0.011 | 0.483 ± 0.036 | 2.43 ± 0.11 |
| J114502.16+191250.4 | 0.4671 | 8949 ± 664 | 111.41 ± 9.90 | 0.01 ± 0.00 | 0.00 ± 0.00 | 0.25 ± 0.03 | 44.636 ± 0.007 | 9.03 ± 0.11 | -1.14 ± 0.18 | -0.042 ± 0.003 | 0.298 ± 0.010 | 2.38 ± 0.31 |
| J115800.24+617706.0 | 0.399 | 3032 ± 97 | 64.62 ± 15.95 | 0.66 ± 0.23 | 4.49 ± 0.23 | 0.43 ± 0.15 | 44.317 ± 0.005 | 7.98 ± 0.05 | -0.36 ± 0.03 | 0.049 ± 0.008 | 0.282 ± 0.024 | 2.47 ± 0.64 |
| J120415.84+560256.4 | 0.09 | 2542 ± 219 | 24.11 ± 9.05 | 0.89 ± 0.47 | 3.53 ± 0.62 | 0.71 ± 0.37 | 43.625 ± 0.001 | 7.74 ± 0.12 | -0.55 ± 0.10 | 0.049 ± 0.008 | 0.282 ± 0.024 | 2.68 ± 0.18 |
| J120452.80-032627.6 | 0.2971 | 7882 ± 506 | 81.42 ± 10.63 | 0.16 ± 0.03 | 2.59 ± 0.15 | 0.43 ± 0.08 | 44.381 ± 0.003 | 8.82 ± 0.09 | -1.16 ± 0.15 | -0.167 ± 0.021 | 0.161 ± 0.010 | 2.46 ± 0.16 |
| J121043.92+390802.4 | 0.6633 | 3833 ± 204 | 75.74 ± 10.83 | 0.35 ± 0.07 | 0.00 ± 0.00 | 0.47 ± 0.09 | 44.782 ± 0.007 | 8.38 ± 0.08 | -0.33 ± 0.04 | -0.025 ± 0.003 | 0.314 ± 0.017 | 2.64 ± 0.70 |
| J121136.00+290748.0 | 0.139 | 5293 ± 487 | 36.91 ± 6.20 | 0.33 ± 0.08 | 2.31 ± 0.16 | 0.56 ± 0.13 | 43.501 ± 0.003 | 8.37 ± 0.08 | -1.25 ± 0.24 | 0.080 ± 0.015 | 0.275 ± 0.021 | 2.46 ± 0.25 |
| J121559.04+362230.0 | 0.5781 | 3971 ± 349 | 57.05 ± 18.80 | 0.00 ± 0.00 | 8.72 ± 18.62 | 0.00 ± 0.00 | 44.459 ± 0.001 | 8.25 ± 0.12 | -0.52 ± 0.09 | 0.334 ± 0.059 | 0.235 ± 0.021 | 1.88 ± 0.25 |
| J121759.76+303256.4 | 0.3626 | 8779 ± 215 | 159.14 ± 11.41 | 0.12 ± 0.01 | 2.24 ± 0.21 | 0.43 ± 0.04 | 44.464 ± 0.004 | 8.94 ± 0.03 | -1.21 ± 0.07 | 0.098 ± 0.005 | 0.278 ± 0.007 | 2.73 ± 0.34 |
| J122001.92-031152.8 | 0.7541 | 3644 ± 222 | 205.83 ± 29.66 | 0.13 ± 0.03 | 0.58 ± 0.12 | 0.43 ± 0.09 | 44.762 ± 0.015 | 8.32 ± 0.09 | -0.30 ± 0.04 | 0.329 ± 0.040 | 0.215 ± 0.013 | 2.32 ± 0.32 |
| J122102.40+155446.8 | 0.2293 | 6895 ± 673 | 81.00 ± 12.81 | 0.11 ± 0.03 | 1.71 ± 0.14 | 0.58 ± 0.12 | 44.276 ± 0.004 | 8.68 ± 0.14 | -0.30 ± 0.04 | -0.029 ± 0.006 | 0.310 ± 0.030 | 2.97 ± 0.39 |
| J122136.72+343542.0 | 0.2993 | 6923 ± 572 | 115.68 ± 28.81 | 0.18 ± 0.06 | 3.25 ± 0.50 | 0.51 ± 0.18 | 44.121 ± 0.004 | 8.66 ± 0.12 | -1.18 ± 0.20 | -0.029 ± 0.008 | 0.315 ± 0.026 | 2.05 ± 0.22 |
| J122137.92+043025.2 | 0.0947 | 4815 ± 352 | 30.58 ± 4.90 | 0.00 ± 0.00 | 2.83 ± 0.16 | 0.45 ± 0.08 | 43.623 ± 0.002 | 8.30 ± 0.10 | -1.25 ± 0.17 | -0.002 ± 0.005 | 0.727 ± 0.053 | 2.33 ± 0.11 |
| J122624.24+324431.2 | 0.2426 | 6967 ± 298 | 123.88 ± 23.61 | 0.66 ± 0.18 | 3.74 ± 0.20 | 0.32 ± 0.09 | 44.090 ± 0.002 | 8.66 ± 0.06 | -1.20 ± 0.17 | -97.615 ± 8.341 | 32.384 ± 1.384 | — — |
| J123733.59+131906.5 | 0.1511 | 2712 ± 752 | 58.79 ± 20.03 | 0.50 ± 0.24 | 3.17 ± 0.60 | 0.70 ± 0.34 | 43.745 ± 0.006 | 7.80 ± 0.39 | -0.42 ± 0.09 | -0.007 ± 0.004 | 0.188 ± 0.052 | 2.58 ± 1.13 |
| J125841.76+271626.4 | 0.3288 | 2456 ± 240 | 22.59 ± 3.75 | 0.00 ± 0.00 | 4.29 ± 0.60 | 1.18 ± 0.28 | 43.821 ± 0.003 | 7.72 ± 0.14 | -0.55 ± 0.09 | -0.032 ± 0.006 | 0.309 ± 0.030 | 2.07 ± 0.53 |
| J132340.32-012750.4 | 0.0769 | 6338 ± 419 | 44.16 ± 4.05 | 0.02 ± 0.00 | 2.13 ± 0.25 | 0.46 ± 0.06 | 43.542 ± 0.001 | 8.53 ± 0.09 | -1.39 ± 0.19 | 0.233 ± 0.031 | 0.193 ± 0.013 | 1.90 ± 0.36 |

| SDSS Name | z | FWHM($H\beta$) | EW($H\beta$) | [OIII]/$H\beta$ | $H\alpha/H\beta$ | $R_{fe}$ | $\log(L_{5100\text{Å}})$ | $\log(M_{BH}/M\odot)$ | $\log(R_{EDD})$ | AI | KI | $\Gamma_X$ |
|---|---|---|---|---|---|---|---|---|---|---|---|---|
| J132447.76+032432.4 | 0.306 | 8917 ± 716 | 79.57 ± 13.97 | 0.33 ± 0.08 | 1.47 ± 0.30 | 0.73 ± 0.18 | 44.752 ± 0.003 | 9.09 ± 0.11 | -1.08 ± 0.18 | 0.043 ± 0.007 | 0.299 ± 0.024 | 2.18 ± 0.19 |
| J132506.96+300834.8 | 0.4811 | 10376 ± 625 | 69.73 ± 9.64 | 0.11 ± 0.02 | 3.10 ± 0.17 | 0.59 ± 0.11 | 44.469 ± 0.002 | 9.09 ± 0.09 | -1.35 ± 0.17 | 0.446 ± 0.054 | 0.560 ± 0.034 | 1.89 ± 0.19 |
| J133103.60+465942.0 | 0.6715 | 5860 ± 231 | 121.07 ± 5.70 | 0.01 ± 0.00 | 0.00 ± 0.00 | 0.41 ± 0.03 | 44.768 ± 0.012 | 8.74 ± 0.06 | -0.71 ± 0.07 | -7.112 ± 0.562 | 2.040 ± 0.081 | 1.80 ± 0.58 |
| J133152.32+111648.0 | 0.0905 | 6664 ± 227 | 46.51 ± 1.36 | 0.00 ± 0.00 | 4.02 ± 0.15 | 0.54 ± 0.05 | 43.280 ± 0.001 | 8.57 ± 0.05 | -1.56 ± 0.07 | 0.097 ± 0.007 | 0.502 ± 0.017 | 1.33 ± 0.69 |
| J133253.28+020046.8 | 0.2158 | 8666 ± 2415 | 39.64 ± 7.94 | 0.21 ± 0.06 | 2.83 ± 0.19 | 0.32 ± 0.09 | 44.180 ± 0.003 | 8.86 ± 0.39 | -1.34 ± 0.75 | -8.582 ± 4.784 | 2.755 ± 0.768 | 2.12 ± 0.18 |
| J133926.16+150758.8 | 0.4318 | 3499 ± 470 | 33.40 ± 7.26 | 1.03 ± 0.32 | 0.00 ± 0.00 | 0.84 ± 0.26 | 44.526 ± 0.003 | 8.17 ± 0.19 | -0.38 ± 0.10 | 0.059 ± 0.016 | 0.298 ± 0.040 | 2.79 ± 0.34 |
| J134351.12+000433.6 | 0.0733 | 3444 ± 1329 | 35.80 ± 4.28 | 0.57 ± 0.10 | 2.13 ± 0.20 | 0.45 ± 0.08 | 43.443 ± 0.003 | 8.00 ± 0.55 | -0.91 ± 0.70 | 0.171 ± 0.132 | 0.180 ± 0.070 | 2.55 ± 0.31 |
| J140320.16+621413.2 | 0.3349 | 6875 ± 729 | 35.84 ± 3.71 | 0.08 ± 0.01 | 2.95 ± 0.18 | 1.12 ± 0.16 | 44.468 ± 0.004 | 8.73 ± 0.15 | -1.00 ± 0.22 | 0.004 ± 0.001 | 0.330 ± 0.035 | 2.36 ± 0.30 |
| J141041.28+531850.4 | 0.3596 | 4389 ± 398 | 109.17 ± 6.94 | 0.51 ± 0.05 | 3.70 ± 0.14 | 0.27 ± 0.02 | 43.715 ± 0.007 | 8.22 ± 0.13 | -0.98 ± 0.19 | 0.419 ± 0.076 | 0.214 ± 0.019 | 2.18 ± 0.24 |
| J141500.48+520657.6 | 0.4242 | 10986 ± 461 | 120.50 ± 21.37 | 0.08 ± 0.02 | 2.11 ± 0.19 | 0.63 ± 0.16 | 44.399 ± 0.004 | 9.12 ± 0.06 | -1.44 ± 0.13 | -0.002 ± 0.000 | 0.274 ± 0.012 | 2.39 ± 0.36 |
| J141625.68+535439.6 | 0.2628 | 7761 ± 891 | 146.46 ± 19.77 | 0.40 ± 0.08 | 2.65 ± 0.27 | 0.39 ± 0.07 | 43.952 ± 0.004 | 8.73 ± 0.16 | -1.36 ± 0.32 | 0.112 ± 0.026 | 0.212 ± 0.024 | 2.25 ± 0.30 |
| J142325.44+384033.6 | 0.2489 | 2024 ± 56 | 42.13 ± 8.85 | 0.00 ± 0.00 | 4.76 ± 0.61 | 0.90 ± 0.27 | 44.224 ± 0.003 | 7.61 ± 0.04 | -0.06 ± 0.00 | -0.039 ± 0.002 | 0.300 ± 0.008 | 2.35 ± 0.21 |
| J142419.11+224631.1 | 0.5917 | 6687 ± 443 | 66.19 ± 7.16 | 0.57 ± 0.08 | 0.00 ± 0.00 | 0.31 ± 0.05 | 44.536 ± 0.003 | 8.73 ± 0.09 | -0.94 ± 0.13 | 0.119 ± 0.016 | 0.309 ± 0.020 | 2.05 ± 0.32 |
| J143030.24-001116.8 | 0.1031 | 2411 ± 401 | 24.83 ± 4.54 | 1.32 ± 0.34 | 4.13 ± 0.31 | 1.45 ± 0.37 | 43.252 ± 0.004 | 7.68 ± 0.24 | -0.69 ± 0.23 | 0.118 ± 0.039 | 0.271 ± 0.045 | 2.71 ± 0.24 |
| J143057.12+324442.0 | 0.1959 | 4204 ± 561 | 26.31 ± 3.59 | 0.46 ± 0.09 | 4.72 ± 0.29 | 0.72 ± 0.14 | 43.858 ± 0.003 | 8.19 ± 0.19 | -0.87 ± 0.24 | -0.038 ± 0.010 | 0.305 ± 0.041 | 1.93 ± 0.19 |
| J143153.76+133314.4 | 0.1596 | 4019 ± 1562 | 50.94 ± 8.62 | 0.38 ± 0.09 | 1.60 ± 0.23 | 0.72 ± 0.17 | 43.619 ± 0.003 | 8.14 ± 0.55 | -0.95 ± 0.75 | -0.085 ± 0.066 | 0.162 ± 0.063 | 2.32 ± 0.12 |
| J143418.48+491236.0 | 0.2485 | 7065 ± 220 | 63.04 ± 2.75 | 0.00 ± 0.00 | 1.36 ± 0.04 | 0.23 ± 0.01 | 44.178 ± 0.005 | 8.68 ± 0.04 | -1.16 ± 0.08 | -0.648 ± 0.040 | 1.223 ± 0.038 | 1.96 ± 0.08 |
| J144415.60-064322.8 | 0.6831 | 15289 ± 517 | 263.28 ± 70.70 | 0.12 ± 0.04 | 0.00 ± 0.00 | 0.26 ± 0.10 | 44.622 ± 0.005 | 9.49 ± 0.05 | -1.61 ± 0.12 | -0.056 ± 0.004 | 0.286 ± 0.010 | 2.15 ± 0.25 |
| J145925.68+493339.6 | 0.5594 | 9030 ± 460 | 110.39 ± 20.55 | 0.00 ± 0.00 | 2.47 ± 0.34 | 0.39 ± 0.10 | 44.446 ± 0.004 | 8.96 ± 0.07 | -1.24 ± 0.13 | 0.154 ± 0.016 | 0.247 ± 0.013 | 2.36 ± 0.26 |
| J145959.28+401803.6 | 0.1239 | 3474 ± 517 | 43.03 ± 5.11 | 0.00 ± 0.00 | 3.21 ± 0.28 | 0.63 ± 0.11 | 43.920 ± 0.002 | 8.03 ± 0.21 | -0.68 ± 0.20 | 0.210 ± 0.062 | 0.165 ± 0.025 | 2.18 ± 0.12 |
| J151106.48+054124.0 | 0.0806 | 2576 ± 288 | 19.02 ± 4.04 | 1.45 ± 0.44 | 6.53 ± 0.25 | 1.12 ± 0.34 | 43.477 ± 0.001 | 7.75 ± 0.16 | -0.64 ± 0.14 | -0.090 ± 0.020 | 0.279 ± 0.031 | — ± — |
| J151133.60-054546.8 | 0.0846 | 4603 ± 616 | 24.79 ± 1.92 | 0.00 ± 0.00 | 2.20 ± 0.10 | 0.21 ± 0.02 | 43.495 ± 0.004 | 8.25 ± 0.19 | -1.13 ± 0.31 | 0.064 ± 0.017 | 0.318 ± 0.043 | 2.20 ± 0.28 |
| J152637.68+165746.8 | 0.1617 | 2601 ± 213 | 18.97 ± 1.33 | 0.07 ± 0.01 | 4.15 ± 0.30 | 1.47 ± 0.15 | 43.900 ± 0.002 | 7.78 ± 0.12 | -0.44 ± 0.07 | -0.025 ± 0.004 | 0.313 ± 0.026 | 2.39 ± 0.81 |
| J153228.80+045400.0 | 0.2182 | 7854 ± 663 | 122.32 ± 3.56 | 0.94 ± 0.04 | 3.03 ± 0.05 | 0.41 ± 0.04 | 44.213 ± 0.004 | 8.78 ± 0.12 | -1.24 ± 0.22 | -0.508 ± 0.086 | 1.074 ± 0.091 | 1.96 ± 0.07 |
| J154213.92+183458.8 | 0.0724 | 4604 ± 459 | 111.59 ± 15.73 | 0.40 ± 0.08 | 3.15 ± 0.21 | 0.57 ± 0.11 | 43.587 ± 0.002 | 8.26 ± 0.14 | -1.09 ± 0.22 | 0.070 ± 0.014 | 0.301 ± 0.030 | 2.43 ± 0.11 |
| J155812.24+345212.0 | 0.3646 | 2781 ± 90 | 64.00 ± 20.21 | 0.01 ± 0.00 | 2.42 ± 0.33 | 0.95 ± 0.42 | 44.466 ± 0.003 | 7.94 ± 0.05 | -0.21 ± 0.01 | 0.023 ± 0.001 | 0.270 ± 0.009 | 3.07 ± 0.65 |
| J161524.00+193813.2 | 0.1564 | 5700 ± 909 | 46.82 ± 2.59 | 0.08 ± 0.01 | 4.25 ± 0.20 | 0.70 ± 0.05 | 43.522 ± 0.003 | 8.44 ± 0.23 | -1.31 ± 0.42 | -0.114 ± 0.036 | 0.141 ± 0.022 | 1.85 ± 0.55 |
| J161745.60+060354.0 | 0.0379 | 7107 ± 112 | 100.01 ± 0.86 | 0.26 ± 0.00 | 2.95 ± 0.13 | 0.42 ± 0.01 | 43.239 ± 0.002 | 8.62 ± 0.02 | -1.64 ± 0.06 | 0.238 ± 0.007 | 0.215 ± 0.003 | 2.81 ± 0.42 |
| J162114.40+181948.0 | 0.1247 | 8840 ± 1266 | 114.47 ± 20.51 | 0.18 ± 0.05 | 1.53 ± 0.23 | 0.42 ± 0.11 | 43.842 ± 0.001 | 8.84 ± 0.20 | -1.53 ± 0.44 | -0.011 ± 0.003 | 0.330 ± 0.047 | 2.65 ± 0.27 |
| J163111.28+404803.6 | 0.2575 | 3511 ± 464 | 95.03 ± 21.87 | 0.00 ± 0.00 | 2.85 ± 0.30 | 0.75 ± 0.25 | 44.439 ± 0.002 | 8.14 ± 0.19 | -0.43 ± 0.11 | 0.038 ± 0.010 | 0.253 ± 0.033 | 2.75 ± 0.22 |
| J163821.84+285902.4 | 0.6719 | 3116 ± 238 | 64.62 ± 20.07 | 0.01 ± 0.00 | 0.00 ± 0.00 | 0.31 ± 0.13 | 44.699 ± 0.006 | 8.15 ± 0.11 | -0.19 ± 0.03 | -0.042 ± 0.006 | 0.295 ± 0.023 | 1.87 ± 0.20 |
| J164823.28+350325.2 | 0.1791 | 6274 ± 881 | 50.11 ± 5.45 | 0.00 ± 0.00 | 3.01 ± 0.22 | 0.29 ± 0.04 | 43.815 ± 0.004 | 8.54 ± 0.20 | -1.24 ± 0.36 | 0.016 ± 0.004 | 0.299 ± 0.042 | 1.78 ± 0.17 |
| J170033.36-355255.2 | 0.1425 | 6436 ± 519 | 64.96 ± 5.39 | 0.39 ± 0.05 | 1.79 ± 0.14 | 0.31 ± 0.04 | 43.993 ± 0.002 | 8.58 ± 0.11 | -1.18 ± 0.19 | -2.880 ± 0.465 | 0.514 ± 0.041 | 2.51 ± 0.27 |
| J170036.48+465533.6 | 0.2668 | 6080 ± 57 | 108.51 ± 2.01 | 0.01 ± 0.00 | 2.84 ± 0.05 | 0.37 ± 0.01 | 44.321 ± 0.002 | 8.58 ± 0.01 | -0.96 ± 0.02 | -0.117 ± 0.002 | 0.867 ± 0.008 | 2.12 ± 0.43 |
| J170227.12-332627.6 | 0.1883 | 3106 ± 395 | 75.98 ± 19.70 | 0.00 ± 0.00 | 3.27 ± 0.26 | 0.64 ± 0.23 | 43.830 ± 0.007 | 7.93 ± 0.18 | -0.62 ± 0.17 | -0.054 ± 0.014 | 0.297 ± 0.038 | 2.29 ± 0.13 |
| J171331.44+433657.6 | 0.1675 | 3004 ± 805 | 18.16 ± 6.54 | 0.62 ± 0.32 | 3.71 ± 0.45 | 0.29 ± 0.15 | 43.644 ± 0.005 | 7.89 ± 0.38 | -0.69 ± 0.37 | -0.034 ± 0.018 | 0.290 ± 0.078 | 2.13 ± 0.47 |
| J213631.68+003151.6 | 0.7958 | 9215 ± 1032 | 211.55 ± 19.61 | 0.18 ± 0.02 | 0.00 ± 0.00 | 0.53 ± 0.07 | 44.666 ± 0.006 | 9.07 ± 0.16 | -1.15 ± 0.27 | -141.964 ± 31.800 | 55.860 ± 6.256 | 1.89 ± 0.20 |
| J221542.24-003610.8 | 0.0991 | 2071 ± 44 | 53.59 ± 22.45 | 1.06 ± 0.63 | 4.49 ± 0.72 | 0.69 ± 0.41 | 43.774 ± 0.002 | 7.57 ± 0.03 | -0.30 ± 0.01 | 0.218 ± 0.009 | 0.255 ± 0.005 | 1.75 ± 0.29 |
| J234718.72+153921.6 | 0.7467 | 10485 ± 644 | 137.14 ± 14.61 | 0.08 ± 0.01 | 0.00 ± 0.00 | 0.28 ± 0.04 | 44.443 ± 0.003 | 9.09 ± 0.09 | -1.37 ± 0.18 | 0.026 ± 0.003 | 0.080 ± 0.005 | 1.76 ± 0.42 |
| J235338.32-095719.1 | 0.1878 | 5225 ± 567 | 48.89 ± 6.35 | 0.65 ± 0.12 | 2.34 ± 0.19 | 0.46 ± 0.08 | 43.947 ± 0.004 | 8.39 ± 0.15 | -1.02 ± 0.23 | 0.435 ± 0.095 | 0.203 ± 0.022 | 2.17 ± 0.71 |

# List of Publications

1. *Accretion Disk Sizes from Continuum Reverberation Mapping of AGN Selected from the ZTF Survey.***Vivek Kumar Jha**, Ravi Joshi, Hum Chand, Xue-Bing Wu, Luis C Ho, Shantanu Rastogi, Quinchun Ma, 2022, Monthly Notices of the Royal Astronomical Society, 511, 2, pp: 3005-3016.

2. *Exploring the AGN Accretion Disks using Continuum Reverberation Mapping.***Vivek Kumar Jha**, Ravi Joshi, Jayesh Saraswat, Hum Chand, Sudhanshu Barway and Amit Kumar Mandal, 2023, *accepted for publication in* the Bulletin of Liège Royal Society of Sciences.

3. *A comparative study of the physical properties for a representative sample of Narrow and Broad-line Seyfert galaxies.* **Vivek Kumar Jha**, Hum Chand, Vineet Ojha, Amitesh Omar, Shantanu Rastogi, 2021, Monthly Notices of the Royal Astronomical Society, 510, 3, pp: 4379-4393.

4. *Properties of Broad and Narrow Line Seyfert galaxies selected from SDSS.* **Vivek Kumar Jha**, Hum Chand, and Vineet Ojha. Communications of the Byurakan Astrophysical Observatory (ComBAO), Volume 67, Issue 2, December 2020.

5. *Evidence of jet induced optical microvariability in radio-loud Narrow Line Seyfert 1 Galaxies* . Vineet Ojha, **Vivek Kumar Jha**, Hum Chand, Veeresh Singh, 2022, Monthly Notices of the Royal Astronomical Society, 514, 4, pp: 5607–5624.

6. *Calibration of the technique of Photometric reverberation mapping*: **Vivek Kumar Jha**, Hum Chand et. al. (in preparation)

7. *Photometric and Spectroscopic Analysis of the Type II Short Plateau SN 2020jfo.* Ailawadhi, Bhavya (et al. including **Vivek Kumar Jha**), 2023, Monthly Notices of the Royal Astronomical Society, 519, 1, pp:248-270.[1]

---

[1]This work is not included in the thesis.

# Work presented at Conferences/Meetings/Seminars:

- **2023** Presented an *online talk* titled: **Exploring the Connection between UV/Optical Variability and Physical Characteristics of X-ray-Selected Type 1 AGN** at the Asia-Pacific Regional IAU Meeting (APRIM) held in Fukushima Prefecture; Japan (07-11 August, 2023).[link]

- **2023** Presented an *online talk* titled: **Tools of optical photometry: data reduction and aperture photometry using Python tools.** at the conference titled: Multidisciplinary Approach to Understand the Mysteries of our Universe, held at National Institute of Technology (NIT) Rourkela, India (17-21 July, 2023).[link]

- **2023** Presented a *poster* titled: **Unveiling the Connection between UV/Optical Variability and Physical Characteristics of X-ray-Selected Type 1 AGN** at the international conference titled: The Restless nature of AGN: 10 years later, held in Naples; Italy (26-30 June, 2023).[link]

- **2023** Presented a *talk* titled: **Accretion disk size measurements for AGN using reverberation mapping** at the 3rd BINA Workshop: Scientific potential of the Indo-Belgian cooperation, held at the Graphic Era Hill University, Bhimtal, India (22-24 March, 2023).[link]

- **2023** Presented a *talk* titled: **New Accretion disk size measurements for reverberation mapped AGN.** at the 41st meeting of the Astronomical Society of India (ASI) held at IIT-Indore, India (01-05 March 2023).[link]

- **2023** Presented an *online talk* titled: **Eyes on the Sky: Current and upcoming telescopes of this decade.** at Deen Dayal Upadhyaya Gorakhpur University, Gorakhpur, India (06 Jan 2023).

- **2022** Presented a *talk* titled: **Tools of Optical Photometry** at ARIES Training School in Observational Astronomy, Nainital, India (16-27 May 2022).[link]

- **2022** Presented a *poster* titled: **Accretion disk sizes for Quasars selected from the Zwicky Transient Facility survey** at the 40th meeting of the Astronomical Society of India (ASI) held at IIT-Roorkee, India (24-29 March 2022).[link]

- **2022** Presented a *talk* titled: **A look into the heart of Quasars: using light echos as a tool** at Central University of Himachal Pradesh (CUHP), Dharamshala; India on 3rd February. [link]

- **2021** Presented an *e-poster* titled: **Correlation analysis on a homogeneous sample of NlSy1 and BlSy1 galaxies** at the workshop titled "Multi-object Spectroscopy for Statistical Measures of Galaxy Evolution" held online by Space Telescope Science Institute (STScI), Baltimore; USA (17-20 May 2021). [link]

- **2020** Presented an *online talk* titled: **A comparative study of Narrow and Broad-line Seyfert galaxies using SDSS** at an international symposium titled "Astronomical Surveys and Big Data 2 (ASBD-2)" held online by the Byurakan Astrophysical Observatory (BAO); Armenia (14-18 September 2020) [link]

- **2019** Presented a *poster* titled: **Devasthal Optical Telescope-AGN Reverberation Monitoring (DOT-ARM): Project strategy and initial results** at the international conference titled "Mapping Central Regions of Active Galactic Nuclei" held in Guilin, Guanxi Province; China (19-24 September 2019). [link]

## Schools/Workshops attended:

- **2021** Attended International Summer School on **The Interstellar Medium of Galaxies, from the Epoch of Reionization to the Milky Way**, held online from July 12-23, 2021. [link]

- **2020** Attended workshop titled: **Less traveled path of dark matter: Axions and primordial black holes**, held online by ICTS-TIFR; India between November 9-13, 2020. [link]

- **2020** Attended **ILMT: International Liquid Mirror Telescope workshop** held online by ARIES, Nainital; India (29 June - 01 July 2020). [link]

- **2020** Attended one day Indo Thai Workshop titled **Investigating the Stellar Variability and Star Formation** held in ARIES, Nainital; India (02 March 2020). [link]

- **2019** Attended **I-TMT (India- TMT) Science and Instruments Workshop** held in ARIES, Nainital; India (17 - 19 October 2019). [link]